\documentclass[10pt,twocolumn,aps,prb,balancelastpage,longbibliography]{revtex4-1}
\usepackage[latin9]{inputenc}
\usepackage{color}
\usepackage{verbatim}
\usepackage{amsmath}
\usepackage{amssymb,ulem,manfnt}
\usepackage{graphicx,cellspace,booktabs}
\usepackage{esint}
\usepackage[unicode=true,pdfusetitle,
 bookmarks=true,bookmarksnumbered=false,bookmarksopen=false,
 breaklinks=false,pdfborder={0 0 1},backref=false,colorlinks=true,citecolor=blue]{hyperref}
\usepackage{breakurl}
\usepackage{epstopdf}
\usepackage{yfonts}

\usepackage{pst-all}
\usepackage{leftidx}
\usepackage[off]{auto-pst-pdf} 

\makeatletter

\@ifundefined{textcolor}{}
{%
 \definecolor{BLACK}{gray}{0}
 \definecolor{WHITE}{gray}{1}
 \definecolor{RED}{rgb}{1,0,0}
 \definecolor{GREEN}{rgb}{0,1,0}
 \definecolor{BLUE}{rgb}{0,0,1}
 \definecolor{CYAN}{cmyk}{1,0,0,0}
 \definecolor{MAGENTA}{cmyk}{0,1,0,0}
 \definecolor{YELLOW}{cmyk}{0,0,1,0}
}

\def\U{\mathrm{U}(1)}

\usepackage[caption=false]{subfig}
\usepackage{bm}

\newcommand{\bra}[1]{\langle #1 |}
\newcommand{\ket}[1]{|#1\rangle}

\newcommand{\im}{\mbox{im}}

\newcommand{\act}[2]{{}^{{\bf #1}} #2}
\newcommand{\acts}[3]{{}^{{\bf #1}_{#2}} #3}

\newcommand{\coho}[1]{\textswab{#1}}
\newcommand{\cohosub}[1]{\protect\scalebox{0.7}{\textswab{#1}}}
\newcommand{\tgh}[2]{\cohosub{t}({\bf #1},{\bf #2})}

\def\l@subsubsection#1#2{}
\makeatother

\begin{document}

\title{Absolute anomalies in (2+1)D symmetry-enriched topological states and exact (3+1)D constructions}

\author{Daniel Bulmash}
\author{Maissam Barkeshli}

\affiliation{Condensed Matter Theory Center and Joint Quantum Institute, Department of Physics, University of Maryland, College Park, Maryland 20472 USA}

\date{\today}
\begin{abstract}
Certain patterns of symmetry fractionalization in (2+1)D topologically ordered phases of matter can be anomalous, which means that they possess an obstruction to being realized in purely (2+1)D. In this paper we demonstrate  how to compute the anomaly for symmetry-enriched topological (SET) states of bosons in complete generality. We demonstrate how, given any unitary modular tensor category (UMTC) and symmetry fractionalization class for a global symmetry group $G$, one can define a (3+1)D topologically invariant path integral in terms of a state sum for a G symmetry-protected topological (SPT) state. We present an exactly solvable Hamiltonian for the system and demonstrate explicitly a (2+1)D $G$ symmetric surface termination that hosts deconfined anyon excitations described by the given UMTC and symmetry fractionalization class. We present concrete algorithms that can be used to compute anomaly indicators in general. Our approach applies to general symmetry groups, including anyon-permuting and anti-unitary symmetries. In addition to providing a general way to compute the anomaly, our result also shows, by explicit construction, that every symmetry fractionalization class for any UMTC can be realized at the surface of a (3+1)D SPT state. As a byproduct, this construction also provides a way of explicitly seeing how the algebraic data that defines symmetry fractionalization in general arises in the context of exactly solvable models. In the case of unitary orientation-preserving symmetries, our results can also be viewed as providing a method to compute the $\mathcal{H}^4(G, U(1))$ obstruction that arises in the theory of $G$-crossed braided tensor categories, for which no general method has been presented to date. 
\end{abstract}

\maketitle

\tableofcontents

\section{Introduction}

In the absence of any symmetries, gapped quantum many-body states of matter in (2+1) space-time dimensions can still form distinct, topologically ordered phases of matter.\cite{wen04,wang2008,nayak2008} Topologically ordered states are characterized by the fusion and braiding properties of topologically non-trivial quasiparticle excitations (anyons), which are described mathematically by an algebraic theory of anyons known as a unitary modular tensor category (UMTC), $\mathcal{C}$.\cite{moore1989b,witten1989,wang2008} It is believed that the pair  ($\mathcal{C}$, $c$), where $c$ is the chiral central charge characterizing the possible gapless degrees of freedom on the boundary, completely characterizes gapped phases of matter in (2+1)D in the absence of symmetry.

In the presence of a symmetry group $G$, the classification of gapped phases is further refined. Symmetry-enriched topological states (SETs) are characterized by a host of additional properties.\cite{barkeshli2019} These include (1) the way that symmetry actions can permute quasiparticle types, (2) the pattern of symmetry fractionalization, which determines the ways in which quasiparticles carry fractional symmetry quantum numbers, and (3) the fusion and braiding properties of symmetry defects. Recently it has been shown how these data can be described in general through the mathematical framework of $G$-crossed braided tensor categories.\cite{barkeshli2019}

The theory of SETs also includes a set of possible anomalies, or obstructions.\cite{ENO2009,barkeshli2019} For example, once it is specified how symmetries permute anyon types in a way that is consistent with the symmetries of the fusion and braiding data, there is an obstruction to defining the notion of symmetry fractionalization. This obstruction, sometimes referred to as a symmetry localization anomaly, takes values in the third cohomology group $\mathcal{H}^3(G, \mathcal{A})$, where $\mathcal{A}$ is the Abelian group formed by fusion of Abelian anyons, and there is an implicit action of $G$ on $\mathcal{A}$ defined by how the symmetry permutes the anyons. Ref. \onlinecite{barkeshli2019} presented a formula for how to compute this obstruction given any possible permutation action of the symmetries. One interpretation of this obstruction is that the associated topological quantum field theory is only compatible with a 2-group symmetry, where $G$ is the ordinary, 0-form, symmetry, $\mathcal{A}$ is the 1-form symmetry,\cite{gaiotto2014} and $[\coho{O}] \in \mathcal{H}^3(G, \mathcal{A})$ characterizes the 2-group.\cite{ENO2009,barkeshli2019, benini2019} 

When the $\mathcal{H}^3$ symmetry localization anomaly vanishes, there are different possible patterns of symmetry fractionalization. Depending on the precise pattern of symmetry fractionalization, there can be a further anomaly, which we refer to as a symmetry fractionalization anomaly, valued in the fourth group cohomology $\mathcal{H}^4(G, U(1))$. In the langauge of high energy physics, this is an example of a 't Hooft anomaly associated with the global symmetry $G$.\cite{kapustin2014b} This is an obstruction to defining a consistent set of braiding and fusion data of symmetry defects; that is, it is an obstruction to defining a full $G$-crossed braided tensor category to describe a consistent set of braiding and fusion of symmetry defects in addition to braiding and fusion of anyons. Stated differently in the context of quantum field theory, this anomaly is an obstruction to consistency of the theory in the presence of background fields associated with $G$. To date, a general method to compute this $\mathcal{H}^4$ anomaly given the data that characterizes symmetry fractionalization has not been presented, aside from some special cases that we review in the following subsection. 

In a variety of examples, it is known that such anomalous symmetry fractionalization classes, while they cannot exist purely in (2+1) space-time dimensions,\footnote{Implicit in this statement is that the symmetry acts in an on-site manner; that is, that the symmetry action in a lattice model decomposes as a tensor product of unitaries acting on each lattice site independently. It is possible to realize anomalous boundary theories in the dimension of the boundary by considering a non-on-site action of the symmetry.\cite{chen2013} The precise extension of the ``on-site'' requirement for space-time symmetries has not yet been formulated, but presumably the requirement is that the symmetry must correspond to an on-site transformation combined with a classical permutation of coordinates and possibly complex conjugation. } can exist at the surface of a (3+1)D symmetry-protected topological (SPT) state.\cite{vishwanath2013,wong2013,Chen2014} SPT states are gapped phases of matter that, if we allow the symmetry to be broken, are adiabatically connected to a trivial direct product state, but are otherwise non-trivial.\cite{senthil2015} In terms of topological quantum field theory (TQFT), SPT states are described by invertible TQFTs with $G$ symmetry, implying that the path integral on all closed manifolds has unit magnitude.\cite{kapustin2014,kapustin2014c,freed2016} A large class of (3+1)D SPT states can be classified by $\mathcal{H}^{4}(G, U(1))$.\cite{chen2013,dijkgraaf1990} The appearance of $\mathcal{H}^4(G, U(1))$ in two different contexts has led to the expectation that all anomalous symmetry fractionalization classes can be realized at the (2+1)D surface of (3+1)D SPT states, although this has not been explicitly demonstrated in general. 

In this paper, we resolve the outstanding issues mentioned above by demonstrating an explicit way to compute the $\mathcal{H}^4$ anomaly in general, and demonstrating by explicit construction how any anomalous symmetry fractionalization class for any UMTC can be realized at the surface of a (3+1)D SPT. Specifically, we show how any UMTC $\mathcal{C}$ and possibly anomalous symmetry fractionalization class can be used to define a (3+1)D topologically invariant path integral. The path integral takes the form of a discrete state sum on a triangulation of the space-time manifold and choice of background $G$-bundle. Topological invariance follows from invariance of the state sum under retriangulation and various other choices made in the construction. We demonstrate that the resulting (3+1)D TQFT describes a bulk (3+1)D SPT state. (3+1)D SPTs can be distinguished by the value of their topological path integral on various $G$-bundles. Our path integral may be computed explicitly on such $G$-bundles, thus allowing us to completely characterize the (3+1)D SPT. In particular this allows us to extract an element of $\mathcal{H}^4(G, U(1))$ given any UMTC and symmetry fractionalization class for arbitrary symmetry groups $G$. 

The path integral on a (3+1)D manifold $M^4$ defines a wave function $|\Psi (\partial M^4)\rangle$ on $\partial M^4$. We further derive a Hamiltonian, which is explicitly $G$ symmetric and acts on a local tensor product Hilbert space, for which the wave function $|\Psi(\partial M^4)\rangle$ is the exact ground state. We then define the Hamiltonian on a three-dimensional space with boundary in a way which preserves the $G$ symmetry, and show that the surface theory is a (2+1)D symmetry-enriched topological state whose quasiparticle excitations are described by the same UMTC and symmetry fractionalization data that was used as input into the (3+1)D state sum. This then proves by explicit construction that any symmetry fractionalization class can appear at the surface of a (3+1)D SPT. Since (2+1)D surfaces of non-trivial (3+1)D SPTs are understood to be anomalous, characterizing the bulk (3+1)D SPT that hosts a given (2+1)D symmetry fractionalization pattern at its surface is equivalent to characterizing the anomaly of the (2+1)D surface. 

We also discuss a method which allows the $\mathcal{H}^4$ anomalies to be computed in an algorithmic way in general. We describe an algorithm that constructs invariants $\mathcal{I}_i$ of $\mathcal{H}^4(G, U(1))$, for $i = 1,\cdots, \text{dim } \mathcal{H}^4(G, U(1))$, in terms of a $U(1)$ valued combination of representative $U(1)$ $4$-cochains. We then show how this combination can be mapped to a state sum and then be used to compute the $\mathcal{H}^4$ invariants by taking as input the UMTC and symmetry fractionalization data. Our procedure thus allows a brute force computation of the $\mathcal{H}^4$ obstruction in complete generality. The complexity of the calculation is exponential in the number of $4$-cochains that appear in $\mathcal{I}_i$ and polynomial in the number of distinct anyons in the UMTC.

Our results apply for arbitrary symmetry groups $G$, whose elements may have either unitary or anti-unitary symmetry action, and may permute the anyons. In particular, to treat anti-unitary symmetries, we generalize the mathematical structures involved in describing symmetry fractionalization for unitary symmetries, making contact with notions in higher category theory.

It is expected that the classification of topological phases with spatial symmetry (in Euclidean space $\mathbb{R}^d$) is equivalent to those with on-site symmetry, and this expectation has been demonstrated explicitly for a wide variety of SPT and SET states.\cite{song2017,huang2017,thorngren2018,cheng2016lsm} Consequently, we expect our methods will apply for computing symmetry fractionalization anomalies associated with spatial symmetries as well. However, not all of our constructions -- in particular our exactly solvable Hamiltonians -- are directly compatible with spatial symmetries. 

\subsection{Relation to prior work}

The fact that a braided tensor category (BTC) can be used to define a (3+1)D topologically invariant path integral state sum was first pointed out by Crane and Yetter \cite{crane1993} (see also Ref. \onlinecite{walker2006}). More recently Walker and Wang \cite{WalkerWang} provided a Hamiltonian realization of these theories. Our work can be viewed as a general symmetry-enriched version of these constructions, which takes as input an arbitrary BTC together with any symmetry fractionalization class, and outputs a topological path integral and a corresponding Hamiltonian realization. If the input BTC is a UMTC, our construction realizes a (3+1)D SPT. Prior to our work, Ref. \onlinecite{Chen2014} studied an example of decorated Walker-Wang models that give SPTs with surface SETs when $G = \mathbb{Z}_2 \times \mathbb{Z}_2$ and the UMTC is the semion theory (denoted $U(1)_2$); a generalization to Abelian anyons not permuted by symmetries and carrying projective representations of $G$ was also sketched. However our general construction is quite different from that of Ref. \onlinecite{Chen2014}.

In (2+1)D, the Turaev-Viro-Barrett-Westbury state sum is well-known to give a topologically invariant path integral for closed 3-manifolds given a spherical fusion category.\cite{turaev1992,barrett1996} The Levin-Wen model provides a Hamiltonian realization of the Turaev-Viro theories associated to unitary fusion categories with a certain tetrahedral symmetry.\cite{levin2005} Recently these (2+1)D theories have been extended to include on-site (internal) G-symmetry, so that a G-graded fusion category can be used to define a (2+1)D state sum\cite{BarkeshliReflection} and also exactly solvable Hamiltonian.\cite{heinrich2016,cheng2016} The work in this paper can be viewed as a (3+1)D generalization of these ideas. 

Recently, Cui presented a state sum construction for a (3+1)D topological quantum field theory that takes as input a $G$-crossed braided tensor category.\cite{CuiTQFT} Its corresponding Hamiltonian realization was studied in Ref. \onlinecite{WilliamsonHamiltonian}. Our construction is closely related to these constructions, but differs in two fundamental ways. First, since Cui's construction takes as input a $G$-crossed braided tensor category, the $\mathcal{H}^4$ obstruction necessarily vanishes; therefore anomalous symmetry fractionalization classes are not included in this construction. Secondly, our construction is capable of taking as input symmetry fractionalization classes associated with anti-unitary symmetries. However, $G$-crossed braided tensor categories require that $G$ correspond to a unitary orientation-preserving symmetry, and therefore such anti-unitary symmetries are not incorporated in the constructions of Ref. \onlinecite{CuiTQFT}. In particular, the anti-unitary symmetry actions allow us to define our path integrals on non-orientable manifolds, generalizing the construction in Ref. \onlinecite{BarkeshliReflection}. The extension to non-orientable manifolds is beyond the most general mathematical constructions for (3+1)D TQFTs in terms of spherical fusion 2-categories defined recently in Ref. \onlinecite{douglas2018}, which assign topological invariants to oriented 4-manifolds. We comment more on the technical similarities and differences between our construction and those of Ref. \onlinecite{CuiTQFT, WilliamsonHamiltonian} in subsequent sections.

While many prior works have studied symmetry fractionalization anomalies, none of them have so far fully solved the problem of computing the $\mathcal{H}^4$ obstruction under completely general circumstances. When the symmetry is finite, unitary, space-time orientation preserving, and does not permute anyon types, Ref. \onlinecite{ENO2009} presented a formula for the $\mathcal{H}^4$ obstruction for bosonic topological phases (where the microscopic constituents are bosons as opposed to fermions; mathematically this corresponds to non-spin TQFTs). This formula was later summarized in Ref. \onlinecite{Chen2014}, and rederived using a more detailed mathematical framework for the data and consistency conditions of $G$-crossed braided tensor categories in Ref. \onlinecite{barkeshli2019} (which is also expected to apply in the cases of continuous and infinite $G$). In the particular case of Abelian symmetry groups (which are unitary, space-time orientation preserving) and Abelian topological orders where the symmetry does not permute the anyon types, Ref. \onlinecite{wang2016b} explicitly studied the bulk-boundary correspondence and a notion of ``anomaly inflow'' to understand the anomaly in more detail. 

On the other hand, for space-time reflection symmetries that square to the identity, denoted $\mathbb{Z}_2^{\bf T}$ or $\mathbb{Z}_2^{\bf r}$, Ref. \onlinecite{BarkeshliReflection} derived a general formula for the anomaly. Note that in this case, the anomaly is mathematically not an obstruction to defining a $G$-crossed braided tensor category, but rather an obstruction to defining a generalized version of a $G$-crossed braided tensor category where the defects can be space-time orientation reversing. Such a mathematical construction was briefly outlined in Ref. \onlinecite{BarkeshliRelativeAnomaly}, but has not yet been fully developed. We note that Ref. \onlinecite{wang2017} also independently conjectured the same formula for the $\mathbb{Z}_2^{\bf T}$ anomaly for both bosonic and fermionic topological orders. The formula for the $\mathbb{Z}_2^{\bf T}$ anomaly for fermionic topological orders was subsequently derived in Ref. \onlinecite{tachikawa2016b}. A number of subsequent works further studied and rederived these anomaly formulae through various other methods.\cite{lee2018,qi2019,kobayashi2019,mao2020}

For symmetries of the form $U(1) \rtimes G'$ that do not permute anyons and for the specific case of $\mathbb{Z}_2$ topological orders, Ref. \onlinecite{hermele2016} provided a method to diagnose symmetry fractionalization anomalies. Subsequently, for $U(1) \times \mathbb{Z}_2^{\bf T}$ and $U(1) \rtimes \mathbb{Z}_2^{\bf T}$ symmetry, Ref. \onlinecite{lapa2019} recently presented formulas for anomaly indicators that apply to general topological orders and allow anyon permutations. 

Aside from these special cases, completely general results have so far been presented for relative anomalies, for bosonic topological phases.\cite{BarkeshliRelativeAnomaly,cui2016} Specifically, given a UMTC and choice of how symmetry permutes the anyons, the difference between symmetry fractionalization patterns is classified by the second group cohomology $\mathcal{H}^2_\rho(G, \mathcal{A})$.\cite{barkeshli2019} Here $\rho$ defines the action of $G$ on $\mathcal{A}$ as specified by how the symmetries permute the anyon types. Given two symmetry fractionalization classes that differ by $[\coho{t}] \in \mathcal{H}^2_\rho(G, \mathcal{A})$, the relative anomaly is the difference in the $\mathcal{H}^4$ anomalies for the two symmetry fractionalization classes. It was shown how these relative anomaly formulae are consistent with all previously derived results.

A main goal of this paper, then, is to complete the story by demonstrating how to compute the absolute anomaly in complete generality for bosonic topological phases of matter. 

\section{Symmetry fractionalization review}

\subsection{Review of UMTC notation}

Here we briefly review the notation that we use to describe UMTCs. For a more comprehensive review
of the notation that we use, see e.g. Ref. \onlinecite{barkeshli2019}. The topologically 
non-trivial quasiparticles of a (2+1)D topologically ordered state are equivalently referred to
as anyons, topological charges, and quasiparticles. In the category theory terminology, they correspond
to isomorphism classes of simple objects of the UMTC. 

A UMTC $\mathcal{C}$ contains splitting spaces $V_{c}^{ab}$, and their dual fusion spaces, $V_{ab}^c$,
where $a,b,c \in \mathcal{C}$ are the anyons. These spaces have dimension 
$\text{dim } V_{c}^{ab} = \text{dim } V_{ab}^c = N_{ab}^c$, where the fusion coefficients $N_{ab}^c$ determine the fusion rules. They are depicted graphically as: 
\begin{equation}
\left( d_{c} / d_{a}d_{b} \right) ^{1/4}
\raisebox{-0.5\height}{
\begin{pspicture}(-0.1,-0.2)(1.5,-1.2)
  \small
  \psset{linewidth=0.9pt,linecolor=black,arrowscale=1.5,arrowinset=0.15}
  \psline{-<}(0.7,0)(0.7,-0.35)
  \psline(0.7,0)(0.7,-0.55)
  \psline(0.7,-0.55) (0.25,-1)
  \psline{-<}(0.7,-0.55)(0.35,-0.9)
  \psline(0.7,-0.55) (1.15,-1)	
  \psline{-<}(0.7,-0.55)(1.05,-0.9)
  \rput[tl]{0}(0.4,0){$c$}
  \rput[br]{0}(1.4,-0.95){$b$}
  \rput[bl]{0}(0,-0.95){$a$}
 \scriptsize
  \rput[bl]{0}(0.85,-0.5){$\mu$}
  \end{pspicture}}
=\left\langle a,b;c,\mu \right| \in
V_{ab}^{c} ,
\label{eq:bra}
\end{equation}
\begin{equation}
\left( d_{c} / d_{a}d_{b}\right) ^{1/4}
\raisebox{-0.5\height}{
\begin{pspicture}[shift=-0.65](-0.1,-0.2)(1.5,1.2)
  \small
  \psset{linewidth=0.9pt,linecolor=black,arrowscale=1.5,arrowinset=0.15}
  \psline{->}(0.7,0)(0.7,0.45)
  \psline(0.7,0)(0.7,0.55)
  \psline(0.7,0.55) (0.25,1)
  \psline{->}(0.7,0.55)(0.3,0.95)
  \psline(0.7,0.55) (1.15,1)	
  \psline{->}(0.7,0.55)(1.1,0.95)
  \rput[bl]{0}(0.4,0){$c$}
  \rput[br]{0}(1.4,0.8){$b$}
  \rput[bl]{0}(0,0.8){$a$}
 \scriptsize
  \rput[bl]{0}(0.85,0.35){$\mu$}
  \end{pspicture}
  }
=\left| a,b;c,\mu \right\rangle \in
V_{c}^{ab},
\label{eq:ket}
\end{equation}
where $\mu=1,\ldots ,N_{ab}^{c}$, $d_a$ is the quantum dimension of $a$, 
and the factors $\left(\frac{d_c}{d_a d_b}\right)^{1/4}$ are a normalization convention for the diagrams. 

We denote $\bar{a}$ as the topological charge conjugate of $a$, for which
$N_{a \bar{a}}^1 = 1$, i.e.
\begin{align}
a \times \bar{a} = 1 +\cdots
\end{align}
Here $1$ refers to the identity particle, i.e. the vacuum topological sector, which physically describes all 
local, topologically trivial excitations. 

The $F$-symbols are defined as the following basis transformation between the splitting
spaces of $4$ anyons:
\begin{equation}
\raisebox{-0.5\height}{
\begin{pspicture}[shift=-1.0](0,-0.45)(1.8,1.8)
  \small
  \psset{linewidth=0.9pt,linecolor=black,arrowscale=1.5,arrowinset=0.15}
  \psline(0.2,1.5)(1,0.5)
  \psline(1,0.5)(1,0)
  \psline(1.8,1.5) (1,0.5)
  \psline(0.6,1) (1,1.5)
   \psline{->}(0.6,1)(0.3,1.375)
   \psline{->}(0.6,1)(0.9,1.375)
   \psline{->}(1,0.5)(1.7,1.375)
   \psline{->}(1,0.5)(0.7,0.875)
   \psline{->}(1,0)(1,0.375)
   \rput[bl]{0}(0.05,1.6){$a$}
   \rput[bl]{0}(0.95,1.6){$b$}
   \rput[bl]{0}(1.75,1.6){${c}$}
   \rput[bl]{0}(0.5,0.5){$e$}
   \rput[bl]{0}(0.9,-0.3){$d$}
 \scriptsize
   \rput[bl]{0}(0.3,0.8){$\alpha$}
   \rput[bl]{0}(0.7,0.25){$\beta$}
\end{pspicture}
}
= \sum_{f,\mu,\nu} \left[F_d^{abc}\right]_{(e,\alpha,\beta)(f,\mu,\nu)}
\raisebox{-0.5\height}{
 \begin{pspicture}[shift=-1.0](0,-0.45)(1.8,1.8)
  \small
  \psset{linewidth=0.9pt,linecolor=black,arrowscale=1.5,arrowinset=0.15}
  \psline(0.2,1.5)(1,0.5)
  \psline(1,0.5)(1,0)
  \psline(1.8,1.5) (1,0.5)
  \psline(1.4,1) (1,1.5)
   \psline{->}(0.6,1)(0.3,1.375)
   \psline{->}(1.4,1)(1.1,1.375)
   \psline{->}(1,0.5)(1.7,1.375)
   \psline{->}(1,0.5)(1.3,0.875)
   \psline{->}(1,0)(1,0.375)
   \rput[bl]{0}(0.05,1.6){$a$}
   \rput[bl]{0}(0.95,1.6){$b$}
   \rput[bl]{0}(1.75,1.6){${c}$}
   \rput[bl]{0}(1.25,0.45){$f$}
   \rput[bl]{0}(0.9,-0.3){$d$}
 \scriptsize
   \rput[bl]{0}(1.5,0.8){$\mu$}
   \rput[bl]{0}(0.7,0.25){$\nu$}
  \end{pspicture}
  }
.
\end{equation}
To describe topological phases, these are required to be unitary transformations, i.e.
\begin{eqnarray}
\left[ \left( F_{d}^{abc}\right) ^{-1}\right] _{\left( f,\mu
,\nu \right) \left( e,\alpha ,\beta \right) }
&= \left[ \left( F_{d}^{abc}\right) ^{\dagger }\right]
  _{\left( f,\mu ,\nu \right) \left( e,\alpha ,\beta \right) }
  \nonumber \\
&= \left[ F_{d}^{abc}\right] _{\left( e,\alpha ,\beta \right) \left( f,\mu
,\nu \right) }^{\ast }
.
\end{eqnarray}

Anyon lines may be ``bent" using the $A$ and $B$ symbols, given diagrammatically by
\begin{equation}
 \raisebox{-0.5\height}{
\begin{pspicture}[shift=-0.65](-0.1,-0.2)(1.5,1.2)
  \small
  \psset{linewidth=0.9pt,linecolor=black,arrowscale=1.5,arrowinset=0.15}
  \psline{->}(0,0)(0,0.55)
  \psline(0,0)(0,0.85)
  \psarc{->}(0.3,0.85){0.3}{25}{115}
  \psarc{-}(0.3,0.85){0.3}{25}{180}
  \psline{->}(1,0)(1,0.45)
  \psline(1,0)(1,0.55)
  \psline(1,0.55) (0.55,1)
  \psline{->}(1,0.55)(0.6,0.95)
  \psline(1,0.55) (1.45,1)	
  \psline{->}(1,0.55)(1.4,0.95)
  \rput[bl]{0}(0.7,0){$c$}
  \rput[br]{0}(1.7,0.8){$b$}
  \rput[bl]{0}(0.4,0.7){$a$}
  \rput[bl]{0}(0.15,0.1){$\bar{a}$}
 \scriptsize
  \rput[bl]{0}(1.15,0.35){$\mu$}
  \end{pspicture}
  }
  = \sum_{\nu} \left[A^{ab}_c\right]_{\mu \nu}
   \raisebox{-0.5\height}{
\begin{pspicture}(-0.1,-0.2)(1.5,-1.2)
  \small
  \psset{linewidth=0.9pt,linecolor=black,arrowscale=1.5,arrowinset=0.15}
  \psline{-<}(0.7,0)(0.7,-0.35)
  \psline(0.7,0)(0.7,-0.55)
  \psline(0.7,-0.55) (0.25,-1)
  \psline{-<}(0.7,-0.55)(0.35,-0.9)
  \psline(0.7,-0.55) (1.15,-1)	
  \psline{-<}(0.7,-0.55)(1.05,-0.9)
  \rput[tl]{0}(0.4,0){$b$}
  \rput[br]{0}(1.4,-0.95){$c$}
  \rput[bl]{0}(0,-0.95){$\bar{a}$}
 \scriptsize
  \rput[bl]{0}(0.85,-0.5){$\nu$}
  \end{pspicture}
  },
  \label{eqn:Abend}
\end{equation}
\begin{equation}
 \raisebox{-0.5\height}{
\begin{pspicture}[shift=-0.65](-0.1,-0.2)(2.15,1.2)
  \small
  \psset{linewidth=0.9pt,linecolor=black,arrowscale=1.5,arrowinset=0.15}
  \psline{->}(2,0)(2,0.55)
  \psline(2,0)(2,0.85)
  \psarc{<-}(1.7,0.85){0.3}{65}{155}
  \psarc{-}(1.7,0.85){0.3}{0}{155}
  \psline{->}(1,0)(1,0.45)
  \psline(1,0)(1,0.55)
  \psline(1,0.55) (0.55,1)
  \psline{->}(1,0.55)(0.6,0.95)
  \psline(1,0.55) (1.45,1)	
  \psline{->}(1,0.55)(1.4,0.95)
  \rput[bl]{0}(0.7,0){$c$}
  \rput[br]{0}(1.55,0.65){$b$}
  \rput[bl]{0}(0.4,0.7){$a$}
  \rput[br]{0}(1.8,0.1){$\bar{b}$}
 \scriptsize
  \rput[bl]{0}(1.15,0.35){$\mu$}
  \end{pspicture}
  }
  = \sum_{\nu} \left[B^{ab}_c\right]_{\mu \nu}
   \raisebox{-0.5\height}{
\begin{pspicture}(-0.1,-0.2)(1.5,-1.2)
  \small
  \psset{linewidth=0.9pt,linecolor=black,arrowscale=1.5,arrowinset=0.15}
  \psline{-<}(0.7,0)(0.7,-0.35)
  \psline(0.7,0)(0.7,-0.55)
  \psline(0.7,-0.55) (0.25,-1)
  \psline{-<}(0.7,-0.55)(0.35,-0.9)
  \psline(0.7,-0.55) (1.15,-1)	
  \psline{-<}(0.7,-0.55)(1.05,-0.9)
  \rput[tl]{0}(0.4,0){$a$}
  \rput[br]{0}(1.4,-0.95){$\bar{b}$}
  \rput[bl]{0}(0,-0.95){$c$}
 \scriptsize
  \rput[bl]{0}(0.85,-0.5){$\nu$}
  \end{pspicture}
  }.
  \label{eqn:Bbend}
\end{equation}
They can be expressed in terms of $F$-symbols by
\begin{align}
\left[A_c^{ab}\right]_{\mu \nu} &= \sqrt{\frac{d_a d_b}{d_c}}\varkappa_a \left[F^{\bar{a}ab}_b\right]^{\ast}_{1,(c,\mu, \nu)}\\
\left[B^{ab}_c\right]_{\mu \nu} &= \sqrt{\frac{d_a d_b}{d_c}}\left[F^{ab\bar{b}}_a\right]_{(c,\mu, \nu),1}
\end{align}
where the phase $\varkappa_a$ is the Frobenius-Schur indicator
\begin{equation}
\varkappa_a = d_a F^{a\bar{a}a}_{a11} .
\end{equation}

The $R$-symbols define the braiding properties of the anyons, and are defined via the the following
diagram:
\begin{equation}
\raisebox{-0.5\height}{
\begin{pspicture}[shift=-0.65](-0.1,-0.2)(1.5,1.2)
  \small
  \psset{linewidth=0.9pt,linecolor=black,arrowscale=1.5,arrowinset=0.15}
  \psline{->}(0.7,0)(0.7,0.43)
  \psline(0.7,0)(0.7,0.5)
 \psarc(0.8,0.6732051){0.2}{120}{240}
 \psarc(0.6,0.6732051){0.2}{-60}{35}
  \psline (0.6134,0.896410)(0.267,1.09641)
  \psline{->}(0.6134,0.896410)(0.35359,1.04641)
  \psline(0.7,0.846410) (1.1330,1.096410)	
  \psline{->}(0.7,0.846410)(1.04641,1.04641)
  \rput[bl]{0}(0.4,0){$c$}
  \rput[br]{0}(1.35,0.85){$b$}
  \rput[bl]{0}(0.05,0.85){$a$}
 \scriptsize
  \rput[bl]{0}(0.82,0.35){$\mu$}
  \end{pspicture}
  }
=\sum\limits_{\nu }\left[ R_{c}^{ab}\right] _{\mu \nu}
\raisebox{-0.5\height}{
\begin{pspicture}[shift=-0.65](-0.1,-0.2)(1.5,1.2)
  \small
  \psset{linewidth=0.9pt,linecolor=black,arrowscale=1.5,arrowinset=0.15}
  \psline{->}(0.7,0)(0.7,0.45)
  \psline(0.7,0)(0.7,0.55)
  \psline(0.7,0.55) (0.25,1)
  \psline{->}(0.7,0.55)(0.3,0.95)
  \psline(0.7,0.55) (1.15,1)	
  \psline{->}(0.7,0.55)(1.1,0.95)
  \rput[bl]{0}(0.4,0){$c$}
  \rput[br]{0}(1.4,0.8){$b$}
  \rput[bl]{0}(0,0.8){$a$}
 \scriptsize
  \rput[bl]{0}(0.82,0.37){$\nu$}
  \end{pspicture}
  }
  .
\end{equation}

Under a basis transformation, $\Gamma^{ab}_c : V^{ab}_c \rightarrow V^{ab}_c$, the $F$ and $R$ symbols change:
\begin{align}
  F^{abc}_{def} &\rightarrow \tilde{F}^{abc}_d = \Gamma^{ab}_e \Gamma^{ec}_d F^{abc}_{def} [\Gamma^{bc}_f]^\dagger [\Gamma^{af}_d]^\dagger
  \nonumber \\
  R^{ab}_c & \rightarrow \tilde{R}^{ab}_c = \Gamma^{ba}_c R^{ab}_c [\Gamma^{ab}_c]^\dagger .
  \end{align}
  where we have suppressed splitting space indices and dropped brackets on the $F$-symbol for shorthand.
  These basis transformations are referred to as vertex basis gauge transformations. Physical quantities correspond to gauge-invariant combinations
  of the data. 
  
The topological twist $\theta_a$ is defined via the diagram:
\begin{equation}
\theta _{a}=\theta _{\bar{a}}
=\sum\limits_{c,\mu } \frac{d_{c}}{d_{a}}\left[ R_{c}^{aa}\right] _{\mu \mu }
= \frac{1}{d_{a}}
\raisebox{-0.5\height}{
\begin{pspicture}[shift=-0.5](-1.3,-0.6)(1.3,0.6)
\small
  \psset{linewidth=0.9pt,linecolor=black,arrowscale=1.5,arrowinset=0.15}
  \psarc[linewidth=0.9pt,linecolor=black] (0.7071,0.0){0.5}{-135}{135}
  \psarc[linewidth=0.9pt,linecolor=black] (-0.7071,0.0){0.5}{45}{315}
  \psline(-0.3536,0.3536)(0.3536,-0.3536)
  \psline[border=2.3pt](-0.3536,-0.3536)(0.3536,0.3536)
  \psline[border=2.3pt]{->}(-0.3536,-0.3536)(0.0,0.0)
  \rput[bl]{0}(-0.2,-0.5){$a$}
  \end{pspicture}
  }
.
\end{equation}
Finally, the modular, or topological, $S$-matrix, is defined as
\begin{equation}
S_{ab} =\mathcal{D}^{-1}\sum%
\limits_{c}N_{\bar{a} b}^{c}\frac{\theta _{c}}{\theta _{a}\theta _{b}}d_{c}
=\frac{1}{\mathcal{D}}
\raisebox{-0.5\height}{
\begin{pspicture}[shift=-0.4](0.0,0.2)(2.6,1.3)
\small
  \psarc[linewidth=0.9pt,linecolor=black,arrows=<-,arrowscale=1.5,arrowinset=0.15] (1.6,0.7){0.5}{167}{373}
  \psarc[linewidth=0.9pt,linecolor=black,border=3pt,arrows=<-,arrowscale=1.5,arrowinset=0.15] (0.9,0.7){0.5}{167}{373}
  \psarc[linewidth=0.9pt,linecolor=black] (0.9,0.7){0.5}{0}{180}
  \psarc[linewidth=0.9pt,linecolor=black,border=3pt] (1.6,0.7){0.5}{45}{150}
  \psarc[linewidth=0.9pt,linecolor=black] (1.6,0.7){0.5}{0}{50}
  \psarc[linewidth=0.9pt,linecolor=black] (1.6,0.7){0.5}{145}{180}
  \rput[bl]{0}(0.1,0.45){$a$}
  \rput[bl]{0}(0.8,0.45){$b$}
  \end{pspicture}
  }
,
\label{eqn:mtcs}
\end{equation}
where $\mathcal{D} = \sqrt{\sum_a d_a^2}$.

A particularly useful quantity for the present discussion is the double braid, which is a phase if either $a$ or $b$ is an Abelian anyon:
\begin{equation}
\raisebox{-0.5\height}{
  \begin{pspicture}[shift=-0.6](0.0,-0.05)(1.1,1.45)
  \small
  \psarc[linewidth=0.9pt,linecolor=black,border=0pt] (0.8,0.7){0.4}{120}{225}
  \psarc[linewidth=0.9pt,linecolor=black,arrows=<-,arrowscale=1.4,
    arrowinset=0.15] (0.8,0.7){0.4}{165}{225}
  \psarc[linewidth=0.9pt,linecolor=black,border=0pt] (0.4,0.7){0.4}{-60}{45}
  \psarc[linewidth=0.9pt,linecolor=black,arrows=->,arrowscale=1.4,
    arrowinset=0.15] (0.4,0.7){0.4}{-60}{15}
  \psarc[linewidth=0.9pt,linecolor=black,border=0pt]
(0.8,1.39282){0.4}{180}{225}
  \psarc[linewidth=0.9pt,linecolor=black,border=0pt]
(0.4,1.39282){0.4}{-60}{0}
  \psarc[linewidth=0.9pt,linecolor=black,border=0pt]
(0.8,0.00718){0.4}{120}{180}
  \psarc[linewidth=0.9pt,linecolor=black,border=0pt]
(0.4,0.00718){0.4}{0}{45}
  \rput[bl]{0}(0.1,1.2){$a$}
  \rput[br]{0}(1.06,1.2){$b$}
  \end{pspicture}
  }
= M_{ab}
\raisebox{-0.5\height}{
\begin{pspicture}[shift=-0.6](-0.2,-0.45)(1.0,1.1)
  \small
  \psset{linewidth=0.9pt,linecolor=black,arrowscale=1.5,arrowinset=0.15}
  \psline(0.3,-0.4)(0.3,1)
  \psline{->}(0.3,-0.4)(0.3,0.50)
  \psline(0.7,-0.4)(0.7,1)
  \psline{->}(0.7,-0.4)(0.7,0.50)
  \rput[br]{0}(0.96,0.8){$b$}
  \rput[bl]{0}(0,0.8){$a$}
  \end{pspicture}
  }
.
  \label{doubleBraid}
\end{equation}

\subsection{Topological symmetry and braided auto-equivalence}

An important property of a UMTC $\mathcal{C}$ is the group of ``topological symmetries,'' which are related to ``braided auto-equivalences'' in the mathematical literature. They are associated with the symmetries of the emergent UMTC description, irrespective of any global symmetries of the microscopic model from which $\mathcal{C}$ emerges as the description of the universal properties of the anyons. 

The topological symmetries consist of the invertible maps
\begin{align}
\varphi: \mathcal{C} \rightarrow \mathcal{C} .
\end{align}
The different $\varphi$, modulo equivalences known as natural isomorphisms, form a group, which we denote as Aut$(\mathcal{C})$.\cite{barkeshli2019}

The symmetry maps can be classified according to a $\mathbb{Z}_2$ grading corresponding to whether $\varphi$ has a unitary or anti-unitary action on the category:
\begin{align}
  \label{qdef}
q(\varphi) = \left\{
\begin{array} {ll}
0 & \text{if $\varphi$ is unitary} \\
1 & \text{if $\varphi$ is anti-unitary} \\
\end{array} \right.
\end{align}
We note that one can consider a more general $\mathbb{Z}_2\times \mathbb{Z}_2$ grading by considering separately transformations that correspond to time-reversal and spatial reflection symmetries.\cite{barkeshli2019,BarkeshliRelativeAnomaly} Here we will not consider spatial parity reversing transformations and thus do not consider this generalization. 
Thus the topological symmetry group can be decomposed as
\begin{align}
\text{Aut}(\mathcal{C}) = \bigsqcup_{\sigma=0,1} \text{Aut}_{\sigma}(\mathcal{C}) .
\end{align}
Aut$_{0}(\mathcal{C})$ is therefore the subgroup corresponding to topological symmetries that are unitary (this is referred to in the mathematical literature as the group of ``braided auto-equivalences''). 

It is also convenient to define
\begin{align}
\sigma(\varphi) = \left\{
\begin{array} {ll}
1 & \text{if $\varphi$ is unitary} \\
* & \text{if $\varphi$ is anti-unitary} \\
\end{array} \right.
\label{eqn:sigmaDef}
\end{align}

The maps $\varphi$ may permute the topological charges:
\begin{align}
\varphi(a) = a' \in \mathcal{C}, 
\end{align}
subject to the constraint that 
\begin{align}
N_{a'b'}^{c'} &= N_{ab}^c
\nonumber \\
S_{a'b'} &= S_{ab}^{\sigma(\varphi)},
\nonumber \\
\theta_{a'} &= \theta_a^{\sigma(\varphi)},
\end{align}
The maps $\varphi$ have a corresponding action on the $F$- and $R-$ symbols of the theory, as well as on the fusion and splitting spaces, which we will discuss in the subsequent section. 

\subsection{Global symmetry}
\label{globsym}

We now consider a system which has a global symmetry group $G$. The global symmetry acts on the anyons and the topological state space through the action of a group homomorphism
\begin{align}
[\rho] : G \rightarrow \text{Aut}(\mathcal{C}) . 
\end{align}
We use the notation $[\rho_{\bf g}] \in \text{Aut}(\mathcal{C})$ for a specific element ${\bf g} \in G$. The square brackets indicate the equivalence class of symmetry maps related by natural isomorphisms, which we define below. $\rho_{\bf g}$ is thus a representative symmetry map of the equivalence class $[\rho_{\bf g}]$. We use the notation
\begin{align}
\,^{\bf g}a \equiv \rho_{\bf g}(a). 
\end{align}
We associate a $\mathbb{Z}_2$ grading $q({\bf g})$ (and related $\sigma({\bf g})$) by defining
\begin{align}
q({\bf g}) &\equiv q(\rho_{\bf g}) 
\nonumber \\
\sigma({\bf g}) &\equiv \sigma( \rho_{\bf g})
\end{align}

$\rho_{\bf g}$ has an action on the fusion/splitting spaces:
\begin{align}
\rho_{\bf g} : V_{ab}^c \rightarrow V_{\,^{\bf g}a \,^{\bf g}b}^{\,^{\bf g}c} .   
\end{align}
This map is unitary if $q({\bf g}) = 0$ and anti-unitary if $q({\bf g}) = 1$. We write this as
\begin{align}
\rho_{\bf g} |a,b;c, \mu\rangle = \sum_{\nu} [U_{\bf g}(\,^{\bf g}a ,
  \,^{\bf g}b ; \,^{\bf g}c )]_{\mu\nu} K^{q({\bf g})} |\,^{\bf g} a, \,^{\bf g} b; \,^{\bf g}c,\nu\rangle,
  \label{eqn:rhoStates}
\end{align}
where $U_{\bf g}(\,^{\bf g}a , \,^{\bf g}b ; \,^{\bf g}c ) $ is a $N_{ab}^c \times N_{ab}^c$ matrix, and 
$K$ denotes complex conjugation.

Under the map $\rho_{\bf g}$, the $F$ and $R$ symbols transform as well:
\begin{widetext}
\begin{align}
\rho_{\bf g}[ F^{abc}_{def}] &= U_{\bf g}(\,^{\bf g}a, \,^{\bf g}b; \,^{\bf g}e) U_{\bf g}(\,^{\bf g}e, \,^{\bf g}c; \,^{\bf g}d) F^{\,^{\bf g}a \,^{\bf g}b \,^{\bf g}c }_{\,^{\bf g}d \,^{\bf g}e \,^{\bf g}f} 
U^{-1}_{\bf g}(\,^{\bf g}b, \,^{\bf g}c; \,^{\bf g}f) U^{-1}_{\bf g}(\,^{\bf g}a, \,^{\bf g}f; \,^{\bf g}d) = K^{q({\bf g})} F^{abc}_{def} K^{q({\bf g})}
\nonumber \\
\rho_{\bf g} [R^{ab}_c] &= U_{\bf g}(\,^{\bf g}b, \,^{\bf g}a; \,^{\bf g}c)  R^{\,^{\bf g}a \,^{\bf g}b}_{\,^{\bf g}c} U_{\bf g}(\,^{\bf g}a, \,^{\bf g}b; \,^{\bf g}c)^{-1} = K^{q({\bf g})} R^{ab}_c K^{q({\bf g})},
\label{eqn:UFURConsistency}
\end{align}
\end{widetext}
where we have suppressed the additional indices that appear when $N_{ab}^c > 1$. 

We demand that composition of $\rho_{\bf g}$ obey the group multiplication law up to a natural isomorphism $\kappa_{\bf g, h}$
\begin{align}
\kappa_{{\bf g}, {\bf h}} \circ \rho_{\bf g} \circ \rho_{\bf h} = \rho_{\bf g h} ,
\end{align}
where the action of $\kappa_{ {\bf g}, {\bf h}}$ on the fusion / splitting spaces is defined as
\begin{align}
\kappa_{ {\bf g}, {\bf h}} ( |a, b;c,\mu \rangle) = \sum_\nu [\kappa_{ {\bf g}, {\bf h}} ( a, b;c )]_{\mu\nu} | a, b;c,\nu \rangle
\end{align}
and, being a natural isomorphism, obeys by definition 
\begin{align}
[\kappa_{ {\bf g}, {\bf h}} (a,b;c)]_{\mu \nu} = \delta_{\mu \nu} \frac{\beta_a({\bf g}, {\bf h}) \beta_b({\bf g}, {\bf h})}{\beta_c({\bf g}, {\bf h}) },
\end{align}
where $\beta_a({\bf g}, {\bf h})$ are $\U$ phases.  The above definitions imply that
\begin{widetext}
\begin{align}
\label{kappaU}
\kappa_{ {\bf g}, {\bf h}} ( a, b;c ) = U_{\bf g}(a,b;c)^{-1} K^{q({\bf g})} U_{\bf h}( \,^{\bar{\bf g}}a, \,^{\bar{\bf g}}b; \,^{\bar{\bf g}}c  )^{-1} K^{q({\bf q})} U_{\bf gh}(a,b;c ),
\end{align}
\end{widetext}
where $\bar{\bf g} \equiv {\bf g}^{-1}$.

\subsection{Symmetry localization and fractionalization}
\label{symmfraclocSec}

Now let us consider the action of a symmetry ${\bf g} \in G$ on the full quantum many-body state of the system, which may correspond, for example, to a lattice model defined on a Hilbert space that decomposes as a local tensor product. We consider systems that are described by a TQFT in the long wave length limit and thus contain many more ``microscopic'' degrees of freedom.

Let $R_{\bf g}$ be the representation of ${\bf g}$ acting on the full Hilbert space of the theory. 
We consider a state $|\Psi_{a_1, \cdots, a_n} \rangle$ in the full Hilbert space of the system, 
which consists of $n$ anyons, $a_1, \cdots a_n$, at well-separated locations, which collectively fuse to the identity topological sector. 
Since the ground state is $G$-symmetric, we expect that the symmetry action $R_{\bf g}$ on this state possesses a property
that we refer to as symmetry localization. This is the property that the symmetry action $R_{\bf g}$ decomposes as
\begin{align}
\label{symloc}
R_{\bf g}  |\Psi_{a_1, \cdots, a_n} \rangle \approx \prod_{j = 1}^n U^{(j)}_{\bf g} U_{\bf g} (\,^{\bf g}a_1, \cdots, \,^{\bf g}a_n ; 1) |\Psi_{\,^{\bf g}a_1, \cdots, \,^{\bf g}a_n} \rangle .
\end{align}
Here, $U^{(j)}_{\bf g}$ are unitary matrices that have support in a region (of length scale set by the correlation length)
localized to the anyon $a_j$. The map $U_{\bf g} (\,^{\bf g}a_1, \cdots, \,^{\bf g}a_n ; 1)$ is the generalization of
$U_{\bf g} (\,^{\bf g}a,\,^{\bf g} b; \,^{\bf g} c)$, defined above, to the case with $n$ anyons fusing to vacuum.
$U_{\bf g} (\,^{\bf g}a_1, \cdots, \,^{\bf g}a_n ; 1)$ only depends on the global topological sector of the system -- that is, on the 
precise fusion tree that defines the topological state -- and not on any other details of the state, in contrast to the local operators $U^{(j)}_{\bf g}$.
The $\approx$ means that the equation is true up to corrections that are exponentially small in the size of $U^{(j)}$ and the distance between the anyons,
in units of the correlation length. 

The choice of action $\rho$ defined above defines an element $[\coho{O}] \in \mathcal{H}^3_{[\rho]}(G, \mathcal{A})$.\cite{barkeshli2019}
If $[\coho{O}]$ is non-trivial, then there is an obstruction to Eq.~(\ref{symloc}) being consistent when considering the
associativity of three group elements.  We refer to this as a symmetry localization anomaly, or
symmetry localization obstruction. See. Ref. \onlinecite{barkeshli2019,fidkowski2015,barkeshli2018} for examples.\footnote{A non-trivial obstruction $[\coho{O}] \in \mathcal{H}^3_{[\rho]}(G, \mathcal{A})$ can be alternatively interpreted as the associated TQFT possessing a non-trivial $2$-group symmetry, consisting of the $0$-form symmetry group $G$ and the $1$-form symmetry group $\mathcal{A}$, with  $[\coho{O}]$ characterizing the $2$-group \cite{ENO2009,barkeshli2019,benini2019}.} 

If $[\coho{O}]$ is trivial, so that symmetry localization as described by Eq.~(\ref{symloc}) is well-defined, then it is possible to define a notion of symmetry
fractionalization.\cite{barkeshli2019}

In general, symmetry fractionalization is characterized by a consistent set of data $\{\eta\}$ and $\{U \}$, where $\{U \}$ was defined
above. $\eta_a({\bf g}, {\bf h})$ is defined as
\begin{align}
  \label{etaDef}
\eta_{a_j}({\bf g}, {\bf h}) U_{\bf gh}^{(j)} |\Psi_{a_1, \cdots, a_n} \rangle = U_{\bf g}^{(j)} \rho_{\bf g} U_{\bf h}^{(j)} \rho_{\bf g}^{-1} |\Psi_{a_1, \cdots, a_n} \rangle
\end{align}
The data $\eta_a({\bf g}, {\bf h})$ characterize the difference in phase obtained when acting ``locally'' on an anyon $a$ by ${\bf g}$ and ${\bf h}$ separately, as compared with acting on $a$ by the product ${\bf gh}$.

This can be captured through a physical process involving symmetry defects, as explained in the next subsection. There are two important consistency conditions for $U$ and $\eta$, which we will use repeatedly later in this paper.\cite{barkeshli2019} The first one is
\begin{align}
  \label{eqn:UEtaConsistency}
\frac{\eta_a({\bf g}, {\bf h}) \eta_b ({\bf g}, {\bf h})}{\eta_c({\bf g}, {\bf h})} = \kappa_{ {\bf g}, {\bf h}}(a, b;c) ,
\end{align}
with $\kappa$ defined in terms of $U$ as in Eq. \eqref{kappaU}. The other one is
\begin{align}
	\eta_a({\bf g},{\bf h})\eta_a({\bf gh}, {\bf k}) =\eta_a({\bf g}, {\bf hk}) \eta^{\sigma({\bf g})}_{\rho_{ {\bf g}^{-1}}(a)}({\bf h}, {\bf k}).
	\label{eqn:etaConsistency}
\end{align}
These data are subject to an additional class of gauge transformations, referred to as symmetry action gauge transformations, which arise by changing $\rho$ by a natural isomorphism: \cite{barkeshli2019}
\begin{align}
  U_{\bf g}(a,b;c) &\rightarrow \frac{\gamma_{a}({\bf g}) \gamma_b({\bf g})}{ \gamma_c({\bf g}) } U_{\bf g}(a,b;c)
\nonumber \\
  \eta_a({\bf g}, {\bf h}) & \rightarrow \frac{\gamma_a({\bf g h}) }{(\gamma_{\,^{\bf g} a}({\bf g}))^{\sigma({\bf g})} \gamma_a({\bf h}) } \eta_a({\bf g}, {\bf h})
\end{align}
We note that $U$ also changes under a vertex basis gauge transformation according to
\begin{equation}
\tilde{U}_{{\bf g}}(a,b,c)_{\mu \nu} = \sum_{\mu', \nu'} [\Gamma^{\act{\bar{g}}{a} \act{\bar{g}}{b}}_{\act{\bar{g}}{c}}]_{\mu, \mu'} U_{{\bf g}}(a,b,c)_{\mu' \nu'}\left[(\Gamma^{ab}_c)^{-1}\right]^{\sigma({\bf g})}_{\nu'\nu},
\end{equation}
with the shorthand $\bar{{\bf g}}={\bf g}^{-1}$. Different gauge-inequivalent choices of $\{\eta\}$ and $\{U\}$ characterize distinct symmetry fractionalization classes.\cite{barkeshli2019} In this paper we will always fix the gauge
\begin{align}
  \eta_1({\bf g},{\bf h})=\eta_a({\bf 1},{\bf g}) = \eta_a({\bf g},{\bf 1})&=1
                                                                             \nonumber \\
  U_{\bf g}(1,b;c)=U_{\bf g}(a,1;c)&=1.
  \end{align}

One can show that symmetry fractionalization forms a torsor over $\mathcal{H}^2_{\rho}(G, \mathcal{A})$. That is, different possible patterns of symmetry fractionalization can be related to each other by elements of $\mathcal{H}^2_{\rho}(G, \mathcal{A})$. In particular, given an element $[\coho{t}] \in \mathcal{H}^2_{\rho}(G, \mathcal{A})$, we can change the symmetry fractionalization class as
\begin{align}
\eta_a({\bf g}, {\bf h}) \rightarrow \eta_a({\bf g}, {\bf h}) M_{a \tgh{g}{h}},
\end{align}
where $\coho{t}({\bf g},{\bf h}) \in \mathcal{A}$ is a representative 2-cocyle for the cohomology class $[\coho{t}]$ and $M_{ab}$ is the double braid (see Eq. \ref{doubleBraid}).

In the case where the permuation $\rho$ is trivial, there is always a canonical notion of a trivial symmetry fractionalization class, where $\eta_a({\bf g},{\bf h}) = 1$ for all ${\bf g}, {\bf h} \in G$. In this case, an element of $\mathcal{H}^2(G, \mathcal{A})$ is sufficient to completely characterize the symmetry fractionalization pattern, as was discussed for the case where the UMTC $\mathcal{C}$ only contains Abelian anyons in Ref. \onlinecite{essin2013}. 

More abstractly, the data $\rho$, $\{U_{\bf g}(a,b;c)\}$ and $\{\eta_a({\bf g}, {\bf h})\rbrace$ defines a categorical $G$ action on $\mathcal{C}$.\cite{barkeshli2019}

\subsection{Symmetry fractionalization and symmetry defects}

\begin{figure}
\includegraphics[width=3.5in]{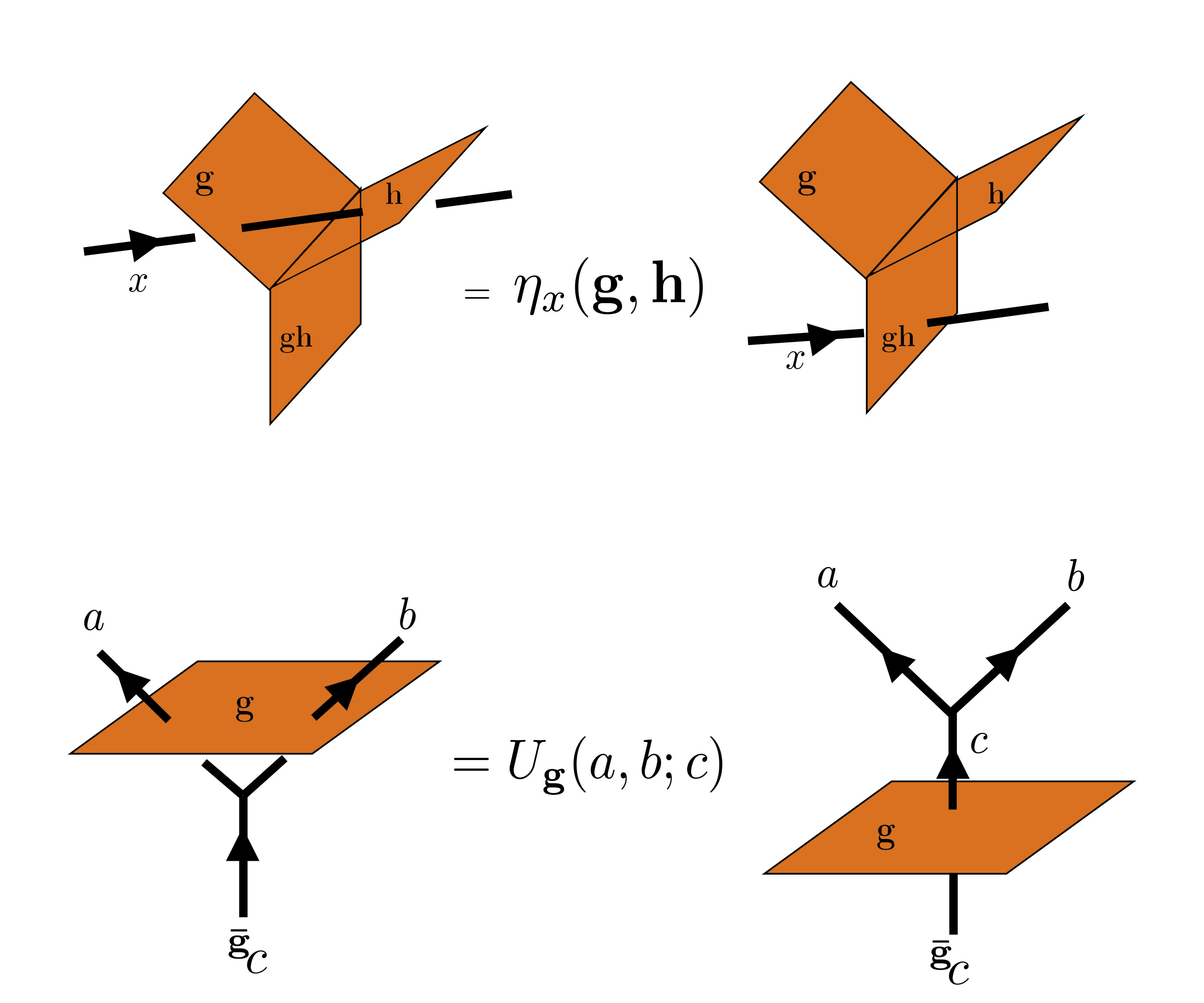}
\caption{Diagrammatic representations of the actions of the $\eta$ and $U$ symbols. Anyon lines are black and branch sheets are orange.}
\label{fig:symfrac}
\end{figure}

A convenient way to understand symmetry fractionalization is through a graphical calculus that incorporates symmetry defects. In 2D space, we can consider a line-like symmetry defect labeled by a group element ${\bf g}$, which we will sometimes refer to as a branch cut. In the (2+1)D space-time, this corresponds to a branch sheet. When an anyon $x$ crosses the ${\bf g}$ defect sheet in space-time, it is permuted to a different anyon, $\,^{\bf g}x$.  $\eta_x( {\bf g}, {\bf h} )$ and $U_{\bf g}(a,b;c)$ can then be understood through the diagrams shown in Fig. \ref{fig:symfrac}.

Note that here we take the symmetry group element to be unitary and orientation-preserving. In Sec.~\ref{sec:antiunitary} we will discuss the generalization to anti-unitary symmetries.

Although we will not do so in the rest of this paper, we may also consider the branch cuts to end at a point in space. In this case, we can have topologically distinct endpoints, labeled as $a_{\bf g}$. If the symmetry fractionalization class is non-anomalous, the symmetry defects $a_{\bf g}$ form a consistent $G$-crossed braided tensor category.\cite{barkeshli2019} It is known for finite $G$ that there is a cohomological obstruction $[\coho{O}] \in \mathcal{H}^4(G, U(1))$ to defining a consistent $G$-crossed braided tensor category, which provides a mathematically precise definition of symmetry fractionalization anomalies, at least for the case of unitary, orientation preserving symmetries.

\section{State-sum construction}
\label{sec:stateSum}

The input data to our construction is a braided tensor category (BTC) $\mathcal{C}$, a symmetry group $G$, the group homomorphism $\rho$ and associated $\{U_{\bf g}(a,b;c)\}$, and the symmetry fractionalization data $\{\eta_a({\bf g}, {\bf h})\}$. The state sum that we define in this section does not require $\mathcal{C}$ to be modular; however when $\mathcal{C}$ is modular, we will see that the resulting state sum defines a (3+1)D SPT and allows a calculation of the symmetry fractionalization anomaly. 

We note that in this section $G$ is discrete and finite and corresponds to an on-site (internal) unitary symmetry. The extension to anti-unitary, spatial, infinite and/or continuous symmetries will be discussed in subsequent sections. 

To simplify the notation, we always assume that the fusion multiplicities $N_{ab}^c$ of simple objects in $\mathcal{C}$ are at most 1; this restriction can be relaxed straightforwardly. 

\subsection{Basic data}

Given a 4-manifold $M^4$, we pick a triangulation and a branching structure (i.e. a local ordering of vertices). We associate to each simplex the following data:
\begin{itemize}
\item 0-simplex $i$: a group element ${\bf g}_i \in G$
\item 2-simplex $ijk$: an anyon $a_{ijk} \in \mathcal{C}$
\item 3-simplex $ijkl$: an anyon $b_{ijkl} \in \mathcal{C}$ obeying certain rules
\end{itemize}

\begin{figure*}
\includegraphics[width=1.7\columnwidth]{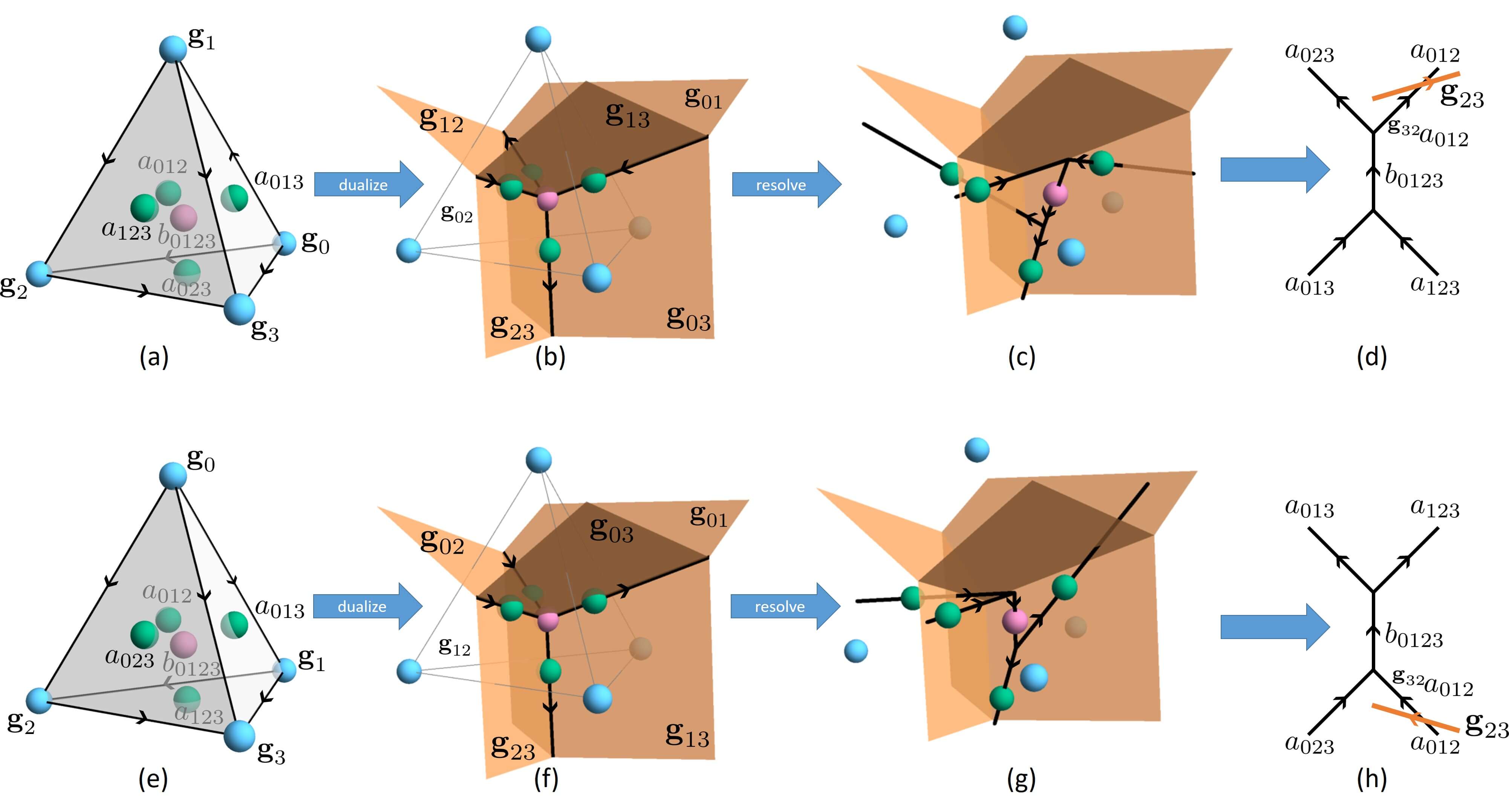}
\caption{Turning state sum data into graphical calculus for a 3-simplex with (a)-(d) positive orientation and (e)-(h) negative orientation. (a) Data associated to 0-, 2-, and 3-simplices in the state sum. Group elements ${\bf g}_i$ (blue spheres) are associated to 0-simplices, anyons $a_{ijk}$ (green spheres) are associated to 2-simplices, and an anyon $b_{0123}$ (purple sphere) is associated to the 3-simplex. (b) Upon dualizing in 3 dimensions, the dual to the 1-simplex $ij$ is a domain wall (orange sheet) acting by the group element ${\bf g}_{ij}= {\bf g}_i {\bf g}_j^{-1}$. The four $a$ anyons fuse into the $b$ anyon, and all the anyons live on domain wall junctions. (c) Our choice of resolution of the fusion channels and symmetry action; fusion is resolved trivalently, and the anyons get pushed towards a 0-simplex (dual 3-simplex). One anyon line, corresponding to $a_{012}$, passes through a domain wall before it reaches a fusion vertex. (d) Representation of (c) in graphical calculus; negatively oriented 2-simplex data are on the bottom of the diagram. Parts (e)-(h) are the same process for a negatively oriented 3-simplex.}
\label{fig:tetrahedronDual}
\end{figure*}

The 3-simplex data is determined as follows. Consider a particular $3$-simplex $ijkl$, where $i<j<k<l$, with group elements $\lbrace {\bf g}_i \rbrace$ on its four 0-simplices and anyons $\lbrace a_{ijk}\rbrace$ on its four 2-simplices. We choose $i,j,k=0,1,2,3$ for concreteness; this is shown in Fig.~\ref{fig:tetrahedronDual}(a). Then we demand that the anyon $b_{0123}$ placed on this 3-simplex obeys
\begin{equation}
N_{a_{023},{}^{{\bf g}_{32}}\!a_{012}}^{b_{0123}} \neq 0 \text{ and } N_{a_{013}, a_{123}}^{b_{0123}} \neq 0
\label{eqn:3simplexRule}
\end{equation} 
where $N_{a,b}^c$ are fusion coefficients and we define
\begin{align}
{\bf g}_{ij} \equiv {\bf g}_i {\bf g}_j^{-1} .
\end{align}

In the language of category theory, the 2-simplex data is a simple object of $\mathcal{C}$, and the data on the 3-simplex is an element of the space Hom$\left(a_{013} \otimes a_{123}, a_{013} \otimes {}^{{\bf g}_{32}}a_{012}\right)$, i.e. an element of $\oplus_b V^b_{013,123} \otimes V_b^{023,\act{32}{012}}$. In other words, if we had allowed $N_{ab}^c > 1$, we would also need to associate elements of a fusion space and a splitting space to the 3-simplices; this generalization is straightforward.

We can interpret this graphically as follows. Consider the 3D dual of the 3-simplex, as in Fig.~\ref{fig:tetrahedronDual}(b). Each dual 1-simplex is now associated to an anyon line, and four anyon lines fuse at the dual 0-simplex. Ignoring the group elements for the moment, the choice of data at the dual 0-simplex is a resolution of those four anyon lines into two trivalent fusion vertices, as in Fig.~\ref{fig:tetrahedronDual}(c). This data is then interpreted in graphical calculus by placing the anyon lines associated to negatively oriented 2-simplices at the bottom of the diagram, as in Fig.~\ref{fig:tetrahedronDual}(d).

To see how the symmetry action in Eq.~\eqref{eqn:3simplexRule} arises, consider how the group elements behave when dualizing, as in Fig.~\ref{fig:tetrahedronDual}(b). The membranes dual to the 1-simplices separate regions of space which are labeled by different group elements (associated to the dual of the 0-simplices). As such, we can think of these membranes as domain walls which have a symmetry action on the anyons. For example, the domain wall between the region associated to ${\bf g}_0$ and ${\bf g}_1$ acts on the anyons by ${\bf g}_{10}$. However, the anyon lines (dual to 2-simplices) naturally live on trijunctions between these domain walls, which leaves the symmetry action ambiguous. To disambiguate the symmetry action, deform all of the data associated to a $k$-simplex towards the highest-numbered 0-simplex (dual 3-simplex) associated with that $k$-simplex. That is, $a_{012}$ is deformed towards the 0-simplex labeled 2, while all other data is deformed towards the 0-simplex labeled 3, as shown in Fig.~\ref{fig:tetrahedronDual}(c). The symmetry action in Eq.~\eqref{eqn:3simplexRule} then appears naturally. At this stage, the data can be interpreted in graphical calculus as in Fig.~\ref{fig:tetrahedronDual}(d). The same process is shown for a negatively oriented 3-simplex in Fig.~\ref{fig:tetrahedronDual}(e)-(h).

There are two types of arbitrary choices in this definition of the data. First, we could have resolved the four-fold fusion in a different channel, related by an $F$-move to our current choice. Second, we could have resolved the symmetry action differently, for example by pushing anyon lines towards the \textit{lowest}-numbered 0-simplex. Once we define our path integral, we will need to show that it does not depend on these arbitrary choices.

Placing ${\bf g}_{23}$ branch cut lines as shown in the graphical calculus of Figs.~\ref{fig:tetrahedronDual}(d) and (h)
is determined as follows. Given a branching structure, each 1-simplex carries an orientation. Upon dualizing in 3 dimensions, this orientation defines a positive normal vector to the domain wall. Consider a domain wall between the ${\bf g}_i$ and ${\bf g}_j$ domains where $i<j$. Given our branching structure the 1-simplex is oriented from $i$ to $j$, so we choose the convention where we associate the domain wall to the group element ${\bf g}_{ij}$ (as opposed to ${\bf g}_{ji}$). An anyon $a$ that passes from the ${\bf g}_i$ region to the ${\bf g}_j$ region is converted to $\,^{ {\bf g}_j{\bf g}_i^{-1}}a$. In other words, an anyon $a$ that crosses a domain wall labeled ${\bf g}_{ij}$ along the positive normal vector to the domain wall converts to $\,^{{\bf g}_{ji}} a$, while an anyon crossing in the direction opposite to the positive normal converts to $\,^{{\bf g}_{ij}} a$.

\begin{figure*}
\includegraphics[width=1.5\columnwidth]{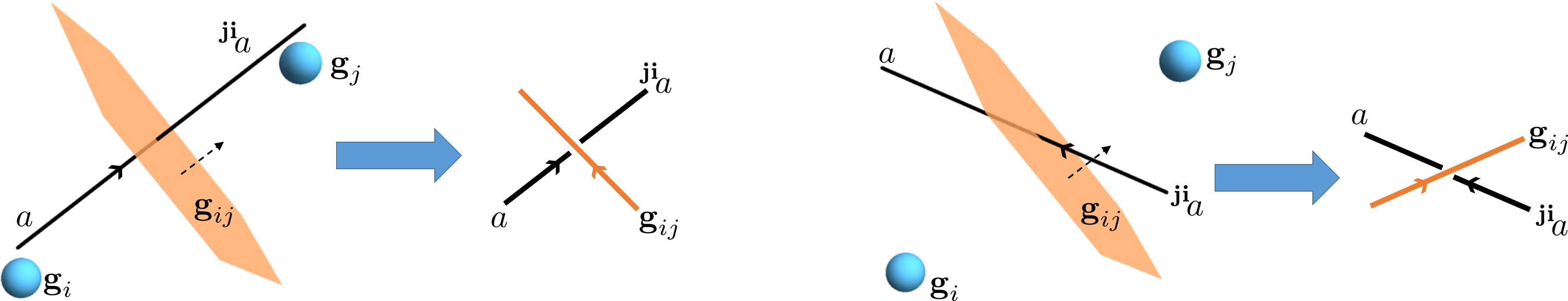}
\caption{Conventions for placing domain wall lines in diagrams in the path integral. Here $i<j$ so that the $1$-simplex $ij$ is oriented from $i$ to $j$, shown by the dashed arrows indicating the outward normal to the ${\bf g}_{ij}$ domain wall.}
\label{fig:pathIntegralDomainWallConventions}
\end{figure*}

In the graphical calculus, when projecting the diagrams to a two-dimensional plane, an oriented domain wall is depicted as an oriented line (colored orange in the figures), where the orientation of the line corresponds to the orientation of the positive normal through the convention depicted in Fig.~\ref{fig:pathIntegralDomainWallConventions}. Therefore, when an anyon passes an oriented domain wall line labeled ${\bf g}_{ij}$ from the right, it transforms to $\,^{{\bf g}_{ij}}a$; when it passes the domain wall from the left, it transforms to $\,^{{\bf g}_{ji}}a$. 

We note that in what follows we will avoid excessive subscripts by labeling simplices and data the same way. For example, we generally notate $R^{{}^{{\bf g}_{32}}a_{012},a_{023}}_{b_{0123}}$ by $R^{{}^{32}012,023}_{0123}$.

\subsection{Definition of the path integral}

\begin{figure*}
\includegraphics[width=1.3\columnwidth]{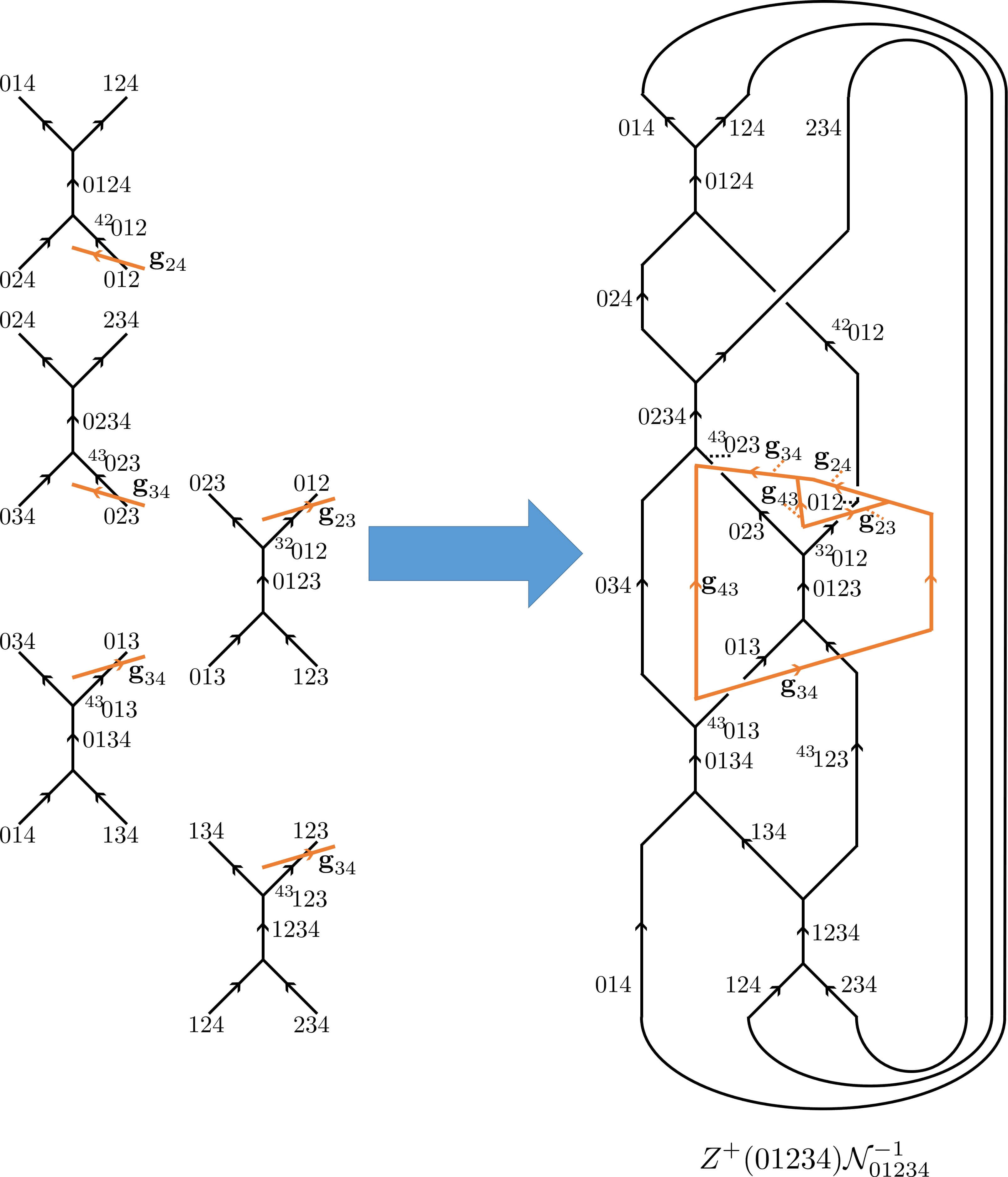}
\caption{Process for defining the diagram used to define the path integral amplitude $Z^{+}(01234)$ for a positively oriented 4-simplex $01234$. Each diagram on the left-hand side comes from a 3-simplex in $01234$; matching lines are connected to obtain the right-hand side from the left-hand side. Orange lines are domain walls. Black anyon lines are labeled by the 2-simplex or 3-simplex to which they are associated. Dashed lines are guides to the eye connecting labels and lines.}
\label{fig:diagramGluing}
\end{figure*}

\begin{figure*}
\includegraphics[width=1.3\columnwidth]{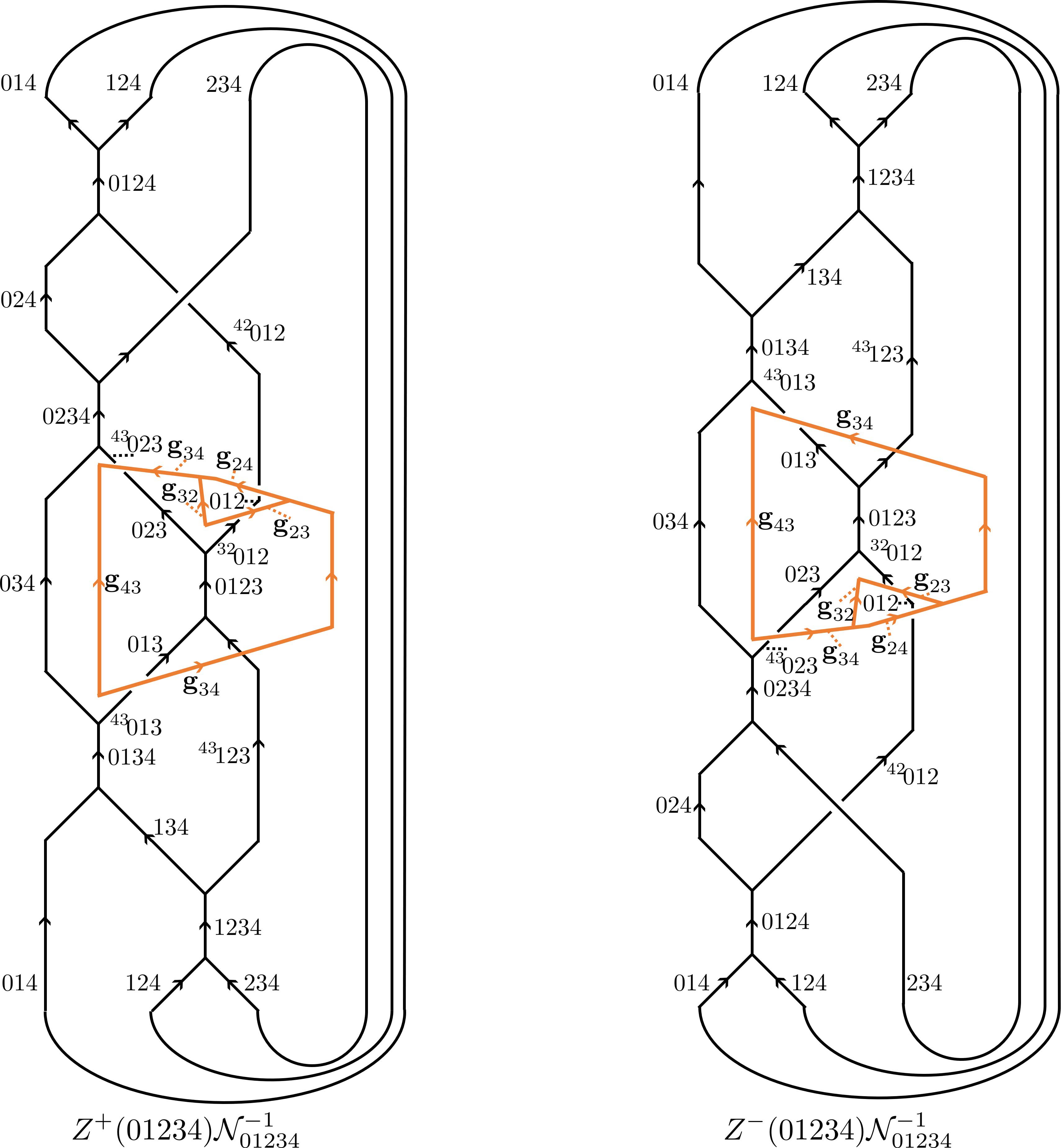}
\caption{Diagrams used to define the path integral amplitudes $Z^{\pm}(01234)$ for a 4-simplex 01234 up to the normalization factor $\mathcal{N}_{01234}$ given in Eq.~\eqref{eqn:normalization}. Orange lines are domain walls. Black anyons lines are labeled by the 2-simplex or 3-simplex to which they are associated. Dashed lines are guides to the eye connecting labels and lines.}
\label{fig:pathIntegralDiagrams}
\end{figure*}

We now define our path integral. Given a labeling of all the simplices, we will compute an anyon diagram for each 4-simplex based on the labeling. We multiply the results, and then sum over all labelings. Let $\Delta^4$ be a 4-simplex, and let $\epsilon(\Delta^4) = \pm$ be its orientation. Then define complex numbers $Z^{\epsilon(\Delta^4)}(\Delta^4)$ in the following diagrammatic way, shown in Fig.~\ref{fig:diagramGluing}. For each 3-simplex in $\Delta^4$, take the diagram obtained in Fig.~\ref{fig:tetrahedronDual} given the induced orientation on that 3-simplex. Lay out these five diagrams in a plane such that anyon lines shared between two 3-simplices are near each other. Then connect up all of the lines and domain walls, sliding domain walls along anyon lines and bending domain walls far from anyon lines as necessary to obtain a closed diagram. This process is shown for positively oriented $\Delta^4$ in Fig.~\ref{fig:diagramGluing}, while the end results are shown for both orientations in Fig.~\ref{fig:pathIntegralDiagrams}. Alternatively, these anyon diagrams arise from projecting the boundary 3-simplices, which form a triangulation of $S^3$, into the plane.

These diagrams can be evaluated explicitly, leading to the following complex number associated to each $4$-simplex:
\begin{widetext}
\begin{align}
Z^+(01234) = \mathcal{N}_{01234} \sum_{d,a \in \mathcal{C}} &F^{024,234,{}^{42}012}_{d,0234,a}\eta^{-1}_{012}(23,34) R_a^{{}^{42}012,234} (F^{024,{}^{42}012,234}_{d})^{-1}_{a,0124}F^{014,124,234}_{d,0124,1234}\times \nonumber \\
&\times (F^{014,134,{}^{43}123}_{d})^{-1}_{1234,0134}F^{034,{}^{43}013,{}^{43}123}_{d,0134,{}^{43}0123}U_{34}^{-1}(023,{}^{32}012,0123)U_{34}(013,123,0123) \times \nonumber \\
&\times (F^{034,{}^{43}023,{}^{42}012}_{d})^{-1}_{^{43}0123,0234} \label{eqn:Zplusdef}
\\
Z^-(01234) = \mathcal{N}_{01234} \sum_{d,a \in \mathcal{C}} &(F^{024,234,{}^{42}012}_{d})^{-1}_{a,0234}\eta_{012}(23,34) (R_a^{{}^{42}012,234})^{-1} F^{024,{}^{42}012,234}_{d,0124,a}\left(F^{014,124,234}_{d}\right)^{-1}_{1234,0124}\times \nonumber\\
&\times F^{014,134,{}^{43}123}_{d,0134,1234} \left(F^{034,{}^{43}013,{}^{43}123}_d\right)^{-1}_{{}^{43}0123,0134}U_{34}(023,{}^{32}012,0123)U_{34}^{-1}(013,123,0123)\times \nonumber \\
&\times F^{034,{}^{43}023,{}^{42}012}_{d,0234,^{43}0123} \label{eqn:Zminusdef}
\end{align}
\end{widetext}

The normalization factor $\mathcal{N}_{01234}$ is given by
\begin{equation}
\mathcal{N}_{01234} = \sqrt{\frac{\prod_{\Delta^3 \in 3\text{-simplices}} d_{b_{\Delta^3}}}{\prod_{\Delta^2 \in 2\text{-simplices}}d_{a_{\Delta^2}}}} ,
\label{eqn:normalization}
\end{equation}
where the products are over $3$-simplices $\Delta^3$ and $2$-simplices $\Delta^2$ of the $4$-simplex $01234$, and $d_a$ is the quantum dimension of anyon $a$. 

We may now define the path integral. On a closed 4-manifold $M^4$, we define
\begin{widetext}
\begin{equation}
Z(M^4) = \sum_{\{a,b,{\bf g}\}} \frac{\mathcal{D}^{2(N_0-N_1)}}{|G|^{N_0}}\frac{\prod_{\Delta^2 \in \mathcal{T}^2}d_{a_{\Delta^2}}\prod_{\Delta^4 \in \mathcal{T}^4}Z_F^{\epsilon(\Delta^4)}(\Delta^4)}{\prod_{\Delta^3 \in \mathcal{T}^3}d_{b_{\Delta^3}}} .
\label{eqn:Zclosed}
\end{equation}
\end{widetext}
Here $\mathcal{T}^k$ denotes the set of $k$-simplices and $N_k = |\mathcal{T}^k|$ is the number of $k$-simplices in the triangulation. The sum is over all possible labelings of the anyons $\{a\}$, $\{b\}$ on the $2$- and $3$-simplices, and group elements on $0$-simplices. 

$Z(M^4)$ defined above can be viewed as a path integral on a 4-manifold $M^4$ equipped with a trivial $G$ bundle. For computing anomaly indicators, it will be crucial to compute path integrals on non-trivial flat $G$ bundles. Therefore, we wish to define the path integral above in the presence of flat background $G$ connections, which modifies the construction as follows. Fix a gauge ${\bf h}_{ij}$ for the background connection, where ${\bf h}_{ij}$ is defined on 1-simplices of the triangulation. Our convention is such that moving from $j$ to $i$ should pick up the action of ${\bf h}_{ij}$ (note that ${\bf h}_{ji}={\bf h}_{ij}^{-1}$, as above). Flatness of the connection requires ${\bf h}_{ij} {\bf h}_{jk} {\bf h}_{ki} = {\bf 1}$ for any 2-simplex $(ijk)$. In every location in the above construction that ${\bf g}_{ij}$ appears, replace ${\bf g}_{ij}$ by $\tilde{{\bf g}}_{ij}$ where 
\begin{equation}
\tilde{{\bf g}}_{ij} = {\bf g}_i{\bf h}_{ij}{\bf g}_j^{-1}
\label{eqn:backgroundGaugeCoupling}
\end{equation}
This modifies the fusion rules used in defining the 3-simplex data and modifies the various transformations and group elements that appear in the definition of $Z^{\pm}$. Otherwise, the construction is completely unchanged. Note in particular that $\tilde{{\bf g}}_{ij}\tilde{{\bf g}}_{jk}=\tilde{{\bf g}}_{ik}$.

Recall that flat $G$-bundles are in one-to-one correspondence with elements of $\text{Hom} (\pi_1(M^4),G)$; that is, the set of group homomorphisms from the fundamental group of $M^4$ into $G$. Therefore the above prescription defines a path integral $Z(M^4, f)$, for $f \in \text{Hom} (\pi_1(M^4), G)$. 

For an open 4-manifold, we fix the $\{a, b, {\bf g} \}$ labels on the boundary of the triangulation, and sum over the rest of the labels. Apart from not summing over surface data, the construction is unmodified. This then defines a wave function on the boundary triangulation. 

We note that when $G$ is trivial, the above path integral reduces to the Crane-Yetter state sum.\cite{crane1993}

Our state sum is closely related to that of Cui.\cite{CuiTQFT} The key difference is that in our construction, the elements of $G$ live on 0-simplices rather than 1-simplices. In a rough sense, we have ``ungauged" Cui's model. This has several consequences. First, and most importantly, this allows us to generalize the construction to allow as input anomalous symmetry fractionalization classes. Cui's construction, on the other hand, is defined by taking as input a $G$-crossed BTC, which necessarily is associated with non-anomalous symmetry fractionalization classes (i.e. trivial $\mathcal{H}^4(G, U(1))$ obstruction). The reason for this difference is that when one allows arbitrary group elements on 1-simplices, there can be a net flux through each 2-simplex. Cui's model associates an object $a_{\bf g}$ of a $G$-crossed BTC to a 2-simplex whenever the flux through a 2-simplex is ${\bf g}$. This construction therefore requires the $G$-crossed BTC to be well-defined and thus correspond to a non-anomalous symmetry fractionalization class. However when group elements are placed on vertices, the net flux through a plaquette is always the identity, so one never has to require the existence of a consistent $G$-crossed BTC. 

Secondly, we will see that for modular $\mathcal{C}$ our model has trivial topological order in the bulk, and therefore corresponds to a $G$ SPT. In contrast, Cui's model is generally topologically ordered (that is, it corresponds to a non-invertible TQFT). Cui's path integral effectively sums over all possible $G$ bundles. This is a crucial distinction that will allow us to extend the construction to include anti-unitary symmetry actions, as described in Sec. \ref{sec:antiunitary}. 

\subsection{Important properties of the state-sum}

The path integral defined in the previous section has a number of important properties. Namely, it
\begin{itemize}
\item possesses a global internal $G$ symmetry,
\item defines a topological invariant for closed $M^4$ and choice of flat $G$ bundle,
\item defines a $G$ SPT (i.e. defines an invertible TQFT) when $\mathcal{C}$ is modular.
\end{itemize}

\subsubsection{Global $G$ symmetry}

Showing that our model has a global $G$ symmetry is straightforward. The only way that the group elements labeling the 0-simplices $\{ {\bf g}_i \}$ enter the state sum is through $\{ {\bf g}_{ij} \}$. Therefore the amplitude in the path integral associated with a given set of $\{{\bf g}_i\}$ labelings on 0-simplices is exactly equal to the amplitude for the labelings $\{{\bf g}_i {\bf g}^{-1} \}$, with the 2- and 3- simplex labelings unchanged.
Note in particular that the wave function defined on the boundary of the triangulation is also symmetric under the operation $\lbrace {\bf g}_i \rbrace \rightarrow \lbrace {\bf g}_i{\bf g}^{-1} \rbrace$.

\subsubsection{Topological invariance}

When $M^4$ is closed, $Z(M^4, f)$ is a topological invariant of $M^4$ and the choice of $G$ bundle, defined by $f \in \text{Hom} (\pi_1(M^4), G)$.

To prove this, we must prove that the construction is independent of the choice of triangulation, branching structure, fusion channel defining the 3-simplex data, and deformation of the anyon labels towards the 0-simplices. 

The retriangularization invariance is proven by using the fact that any two triangulations can be related by a finite sequence of Pachner moves. Therefore to prove retriangularization invariance, we directly compute the product of diagrams associated with each Pachner move to demonstrate invariance under the $3-3$, $2-4$, and $1-5$ Pachner moves. The calculations are essentially the same as those of Ref.~\onlinecite{CuiTQFT} and are reproduced in Appendix~\ref{app:Pachner}.

In Appendix~\ref{app:invariances}, we further prove invariance under the choice of fusion channel for the 3-simplex data and choice of deformation of the anyon labels towards the 0-simplices. The proof of independence of the choice of branching structure follows from the results of Ref. ~\onlinecite{CuiTQFT} and is not reproduced here.

Finally, we must prove that the construction is independent of the particular choice of flat connection ${\bf h}_{ij}$ used to describe the flat $G$ bundle. To do so, we must prove gauge invariance of the path integral under the gauge transformation
\begin{align}
{\bf h}_{ij} \rightarrow {\bf s}_i {\bf h}_{ij} {\bf s}_j^{-1} .
\end{align}
This follows from the fact that given a set of labelings, the amplitude in the state sum only depends on $\tilde{{\bf g}}_{ij} = {\bf g}_i {\bf h}_{ij} \bf {g}_j^{-1}$. Therefore, under the gauge transformation, $\tilde{{\bf g}}_{ij} \rightarrow  {\bf g}_i {\bf s}_i {\bf h}_{ij} {\bf s}_j^{-1} \bf {g}_j^{-1}$. Then the relabeling ${\bf g}_i \rightarrow {\bf g}_i {\bf s}_i$ gives back the original path integral due to the sum over all group element labels of the 0-simplices.

\subsubsection{Effect of changing projection}

The precise braided fusion diagram that we associate to a 4-simplex depends on how we project the dual 1-skeleton of the 3-simplices on the boundary of the 4-simplex onto the plane. There are a number of different choices we can consider associated with this projection. We could consider (1) flipping the fusion diagrams associated to 3-simplices about a vertical axis, (2) flipping the fusion diagrams associated to 3-simplices about a horizontal axis, and (3) changing the over-crossing to an under-crossing. Note that applying (1) and (3) together is equivalent to rotating the diagram about a vertical axis by $\pi$, which as we discuss in Appendix~\ref{vertMirror} does not change the value of the path integral on a closed manifold. Therefore (1) and (3) are equivalent choices for the path integral on a closed manifold. 

Applying (2) is equivalent to changing the convention for positive and negative orientation, which complex conjugates the path integral. This convention can be fixed by comparing to a known example, which we do by considering the case with $G = \mathbb{Z}_3 \times \mathbb{Z}_3$ and the unique UMTC of rank 3 with Abelian $\mathbb{Z}_3$ fusion rules. 

In Appendix~\ref{crossChange} we discuss the effect of applying (3) alone. In the Crane-Yetter theories with a UMTC taken as input, one can prove that changing the over- to under-crossing simply complex conjugates the path integral.\footnote{We thank Shawn Cui for helpful discussions regarding this point.} We show under a mild assumption that the same holds in our theory, that is, applying (3) complex conjugates the path integral on a closed manifold. The proof is rather indirect, relying on the relationship between the path integral and SET anomaly in order to use the relative anomaly formula\cite{BarkeshliRelativeAnomaly}. We have verified this by computerized computation of the path integral for the examples in Sec. \ref{exampleSec} along with the example of $G = \mathbb{Z}_3 \times \mathbb{Z}_3$ and the unique UMTC of rank 3 with Abelian $\mathbb{Z}_3$ fusion rules mentioned above.

We note that in Cui's construction where a $G$-crossed BTC is used to define the path integral, there is a unique choice of over- or under-crossing that allows the path integral to be well-defined. It is thus natural for us to use this convention as it is compatible with the natural generalization to Cui's state sum.

\subsection{SPT for modular $\mathcal{C}$}

It has been shown that every (at least once-extended) $(d+1)$-dimensional TQFT which assigns one-dimensional vector spaces to both $T^d$ and $S^d$ is invertible.\cite{ToriInvertibleTQFTs}\footnote{Note that a $(d+1)$-dimensional TQFT is once-extended if it assigns data to every closed $(d+1)$-, $d$- and $(d-1)$-manifold. Our model assigns a complex number to each closed $4$-manifold and a quantum state to the boundary of every $4$-manifold with boundary. Since every 3-manifold can exist at the boundary of a 4-manifold, it follows that our construction assigns a vector space to every closed $3$-manifold. The Hamiltonian formulation we study later shows how our formalism also assigns states to the boundaries of 3-manifolds as well, which determines a UMTC with $G$ symmetry fractionalization data.} Equivalently, if $Z(T^4)=Z(S^3\times S^1)=1$, the theory does not have intrinsic bulk topological order. We demonstrate for our model that if $\mathcal{C}$ is modular, then $Z(T^4)=Z(S^3\times S^1)=1$ (for trivial background $G$-bundle). This then shows that our path integral defines a $G$ SPT. 

\begin{figure}
\includegraphics[width=0.4\columnwidth]{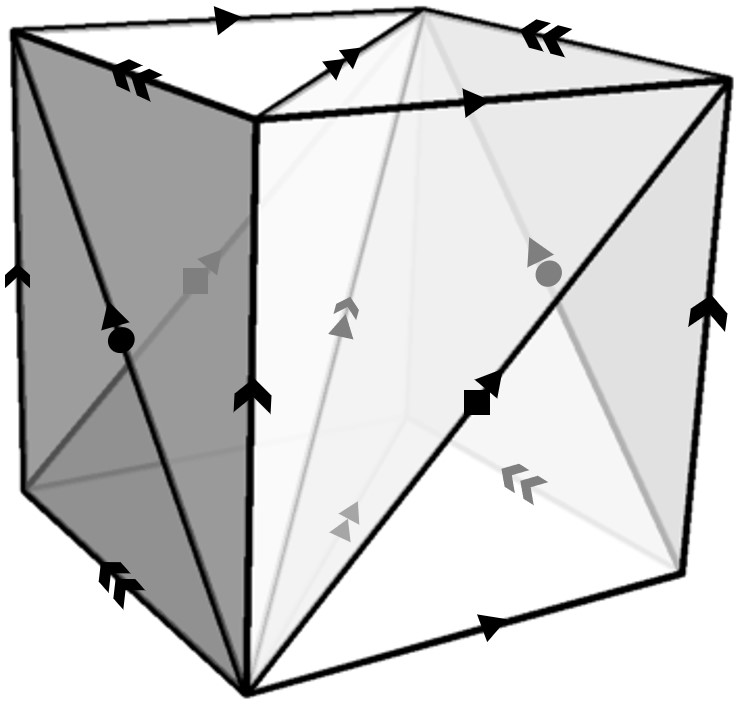}
\caption{Cellulation of $T^3$ analogous to the one used in the main text for $T^4$. Opposite faces are identified, which identifies all $0$-simplices and any 1- and 2-simplices with matching markings. This cellulation contains six 4-simplices.}
\label{fig:T3cellulation}
\end{figure}

Given that our state sum is topologically invariant, we may compute $Z(T^4)$ using the simplest possible triangulation. In fact, we choose to use a cellulation rather than a triangulation.\footnote{A $d$-dimensional triangulation strictly speaking requires two $d$-simplices to share only one $d-1$ simplex. A cellulation does not require this stringent condition and therefore allows more efficient ways to construct manifolds by gluing together simplices. Our state sum is still well-defined when we relax this condition.} The simplest cellulation of $T^4$ consists of twenty-four 4-simplices, but only one 0-simplex which we label $0$. See Fig.~\ref{fig:T3cellulation} for the analogous cellulation of $T^3$. The details of the cellulation are described in Appendix~\ref{app:cellulations}, but given that it exists, we can use it to compute $Z(T^4)$. Because there is only a single 0-simplex and we choose the background $G$ bundle to be trivial, there are no domain walls; all the ${\bf g}_{ij}$ are the identity. Therefore, for this particular cellulation, every term is independent of ${\bf g}_0$, so we may as well choose the term where ${\bf g}_0$ is the identity. It immediately follows that
\begin{equation}
Z(T^4)=Z_{CY}(T^4) = 1
\end{equation}
where $Z_{CY}$ is the Crane-Yetter path integral for $\mathcal{C}$. This is the desired result because $Z_{CY}(T^4)=1$ for modular $\mathcal{C}$ (see e.g. Ref. \onlinecite{walker2006}). 

The partition function $Z(S^3 \times S^1)$ in the absence of a background gauge field can also be calculated explicitly. The simplest cellulation of this manifold involves eight 4-simplices and is detailed in Appendix~\ref{app:cellulations}. Once the cellulation is written down, one can explicitly show that the symmetry factors $U$ and $\eta$ cancel out in the absence of a background gauge field, reducing the calculation to the Crane-Yetter path integral for $\mathcal{C}$ on $S^3 \times S^1$:
\begin{equation}
Z(S^3 \times S^1)=Z_{CY}(S^3 \times S^1) = 1.
\end{equation}
Since describing the cellulation is rather tedious, this calculation is relegated to Appendix~\ref{app:ZS3xS1}.

For reference, we note that it is also straightforward to check that
\begin{align}
Z(S^4) = 1.
\end{align}

As a technical aside, in order to use such simple decompositions of the manifolds in question, we are using the fact that our path integral is independent of the \textit{cellulation} of the manifold in question. That this is the case for the Crane-Yetter state sum was proven in Ref. \onlinecite{walker2006}, and we similarly expect it to be true for our construction. Alternatively, one can check whether Pachner moves can convert the cellulations used here to triangulations, which we also expect to be true (it is a simple exercise to check the lower dimensional analog for $T^2$). 

\subsection{Conjecture for removing sum over group elements}

The state sum defined above contains a sum over all group labels on the $0$-simplices, $\frac{1}{|G|^{N_0}} \sum_{\{{\bf g}_i \}}$. We conjecture that, at least for modular $\mathcal{C}$, every term in this sum is equal and therefore the sum is actually unnecessary. In other words, the amplitude associated to any given set of labelings is actually independent of the choice of $G$-defect networks present in the (3+1)D space-time. In this case, the symmetry fractionalization data only contributes when the $G$-bundle is non-trivial, in which case the braided fusion diagrams that define the 4-simplex amplitudes necessarily have domain walls due to the non-trivial holonomies around non-contractible cycles. This conjecture is borne out in the examples that we study in Sec. \ref{exampleSec}.

We note that the analogous statement is also true for the SPT state sum based on group cohomology.\cite{chen2013}

It is clear that if such a conjecture is true, then it implies that the state sum is a (3+1)D G-SPT. The reason is that the computation of the state sum on trivial $G$ bundles necessarily reduces to the Crane-Yetter state sum for which it is known that the path integral is unity on every trivial closed $G$-bundle. We believe that the converse is also true: the fact that modular $\mathcal{C}$ with the symmetry fractionalization data determines a $G$-SPT should imply the aforementioned conjecture. 

\section{Extension to anti-unitary on-site symmetries}
\label{sec:antiunitary}

In this section we extend our state sum construction to symmetry groups $G$ that contain anti-unitary symmetry actions. This will allow us to extend our results to arbitrary time-reversing or spatial parity reversing symmetries. In particular, theories with such anti-unitary actions will allow us to formulate path integrals on non-orientable manifolds for theories that have anti-unitary symmetry actions. 

We note that the symmetry fractionalization formalism discussed in this section is closely related to previous discussions in Ref. \onlinecite{barkeshli2019,BarkeshliReflection} although the discussion here proceeds through a somewhat different perspective in light of our model. 

\subsection{Reformulation of unitary symmetry fractionalization}

In order to extend our construction to allow $G$ to contain anti-unitary symmetry actions, we will first consider the unitary case from a different perspective. Formally this amounts to packaging the UMTC and symmetry fractionalization data into the structure of a 3-category where the objects ($0$-morphisms) are group elements of $G$. 

The state sum construction we defined in Sec.~\ref{sec:stateSum} required us to input a consistent graphical calculus involving the data
$\{ F^{abc}_{def} , R^{ab}_c , G, \rho, U_{{\bf g}_{ij}}(a,b;c), \eta_a({\bf g}_{ij}, {\bf g}_{jk})\}$. The domain walls ${\bf g}_{ij}$ are essentially a shorthand tracking the group elements ${\bf g}_i$ associated to regions of space in the graphical calculus (recall that in passing from simplices to the graphical calculus, we consider the dual of the triangulation, so group elements on the $0$-simplices are associated to regions of space in the graphical calculus). We can do this because the state sum does not explicitly depend on any of the ${\bf g}_i$, but rather just on the domain walls. 

\begin{figure*}
\centering
\includegraphics[width=1.5\columnwidth]{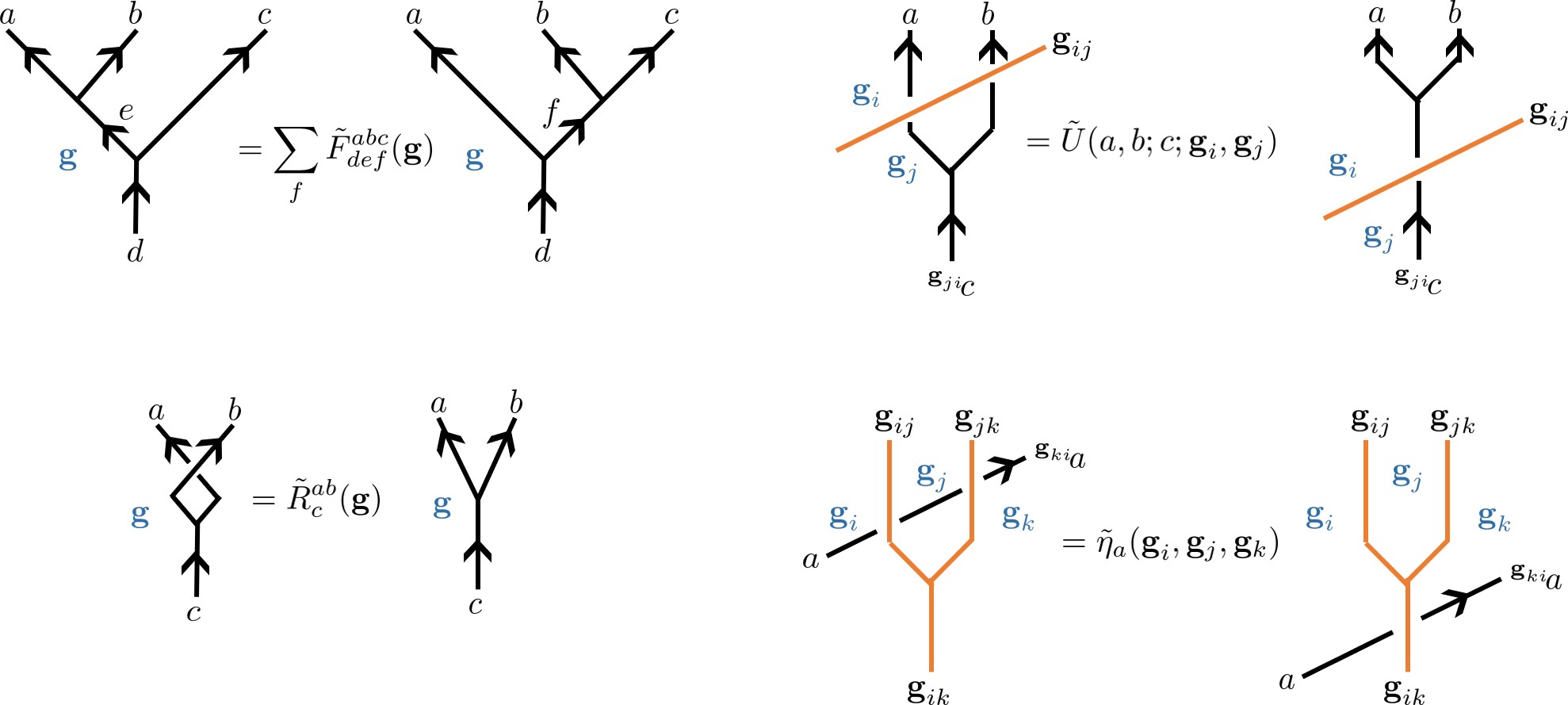}
\caption{Graphical calculus defining $\tilde{F}$, $\tilde{R}$, $\tilde{U}$, and $\tilde{\eta}$ so that they depend on the group elements in the relevant regions of space. Anyon lines are black, and group elements label regions of space or domain walls (orange). Note that anyon lines do not change the group element associated to a region of space.}
\label{fig:tildeGraphicalCalculus}
\end{figure*}

However, we could have reasonably defined the same type of state sum by having the fusion spaces explicitly depend on the group element of the domain in which they sit. That is, we define the fusion and splitting spaces $\tilde{V}_{ab}^c({\bf g})$, $\tilde{V}_c^{ab}({\bf g})$. Then the $c-ab$ splitting vertex that has been deformed towards the 0-simplex $i$ is an element of $\tilde{V}_c^{ab}({\bf g}_i)$; diagrammatically, this is represented as
\begin{equation}
\left( d_{c} / d_{a}d_{b}\right) ^{1/4}
\raisebox{-0.5\height}{
\begin{pspicture}[shift=-0.65](-0.1,-0.2)(1.5,1.2)
  \small
  \psset{linewidth=0.9pt,linecolor=black,arrowscale=1.5,arrowinset=0.15}
  \psline{->}(0.7,0)(0.7,0.45)
  \psline(0.7,0)(0.7,0.55)
  \psline(0.7,0.55) (0.25,1)
  \psline{->}(0.7,0.55)(0.3,0.95)
  \psline(0.7,0.55) (1.15,1)	
  \psline{->}(0.7,0.55)(1.1,0.95)
  \rput[bl]{0}(0.4,0){$c$}
  \rput[br]{0}(1.4,0.8){$b$}
  \rput[bl]{0}(0,0.8){$a$}
  \rput[bl]{0}(0.2,0.35){$\textcolor{blue}{{\bf g}}$}
 \scriptsize
  \rput[bl]{0}(0.85,0.35){$\mu$}
  \end{pspicture}
  }
=\left| a,b;c,\mu \right\rangle_{{\bf g}} \in
\tilde{V}_{c}^{ab}({\bf g}).
\label{eq:ketWithG}
\end{equation}
The data defining the theory can then depend on these group elements: $\tilde{F}^{abc}_{def}({\bf g}_i)$, $\tilde{R}^{ab}_c({\bf g}_i)$, $\tilde{\eta}_a({\bf g}_i,{\bf g}_j,{\bf g}_k)$ and $\tilde{U}(a,b;c;{\bf g}_i,{\bf g}_j)$, where the $\lbrace {\bf g}_i \rbrace$ label the regions involved in the graphical calculus operation. For example, the ${\bf g}_i$ in $\tilde{R}$ is associated to the spatial region in which the $abc$ fusion vertex being acted on by the $R$-move is contained. Diagrammatic representations of these generalized data are shown in Fig.~\ref{fig:tildeGraphicalCalculus}.

In particular, $\tilde{U}$ is now defined via the unitary maps $\tilde{\rho}^L$:
\begin{align}
  \tilde{\rho}_{\bf h}^L : \tilde{V}_{ab}^c({\bf g}) \rightarrow \tilde{V}_{\,^{\bf h}a \,^{\bf h} b}^{\,^{\bf h} c} ({\bf h g })
\end{align}
\begin{align}
  \tilde{\rho}_{ {\bf h} }^L &|a,b;c, \mu \rangle_{\bf g} =
                             \nonumber \\
&  \sum_{\nu} [\tilde{U}(\,^{\bf h} a, \,^{\bf h} b; \,^{ \bf h}c, {\bf hg}, {\bf g})]_{\mu\nu} |\,^{\bf h}a,\,^{\bf h}b;\,^{\bf h}c, \mu \rangle_{\bf h g }
\end{align}
where the $L$ superscript refers to the fact that group elements are left-multiplied under this map.

\begin{figure*}
\centering
\includegraphics[width=1.3\columnwidth]{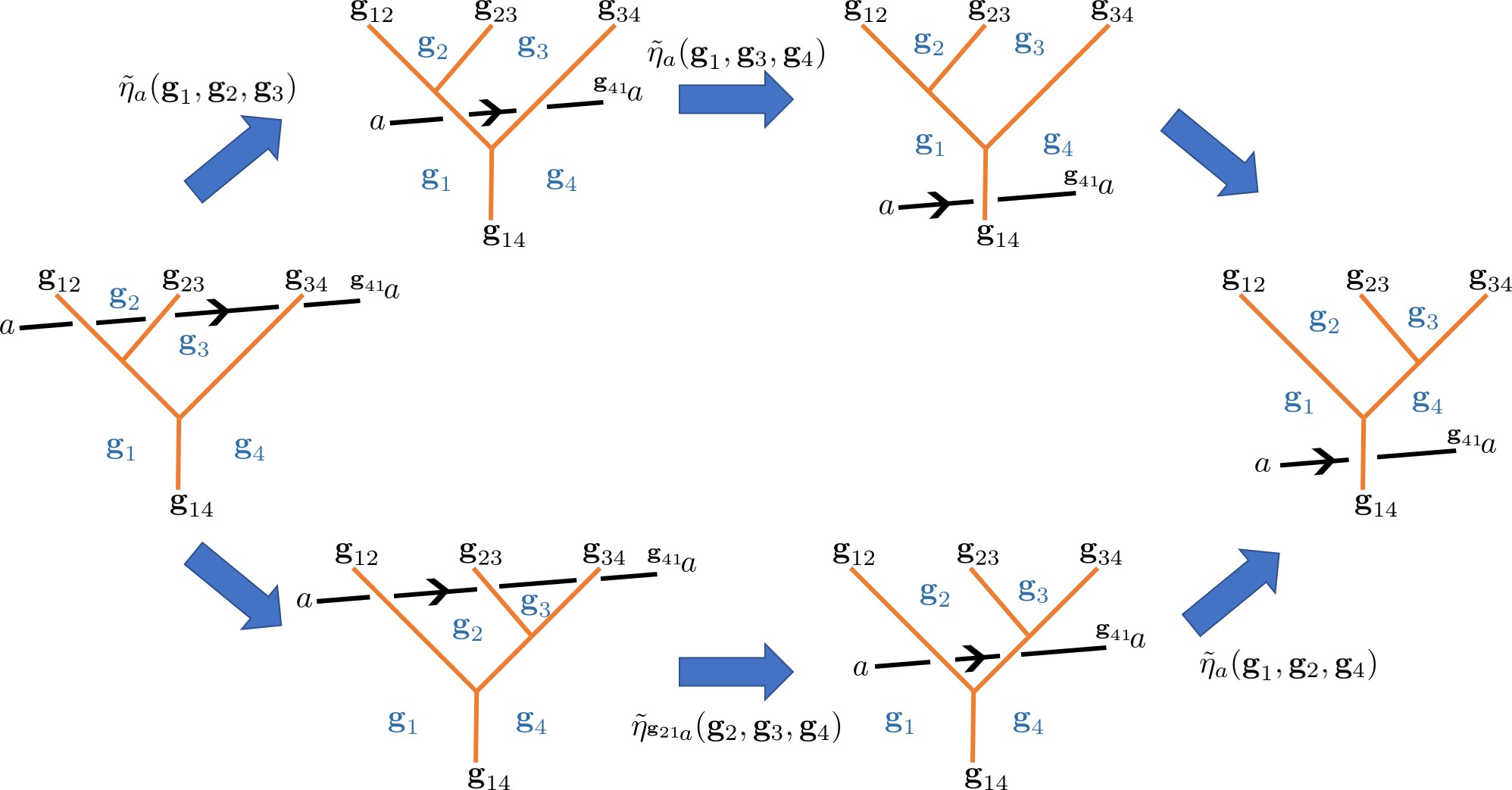}
\caption{Consistency condition for $\tilde{\eta}$, where $\tilde{\eta}$ is defined graphically in Fig.~\ref{fig:tildeGraphicalCalculus}. Domain walls are orange, anyon lines are black.}
\label{fig:tildeEtaConsistency}
\end{figure*}

The consistency of the graphical calculus then leads to consistency conditions analogous to the ones we had previously: 
\begin{widetext}
\begin{align}
\tilde{F}^{fcd}_{egl}({\bf g})\tilde{F}^{abl}_{efk}({\bf g}) &= \sum_h \tilde{F}^{abc}_{gfh}({\bf g})\tilde{F}^{ahd}_{egk}({\bf g})\tilde{F}^{bcd}_{khl}({\bf g}) \label{eqn:tildePentagon}\\
\tilde{R}^{ca}_e({\bf g})\tilde{F}^{acb}_{deg}({\bf g})\tilde{R}^{cb}_g({\bf g}) &= \sum_f \tilde{F}^{cab}_{def}({\bf g})\tilde{R}^{cf}_d({\bf g})\tilde{F}^{abc}_{dfg}({\bf g})\\
(\tilde{R}^{ac}_e({\bf g}))^{-1}\tilde{F}^{acb}_{deg}({\bf g})(\tilde{R}^{bc}_g({\bf g}))^{-1} &= \sum_f \tilde{F}^{cab}_{def}({\bf g})(\tilde{R}^{fc}_d({\bf g}))^{-1}\tilde{F}^{abc}_{dfg}({\bf g})\\
\tilde{U}(\acts{g}{21}{a},\acts{g}{21}{b} ;\acts{g}{21}{c} ,{\bf g}_2,{\bf g}_3)\tilde{U}(a,b;c,{\bf g}_1, {\bf g}_2) &= \tilde{U}(a, b;c, {\bf g}_1, {\bf g}_3)\frac{\eta_c({\bf g}_1, {\bf g}_2, {\bf g}_3)}{\eta_a({\bf g}_1, {\bf g}_2, {\bf g}_3)\eta_b({\bf g}_1, {\bf g}_2, {\bf g}_3)}\\
\tilde{U}(\acts{g}{12}{a},\acts{g}{12}{b} ;\acts{g}{12}{e},{\bf g}_1,{\bf g}_2)\tilde{U}(\acts{g}{12}{e},\acts{g}{12}{c} ;\acts{g}{12}{d},{\bf g}_1,{\bf g}_2)&\tilde{F}^{\acts{g}{12}{a}\act{12}{b}\act{12}{c}}_{\act{12}{d}\act{12}{e}\act{12}{f}}({\bf g}_1) \times \nonumber \\
\times \tilde{U}^{-1}(\acts{g}{12}{b},\acts{g}{12}{c} ;\acts{g}{12}{f},{\bf g}_1,{\bf g}_2)&\tilde{U}^{-1}(\acts{g}{12}{a},\acts{g}{12}{f} ;\acts{g}{12}{d},{\bf g}_1,{\bf g}_2) = \tilde{F}^{abc}_{def}({\bf g}_2)\\
  \tilde{U}(\acts{g}{12}{b},\acts{g}{12}{a} ;\acts{g}{12}{c},{\bf g}_1,{\bf g}_2)\tilde{R}^{\act{12}{a}\act{12}{b}}_{\act{12}{c}}({\bf g}_1)\tilde{U}^{-1}(\acts{g}{12}{a},\acts{g}{12}{b} ;\acts{g}{12}{c},{\bf g}_1,{\bf g}_2) &= \tilde{R}^{ab}_c({\bf g}_2) \label{eqn:tildeURConsistency}
\\
  \tilde{\eta}_{{}^{{\bf g}_{21}}\!a}({\bf g}_2,{\bf g}_3,{\bf g}_4) \tilde{\eta}_a({\bf g}_1, {\bf g}_3, {\bf g}_4)
                                                             &= \tilde{\eta}_a({\bf g}_1,{\bf g}_2,{\bf g}_3) \tilde{\eta}_a({\bf g}_1, {\bf g}_2, {\bf g}_4)
                                                              \label{eqn:tildeEtaConsistency}
\end{align}
\end{widetext}
For example, in deriving the analogue of Eq.~\eqref{eqn:etaConsistency}, one considers the equality of the two diagrammatic processes  shown in Fig.~\ref{fig:tildeEtaConsistency}, producing Eq.~\eqref{eqn:tildeEtaConsistency}. 
Internal indices for the fusion spaces can be added straightforwardly at the cost of additional notation.

One can check that the consistency of this graphical calculus yields a topologically invariant path integral that is independent of the various arbitrary choices made in the definition. However, it does not yet have a $G$ symmetry, as everything depends explicitly on the ${\bf g}_i$. Therefore it cannot be consistently coupled to a background $G$ connection and defined on non-trivial $G$ bundles. 

To ensure that the theory is $G$ symmetric, we demand that the wavefunction on any closed 3-manifold $\partial M^4$ be $G$-invariant. That is, we demand invariance under the unitary $G$ action which right-multiplies the $G$ degrees of freedom by $\bar{{\bf g}}$ and does not modify any of the other degrees of freedom. Since the wave function is given by the partition function on $M^4$ with fixed boundary data, this means that
\begin{align}
Z[\partial M^4; &\lbrace {\bf g}_i{\bf \bar{g}} \rbrace, \lbrace a_{ijk} \rbrace , \lbrace b_{ijkl} \rbrace] \nonumber \\
&= Z[\partial M^4; \lbrace {\bf g}_i \rbrace, \lbrace a_{ijk} \rbrace ,\lbrace b_{ijkl}\rbrace]
\label{eqn:unitaryGAction}
\end{align}
where the $ \lbrace {\bf g}_i \rbrace, \lbrace a_{ijk} \rbrace ,\lbrace b_{ijkl}\rbrace$ are the labels on $\partial M^4$, and $i,j,k,l$ label boundary vertices. It is important that we have chosen right-multiplication for the $G$ action because it preserves the combinations ${\bf g}_i {\bf \bar{g}}_j$ that naturally act on anyons in the path integral.

Specializing to the case where $M^4$ is a single 4-simplex $01234$, (meaning $\partial M^4 = S^3$), we thus require
\begin{align}
Z^+[01234; &\lbrace {\bf g}_i{\bf \bar{g}} \rbrace, \lbrace a_{ijk} \rbrace ,\lbrace b_{ijkl}\rbrace] \nonumber\\
&= Z^+\left[01234;  \lbrace {\bf g}_i \rbrace, \lbrace a_{ijk} \rbrace ,\lbrace b_{ijkl}\rbrace\right]
\label{eqn:symmetryOnD4}
\end{align}
which equates two anyon diagrams, with no bulk data to sum over, for every choice of anyons and group elements. A sufficient condition to satisfy Eq.~\eqref{eqn:symmetryOnD4} is to demand $\tilde{F}$, $\tilde{R}$, $\tilde{\eta}$ and $\tilde{U}$ are individually invariant under this $G$-action:
\begin{align}
\tilde{F}^{abc}_{def}({\bf g}_i) &= \tilde{F}^{abc}_{def}({\bf g}_i \bar{{\bf g}}) \label{eqn:unitaryTildeF}\\
\tilde{R}^{ab}_c({\bf g}_i) &= \tilde{R}^{ab}_c({\bf g}_i \bar{{\bf g}})\\
\tilde{\eta}_a({\bf g}_i,{\bf g}_j,{\bf g}_k) &= \tilde{\eta}_a({\bf g}_i\bar{{\bf g}},{\bf g}_j\bar{{\bf g}},{\bf g}_k\bar{{\bf g}})\\
\tilde{U}(a,b;c;{\bf g}_i, {\bf g}_j) &= \tilde{U}(a,b;c;{\bf g}_i \bar{{\bf g}}, {\bf g}_j \bar{{\bf g}}) \label{eqn:unitaryTildeU}
\end{align}

Another way to say this is the following. For fixed $a,b,c$, all the fusion spaces $\tilde{V}_{ab}^c({\bf g}_i)$ are isomorphic. In order to impose $G$ symmetry, we should identify all of these fusion spaces using isomorphisms
\begin{equation}
\alpha_{\bf g}^R: \tilde{V}_{ab}^c({\bf g}_i) \rightarrow \tilde{V}_{ab}^c({\bf g}_i\bar{{\bf g}}) \label{eqn:alphaDef},
\end{equation}
and then demand $G$-equivariance. The $R$ superscript refers to the fact that group elements are right-multiplied. That is, the maps $\alpha_{\bf g}^R$ should be unitary and strictly preserve all of the UMTC plus symmetry fractionalization data of the theory. In particular we demand that
\begin{equation}
\tilde{\rho}_{\bf g}^L \circ \alpha_{\bf h}^R = \alpha_{\bf h}^R \circ \tilde{\rho}_{\bf g}^L .
\label{eqn:equivariance}
\end{equation}

It follows that
\begin{align}
\tilde{F}^{abc}_{def}({\bf g}_i) &= \tilde{F}^{abc}_{def}({\bf 1}) \equiv F^{abc}_{def}\\
\tilde{R}^{ab}_c({\bf g}_i) &= \tilde{R}^{ab}_c({\bf 1}) \equiv R^{ab}_c\\
\tilde{\eta}_a({\bf g}_i,{\bf g}_j,{\bf g}_k) &= \tilde{\eta}_a({\bf g}_{ik}, {\bf g}_{jk}, {\bf 1}) \equiv \eta_a({\bf g}_{ij},{\bf g}_{jk}) \\
\tilde{U}(a,b;c;{\bf g}_i, {\bf g}_j) &= \tilde{U}(a,b;c;{\bf g}_{ij}, {\bf 1}) \equiv U_{{\bf g}_{ij}}(a,b;c) .
\end{align}
The last equalities define the data $F$, $R$, $\eta$ and $U$ (without tildes), which are what we usually use to define symmetry fractionalization on a BTC. It is straightforward to check that the consistency conditions Eqs.~\eqref{eqn:tildeEtaConsistency}-\eqref{eqn:tildeURConsistency} for the data with tildes, when recast in terms of the data without tildes, become the usual consistency conditions Eqs.~\eqref{eqn:UFURConsistency} and \eqref{eqn:UEtaConsistency}-\eqref{eqn:etaConsistency}.

Another way to say this is that the data without tildes that appears in Sec.~\ref{globsym} should be identified as the identity sector of the data with tildes, that is,
\begin{equation}
V_{ab}^c = \tilde{V}_{ab}^c({\bf 1}).
\end{equation}
We can then define a map $\rho_{\bf g}$ which acts only on the spaces without tildes and also has the properties given in Sec.~\ref{globsym} by 
\begin{align}
\rho_{{\bf g}} = \alpha_{{\bf g}}^R \circ \tilde{\rho}_{{\bf g}}^L\bigg|_{V}
\label{eqn:rhodef}
\end{align}
where the restriction  means that we are restricting $\tilde{\rho}$ to act only on vector spaces without tildes, i.e. those in the identity domain.

At this stage, we have simply added a lot of extra notation for very little gain, as it is clear from Eqs.~\eqref{eqn:unitaryTildeF}-\eqref{eqn:unitaryTildeU} that the data without tildes characterizes the data of the theory with tildes. However, this machinery will make the extension to antiunitary symmetries much more natural.

Before proceeding, we make two technical comments. First, in principle both the vertex basis and symmetry action gauge transformations can depend on the domain in question, and furthermore one could imagine changing $\alpha^R$ and $\tilde{\rho}^L$ by natural isomorphisms separately. However, imposing the equivariance conditions, Eq.\eqref{eqn:unitaryTildeF}-\eqref{eqn:unitaryTildeU}, which arose from demanding symmetry of the wave function, Eq.~\eqref{eqn:symmetryOnD4}, constrains the gauge transformations so that the transformations must act the same way on all the domains. For example, if $\tilde{\Gamma}_{ab}^c({\bf g}_i)$ is a vertex basis gauge transformation on $\tilde{V}_{ab}^c({\bf g}_i)$, then
\begin{equation}
\tilde{\Gamma}_{ab}^c({\bf g}_i) = \tilde{\Gamma}_{ab}^c({\bf g}_i {\bf \bar{g}}) ,
\end{equation}
and similarly for the symmetry action gauge transformations. 
Second, in the equivariance condition Eq.~\eqref{eqn:equivariance}, it is crucial that if $\tilde{\rho}$ left-multiplies the group elements, $\alpha$ must right-multiply. If we instead used a map
\begin{equation}
\alpha_{\bf g}^L : V_{ab}^c({\bf g}_i) \rightarrow V_{ab}^c({\bf gg}_i),
\end{equation}
then the equivariance condition would not even make any sense if $G$ is non-Abelian because the codomains of the two maps are different spaces:
\begin{align}
\tilde{\rho}_{\bf g}^L \circ \alpha_{\bf h}^L &: V_{ab}^c({\bf g}_i) \rightarrow V_{\act{g}{a} \act{g}{b}}^{\act{g}{c}}({\bf g h g}_i)\\
\alpha_{\bf h}^L \circ \tilde{\rho}_{\bf g}^L  &: V_{ab}^c({\bf g}_i) \rightarrow V_{\act{g}{a} \act{g}{b}}^{\act{g}{c}}({\bf h g g}_i)
\end{align}

\subsection{Anti-unitary symmetry actions}

We now generalize the above discussion to the case where $G$ contains anti-unitary symmetry actions. Then the natural $G$ action Eq.~\eqref{eqn:unitaryGAction} on the wavefunction involves complex conjugations:
\begin{align}
Z[\partial M^4; &\lbrace {\bf g}_i{\bf \bar{g}} \rbrace, \lbrace a_{ijk} \rbrace , \lbrace b_{ijkl} \rbrace] = \nonumber \\
&= Z^{\sigma({\bf g})}[\partial M^4; \lbrace {\bf g}_i \rbrace, \lbrace a_{ijk} \rbrace ,\lbrace b_{ijkl}\rbrace]
\label{eqn:TReversingAction}
\end{align}
where $\sigma$ was defined in Eq.~\eqref{eqn:sigmaDef}. Following the same logic, the desired symmetry transformation rules change:
\begin{align}
\tilde{F}^{abc}_{def}({\bf g}_i) &= [\tilde{F}^{abc}_{def}({\bf g}_i {\bf {\bar{g}}})]^{\sigma({\bf g})}\\
\tilde{R}^{ab}_c({\bf g}_i) &= [\tilde{R}^{ab}_c({\bf g}_i {\bf \bar{g}})]^{\sigma({\bf g})}\\
\tilde{\eta}_a({\bf g}_i,{\bf g}_j,{\bf g}_k) &= [\tilde{\eta}_a({\bf g}_i{\bf \bar{g}},{\bf g}_j{\bf \bar{g}},{\bf g}_k\bar{{\bf g}})]^{\sigma({\bf g})}\\
\tilde{U}(a,b;c;{\bf g}_i, {\bf g}_j) &= [\tilde{U}(a,b;c;{\bf g}_i {\bf \bar{g}}, {\bf g}_j {\bf \bar{g}})]^{\sigma({\bf g})}
\end{align}
In particular, we now take the isomorphisms $\alpha_{{\bf g}}^R$ (see Eq.~\eqref{eqn:alphaDef}) to be \textit{anti-unitary} maps when ${\bf g}$ is an anti-unitary symmetry action. 

With this in mind, we may now generalize the definitions of the data $F$, $R$, $\eta$ and $U$ (without tildes) via
\begin{align}
\tilde{F}^{abc}_{def}({\bf g}_i)^{\sigma({\bf g})} &= \tilde{F}^{abc}_{def}({\bf 1}) \nonumber\\
&\equiv F^{abc}_{def} \label{eqn:FNoTilde}\\
\tilde{R}^{ab}_c({\bf g}_i)^{\sigma({\bf g})} &= \tilde{R}^{ab}_c({\bf 1}) \nonumber\\
&\equiv R^{ab}_c  \label{eqn:RNoTilde}\\
\eta_a({\bf g}_i,{\bf g}_j,{\bf g}_k)^{\sigma({\bf g}_i)} &= \tilde{\eta}^{\sigma({\bf g}_{ik})}_a({\bf g}_{ik}, {\bf g}_{jk}, {\bf 1}) \nonumber\\
&\equiv \eta_a({\bf g}_{ij},{\bf g}_{jk}) \label{eqn:etaNoTilde}\\
\tilde{U}(a,b;c;{\bf g}_i, {\bf g}_j)^{\sigma({\bf g}_i)} &= \tilde{U}^{\sigma({\bf g}_{ij})}(a,b;c;{\bf g}_{ij}, {\bf 1}) \nonumber \\
&\equiv U_{{\bf g}_{ij}}(a,b;c) \label{eqn:UNoTilde}
\end{align}
Another way to think of Eqs.~\eqref{eqn:FNoTilde}-\eqref{eqn:UNoTilde} is that they allow us to take symmetry fractionalization data given in the language of Sec.~\ref{globsym} and convert it into data ``with tildes" which can be inserted into our path integral.

Following the above, we can now write the consistency conditions for the data $F, R, \eta, U$ in the presence of anti-unitary symmetry actions by simply substituting Eqs.~\eqref{eqn:FNoTilde}-\eqref{eqn:UNoTilde} into Eqs.~\eqref{eqn:tildePentagon}-\eqref{eqn:tildeEtaConsistency}
\begin{widetext}
\begin{align}
\eta_a({\bf g},{\bf h})\eta_a({\bf gh},{\bf k})&=\eta_{{}^{\bar{{\bf g}}}\!a}^{\sigma({\bf g})}({\bf h}, {\bf k}) \eta_a({\bf g}, {\bf hk}) \label{eqn:etaConsistencyAnti}\\
U_{{\bf k}}({}^{{\bf k}}\!a,{}^{{\bf k}}\!b;{}^{{\bf k}}\!e)U_{{\bf k}}({}^{{\bf k}}\!e,{}^{k}\!c,{}^{k}\!d)F^{{}^{{\bf k}}\!a {}^{{\bf k}}\!b {}^{{\bf k}}\!c}_{{}^{{\bf k}}\!d {}^{{\bf k}}\!e {}^{{\bf k}}\!f}U_{{\bf k}}^{-1}({}^{{\bf k}}\!b,{}^{{\bf k}}\!c;{}^{{\bf k}}\!f)U_{{\bf k}}^{-1}({}^{{\bf k}}\!a,{}^{{\bf k}}\!f;{}^{{\bf k}}\!d)&= \left(F^{abc}_{def}\right)^{\sigma({\bf k})} \label{eqn:UFconsistencyAnti}\\
U_{{\bf \ell}}^{\sigma({\bf g})}({}^{\bf {\bar{k}}}a,{}^{\bf {\bar{k}}}b;{}^{\bf {\bar{k}}}c)U_{{\bf k}}(a, b;c) &= U_{{\bf kl}}(a,b;c)\frac{\tilde{\eta}_c({\bf k}, {\bf \ell})}{\tilde{\eta}_a({\bf k}, {\bf \ell})\tilde{\eta}_b({\bf k}, {\bf \ell})}\\
U_{{\bf g}}({}^{{\bf g}}\!b,{}^{{\bf g}}\!a;{}^{{\bf g}}\!c)R^{{}^{{\bf g}}\!a {}^{{\bf g}}\!b}_{{}^{{\bf g}}\!c}U^{-1}_{{}^{{\bf g}}\!}({}^{{\bf g}}\!a, {}^{{\bf g}}\!b, {}^{{\bf g}}\!c)&=\left(R^{ab}_c\right)^{\sigma({\bf g})} \label{eqn:URconsistencyAnti}
\end{align}
\end{widetext}
These exactly reproduce Eqs.~\eqref{eqn:UFURConsistency} and \eqref{eqn:UEtaConsistency}-\eqref{eqn:etaConsistency} in the presence of anti-unitary symmetry actions. Because $F$- and $R$-moves need only involve a single domain, the pentagon and hexagon equations take their usual forms.

To make contact with Sec.~\ref{globsym}, we again note that all of the data without tildes act on vector spaces $\tilde{V}_{ab}^c({\bf 1})$ in the identity domain. We claim that the formalism in Sec.~\ref{globsym} arises if we define $V_{ab}^c = \tilde{V}_{ab}^c({\bf 1})$ and define the topological symmetry operator $\rho_{\bf g}$, without a tilde, again by Eq.~\eqref{eqn:rhodef}. The map $\rho_{\bf g}$ is anti-unitary if the action of ${\bf g}$ is anti-unitary (since $\tilde{\rho}$ is always unitary and $\alpha_{\bf g}$ is anti-unitary when the action of ${\bf g}$ is anti-unitary). We can check that this $\rho_{\bf g}$ behaves in the desired way:
\begin{align}
\rho_{\bf g}\ket{a,b;c}_{\bf 1} &= \alpha_{\bf g}^R \tilde{U}(\act{g}{a},\act{g}{b};\act{g}{c},{\bf g},{\bf 1})\ket{\act{g}{a}, \act{g}{b}; \act{g}{c}}_{\bf g}\\
&=\tilde{U}^{\sigma({\bf g})}(\act{g}{a},\act{g}{b};\act{g}{c},{\bf g},{\bf 1})K^{q({\bf g})}\ket{\act{g}{a}, \act{g}{b}; \act{g}{c}}_{\bf 1}\\
&= U_{\bf g}(\act{g}{a},\act{g}{b};\act{g}{c})K^{q({\bf g})}\ket{\act{g}{a}, \act{g}{b}; \act{g}{c}}_{\bf 1}
\end{align}
where the anti-unitarity of $\alpha_{\bf g}^R$ played an important role by complex conjugating $\tilde{U}$.

A key point to notice is that the derivations of the consistency equations for the data with tildes, Eqs.~\eqref{eqn:tildePentagon}-\eqref{eqn:tildeEtaConsistency}, do \textit{not} depend on whether the symmetry group elements have unitary or anti-unitary action. 

We comment at this point that, in terms of symmetry fractionalization data, the maps $\rho_{{\bf g}}$ defined in Sec.~\ref{globsym} and Eq.~\eqref{eqn:rhodef} are either unitary or anti-unitary according to the action of ${\bf g}$. Importantly, in our present formalism, $\rho_{{\bf g}}$ does not describe sweeping a domain wall over a fusion vertex in the graphical calculus defined in Fig.~\ref{fig:tildeGraphicalCalculus}. Instead, the action of a domain wall in the graphical calculus is related to the map $\tilde{\rho}^L_{\bf g}$. 
We take $\tilde{\rho}^L_{{\bf g}}$ to always be unitary. Then $\rho_{\bf g} = \alpha_{\bf g}^R \circ \tilde{\rho}^L_{\bf g}$ and $\alpha^R_{\bf g}$ are either both unitary or anti-unitary.\footnote{It is interesting to consider a theory where $\tilde{\rho}$ is taken to be anti-unitary, which can presumably lead to a consistent graphical calculus and may potentially be of mathematical interest. However, the natural generalization of our Hamiltonian construction in Sec. \ref{HamiltonianSec}-\ref{sec:surfaceTO} to this case is non-local and thus we do not consider this possibility here.}. 

We note that the results of this section highlight the fact that to properly describe anti-unitary symmetry actions, it is helpful to view symmetry fractionalization in terms of the language of 3-categories. Note that any UMTC can be interpreted as a 3-category with trivial $0$ and $1$ morphisms. The objects of the UMTC, which are the anyons, then correspond to $2$-morphisms while the $1$-morphisms of the UMTC are promoted to $3$-morphisms in the 3-category. Symmetry fractionalization can then be described by a 3-category where the objects (0-morphisms) correspond to group elements of $G$ and the 1-morphisms correspond to the domain walls. The 3-category is equipped with $G$ actions, which define the data $\tilde{\rho}$, $\tilde{U}$, $\tilde{\eta}$. Finally, the 3-category has a $G$-equivariance condition, where the possible anti-unitary action of symmetry group elements complex conjugates the data.

\subsection{Defining the state sum on non-orientable manifolds}

For unitary spatial parity-preserving symmetries, SPTs are distinguished by their topological path integrals on non-trivial $G$ bundles. For anti-unitary symmetries such as time-reversal symmetry, or unitary spatial parity-reversing symmetries, it is well-known that one must also consider the path integral on non-orientable manifolds. Equivalently, certain anomaly indicators for (2+1)D SETs involving time-reversal or spatial reflection symmetries are associated with (3+1)D path integrals on non-orientable manifolds.\cite{BarkeshliReflection} For example, for $G = \mathbb{Z}_2^{\bf T}$ time-reversal symmetry, SPTs are distinguished by their path integrals on $\mathbb{RP}^4$ and $\mathbb{CP}^2$. We therefore should ensure that our path integral is well-defined on non-orientable manifolds when $G$ contains group elements with anti-unitary symmetry action. 

Suppose that we wish to calculate the path integral on some non-orientable manifold $M^4$. 
First let $W^3$ be a closed submanifold dual to the first Stieffel-Whitney class $w_1(M^4)$. To perform the calculation, we cut open $M^4$ along some set of 3-simplices that triangulates $W^3$ in order to form an open, orientable $4$-manifold $\check{M}^4$, such that $\partial \check{M}^4 = W^3 \cup \bar{W}^3$. Assign $\check{M}^4$ a global orientation, and for a 4-simplex $\Delta^4$ let $s(\Delta^4)=+1$ (respectively $-1$) if the orientation of $\Delta^4$ given by the ordering of its vertices matches (respectively fails to match) the orientation of $\check{M}^4$. This allows us to compute a path integral on $\check{M}^4$ as a function of the labels on the boundary simplices. 

Next, to construct the path integral on $M^4$, we glue along the boundary 3-simplices with a twist by an anti-unitary symmetry group element as follows. Group elements on opposite sides of the cut are identified up to an action of an anti-unitary symmetry group element. The data assigned to 2- and 3- simplices are however unaffected by the twist. The path integral on $M^4$ is then obtained by summing over the labels on $\partial \check{M}^4$, including the appropriate normalization factors for the boundary data. 

The above procedure may alternatively be thought of as adding a flat background $G$ gauge field to $M^4$. The important point is that the holonomy of this background gauge field must agree with the orientation bundle of $M^4$; that is, the $G$-holonomy around any cycle corresponds to a symmetry group element ${\bf g}$ with anti-unitary action if and only if the local orientation reverses when traveling around that cycle. As we show in Appendix~\ref{app:invariances}, this requirement, together with the anti-unitary action of the isomorphisms $\alpha^R_{{\bf g}}$, ensures that the path integral is invariant under vertex basis and symmetry action gauge transformations. Importantly, gauge invariance cannot be maintained on non-orientable manifolds if $\alpha^R_{\bf g}$ is unitary. 

From this perspective we see why ``gauging $G$'' is not possible for anti-unitary symmetries. Gauging $G$ corresponds to summing over all possible $G$ bundles, which is not possible when $G$ contains anti-unitary symmetries because of the requirement that anti-unitary holonomies must be compatible with the orientation bundle of $M^4$, which is fixed. 

\section{Constructing anomaly indicators}
\label{sec:anomalyIndicators}

We have seen in Sec. \ref{sec:stateSum} that a UMTC $\mathcal{C}$, symmetry group $G$, and symmetry fractionalization data specified by $\{\rho, U, \eta \}$ defines a path integral for a $(3+1)D$ G-SPT. $G$-SPTs are characterized completely by the values of the TQFT path integral on an appropriate set of 4-manifolds and non-trivial $G$ bundles. The specific set of 4-manifolds and $G$ bundles necessary to fully specify the (3+1)D SPT depends on $G$ and correspond to generators of cobordism groups. 

$G$-SPTs in (3+1)D form an Abelian group $\mathcal{S}_G$. A given $G$-SPT is therefore characterized by a set of roots of unity $\{ \mathcal{I}_i \}$, for $i  =1, \cdots, n$, where $n$ is the number of generators of $\mathcal{S}_G$, and where
\begin{align}
\mathcal{I}_i \equiv Z( M^4_i, f_i) . 
\end{align}
Here $M^4_i$ is a 4-manifold and $f_i \in \text{Hom}(\pi_1(M^4), G)$ specifies a $G$ bundle. 

The set $\{\mathcal{I}_i\}$ are referred to as the anomaly indicators associated with the data $\{\mathcal{C}, G, \rho, U, \eta\}$. The reason is as follows. As we will show in Sec.~\ref{sec:surfaceTO}, the (3+1)D $G$ SPT that we have defined hosts a (2+1)D surface termination that respects the $G$ symmetry and that is characterized by the data $\{ \mathcal{C}, G, \rho, U, \eta\}$. Therefore, $\{\mathcal{I}_i\}$ specifies the bulk (3+1)D SPT that hosts the given symmetry fractionalization class at its $(2+1)$D surface. 

Since (3+1)D $G$-SPTs are partially classified by $\mathcal{H}^4(G, U(1))$, we have that $\mathcal{H}^4(G, U(1)) \subset \mathcal{S}_G$. The $\mathcal{H}^4(G, U(1))$ part is therefore determined by a subset of the anomaly indicators $\{ \mathcal{I}_i\}$. In this section, we provide a general algorithm that can determine this subset of anomaly indicators, which thus specifies an element $[\coho{S}] \in \mathcal{H}^4(G, U(1))$. This converts the problem of computing the $\mathcal{H}^4(G, U(1))$ anomalies from the data of $\{\mathcal{C}, G, \rho, U, \eta \}$ into a concrete algorithm that can be run by computer.

To do this, we start with a finite group $G$ and present an algorithm for finding cohomology invariants, that is, functionals of a cocycle whose value depends only on the cohomology class of the cocycle. From a given $\mathcal{H}^4(G, U(1))$ invariant, we construct a cellulation of a 4-manifold with $G$ bundle. Our path integral, evaluated on that 4-manifold with $G$ bundle, is an anomaly indicator for a $G$-SPT. We show that our procedure produces a full set of anomaly indicators; that is, the anomaly indicators are sufficient to fully extract the element of $\mathcal{H}^4(G,U(1))$ that characterizes the anomaly. The approach in this section works both for unitary and anti-unitary symmetry actions.

\subsection{Generating cohomology invariants}
\label{subsec:generateInvariants}

We will first show how to find a complete set of cohomology invariants given a group $G$. We do this by recasting the cocycle condition in a convenient form, after which some simple manipulations yield the desired invariants.

Consider a representative $n$-cocycle $\omega=e^{2\pi i\phi}$ of a class $[\omega] \in \mathcal{H}^n(G,U(1))$. Then $\omega$ is a map from $G^n \rightarrow U(1)$. We can choose to think of $\omega$ instead as an element of $(U(1))^{|G|^n}$; that is, as a vector of $U(1)$ phases associated to each of the $|G|^n$ possible inputs. In a similar fashion, $\phi$ is a map from $G^n \rightarrow \mathbb{R}$ (ignoring for the moment the integer ambiguity) and can therefore be written as a real-valued vector $v^{(n)}$ of length $|G|^n$. For example, in the $n=1$ case,
\begin{equation}
v^{(1)} = (\phi({\bf g}_1), \phi({\bf g}_2), \ldots, \phi({\bf g}_{|G|})
\end{equation}
where we have chosen an arbitrary ordering ${\bf g}_i$ of the elements of $G$.

From now on we fix $n$ and suppress the $n$ superscript on $v$ to make the notation less cumbersome.

The coboundary operator $d$ takes $n$-cochains to $(n+1)$-cochains; in the above language, it maps elements of $(U(1))^{|G|^n}$ to elements of $(U(1))^{|G|^{n+1}}$. Crucially, the coboundary operator can be written as a \textit{linear} map on $v$ because $d\omega$ is a product of $\omega$'s. For example, for $n=1$,
\begin{equation}
d\omega({\bf g}, {\bf h}) = \frac{\omega(\bf{gh})}{\omega({\bf g})\omega({\bf h})} = e^{2\pi i \left(\phi(\bf{gh})-\phi({\bf g})-\phi({\bf h})\right)}
\label{eqn:domega}
\end{equation}
which defines the map $\tilde{d}\phi({\bf g}, {\bf h}) \equiv \phi({\bf gh})-\phi({\bf g})-\phi({\bf h})$. We have used the notation $\tilde{d}\phi$ to indicate that the coboundary operator $d$ has induced an operator $\tilde{d}$ acting on $\phi$. ``Vectorizing" $\tilde{d}\phi$ into a vector $\tilde{d}v$ of length $|G|^{n+1}$, we see that
\begin{equation}
\tilde{d}v = M^G_n v
\end{equation}
where $M^G_n$ is a $|G|^{n+1}\times |G|^n$ integer matrix which depends on $n$ and on $G$. For example, for $G=\mathbb{Z}_2$ and $n=1$, we have $\tilde{d}\phi({\bf 0},{\bf 0}) = \tilde{d}\phi({\bf 0},{\bf 1})= \tilde{d}\phi({\bf 1},{\bf 0}) = -\phi({\bf 0})$ and $\tilde{d}\phi({\bf 1},{\bf 1})=\phi({\bf 0})-2\phi({\bf 1})$. Hence
\begin{equation}
M^{\mathbb{Z}_2}_1 = \begin{pmatrix}
-1 & 0\\
-1 & 0\\
-1 & 0\\
1 & -2
\end{pmatrix}
\label{eqn:exampleM}
\end{equation}

In this language, the cocycle condition $d\omega = 1$ becomes
\begin{equation}
M^G_n v = \Lambda
\label{eqn:cocycleM}
\end{equation}
where $\Lambda$ is any integer vector with $|G|^{n+1}$ components.

Next, we write $M^G_n$ in terms of the Smith normal form:
\begin{equation}
M^G_n = ADB
\end{equation}
where $A,D,B$ are all integer-valued matrices, $D$ is diagonal and $A,B$ are invertible over the integers. Multiplying Eq.~\eqref{eqn:cocycleM} on the left by $A^{-1}$ means
\begin{equation}
\sum_j D_{ii}B_{ij} v_j = \sum_k (A^{-1})_{ik}\Lambda_k
\end{equation}
for each $i$. For each nonzero $D_{ii}$, then,
\begin{equation}
\sum_j B_{ij} v_j = \frac{1}{D_{ii}}\sum_k (A^{-1})_{ik}\Lambda_k
\label{eqn:rational}
\end{equation}
Let us recall where each term comes from. The matrices $A,B,D$ are determined entirely by $G$ and $n$, while $v$ is a reparametrization of the cocycle $\omega$. Provided $\omega$ is indeed a cocycle, then $\Lambda$ exists and is determined by $v$. Note that $v$ is ambiguous up to an integer change of each of its entries, $v \rightarrow v + X$ for any integer vector $X$. Such a change will shift $\Lambda$ by an integer vector: $\Lambda \rightarrow \Lambda + M_n^G  X$. 

Suppose we change the cocycle $\omega$ by a coboundary. Then $v$ changes to $v+M^G_{n-1}e$ for some element $e$ of $\mathbb{R}^{|G|^{n-1}}$. Varying $\omega$ over all representatives of its cohomology class corresponds to continuously varying $e$ over all its values. In this process, the left-hand side of Eq.~\eqref{eqn:rational} varies continuously. But the right-hand side of Eq.~\eqref{eqn:rational} is rational and thus can only vary discretely. The only way that this is possible is if the left-hand-side is completely invariant when $\omega$ changes by a coboundary, that is, if 
\begin{equation}
\mathcal{I}_i = \exp\left(2\pi i \sum_j B_{ij} v_j\right)
\label{eqn:cohomologyInvariants}
\end{equation}
is a cohomology invariant when $D_{ii} \neq 0$, where we have re-exponentiated to avoid the integer ambiguity in $v$. Furthermore, since $\sum_k (A^{-1})_{ik}\Lambda_k$ is an integer, $\mathcal{I}_i$ must be a $(D_{ii})$-th root of unity.

Of course, many of these invariants are trivial. For example, $d\omega$ evaluated on any set of group elements is 1 and therefore an invariant. In particular, when $D_{ii}=1$, the right-hand side of Eq.~\eqref{eqn:rational} is always an \textit{integer}, so multiplying both sides by $2\pi i$ and exponentiating tells us that some product of $\omega$'s is 1, independent of the cohomology class. While this is an invariant, it is not useful for distinguishing different cohomology classes. Therefore, nontrivial invariants only occur when $D_{ii}>1$. Recalling that $B$ has an integer inverse, it is straightforward to check that choosing 
\begin{equation}
v_j = \sum_m k_m \frac{(B^{-1})_{jm}}{D_{mm}}
\label{eqn:explicitCocycle}
\end{equation}
for $k_m \in \mathbb{Z}$ produces an explicit cocycle with anomaly indicator
\begin{align}
  \label{indicatorFormula}
\mathcal{I}_m=\exp(2\pi i k_m/D_{mm}).
\end{align}
 Note that $k_m$ must be an integer in order to satisfy the cocycle condition \eqref{eqn:cocycleM}. Hence all such invariants are nontrivial, and $\mathcal{H}^n(G,U(1))$ contains a subgroup $\mathbb{Z}_{D_{ii}}$ for each $D_{ii}>1$. We prove in Appendix~\ref{app:exhaustiveInvariants} that these invariants provide a full characterization of $\mathcal{H}^n(G,U(1))$.

There is no conceptual change to this procedure in the presence of anti-unitary symmetries; all that happens is that the cocycle condition changes slightly. For example, for $n=1$, the boundary operator Eq.~\eqref{eqn:domega} changes to
\begin{equation}
d\omega({\bf g},{\bf h}) = \frac{\omega(\bf{gh})}{\omega({\bf g})\omega({\bf h})^{\sigma({\bf g})}} = e^{2\pi i\left(\phi(\bf{gh})-\phi({\bf g})-(-1)^{q({\bf g})}\phi({\bf h})\right)}
\end{equation}
which, for $G=\mathbb{Z}_2^{\bf T}$, modifies Eq.~\eqref{eqn:exampleM} to
\begin{equation}
M_1^{\mathbb{Z}_2^{\bf T}}=\begin{pmatrix}
-1 & 0\\
-1 & 0\\
1 & 0\\
1 & 0
\end{pmatrix}
\end{equation}
In this case, $M_1^{\mathbb{Z}_2^{\bf T}}$ has no $D_{ii}>1$, corresponding to the fact that $\mathcal{H}^1(\mathbb{Z}_2^{\bf T},U(1))$ is trivial. 

\subsection{Generating cellulations}
\label{subsec:generateCellulations}

Suppose that we have a cohomology invariant for $\mathcal{H}^d(G, U(1))$: $\mathcal{I} = \omega({\bf g}_1,{\bf g}_2,\ldots,{\bf g}_d)^{\epsilon_g}\omega({\bf h}_1,{\bf h}_2,\ldots,{\bf h}_d)^{\epsilon_h}\cdots$ where the $\epsilon$ are $\pm 1$. We can associate a labeled $d$-simplex to each factor as follows. Consider the factor $\omega({\bf g}_1,{\bf g}_2,\ldots,{\bf g}_d)^{\epsilon_g}$. We consider a $d$-simplex with orientation given by $\epsilon_g$ and order its vertices. Suppose its vertices are, in order, $0123\cdots d$. For each $i$, label the 1-simplex connecting vertices $(i-1)$ and $i$ by ${\bf g}_i$. This labels $d$ of the $d(d+1)/2$ 1-simplices. To label the remaining 1-simplices, we demand that the net flux through each 2-simplex is trivial. This uniquely specifies the labels of the remaining 1-simplices.

We claim that there is always a way to glue together all of the $d$-simplices associated to a given invariant $\mathcal{I}$ such that the 1-simplex group elements and induced orientations match to form a closed $d$-manifold with some nontrivial flat $G$-bundle. 

To see this, suppose that we modify $\omega$ by a coboundary $d\alpha$. Just as we can interpret $\omega$ as a $d$-simplex, we can also interpret $\alpha({\bf k}_1, \ldots, {\bf k}_{d-1})^{\epsilon}$ as corresponding to a $(d-1)$-simplex with orientation $\epsilon$ and a flat $G$-connection, where the group elements ${\bf k}_i$ are placed on $(d-1)$ 1-simplices and determine (via flatness) group elements on the other 1-simplices. Then, by definition, $d\alpha({\bf g}_1,{\bf g}_2,\ldots,{\bf g}_d)$ is a product of one factor of $\alpha$ for each $(d-1)$-simplex of $\omega({\bf g}_1,{\bf g}_2,\ldots,{\bf g}_d)$, where the orientation of the $(d-1)$-simplex corresponding to each factor of $\alpha$ is given by the $(d-1)$-simplex's orientation relative to $\omega$. In order for $\mathcal{I}$ to be a cohomology invariant, every factor of $\alpha$ must cancel, that is, each factor of $\alpha$ must appear in $\mathcal{I}$ the same number of times with positive relative orientation as it appears with negative relative orientation. That is, each $(d-1)$-simplex can be paired with an identical simplex, but with opposite orientation, which is all that is needed to be able to glue the $d$-simplices together to form a closed manifold.

For example, for $G=\mathbb{Z}_2^2$, $d=4$, our algorithm generates the cohomology invariants
\begin{widetext}
\begin{align}
\mathcal{I}_1 &= \frac{\omega({\bf 1},{\bf ZX},{\bf X},{\bf ZX})\omega({\bf 1},{\bf ZX},{\bf Z},{\bf ZX})\omega({\bf ZX},{\bf X},{\bf Z},{\bf ZX})\omega({\bf ZX},{\bf ZX},{\bf X},{\bf ZX})\omega({\bf ZX},{\bf ZX},{\bf Z},{\bf ZX})}{\omega({\bf X},{\bf ZX},{\bf X},{\bf ZX})\omega({\bf X},{\bf ZX},{\bf Z},{\bf ZX})\omega({\bf ZX},{\bf X},{\bf ZX},{\bf Z})\omega({\bf ZX},{\bf Z},{\bf X},{\bf ZX})\omega({\bf ZX},{\bf Z},{\bf ZX},{\bf Z})}
\label{eqn:z2z2Invariant1} \\
\mathcal{I}_2 &= \frac{\omega({\bf 1},{\bf ZX},{\bf X},{\bf ZX})\omega({\bf 1},{\bf ZX},{\bf Z},{\bf ZX})\omega({\bf Z},{\bf Z},{\bf Z},{\bf ZX})\omega({\bf Z},{\bf ZX},{\bf X},{\bf ZX})\omega({\bf Z},{\bf ZX},{\bf Z},{\bf Z})\omega({\bf Z},{\bf ZX},{\bf Z},{\bf ZX})}{\omega({\bf X},{\bf ZX},{\bf X},{\bf ZX})^2\omega({\bf X},{\bf ZX},{\bf Z},{\bf ZX})^2 \omega({\bf Z},{\bf Z},{\bf ZX},{\bf X})\omega({\bf Z},{\bf ZX},{\bf ZX},{\bf ZX})\omega({\bf ZX},{\bf X},{\bf ZX},{\bf Z})} \times \nonumber \\
&\hspace{1cm}\times \frac{\omega({\bf ZX},{\bf X},{\bf X},{\bf ZX})\omega({\bf ZX},{\bf X},{\bf Z},{\bf ZX})^2\omega({\bf ZX},{\bf Z},{\bf ZX},{\bf ZX})\omega({\bf ZX},{\bf ZX},{\bf X},{\bf ZX})\omega({\bf ZX},{\bf ZX},{\bf ZX},{\bf Z})}{\omega({\bf ZX},{\bf Z},{\bf X},{\bf ZX})^2\omega({\bf ZX},{\bf Z},{\bf Z},{\bf Z})\omega({\bf ZX},{\bf Z},{\bf Z},{\bf ZX})\omega({\bf ZX},{\bf Z},{\bf ZX},{\bf Z})} \label{eqn:z2z2Invariant2}
\end{align}
\end{widetext}
 where ${\bf Z}$ and ${\bf X}$ generate $\mathbb{Z}_2^2$ in multiplicative notation and obey ${\bf XZ} = {\bf ZX}$. We focus on $\mathcal{I}_1$ since it has many fewer factors of $\omega$. For example, the factor of $\omega({\bf 1},{\bf ZX},{\bf X},{\bf ZX})$ in Eq.~\eqref{eqn:z2z2Invariant1} is associated to a 4-simplex $01234$ with group elements associated to the links via $01={\bf 1},12={\bf ZX},23={\bf X},34={\bf ZX}$, and the rest of the group elements determined by requiring that the 2-simplices all have trivial flux through them. To determine how to make a closed manifold, from left to right, denote the five $4$-simplices in the numerator of Eq.~\eqref{eqn:z2z2Invariant1}  $1,2,3,4,5$ and the ones in the denominator $6,7,8,9,10$. Then, for example, the 4-simplices $1$ and $2$ can be glued together along two 3-simplices, one with elements ${\bf 1},{\bf ZX},{\bf X}$ on three successive edges (+ and - orientation for 1 and 2 respectively) and one with elements ${\bf 1},{\bf ZX},{\bf Z}$ on three successive edges (- and + orientation for 1 and 2 respectively). 

\begin{figure}
\includegraphics[width=0.4\columnwidth]{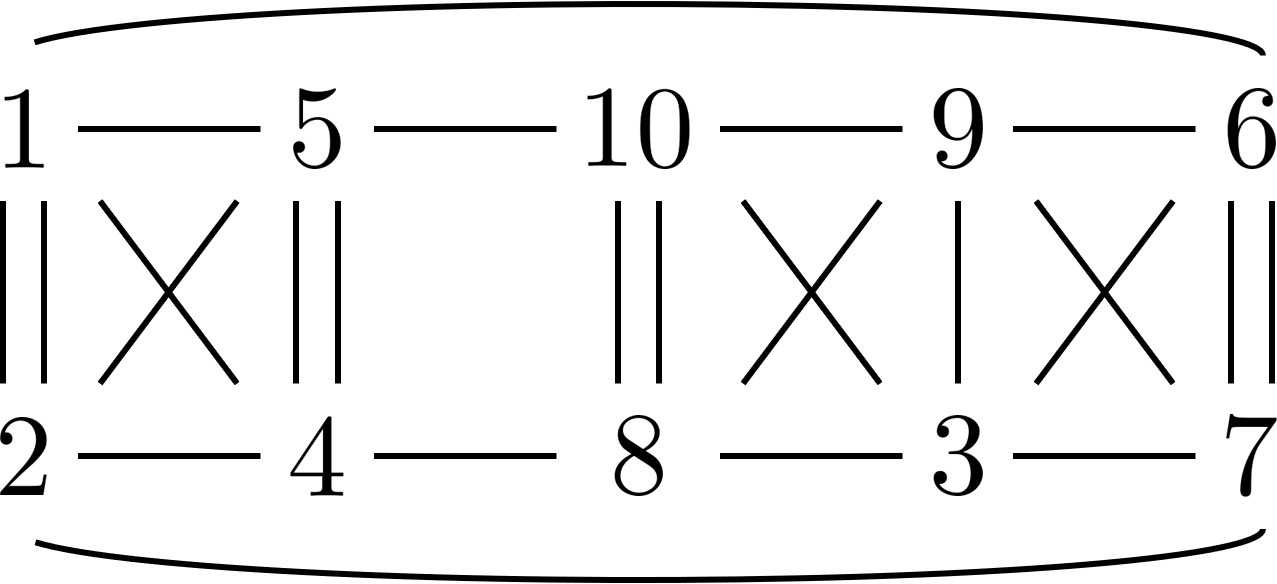}
\caption{Gluing 4-simplices generated by the $\mathbb{Z}_2\times \mathbb{Z}_2$ cohomology invariant $\mathcal{I}_1$, given in Eq.~\eqref{eqn:z2z2Invariant1}, to create a closed 4-manifold. Each number represents 4-simplices corresponding to factors in Eq.~\eqref{eqn:z2z2Invariant1} as described in the main text, and a line between 4-simplices $i$ and $j$ represents a 3-simplex on which $i$ and $j$ have been glued together.}
\label{fig:I1gluing}
\end{figure}

In fact, it can be checked that a closed manifold with a consistent $G$ connection can be obtained by gluing the 4-simplices in the way shown in Fig.~\ref{fig:I1gluing}, where a line between two 4-simplices represents a 3-simplex shared between those 4-simplices.

There are, in general, multiple valid ways to glue together the 4-simplices to produce a closed manifold. Presumably these different gluings correspond to different cellulations of the same manifold with the same holonomies, although we have not studied this in detail. 

This procedure needs a small modification in the presence of anti-unitary symmetries because of the non-trivial group action that appears in the coboundary operator. In particular, let us re-run the argument that the $d$-simplices can be glued together. The key difference is that now group actions can appear: the $d$-simplex $\omega({\bf g}_1, {\bf g}_2, \ldots, {\bf g}_d)$ now picks up a $(d-1)$-simplex $\alpha^{\sigma({\bf g}_1)}({\bf g}_2, \ldots, {\bf g}_d)$. That is, if ${\bf g}_1$ is antiunitary, then we should interpret the 3-simplex $\alpha({\bf g}_2, \ldots, {\bf g}_d)$ as having a \textit{negative} orientation relative to $\omega({\bf g}_1, {\bf g}_2, \ldots, {\bf g}_d)$.

\begin{figure}
\centering
\subfloat[\label{fig:2dZ2TSimplices}]{\includegraphics[width=0.5\columnwidth]{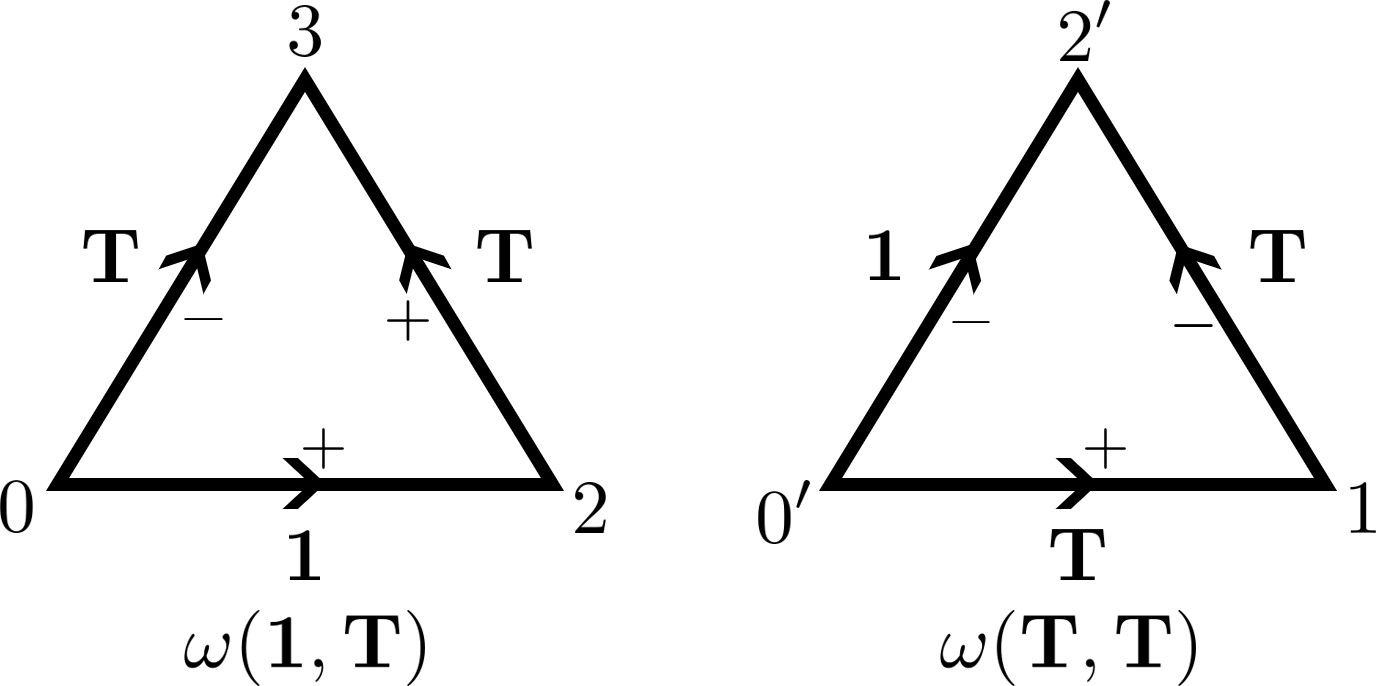}} \hspace{0.1\columnwidth}
\subfloat[\label{fig:2dZ2TSquare}]{\includegraphics[width=0.2\columnwidth]{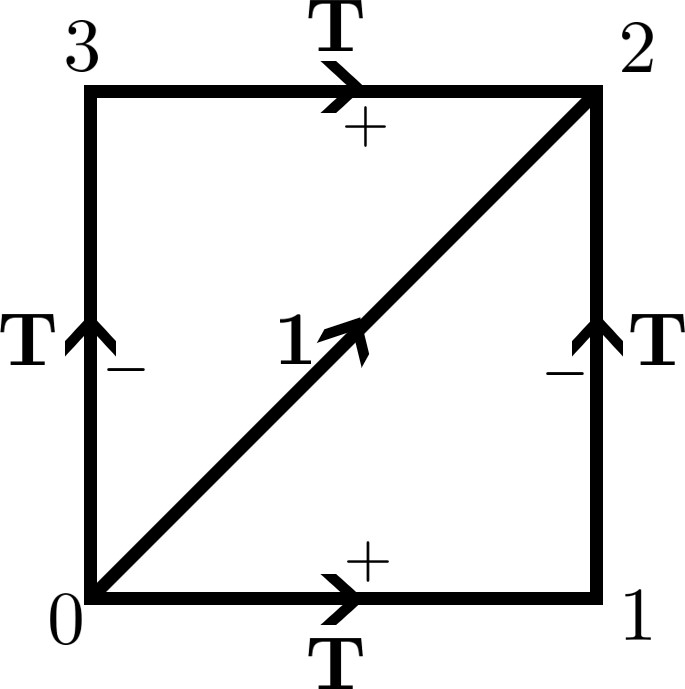}}
\caption{Constructing a cellulation of a manifold from the anomaly indicator Eq.~\eqref{eqn:2dZ2TIndicator} for $\mathbb{Z}_2^{\bf T}$ in $d=2$. (a) Each factor of $\omega$ in the anomaly indicator corresponds to a 2-simplex. The relative orientations of the 1-simplices are shown next to them; the 1-simplex $12'$ is interpreted as having the ``wrong" relative orientation because of the group action arising from ${\bf g}_{0'1}={\bf T}$. (b) The 1-simplices $02$ and $0'2'$ are glued together. There are two valid ways to glue the remaining 1-simplices together, either by gluing $01$ to $03$ or $01$ to $12$. By inspection, both produce $\mathbb{RP}^2$.}
\end{figure}

As an example, take $G=\mathbb{Z}_2^{\bf T}$ with $d=2$. The above procedure produces the cohomology invariant
\begin{equation}
\mathcal{I}^2_{\mathbb{Z}_2^{\bf T}} = \omega({\bf 1}, {\bf T})\omega({\bf T}, {\bf T})
\label{eqn:2dZ2TIndicator}
\end{equation}
The 2-simplices that arise from this invariant are shown in Fig.~\ref{fig:2dZ2TSimplices}, along with the relative orientations on the 1-simplices accounting for the group action of ${\bf g}_1$. Crucially, the 1-simplex $12'$ should be interpreted as having negative relative orientation thanks to the fact that ${\bf g}_{0'1}={\bf T}$ has a group action on $12'$. The 1-simplices $02$ and $0'2'$ with the group element ${\bf 1}$ on them can be glued to produce Fig.~\ref{fig:2dZ2TSquare}, but there are two valid ways to glue the 1-simplices with ${\bf T}$ on them to produce a closed manifold. It is easy to see by inspection that both produce the real projective plane, $\mathbb{RP}^2$. It is well-known that $Z(\mathbb{RP}^2)$ is the anomaly indicator for a $1+1$D $\mathbb{Z}_2^{\bf T}$ SPT.\cite{kapustin2014}

\subsection{Algorithm for determining anomaly indicators}

We can now combine the above results to obtain anomaly indicators, which fully specify an element of $\mathcal{H}^4(G, U(1))$ from the symmetry fractionalization data.  
Given a group $G$, UMTC $\mathcal{C}$, and symmetry fractionalization data $\{\rho, U,\eta\}$ we proceed by the following algorithm:
\begin{enumerate}
\item Compute cohomology invariants $\{\mathcal{I}_i\}$ for $\mathcal{H}^4(G,U(1))$ in the way described in Sec.~\ref{subsec:generateInvariants}.
\item Construct a cellulation for a closed $4$-manifold $M_i$ with a background $G$ connection $A_i$ for each such invariant as described in Sec.~\ref{subsec:generateCellulations}.
\item Input $\mathcal{C}$, $G$, and the given symmetry fractionalization data to our state sum construction to compute the path integral $Z(M_i;A_i)$ on each $M_i$ equipped with its $G$ connection $A_i$ by using the cellulation obtained in the previous step. 
\end{enumerate}
The above algorithm therefore allows explicit computation of a set of anomaly indicators $\{\mathcal{I}_i\}$ given the data $\{\mathcal{C}, G, \rho, U, \eta\}$.  These anomaly indicators may be used in Eq.~\eqref{eqn:explicitCocycle} to compute an explicit representative cocycle in the anomaly's cohomology class.

For the $G = \mathbb{Z}_2^2$ example introduced above, it is already known that one choice of a full set of SPT invariants consists of the SPT path integral on the manifold $L(2,1)\times S^1$ with flux ${\bf ZX}$ through one cycle and flux ${\bf Z}$ through the other cycle\cite{TantivasadakarnSPT,TiwariSPT}. Here $L(p,q)$ is Lens space and $L(2,1) = \mathbb{RP}^3$. Explicit formulas for these path integrals in terms of representative cocycles are known\cite{WangLevinPRB,TantivasadakarnSPT,TiwariSPT}, so given a representative cocycle, we can evaluate the SPT path integrals and compare the results to explicitly evaluating $\mathcal{I}_1$ and $\mathcal{I}_2$ for the same cocycle. We find that
\begin{align}
\mathcal{I}_1 = Z_{\text{SPT}}(L(2,1)\times S^1;{\bf ZX},{\bf Z}) \label{eqn:I1Z}\\
\mathcal{I}_2 = Z_{\text{SPT}}(L(2,1)\times S^1;{\bf Z},{\bf ZX})\label{eqn:I2Z}
\end{align}
where the first and second group elements label the fluxes through the $L(2,1)$ and $S^1$ cycles, respectively. As such, we believe that the cellulations produced by the above algorithm applied to $\mathcal{I}_1$ and $\mathcal{I}_2$ actually yield cellulations of $L(2,1) \times S^1$ with the appropriate holonomies, although we have not verified this directly.

\section{Examples of absolute anomalies}
\label{exampleSec}

In this section we compute a number of examples. Where there is overlap, we find consistency with all previous calculations for anomalies. In particular our computation of absolute anomalies for $G = \mathbb{Z}_2 \times \mathbb{Z}_2$ anyon permuting symmetries for the $U(1)_4$ topological order has not been performed before through any method and is consistent with the relative anomaly calculation of Ref. \onlinecite{BarkeshliRelativeAnomaly}. 

We note that we are limited in the number of examples that we can easily calculate due to the large number of terms in the state sum. In principle the calculations can be parallelized on the computer, and the number of terms in the state sum scales polynomially with the number of anyons in the UMTC. However we expect that the complexity of the cellulations involved increases with $|G|$ such that in general the complexity of the computation would be exponential in $|G|$.

\subsection{$G=\mathbb{Z}_2 \times \mathbb{Z}_2$}

Here we compute symmetry fractionalization anomalies for a series of topological orders: $U(1)_2$ and $U(1)_4$ with $G = \mathbb{Z}_2 \times \mathbb{Z}_2$ symmetry, with both permuting and non-permuting symmetry actions. The case $U(1)_2$ and $U(1)_4$ with no anyon permutations reproduces previous known results, while the case $U(1)_4$ includes anyon-permuting symmetries and is consistent with the relative anomaly calculation of Ref. \onlinecite{BarkeshliRelativeAnomaly}. 

\subsubsection{(3+1)D SPTs}

For $G=\mathbb{Z}_2 \times \mathbb{Z}_2$, we have
\begin{align}
\mathcal{H}^4(G, U(1)) = \mathbb{Z}_2 \times \mathbb{Z}_2 .
  \end{align}
  In principle we could use the computer-generated invariants in Eqs.~\eqref{eqn:z2z2Invariant1} and ~\eqref{eqn:z2z2Invariant2} to construct cellulations. However, they contain a
  large number of 4-simplices. It will be simpler to use the known result that for a $G=\mathbb{Z}_2 \times \mathbb{Z}_2$ SPT, the path integral
  on $M^4 =L(2,1)\times S^1$ provides a set of anomaly indicators for certain choices of background $G$ connections.\cite{TantivasadakarnSPT} Recall that
  here $L(p,q)$ is a Lens space, and $L(2,1) =\mathbb{RP}^3$ is real projective space. This manifold has two nontrivial cycles; one which appears in the Lens space
  and one around the $S^1$. We call the flux threading those cycles ${\bf g}$ and ${\bf h}$ respectively. More precisely, the fundamental group
  $\pi_1(L(2,1)\times S^1)=\mathbb{Z}_2 \times \mathbb{Z}$; the ${\bf g}$ (respectively, ${\bf h}$) flux is the one through the loop generating the factor of $\mathbb{Z}_2$ (respectively, $\mathbb{Z}$). We can then define the anomaly indicators
\begin{equation}
\mathcal{I}_{{\bf g}, {\bf h}} \equiv Z(M^4;{\bf g}, {\bf h})
\end{equation}

Of course, only two such indicators are independent. We take as a generating set $\mathcal{I}_{{\bf X},{\bf Z}}$ and $\mathcal{I}_{{\bf Z},{\bf X}}$, where ${\bf X}$ and ${\bf Z}$ generate $\mathbb{Z}_2 \times \mathbb{Z}_2$. That is, the values of $\mathcal{I}_{{\bf X},{\bf Z}}$, $\mathcal{I}_{{\bf Z},{\bf X}}$ characterize the four distinct possible SPTs in this case. Note that this is a slightly different set of generators than we found algorithmically in Sec.~\ref{sec:anomalyIndicators}; one can check that, as defined in Eqs.~\eqref{eqn:I1Z} and \eqref{eqn:I2Z}, we have
\begin{align}
\mathcal{I}_{{\bf X},{\bf Z}} = \mathcal{I}_1 \mathcal{I}_2\\
\mathcal{I}_{{\bf Z},{\bf X}} =  \mathcal{I}_2
\end{align}

We can make contact with previous parameterizations\cite{TantivasadakarnSPT,TiwariSPT} of the (3+1)D path integrals as follows. We
use the following set of representative cocycles for $\mathcal{H}^4(\mathbb{Z}_2 \times \mathbb{Z}_2,U(1)) = \mathbb{Z}_2 \times \mathbb{Z}_2$:
\begin{equation}
  \omega_q({\bf g},{\bf h},{\bf k},{\bf l})=
  e^{i \frac{\pi}{2} \sum_{I,J=1}^2 q_{IJ} {\bf g}_I {\bf h}_J\left({\bf k}_J + {\bf l}_J -\left[{\bf k}_J + {\bf l}_J \right] \right)}
\label{eqn:z2z2Cocycles}
\end{equation}
where group elements ${\bf g}=({\bf g}_1,{\bf g}_2)$ with $(1,0) = {\bf X}$ and $(0,1) = {\bf Z}$, and where $\left[ {\bf k}_J+ {\bf l}_J \right] = ({\bf k}_J+{\bf l}_J) \mod 2$. Here $q_{IJ}$ is an off-diagonal matrix with entries in $\mathbb{Z}_2$ which labels the cohomology class in question. If, on $M^4$, there is a flux ${\bf g}$ through the non-trivial Lens space cycle and a flux ${\bf h}$ through $S^1$, then the SPT partition function is
\begin{equation}
Z(M^4;{\bf g,h}) = e^{\pi i \sum_{IJ} q_{IJ} {\bf g}_J\left({\bf h}_I {\bf g}_J-{\bf g}_I{\bf h}_J \right)}
\end{equation}
This results in the path integrals:
\begin{align}
  \mathcal{I}_{{\bf X},{\bf Z}} &= (-1)^{q_{21}},
                                  \nonumber \\
  \mathcal{I}_{{\bf Z},{\bf X}} &= (-1)^{q_{12}}.
  \end{align}

We then numerically explicitly evaluate the path integral on $L(2,1)\times S^1$ using a cellulation given in Appendix~\ref{app:cellulations}. The aforementioned cellulation contains eight 4-simplices, which is much simpler than the ten and twenty-four 4-simplices produced by Eqs.~\eqref{eqn:z2z2Invariant1} and ~\eqref{eqn:z2z2Invariant2}, respectively. Even so, our implementation of the numerical computation has a runtime which scales as $N_a^9$ for Abelian $\mathcal{C}$, where $N_a$ is the number of anyons in $\mathcal{C}$, which is computationally expensive. As such, we only examined topological orders with at most four anyon types.

\subsubsection{Gauge-invariant characterization of $\mathbb{Z}_2 \times \mathbb{Z}_2$ symmetry fractionalization}

For a given symmetry action, one can solve the consistency conditions to determine all possible fractionalization patterns up to gauge equivalence. In the tables that follow, we specify symmetry fractionalization patterns by a set of gauge-invariant quantities $\lambda^{\bf g}_c$, $\tilde{\lambda}^{\bf g}_c$ and $\tau_c$ where ${\bf g} \in G$ and $c \in \mathcal{C}$. 

The $\lambda^{\bf g}_c$ and $\tilde{\lambda}^{\bf g}_c$ invariants are defined following Ref.~\onlinecite{BarkeshliRelativeAnomaly}:

First suppose that $c={}^{\bf g}\!c$ for some fixed ${\bf g}$, and $c^n=1$ for some even $n$. Further suppose that there exists a sequence of ${\bf g}$-invariant anyons $c_0=c, c_1,c_2,\ldots, c_{n-2},c_{n-1}=1$ such that $c \times c_k$ can fuse to $c_{k+1}$, that is, $N_{c_{k+1}}^{c,c_k} > 0$. Then we can define
\begin{equation}
\lambda_c^{\bf g} = \eta_c({\bf g},{\bf g})^{n/2}\prod_{k=0}^{n-2}U_{\bf g}(c,c_k;c_{k+1}).
\end{equation}
If instead $\bar{c}={}^{\bf g}\!c$ for some fixed ${\bf g}$, then define
\begin{equation}
\tilde{\lambda}_c^{\bf g} = \eta_c({\bf g}, {\bf g})U_{\bf g}(c,\bar{c};1)R_1^{\bar{c},c}\theta_c.
\end{equation}

Finally, if $c$ is $G$-invariant, that is, ${}^{\bf g}c=c$ for all ${\bf g} \in \mathbb{Z}_2 \times \mathbb{Z}_2$, then the quantity
\begin{equation}
\tau_c \equiv \frac{\eta_c({\bf X}, {\bf Z})}{\eta_c({\bf Z}, {\bf X})} = \frac{\eta_c({\bf X}, {\bf ZX})}{\eta_c({\bf ZX}, {\bf X})} = \frac{\eta_c({\bf Z},{\bf ZX})}{\eta_c({\bf ZX}, {\bf Z})}
\end{equation}
is also gauge-invariant. The equalities follow from the cocycle-like condition on $\eta$. 

\subsubsection{$U(1)_2$ (semion) topological order}

Semion topological order has anyon types $\lbrace 1, s \rbrace$ with one non-trivial $F$-symbol $F^{sss}=-1$ and one non-trivial $R$-symbol $R^{ss}=i$. Any symmetry action is non-permuting, so we can gauge-fix all the $U$s to be 1. In this gauge, $\lambda_s^{\bf g} = \eta_s({\bf g},{\bf g})$, which is invariant under any additional gauge transformations. One can check that these quantities can be independently chosen to be $\pm 1$ for each ${\bf g} \neq {\bf 1}$, and that this is an exhaustive list of symmetry fractionalization classes. Our numerics produce the absolute anomalies shown in Table~\ref{tab:semionAnomalies}.

\begin{table}
\centering
\renewcommand{\arraystretch}{1.3}
\begin{tabular}{@{}llrlr}
\toprule[2pt]
$(\lambda_s^{\bf X},\lambda_s^{\bf Z}\lambda_s^{{\bf ZX}})$&&$\mathcal{I}_{{\bf X},{\bf Z}}$ && $\mathcal{I}_{{\bf Z},{\bf X}}$ \\ \hline
$(1,1,1)$ && 1 && 1\\
$(1,1,-1)$ && -1 && -1\\
$(1,-1,1)$ && -1 && 1\\
$(1,-1,-1)$ && 1 && 1\\
$(-1,1,1)$ && 1 && -1\\
$(-1,1,-1)$ && 1 && 1\\
$(-1,-1,1)$ && 1 && 1\\
$(-1,-1,-1)$ && 1 && 1
\\	\bottomrule[2pt]	
\end{tabular}
\caption{Absolute anomalies $\mathcal{I}_{{\bf X},{\bf Z}}$ and $\mathcal{I}_{{\bf Z},{\bf X}}$ for semion topological order with $\mathbb{Z}_2 \times \mathbb{Z}_2$ symmetries.}
\label{tab:semionAnomalies}
\end{table}

\subsubsection{$U(1)_4$ topological order with permutations}

We label anyons in $U(1)_4$ by elements $[k] \in \mathbb{Z}_4$ with fusion given by addition modulo 4. The $F$-symbols are given by
\begin{equation}
F^{abc}=e^{\frac{i\pi}{4}a(b+c-[b+c])}
\end{equation}
where $[b+c]=(b+c) \mod 4$, and the $R$-symbols are
\begin{equation}
R^{ab}=e^{\frac{i\pi}{4}ab}
\end{equation}

We choose ${\bf X}$ to act by charge conjugation $[k] \rightarrow [4-k]$ and ${\bf Z}$ to not permute the anyons. Then we can gauge fix
\begin{align}
U_{\bf X}([c],[d];[c+d]) &= U_{\bf Z X}([c],[d];[c+d])\nonumber\\
&= \begin{cases}
(-1)^c & [d]>0\\
1 & [d]=0
\end{cases}
\end{align}
and $U_{\bf Z}=1$. After solving the consistency conditions and accounting for gauge redundancy, one finds that $\tilde{\lambda}^{\bf X}_{[1]}$, $\tilde{\lambda}_{[1]}^{\bf ZX}$, and $\lambda^{\bf Z}_{[1]}$ are sufficient to distinguish the eight fractionalization classes. Our numerics lead to Table~\ref{tab:U14PermutingAnomalies}.

\begin{table}
\centering
\renewcommand{\arraystretch}{1.3}
\begin{tabular}{@{}llrlr}
\toprule[2pt]
$(\tilde{\lambda}^{\bf X}_{[1]},\lambda^{\bf Z}_{[1]},\tilde{\lambda}^{{\bf ZX}}_{[1]})$&&$\mathcal{I}_{{\bf X},{\bf Z}}$ && $\mathcal{I}_{{\bf Z},{\bf X}}$ \\ \hline
$(1,1,1)$ && 1 && 1\\
$(1,-1,1)$ && 1 && 1\\
$(1,1,-1)$ && 1 && 1\\
$(1,-1,-1)$ && 1 && -1\\
$(-1,1,1)$ && 1 && 1\\
$(-1,1,-1)$ && 1 && 1\\
$(-1,-1,1)$ && -1 && -1\\
$(-1,-1,-1)$ && -1 && 1
\\	\bottomrule[2pt]	
\end{tabular}
\caption{Absolute anomalies $\mathcal{I}_{{\bf X},{\bf Z}}$, $\mathcal{I}_{{\bf Z},{\bf X}}$ for $U(1)_4$ topological order with anyon-permuting $\mathbb{Z}_2 \times \mathbb{Z}_2$ symmetry, where the element ${\bf X}$ of $\mathbb{Z}_2 \times \mathbb{Z}_2$ acts as charge conjugation and ${\bf Z}$ does not permute anyons.}
\label{tab:U14PermutingAnomalies}
\end{table}

\subsubsection{$U(1)_4$ topological order with no permutations}

Since the symmetry is non-permuting in this case, we can gauge-fix $U=1$. In this gauge $\lambda^{\bf g}_{[1],[3]}=\eta_{[1],[3]}({\bf g},{\bf g})^2$ and $\lambda^{\bf g}_{[2]}=\eta_{[2]}({\bf g},{\bf g})$. One finds that there are eight solutions to the consistency equations. One can check straightforwardly from Eq.~\eqref{eqn:UEtaConsistency} that only $\eta_{[1]}({\bf g},{\bf g})$ is independent, and further that $\lambda^{\bf Z}_{[1]}\lambda^{\bf X}_{[1]}=\lambda^{{\bf ZX}}_{[1]}$. However, $\tau_{[1]}$ is well-defined and independent of the $\lambda^{\bf g}_{[1]}$. Our numerics lead to the absolute anomalies in Table~\ref{tab:U14nonPermutingAnomalies}.

\begin{table}
\centering
\renewcommand{\arraystretch}{1.3}
\begin{tabular}{@{}llrlr}
\toprule[2pt]
$(\lambda^{\bf X}_{[1]},\lambda^{\bf Z}_{[1]},\tau_{[1]})$&&$\mathcal{I}_{{\bf X},{\bf Z}}$ && $\mathcal{I}_{{\bf Z},{\bf X}}$ \\ \hline
$(1,1,1)$ && 1 && 1\\
$(1,1,-1)$ && 1 && 1\\
$(1,-1,1)$ && 1 && 1\\
$(1,-1,-1)$ && 1 && -1\\
$(-1,1,1)$ && 1 && 1\\
$(-1,1,-1)$ && -1 && 1\\
$(-1,-1,1)$ && 1 && 1\\
$(-1,-1,-1)$ && -1 && -1
\\	\bottomrule[2pt]	
\end{tabular}
\caption{Absolute anomalies $\mathcal{I}_{{\bf X},{\bf Z}}$, $\mathcal{I}_{{\bf Z},{\bf X}}$ for $U(1)_4$ topological order with non-permuting $\mathbb{Z}_2 \times \mathbb{Z}_2$ symmetry.}
\label{tab:U14nonPermutingAnomalies}
\end{table}

\subsection{$G=\mathbb{Z}_2^{\bf T}$}

It is known that $\mathcal{H}^4(\mathbb{Z}_2^{\bf T},U(1))=\mathbb{Z}_2$, with the anomaly indicator given by the path integral of the (3+1)D SPT on $\mathbb{RP}^4$. There is also a beyond cohomology SPT, with the anomaly indicator given by the (3+1)D path integral evaluated on $\mathbb{CP}^2$, but this is not expected to be related to the SET anomaly. Hence here we study only $Z(\mathbb{RP}^4)$. 

An explicit formula for the anomaly indicator corresponding to $Z(\mathbb{RP}^4)$ is already known:\cite{BarkeshliReflection}
\begin{equation}
Z(\mathbb{RP}^4) = \frac{1}{\mathcal{D}}\sum_{a|a={}^{\bf T}\!a}\eta_{a}({\bf T},{\bf T})\theta_a d_a
\label{eqn:z2TAnalyticIndicator}
\end{equation}
This indicator is easy to compute and serves as a convenient cross-check for our results.

\begin{figure}
\includegraphics[width=0.13\columnwidth]{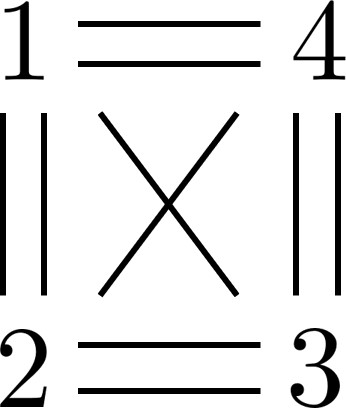}
\caption{Gluing together the 4-simplices generated by the $\mathbb{Z}_2^{{\bf T}}$ cohomology invariant $\mathcal{I}$, given in Eq.~\eqref{eqn:Z2TInvariant}, to create a closed 4-manifold. The 4-simplices correspond to factors in Eq.~\eqref{eqn:Z2TInvariant} as described in the main text, and a line between two 4-simplices $i$ and $j$ represents a 3-simplex on which $i$ and $j$ have been glued together.}
\label{fig:Z2TInvariantGluing}
\end{figure}

The procedure in Sec.~\ref{sec:anomalyIndicators} produces one invariant
\begin{align}
\mathcal{I} = \omega({\bf 1},{\bf T},{\bf 1},{\bf T})&\omega({\bf 1},{\bf T},{\bf T},{\bf T})\omega({\bf T},{\bf T},{\bf 1},{\bf T}) \times \nonumber \\
&\times \omega({\bf T},{\bf T},{\bf T},{\bf T}) \label{eqn:Z2TInvariant}
\end{align}
which can be turned into a cellulation with four 4-simplices. Let the 4-simplices corresponding to the four factors in Eq.~\eqref{eqn:Z2TInvariant} be labeled 1,2,3,4 in the order in which they appear. Then, accounting for the matching of the background connections, these 4-simplices can be glued together in the manner shown in Fig.~\ref{fig:Z2TInvariantGluing}, where each number represents a 4-simplex and each line represents a 3-simplex on which two 4-simplices have been glued together.

One can check that for Abelian $\mathcal{C}$, the number of terms in the sum scales as $N_a^4$, where $N_a$ is the total number of anyons in $\mathcal{C}$.

\subsubsection{Results: $\mathbb{Z}_N$ toric code}

Let the excitations of the $\mathbb{Z}_N$ toric code be labeled by $(a_1,a_2)$ where $a_1,a_2 \in \mathbb{Z}_N$ represent the charge and flux respectively and fusion is given by addition mod $N$. The $F$-symbols can be chosen to all be 1 and the $R$-symbols are
\begin{equation}
R^{ab}=e^{2\pi i a_2b_1/N}
\end{equation}

We take the $\mathbb{Z}_2^{\bf T}$ action on the anyons to be ${}^{\bf T}\!(a_1,a_2) = (a_1,-a_2)$, taken mod $N$; this symmetry action leaves $(m,0)$ and $(n,N/2)$ invariant for any $m,n\in \mathbb{Z}_N$.

Since all the $F$-symbols are 1, we can take all the $U=1$. The only non-trivial $\eta$s are $\eta_x^{\bf T} \equiv \eta_x({\bf T},{\bf T})$, for $\,^{\bf T}x = x$. The consistency conditions require:
\begin{align}
\eta_{{}^{\bf T}\!x}^{\bf T} &= (\eta_x^{\bf T})^{\ast}\\
\eta_x^{\bf T}\eta_y^{\bf T} &= \eta_{x \times y}^{\bf T}
\end{align}
One can check straightforwardly that if ${}^{\bf T}\!x=x$, then $\eta_x^{\bf T}$ is gauge-invariant and real. In our gauge $\eta_{(0,0)}^{\bf T}=1$, which can be used to show that $\eta_{(1,0)}^{\bf T} = \pm 1$ and $\eta_{(0,N/2)}^{\bf T} = \pm 1$, and that up to gauge transformations, these specify all fractionalization classes. Physically, $\eta_x^{\bf T}$ (for a ${\bf T}$-invariant anyon $x$) specifies whether $x$ carries a local Kramers degeneracy, that is, roughly speaking, locally ``${\bf T}^2=\eta_x^{\bf T}$'' for the anyon $x$.

We numerically evaluated the path integral for $N=2,4,6$ and obtained
\begin{equation}
Z(M) = \begin{cases}
-1 & \eta_{(1,0)}^{\bf T}=\eta_{(0,N/2)}^{\bf T}=-1\\
1 & \text{else}
\end{cases}
\end{equation}
This says that of the four distinct $\mathbb{Z}_2^{\bf T}$ symmetry fractionalization patterns, only the state with $(\eta_{(1,0)}^{\bf T},\eta_{(0,N/2)}^{\bf T}) = (-1,-1)$ is anomalous, which matches the results from
Eq.~\eqref{eqn:z2TAnalyticIndicator} and obtained previously.\cite{BarkeshliReflection} For the case of the $\mathbb{Z}_2$ toric code, the anomalous symmetry fractionalization pattern is referred to as eTmT and was found to be anomalous in Ref. \onlinecite{vishwanath2013}.

\section{Hamiltonian formulation}
\label{HamiltonianSec}

Here we present a Hamiltonian formulation of the system described by the path integral state sum of the previous sections. In particular, we provide an exactly solvable Hamiltonian whose exact ground state on a closed 3-manifold corresponds to the wave function obtained by evaluating the path integral of Sec.~\ref{sec:stateSum} on a 4-manifold with boundary. The relation between the Hamiltonian and the path integral state sum here is similar to the approach of previous works: the Levin-Wen model gives an exactly solvable Hamiltonian formulation of the (2+1)D Turaev-Viro state sum, while the Walker-Wang model provides an exactly solvable Hamiltonian for the Crane-Yetter state sum. In particular our formulation is a generalization of the Walker-Wang and Williamson-Wang models\cite{WalkerWang,WilliamsonHamiltonian} to allow $G$ domain walls and anomalous $G$ symmetry actions. 

Our primary motivation for studying the Hamiltonian formulation is to be able to study the model on a 3D space with boundary. This allows us to demonstrate explicitly that there exists a (2+1)D surface termination for the Hamiltonian that preserves the $G$ symmetry, and that the surface hosts a (2+1)D SET described by the same UMTC $\mathcal{C}$ and symmetry fractionalization data $\{\rho, U, \eta\}$ that was used to define the model. In particular, this therefore allows us to demonstrate explicitly that (1) every symmetry fractionalization class for any UMTC can be realized at the surface of a (3+1)D SPT, and (2) to see the data describing symmetry fractionalization at the surface $\{\rho, U, \eta\}$ emerge explicitly via an exactly solvable model. 

Schematically, the ground state of our 3D Hamiltonian will consist of a superposition of all braided fusion diagrams in $\mathcal{C}$ and all possible networks of $G$-domain walls. The relative weights of the different terms in the superposition are obtained by evaluating the corresponding braided fusion diagrams with $G$ actions, which are obtained using the data of the UMTC $\mathcal{C}$ and the symmetry fractionalization data. With open boundary conditions we will see how the $G$-symmetric (2+1)D surface possesses anyonic excitations described by $\mathcal{C}$ and symmetry fractionalization described by $\{\rho, U, \eta\}$. 

We note one technicality in relating the Hamiltonian formulation to the path integral formulation. Our Hamiltonian will be defined on a certain 3D trivalent lattice and a preferred direction in space -- here for concreteness we will take a trivalently-resolved cubic lattice, as described below. On the other hand, the wave function obtained from the path integral construction is defined on a 3D triangulation. We assert that the ground state wave function obtained from our Hamiltonian can be related to the wave function obtained from the path integral construction by a local constant depth circuit, and therefore realizes the same phase of matter. This assertion is on a firm footing given our understanding of the relation between Levin-Wen models and Turaev-Viro path integrals, and between Walker-Wang models and Crane-Yetter path integrals, for which the analogous statements have also not been proven rigorously in general. In this spirit we therefore leave a rigorous proof of this assertion to future work. 

\subsection{Setup}

We start with a trivalently resolved cubic lattice. As usual, we choose a particular projection of this lattice to 2D (similar to how braided fusion diagrams are evaluated according to a choice of projection from 3D to 2D). This trivalent resolution breaks the rotational and reflection symmetries of the cubic lattice. 

\begin{figure*}
\centering
\subfloat[\label{fig:domainWallUnshifted}]{\includegraphics[width=0.9\columnwidth]{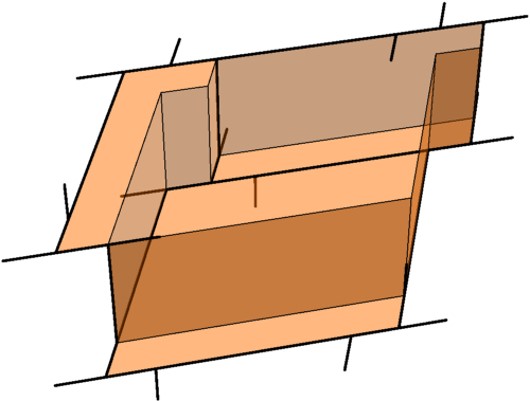}}
 \hspace{0.1\columnwidth}
\subfloat[\label{fig:domainWallShift}]{\includegraphics[width=0.9\columnwidth]{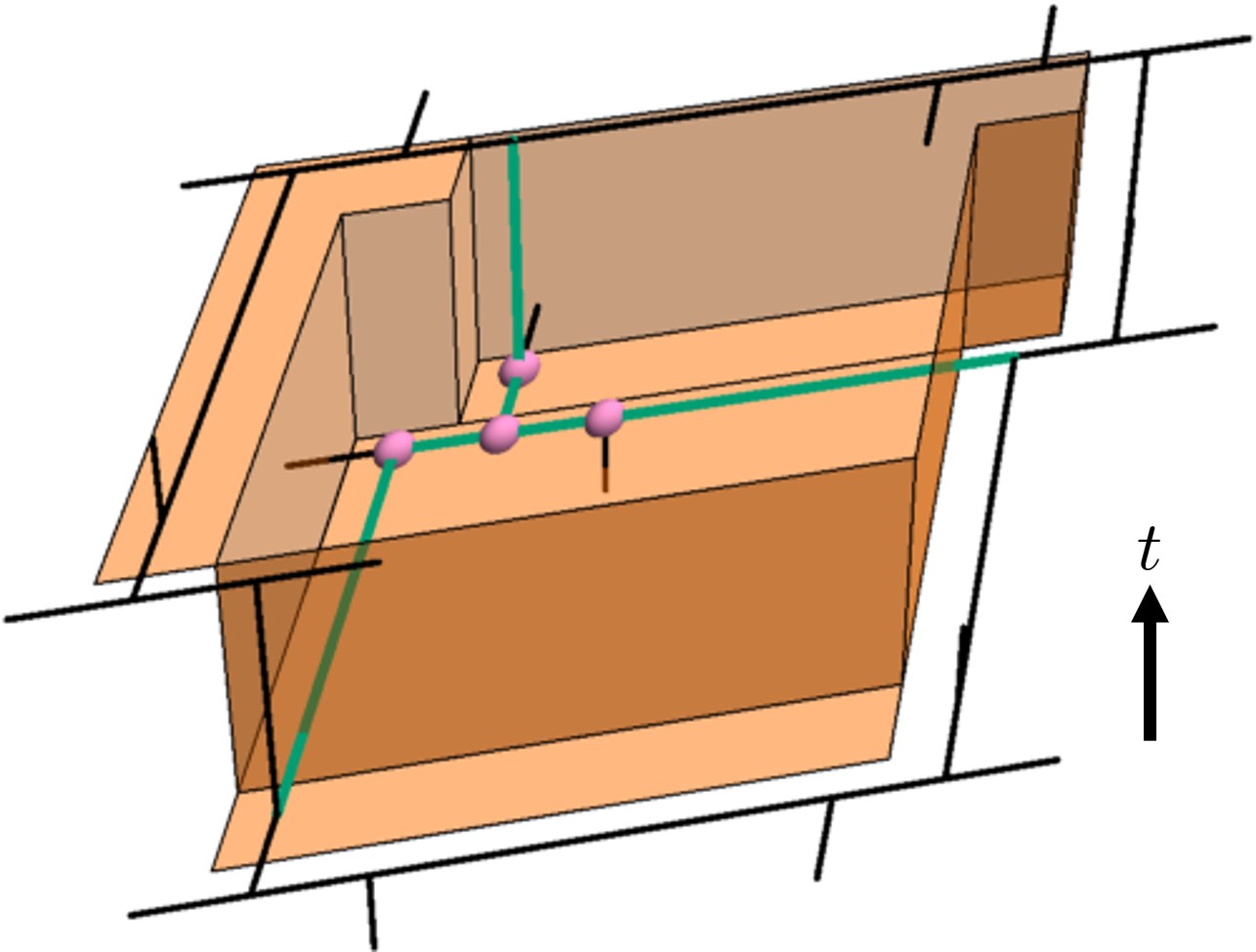}}
\caption{(a) Trivalently resolved cubic lattice. Domain walls (orange) live on the faces, which are deformed from their shape on the cubic lattice. All edges have anyon degrees of freedom on them and are oriented upwards (not explicitly illustrated), but also overlap with domain wall junctions. $G$ degrees of freedom live at the cube centers and are not shown here, so a $G$ domain consists of the deformed cube bounded by the domain walls shown here and a domain wall at the top. (b) Domain walls in (a) are shifted left, back, and down in order to move domain wall junctions away from anyon lines. After this resolution, the green anyon lines and the purple vertices now live inside the domain bounded by the given domain walls, while other lines and vertices live outside said domain.  The lattice lives in three-dimensional space, while the ``arrow of time" indicated in the figure is used exclusively for interpreting states as anyon diagrams in $\mathcal{C}$.
}
\end{figure*}

Next, we choose one direction of the space that will be interpreted as a ``time'' direction in the braided fusion diagrams that will appear in the calculations. Importantly, the ``time'' direction must be picked in such a way that all edges of the lattice have some non-zero projection in the ``time'' direction. The edges are then all oriented along the positive ``time'' direction, as in Fig. ~\ref{fig:domainWallShift}.  These choices are made  to obtain a well-defined graphical calculus for completely general BTCs in order to define the terms in the Hamiltonian. Choosing different ``time'' directions will produce a slightly different Hamiltonian with the same general properties.

We note that the above choice of ``time'' direction is also required to obtain a Hamiltonian realization for general Crane-Yetter models. The Walker-Wang models\cite{WalkerWang} also require these choices unless additional restrictions are made on the type of anyons theories that can be used as input into the construction (although Ref. \onlinecite{WalkerWang} did not explicitly describe these choices). Models that do not actually depend on this choice of ``time'' direction presumably possess some notion of ``hypertetrahedral'' symmetry, which means that the evaluation of braided fusion diagrams obtained from a labeled 4-simplex should be, in an appropriate sense, independent of the branching structure on the 4-simplex. This notion of hypertetrahedral symmetry does not appear to have been formally defined for BTCs. This restriction is analogous to how Levin-Wen models were originally defined\cite{levin2005} to take as input fusion categories with a tetrahedral symmetry, even though the path integral construction can be extended to any spherical fusion category.\cite{barrett1996}

The Hilbert space of the model consists of the following. On each cube center, we place a qudit where a basis of orthonormal states are labeled $\ket{{\bf g}}$, for ${\bf g} \in G$. At the center of each link of the trivalently resolved lattice, we place a qudit with an orthonormal basis labeled by anyons in $\mathcal{C}$.

On the cubic lattice, cubes are naturally $G$ domains with domain walls naturally living on the faces of the lattice; on the trivalently resolved cubic lattice, the domain (cube) and domain walls (faces) are deformed as shown in Fig.~\ref{fig:domainWallUnshifted}. If an anyon line extends from a cube with ${\bf g}_0$ in its center to a cube with ${\bf g}_1$ in its center, the anyon should be acted on by ${\bf g}_1{\bf g}_0^{-1} \equiv {\bf g}_{10}$. Note that 
\begin{equation}
{\bf g}_{ij}{\bf g}_{jk}={\bf g}_{ik}.
\label{eqn:flatness}
\end{equation} 

Furthermore, in the trivalently resolved cubic lattice, some of the domain wall junctions lie along edges of the lattice. Just as in the path integral construction, we will not want anyon lines to be coincident with domain wall junctions. Thus we choose to resolve this ambiguity by shifting the domain walls slightly, as shown in Fig.~\ref{fig:domainWallShift}; different choices of this shift will not change the topological properties, although the model will be slightly different (just as it would change if we changed the projection to 2D). This deformation naturally associates, in a translation-invariant way, six links and four vertices (shown in green and purple, respectively, in Fig.~\ref{fig:domainWallShift}) to a domain (deformed cube).  One can further choose a trivalent resolution of the domain wall junctions, but this will have no effect on the model. 

As a technical comment, since we will be demanding that the anyon degrees of freedom satisfy the fusion rules at the vertices, we would need to add degrees of freedom on the vertices to track elements of the fusion spaces $V_{ab}^c$ if we allowed $N_{ab}^c>1$. Because $\dim V_{ab}^c$ depends in general on $a,b,c$, additional care must be taken to ensure that the Hilbert space remains a tensor product of local Hilbert spaces. While this is possible, such a generalization has not been described in the literature for the Levin-Wen or Walker-Wang models. For the models described in this paper, we also restrict to the cases with $N_{ab}^c \leq 1$ and leave the generalization for future consideration. 

\subsection{Constructing the Hamiltonian}

In order to have the desired ground state, the Hamiltonian has the form
\begin{equation}
H = -\sum_v A_v - \sum_{p} B_p - \sum_{c} D_c
\end{equation}
where $v,p,c$ are vertices, plaquettes, and cubes of the trivalently resolved cubic lattice, respectively. The vertex term $A_v$ enforces the fusion rules at each vertex, paying attention to any domain walls that are crossed. The plaquette term has the form
\begin{equation}
B_p = \sum_{s \in \mathcal{C}} B_p^s
\end{equation}
where $B_p^s$ creates a closed anyon loop on plaquette $p$ and fuses it into the edges of $p$. For comparison, $A_v$ and $B_p$ are analogous to the terms of the Walker-Wang model but are modified in the presence of the domain walls. The cube term $D_c$ has the form
\begin{equation}
D_c = \frac{1}{|G|}\sum_{{\bf g} \in G} D_c^{{\bf g}}
\end{equation}
where $D_c^{{\bf g}}$ creates a closed domain wall of type ${\bf g}$ at the center of cube $c$, then fuses that domain wall into the domain walls on the plaquettes. The factor of $|G|^{-1}$ is for later convenience.

In the rest of this section, we explain how we construct each of these terms. Before doing so, we make some notational comments. Because $B_p^s$ and $D_c^{{\bf g}}$ have very complicated formulas, we will use compact notation where no confusion results: we use the notation $^{ij}a$ to mean $^{{\bf g}_{ij}}a$, and use bars to refer to inverses of matrices (e.g. $\bar{F} = F^{-1}$), inverses of group elements ($\bar{{\bf g}} = {\bf g}^{-1}$), and antiparticles. We will also slightly overload  the variable $v$, which will refer either to a vertex or to an anyon on a specific link. The usage should be clear from context.

\subsubsection{Vertex term}

\begin{figure*}
\centering
\subfloat[\label{fig:plaquetteLabelings}]{\includegraphics[width=0.8\columnwidth]{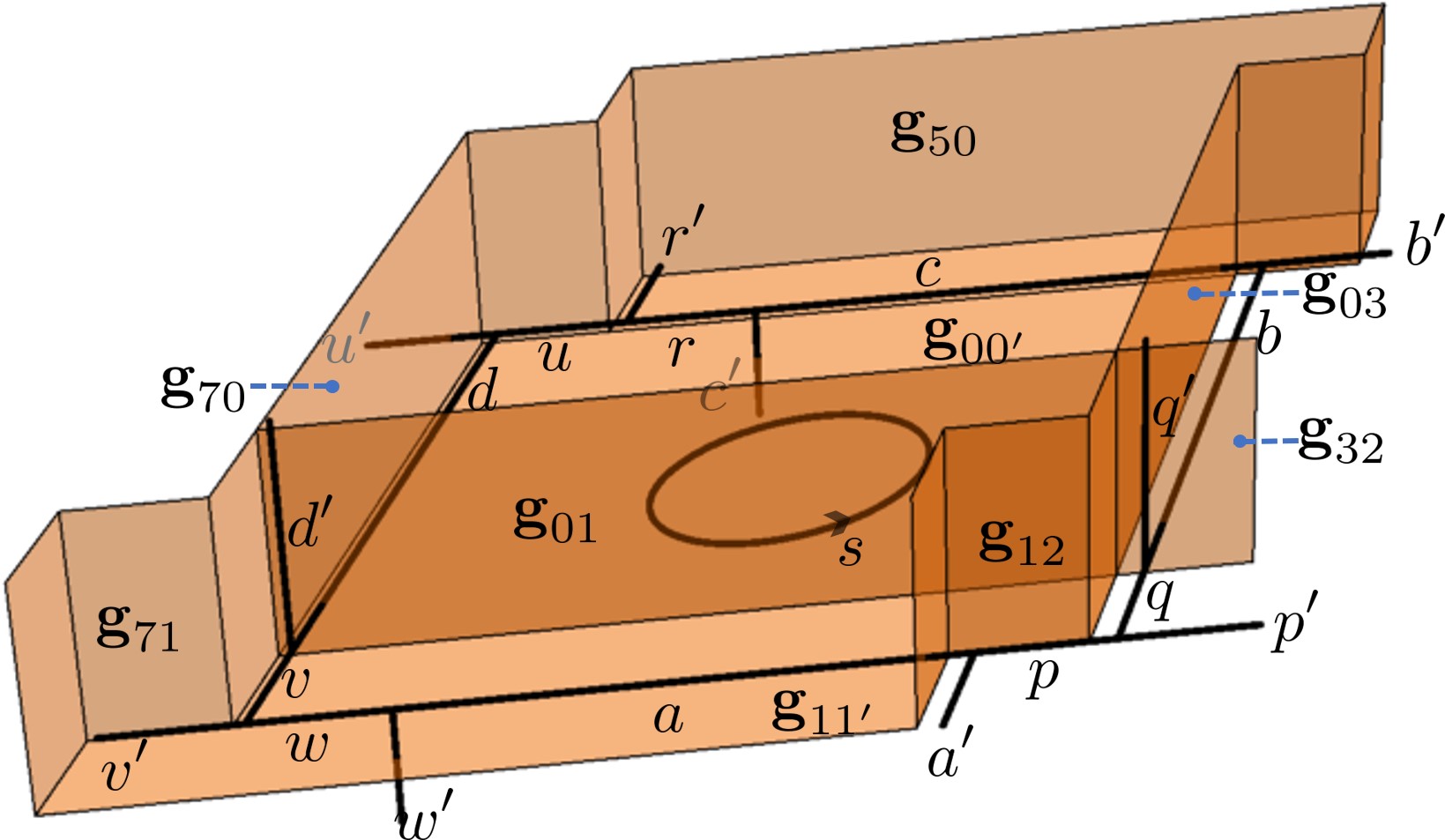}}\hspace{0.1\columnwidth}
\subfloat[\label{fig:domainNumbering}]{\includegraphics[width=0.4\columnwidth]{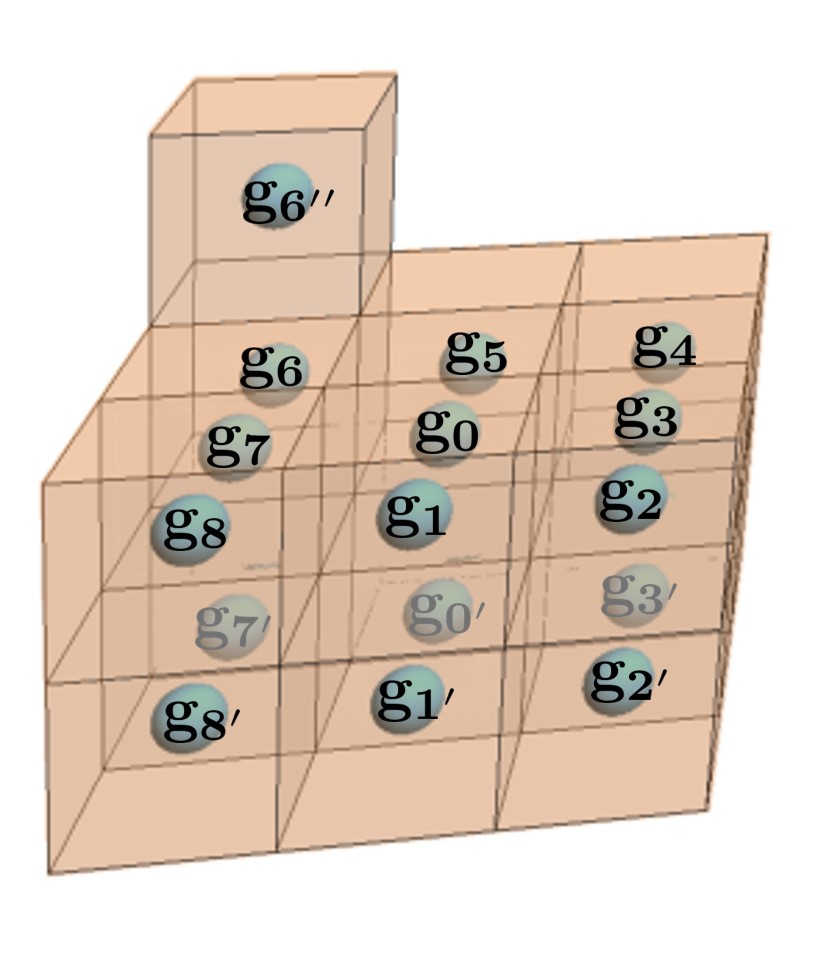}}
\caption{Labeling scheme for the vertex terms and setup for calculating the plaquette term for our Hamiltonian model. (a) Domain walls and anyon lines for the plaquette at the bottom of domain ${\bf g}_0$. Black lines are anyon lines, oriented upwards in the plane of the page and labeled by the nearest letter. Orange sheets are domain walls corresponding to the group element labeling them (dashed blue lines are guides for labeling). The operator $B_p^s$ creates the loop $s$ and fuses it into the anyon lines. (b) Numbering scheme for the domains surrounding ${\bf g}_0$. Primes denote domains below ${\bf g}_0$, double primes denote domains above ${\bf g}_0$. }
\end{figure*}

The term $A_v$ enforcing the fusion rules is straightforward. It acts as the identity on the $G$ degrees of freedom and can be written
\begin{equation}
A_v = \delta_{{}^{07}\!u',d;u} \ket{u',d,u,\lbrace {\bf g}_i\rbrace}\bra{u',d,u,\lbrace {\bf g}_i \rbrace}
\end{equation}
for, e.g., the vertex where $u',u,d$ meet in Fig.~\ref{fig:plaquetteLabelings}. The domains are numbered using the scheme in Fig.~\ref{fig:domainNumbering}. The delta function here means
\begin{equation}
\delta_{{}^{07}\!u',d;u} = \begin{cases} 1 & N^{u}_{{}^{07}\!u',d} >0\\ 0 &\text{else}
\end{cases}
\end{equation}
where $N_{ab}^c$ is a fusion coefficient. The symmetry action on $u'$ comes from the fact that $u'$ lives on a link associated to the ${\bf g}_7$ domain, while the fusion vertex is associated to the ${\bf g}_0$ domain. Hence the anyon line passes through the domain wall ${\bf g}_{70}$ and is thus acted on by the symmetry. The terms for the other vertices are straightforward generalizations.

\subsubsection{Plaquette term}

The matrix elements of the plaquette term are obtained by evaluating the amplitude obtained in the braided fusion diagram associated with starting with a closed loop labeled by an anyon $s$ at the center of the plaquette $p$ and fusing it into the edges. This must be done carefully to ensure that we capture all of the group actions correctly.

The setup is as in Fig.~\ref{fig:plaquetteLabelings}. The procedure is similar to that for the Walker-Wang models\cite{WalkerWang}, where we take that diagram and fuse the anyon line $s$ into the lattice anyon lines. This is done by first doing an $R$ move on the $q-q'-{}^{23}\!b$ and $r-{}^{00'}\!c'-c$ vertices, then fusing $s$ into $a$, then working our way around counterclockwise, then undoing the $R$-moves, all while tracking the domain walls. In the end, all unprimed lines (e.g. $a$) will be replaced with their double-primed counterparts (e.g. $a''$).

The final result is a product of four $R$s, twelve $F$s, two $\eta$s, five $U$s, two bending moves, and some factors of quantum dimension:
\begin{widetext}
  \begin{equation}
    \label{Bps}
B_p^s = \sum_{\text{anyons}} \mathcal{R}\mathcal{F}\mathcal{E}\mathcal{U}\mathcal{D} \ket{a'',p'',q'',b'',c'',r'',u'',d'',v'',w''}\bra{a,p,q,b,c,r,u,d,v,w}
\end{equation}
where $B_p^s$ acts diagonally on all other degrees of freedom and the various factors are
\begin{align}
\mathcal{R} &= R^{{}^{23}\!bq'}_q \bar{R}^{{}^{23}\!b''q'}_{q''}\bar{R}^{{}^{00'}\!c'r}_cR^{{}^{00'}\!cr''}_{c''} \label{eqn:BpR}\\
\mathcal{E} &= \eta_{s}^{-1}({\bf g}_{01}, {\bf g}_{12})\eta_{s}({\bf g}_{03}, {\bf g}_{32})\\
\mathcal{U} &= U^{-1}_{01}(\bar{s},s;1)
U^{-1}_{12}({}^{10}\!s,a;a'')
U_{32}({}^{30}\!s,b;b'')
U_{03}(\bar{s},s;1)
U_{03}(c,\bar{s},c'')
U_{01}(d,\bar{s};d'')\\
\mathcal{D} &= \sqrt{\frac{d_{p''}d_rd_{u''}d_dd_{v''}d_wd_{a''}}{d_pd_{r''}d_ud_{d''}d_vd_{w''}d_a}} A^{{}^{30}\!s b}_{b''}\bar{B}^{v''{}^{10}\!s}_v \label{eqn:BpD}\\
\mathcal{F} &= F^{{}^{20}\!s{}^{21}\!aa'}_{p''{}^{21}\!a''p} 
\left(F^{{}^{20}\!sqp'}_{p''}\right)^{-1}_{pq''} 
\left(F^{{}^{20}\!s{}^{23}\!bq'}_{q''}\right)^{-1}_{q{}^{23}\!b''}
F^{{}^{30}\!c{}^{30}\!\bar{s}b''}_{b'{}^{30}\!c''b}
\left(F^{{}^{00'}\!c'r\bar{s}}_{c''}\right)^{-1}_{r''c}
F^{r'r\bar{s}}_{u''ur''}
\left(F^{{}^{07}\!u'd\bar{s}}_{u''}\right)^{-1}_{d''u}
F^{d'{}^{10}\!d{}^{10}\!\bar{s}}_{v''v{}^{10}\!d''}
F^{v''{}^{10}\!sw}_{v'vw''}
F^{{}^{10}\!sw{}^{11'}\!w'}_{a''w''a} \label{eqn:BpF}
\end{align}
\end{widetext}

The sum over anyons is over all of the anyons that appear in the bra and ket in Eq. \ref{Bps}. Note that $A_{a}^{bc}$ and $B_{a}^{bc}$ are the bending moves defined in Eqs.~\eqref{eqn:Abend} and \eqref{eqn:Bbend} (not to be confused with the vertex and plaquette terms $A_v$ and $B_p$). One can check at this stage that when $G$ is trivial, our model reduces precisely to the Walker-Wang model (in a slightly more general form than the usual presentation).

We have assumed in the above that the symmetry actions are unitary. To include both unitary and anti-unitary symmetries, the above calculation can be repeated using the graphical calculus involving the data $\lbrace \tilde{F},\tilde{R},\tilde{U},\tilde{\eta}\rbrace$ defined in Sec. \ref{sec:antiunitary}. In this case the group elements (not just the domain walls) need to be carefully tracked. For example, in the graphical calculus derivation, the $F$-move producing the term $F^{r'r\bar{s}}_{u''ur''} $ in $\mathcal{F}$ occurs in the ${\bf g}_0$ domain. More generally this term should thus be replaced by $\tilde{F}^{r'r\bar{s}}_{u''ur''}({\bf g}_0)$. The relation between the data with tildes and the data without tildes, Eqs. \ref{eqn:FNoTilde}-\ref{eqn:UNoTilde}, determines whether the model will be symmetric under unitary or anti-unitary symmetry actions. 

\subsubsection{Cube term}

\begin{figure*}
\centering
\includegraphics[width=0.8\columnwidth]{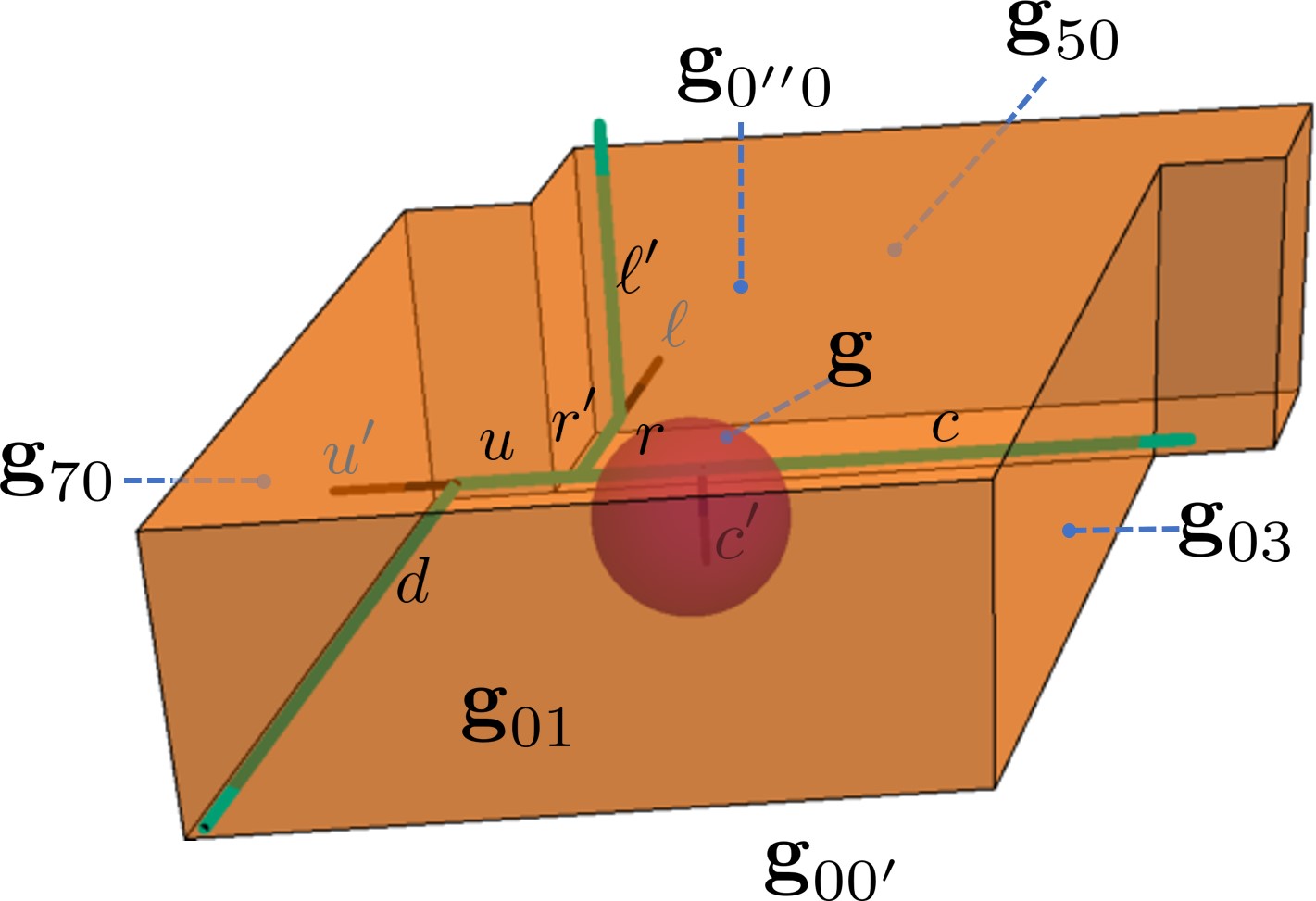}
\caption{Setup for determining the cube term $D_c^{{\bf g}}$ by graphical calculus. A domain wall of type ${\bf g}$ (dark purple) is nucleated at the center of the ${\bf g}_0$ domain, expanded, and fused into the preexisting domain walls (orange). Green anyon lines are transformed by ${\bf g}$, while all anyons shown participate in factors of $U$ or $\eta$ resulting from this process.}
\label{fig:cubeTerm}
\end{figure*}

The cube term $D_c^{{\bf g}}$ fluctuates the domain wall configuration by changing ${\bf g}_i \rightarrow {\bf g}{\bf g}_i$ for ${\bf g} \in G$. Schematically, this creates a domain wall of type ${\bf g}$ at cube $c$ and fuses it into the domain wall configuration, as in Fig.~\ref{fig:cubeTerm}. Using the notation of Fig.~\ref{fig:cubeTerm}, we obtain
\begin{align}
D_c^{{\bf g}} = &\sum_{{\bf g}_0, \text{anyons}} e^{i\phi_{{\bf g}}(\text{anyons},\lbrace {\bf g}_{ij}\rbrace)}\prod_{v \in c} \mathcal{P}_v\ket{{\bf g}{\bf g}_0}\bra{{\bf g}_0} \otimes \nonumber\\
&\hspace{0.05\columnwidth}\ket{{}^{{\bf g}}\!c,{}^{{\bf g}}\!r,{}^{{\bf g}}\!r',{}^{{\bf g}}\!\ell',{}^{{\bf g}}\!u,{}^{{\bf g}}\!d}\bra{c,r,r',\ell',u,d}
\end{align}
where $e^{i\phi_{{\bf g}}}$ is a phase factor which depends on nearby anyons and group elements in a way which we will determine shortly. Here $\mathcal{P}_v$ projects onto the $A_v=+1$ subspace of vertex $v$, ensuring that the fusion rules are satisfied.  Although $\phi$ depends on more anyons and group elements, $D_c^{{\bf g}}$ acts diagonally on all anyons apart from $c,r,r',\ell',u,d$ and group elements apart from ${\bf g}_0$.

The result is obtained similarly to the plaquette term using diagrammatics. Since there are no anyon lines in the way, the ${\bf g}$ domain wall may be immediately fused into the front right corner of Fig.~\ref{fig:cubeTerm}, where the ${\bf g}_{01}$, ${\bf g}_{03}$, and ${\bf g}_{00'}$ branch sheets meet. We then slide the domain wall around, crossing it over various anyon lines in the process.

The final result is
\begin{widetext}
\begin{align}
e^{i \phi_{{\bf g}}} = \eta^{-1}_c(\bar{{\bf g}},{\bf g}{\bf g}_{03})&U_{\bar{{\bf g}}}(r,{}^{00'}\!c';c)\eta_{{}^{00'}\!c'}({\bf \bar{g}},{\bf gg}_{00'})\eta_{\ell}^{-1}({\bf g}_{50},\bar{{\bf g}})U^{-1}_{\bar{{\bf g}}}(\ell',{}^{05}\!\ell;r')\eta_{{}^{0''0}\!\ell'}^{-1}({\bf g}_{0''0},\bar{{\bf g}})\times \nonumber \\
&\times U^{-1}_{\bar{{\bf g}}}(r',r;u)U_{\bar{{\bf g}}}({}^{07}\!u',d;u)\eta_{u'}({\bf g}_{70},\bar{{\bf g}})\eta_d(\bar{{\bf g}},{\bf g}{\bf g}_{01})
 \label{eqn:cubePhase}
\end{align}
\end{widetext}

One can check by a tedious computation that the $D_c^{{\bf g}}$ obey $G$ multiplication rules, that is,
\begin{equation}
D_c^{{\bf g}} D_c^{{\bf h}} = D_c^{{\bf gh}}
\end{equation}

Eq.~\eqref{eqn:cubePhase} assumes that the symmetry actions are unitary. Just as for the plaquette operator, anti-unitary symmetry actions may be taken into account using the graphical calculus of $\lbrace \tilde{U},\tilde{\eta}\rbrace$ defined in Sec. \ref{sec:antiunitary}, tracking the group elements carefully. It follows that, physically speaking, $D_c^{{\bf g}}$ implements the map $\tilde{\rho}_{{\bf g}}$ defined in Sec.~\ref{sec:antiunitary} on all of the fusion and splitting spaces contained inside the domain $c$.

Recall that in Sec. \ref{sec:antiunitary} we defined $\tilde{\rho}_{\bf g}^L$ to always be a unitary map. If we instead considered a theory with an anti-unitary $\tilde{\rho}_{{\bf g}}^L$, the operator $D_c^{{\bf g}}$ would need to be anti-unitary, leading to a non-local Hamiltonian. Therefore, to allow for a local Hamiltonian, we require $\tilde{\rho}_{{\bf g}}^L$ to be unitary. We account for the possibility of anti-unitary symmetry actions by allowing the map $\alpha_{{\bf g}}^R$ defined in Sec. \ref{sec:antiunitary}  to be anti-unitary.

\subsection{Bulk properties}

The graphical interpretation of the Hamiltonian terms make it clear that the (unnormalized) ground state wavefunction is
\begin{equation}
\ket{\Psi_0} = \sum_{\lbrace {\bf g}_i \rbrace} \sum_{D} V(D)\ket{D,\lbrace {\bf g} \rbrace}
\end{equation}
where $\lbrace {\bf g}_i \rbrace$ is a group element configuration on the cube centers, $D$ is a braided fusion diagram in the presence of the domain walls for the configuration $\lbrace {\bf g_i} \rbrace$, and $V(D)$ is the value of the diagram $D$ evaluated by the usual graphical calculus.

Since this ground state corresponds to the state sum constructed in Sec.~\ref{sec:stateSum}, the model is gapped and there is a unique ground state on a closed 3-manifold when $\mathcal{C}$ is modular. We do not know a general proof of this fact directly from the Hamiltonian formalism.

The model also has a global $G$ symmetry
\begin{equation}
R_{{\bf g}} = \prod_{c\in \text{cubes}} R^{(c)}_{\bf g}
\label{eqn:globalG}
\end{equation}
where $R^{(c)}_{\bf g}$ is the right multiplication operator on cube $c$ 
\begin{equation}
R^{(c)}_{\bf g} = \sum_{{\bf g}_c}\ket{{\bf g}_c\bar{{\bf g}}}\bra{{\bf g}_c}.
\end{equation}
It is immediately obvious that $R_{{\bf g}}R_{{\bf h}}=R_{{\bf gh}}$. 

The symmetry action $R_{\bf g}$ commutes with $A_v$, $B_p^s$, and $D_c^{\bf g}$. This follows immediately from the fact that $A_v$, $B_p^s$, and $D_c^{\bf g}$ only depend on the domain walls, ${\bf g}_{ij}$ and not on the ${\bf g}_i$ individually. Note that the symmetry acts by right multiplication because the domain walls act on anyons by group elements of the form ${\bf g}_{ij} = {\bf g}_i {\bf g}_j^{-1}$. This action could be changed to make the symmetry act by left multiplication, but both the domain wall action on anyons and the interpretation of the $D_c^{\bf g}$ operator as fusing in a domain wall would be altered to be considerably less natural.

It is straightforward to check that when $R_{\bf g}$ has an anti-unitary action, then the model is also symmetric when the modifications described in the preceding section for anti-unitary symmetry actions are made.

\section{(2+1)D surface topological order}
\label{sec:surfaceTO}

In this section, we explain how, with appropriate boundary conditions, the (2+1)D surface of our model realizes a symmetry-enriched topological phase where the topological order is described by the UMTC $\mathcal{C}$ and the symmetry fractionalization by the data $\{\rho, U, \eta\}$ that was used to define the model.

\subsection{$G$-symmetric boundary Hamiltonian}

\begin{figure*}
\subfloat[\label{fig:surfaceTermination}]{\includegraphics[width=0.8\columnwidth]{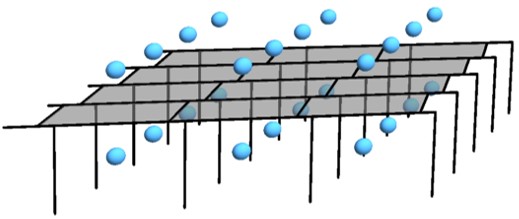}}
\hspace{0.1\columnwidth}
\subfloat[\label{fig:surfaceHilbertSpace}]{\includegraphics[width=0.8\columnwidth]{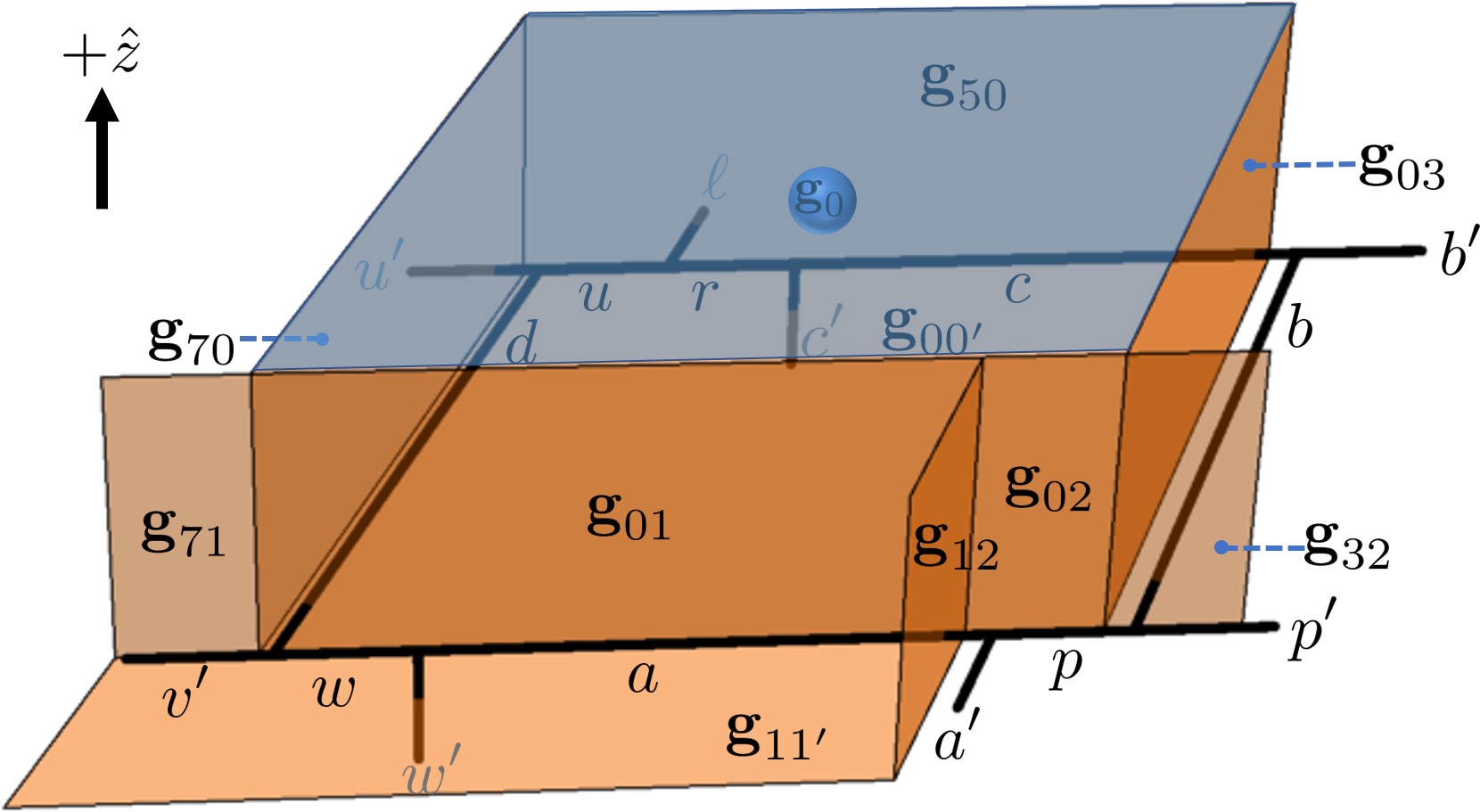}}
\caption{Surface termination for a system occupying $z<0$. (a) Degrees of freedom near the surface (grey); each black link carries an anyon degree of freedom, and each blue sphere is a $G$ degree of freedom. Note that the surface is smooth for the anyon degrees of freedom but in some sense rough for the $G$ degrees of freedom. (b) Domain wall configurations and labelings of degrees of freedom near a single surface plaquette. The ${\bf g}_{ij}={\bf g}_i{\bf g}_j^{-1}$ label domain walls (orange) and are determined by the $G$ degrees of freedom, with the same numbering scheme for nearby cubes as in the bulk (see Fig.~\ref{fig:domainNumbering}, though the double-primed domains do not exist at the surface). Black lines are anyon lines and are oriented upward. The blue sheet is a fictitious domain wall between ${\bf g}_0$ and the vacuum; it plays no role in the Hamiltonian but is useful for defining string operators.}
\end{figure*}

For concreteness we choose a boundary normal to the arrow of time direction (which we call $\hat{z}$ in this section) in Fig.~\ref{fig:domainWallShift} (strictly speaking the ``time'' direction is tilted slightly from the normal of the boundary to ensure that all links can be oriented in the positive ``time'' direction with no ambiguity). For the links in the trivalently resolved cubic lattice, choose smooth boundary conditions, where links which protrude above the surface are not included. In addition, we include the $G$ degrees of freedom in the plane above the surface as well. This surface termination is shown in Fig.~\ref{fig:surfaceTermination}, and the domain walls on the surface are shown in Fig.~\ref{fig:surfaceHilbertSpace}.

The reason that we include the $G$ degrees of freedom in the plane above the surface is so that the model possesses domain walls lying along the $\hat{z}$ direction at the surface. The anyon lines at the surface (which are orthogonal to the $\hat{z}$ direction) will then pierce these domain walls and then be acted on by the corresponding symmetry group elements. Note that, as discussed in the previous section, we have defined the model by deforming the domain walls downward so that the domain wall junctions are not coincident with the links of the lattice. If we had instead chosen to deform the domain walls upward, then we would not need the dangling $G$ degrees of freedom at the surface, as domain walls that affect the anyon lines at the surface would already exist. 

The surface Hamiltonian terms are obtained straightforwardly from the bulk terms by accounting for the erased links. Using the labelings of Fig.~\ref{fig:surfaceHilbertSpace}, this amounts to removing any terms involving $d'$ and $q'$ from $B_p$, setting $v={}^{10}\!d$, $q={}^{32}b$, and $r'={}^{05}\ell$ (since those links are identified on the surface), and removing terms involving $\ell'$ from $D_c$. Explicitly, on the surface, we modify Eqs.~\eqref{eqn:BpR},\eqref{eqn:BpD},\eqref{eqn:BpF},\eqref{eqn:cubePhase} to
\begin{widetext}
\begin{align}
\mathcal{R}_{\text{surface}} &= \bar{R}^{{}^{00'}\!c'r}_cR^{{}^{00'}\!cr''}_{c''}\\
\mathcal{D}_{\text{surface}} &= \sqrt{\frac{d_{p''}d_rd_{u''}d_wd_{a''}}{d_pd_{r''}d_ud_{w''}d_a}} A^{{}^{30}\!s b}_{b''}\bar{B}^{{}^{01}\!d''{}^{10}\!s}_{{}^{01}\!d}\\
\mathcal{F}_{\text{surface}} &= F^{{}^{20}\!s{}^{21}\!aa'}_{p''{}^{21}\!a''p} 
\left(F^{{}^{20}\!s{}^{32}\!bp'}_{p''}\right)^{-1}_{p{}^{32}\!b''} 
F^{{}^{30}\!c{}^{30}\!\bar{s}b''}_{b'{}^{30}\!c''b}
\left(F^{{}^{00'}\!c'r\bar{s}}_{c''}\right)^{-1}_{r''c}
F^{{}^{05}\!\ell r \bar{s}}_{u''ur''}
\left(F^{{}^{07}\!u'd\bar{s}}_{u''}\right)^{-1}_{d''u}
F^{{}^{10}\!d''''{}^{10}\!sw}_{v'{}^{10}\!dw''}
F^{{}^{10}\!sw{}^{11'}\!w'}_{a''w''a}\\
\left(e^{i \phi_{\bf g}}\right)_{\text{surface}} &= \eta^{-1}_c(\bar{{\bf g}},{\bf gg}_{03})U_{\bar{{\bf g}}}(r,{}^{00'}\!c';c)\eta_{\ell}^{-1}({\bf g}_{50},\bar{{\bf g}})U^{-1}_{\bar{{\bf g}}}({}^{05}\!\ell,r;u)U_{\bar{{\bf g}}}({}^{07}\!u',d;u)\eta_{u'}({\bf g}_{70},\bar{{\bf g}})\eta_d(\bar{{\bf g}},{\bf gg}_{01})
\end{align}
\end{widetext}
with $\mathcal{E}$ and $\mathcal{U}$ unmodified. The labelings are given in Fig.~\ref{fig:surfaceHilbertSpace}.

It is not hard to check that the global $G$ symmetry with generators $R_{{\bf g}}$ given in Eq.~\eqref{eqn:globalG} is still present with this choice of boundary conditions.

The ground state wave function is still a superposition of domain walls and closed braided fusion diagrams, as in the bulk. Although it plays no role in the Hamiltonian or the ground state wavefunction, it will be useful to think of the domains as ``closed" on top as if the vacuum is assigned to an identity domain; this is indicated by the blue sheet in Fig.~\ref{fig:surfaceHilbertSpace}.

\subsection{Surface topological order}

Here we will see how the surface possesses deconfined anyon excitations, whose algebraic properties are described by the same UMTC $\mathcal{C}$ that was used to define the model. 

In the case that $G$ is trivial, the model reduces to the Walker-Wang model and the boundary conditions we consider are the standard smooth boundary conditions. It has been found\cite{vonKeyserlingkSurfaceAnyons} in that case that such a boundary has surface topological order described by the UMTC $\mathcal{C}$ that was used to define the theory.

\subsubsection{Definition of string operators}

\begin{figure*}
\includegraphics[width=1.2\columnwidth]{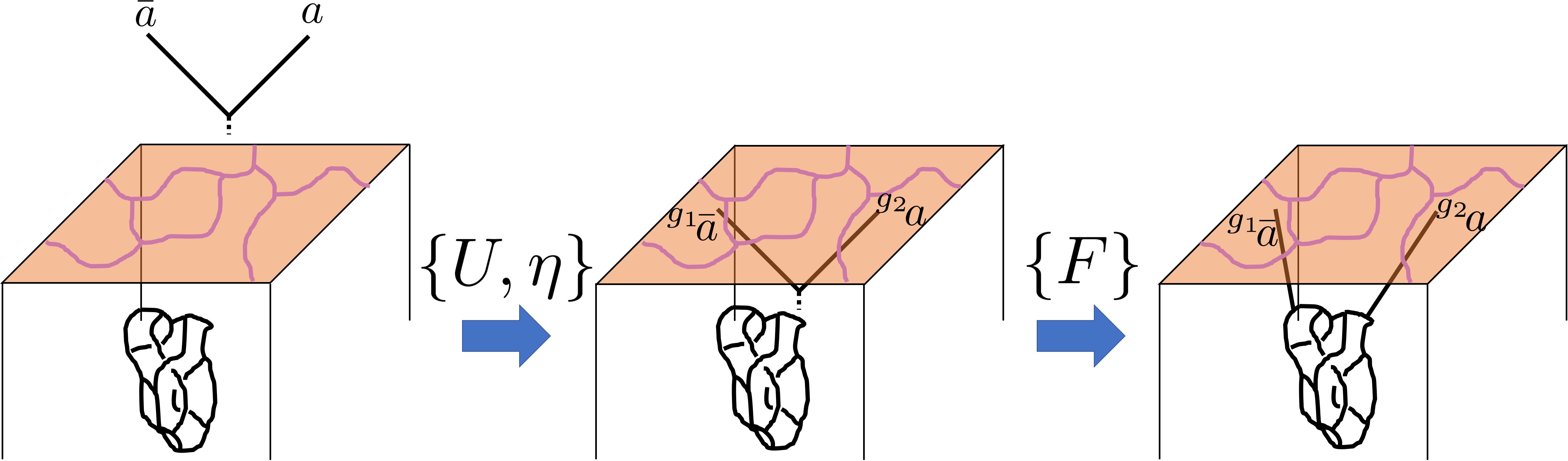}
\caption{Schematic process for the action of the operator $\mathcal{S}^{{\bf k},{\pmb{\ell}}}_a$ which creates a pair of anyons $\bar{a},a$ on the surface. Anyon lines of type $a$ or $\bar{a}$ are created far above the surface, passed through the domain walls on the surface (light grey, orange are junctions of domain walls including those in the bulk), and fused into the bulk braided fusion diagram.}
\label{fig:surfaceStringOp}
\end{figure*}

We now generalize the techniques from the Walker-Wang model to describe the string operator $\mathcal{S}^{{\bf k},{\pmb{\ell}}}_a$ for our model, which creates a pair of topologically non-trivial excitations at the surface corresponding to the anyon pair $\bar{a}$ and $a$. Ignoring the group elements ${\bf k}$ and ${\pmb{\ell}}$ for the moment, the action of the string operators $\mathcal{S}^{{\bf k},{\pmb{\ell}}}_a$ away from the string endpoints can be determined by following the graphical calculus for evaluating braided fusion diagrams. Consider an open string of type $a$ far above the surface. Then, for a given configuration of domain walls and fusion diagrams, fuse that string into the braided fusion diagram, acting on it with any domain walls it passes through (and in particular acting with the blue domain wall in Fig.~\ref{fig:surfaceHilbertSpace}). This process is shown in Fig.~\ref{fig:surfaceStringOp}. The amplitude that relates the original configuration to the final configuration then gives the matrix elements of $\mathcal{S}^{{\bf k},{\pmb{\ell}}}_a$.

This schematic is in principle sufficient to determine, using the graphical calculus, the action of the string operators $\mathcal{S}^{{\bf k},{\pmb{\ell}}}_a$ everywhere except at the string endpoints, which is where the ${\bf k}$ and ${\pmb{\ell}}$ labels come in. To further fix the action at the endpoints, recall that, after the domain walls are resolved away from the anyon degrees of freedom, each vertex is contained in a cubical domain $c$, see Fig.~\ref{fig:domainWallShift}. Suppose that the string ends at the vertices $v_1$ and $v_2$ inside cubes $c_1$ and $c_2$, respectively. Then we fix the local action of $\mathcal{S}^{{\bf k},{\pmb{\ell}}}_a$ on the $G$ degrees of freedom in $c_1$ and $c_2$ by declaring
\begin{equation}
\mathcal{S}^{{\bf k},{\pmb{\ell}}}_a \propto \mathcal{P}_{c_1}^{{\bf k}}\mathcal{P}_{c_2}^{{\pmb{\ell}}}
\end{equation}
where $\mathcal{P}_{c}^{{\bf g}}$ projects the $G$ degree of freedom in cube $c$ to the state $\ket{{\bf g}}$. We will fix this local action a little bit more precisely in what follows but will not describe in more detail the string operators along the length of the string.

We claim that the $\mathcal{S}^{{\bf k},{\pmb{\ell}}}_a$ commute with every term in the Hamiltonian that does not involve $v_1$ or $v_2$, which implies that they create deconfined point-like excitations at their endpoints. This is obviously true for the vertex term $A_v$. To see this for the $B_p$ operators, note that $B_p^s$ is obtained by fusing in a loop labeled $s$ into a plaquette, while the string operator is defined by fusing a string labeled by $a$ from ``above'' the surface. Since the $a$ and $s$ strings are unlinked, the graphical calculus makes it clear that the order in which these are done does not affect the final result, which implies that the string operator must commute with $B_p^s$. 
Finally, the consistency of the graphical calculus ensures that the string operator commutes with the cube operators $D_c$ away from the endpoints since both $D_c$ and $\mathcal{S}^{{\bf k},\pmb{\ell}}_a$ only involve sliding domain walls over anyon lines.

Consider now the state 
\begin{align}
\ket{\bar{a},a;1;{\bf 1},{\bf 1}} = \mathcal{S}^{{\bf 1},{\bf 1}}_a |\Psi_0 \rangle ,
\label{eqn:S11}
\end{align}
where the $a$-type string operator $\mathcal{S}^{{\bf 1},{\bf 1}}_a$ stretching from vertex $v_1$ to vertex $v_2$ has been applied to the ground state $|\Psi_0 \rangle$. (The reason for the notation will be clear in the subsequent section.) Clearly this state has $A_{v_i}=0$ where $v_1,v_2$ are the string endpoints. Any plaquette $p$ involving a $v_i$ and the cubes $c_i$ containing $v_i$ (recall that each vertex is contained in a single cube after the domain walls are shifted, see Fig.~\ref{fig:domainWallShift}) automatically obey
\begin{equation}
B_p\ket{\bar{a},a;1;{\bf 1},{\bf 1}} = D_{c_i}\ket{\bar{a},a;1;{\bf 1},{\bf 1}} = 0
\end{equation}
because the $F$-symbols are zero if the fusion rules are not obeyed and because $D_c$ projects onto states which obey the fusion rules inside cube $c$. Hence we have constructed an eigenstate of $H$, even though $\mathcal{S}^{{\bf 1},{\bf 1}}_a$ contains projectors onto certain group elements in $c_1,c_2$.

\subsubsection{Excited state wavefunction}

\begin{figure*}
\includegraphics[width=1.8\columnwidth]{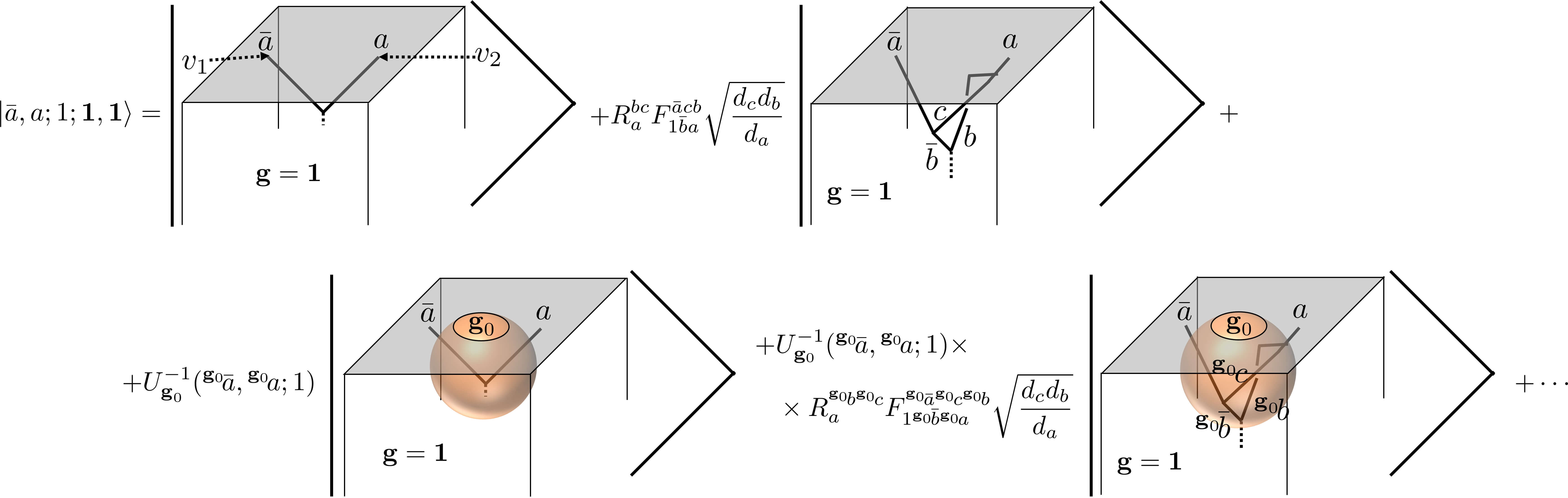}
\caption{Wavefunction of the state $\ket{\bar{a},a;1;{\bf 1},{\bf 1}}$. Black are anyon lines, orange are domain walls, $G$ labels label domains. The first term is the ``reference" state with an $a$-type string running from $v_1$ to $v_2$. The Hamiltonian term $B_p$ causes fluctuations from the first and third terms to the second and fourth, while $D_c$ causes fluctuations between the first two terms and the last two.}
\label{fig:wavefunctionWithAnyons}
\end{figure*}

We now describe the wavefunction of $\ket{\bar{a},a;1;{\bf 1},{\bf 1}}$. We begin by applying $\mathcal{S}^{{\bf 1},{\bf 1}}_a$ to the particular term in the ground state superposition where all of the group element degrees of freedom ${\bf g}_i$ and all the respective anyon labels are the identity. This creates a string of anyon labels $a$ from $v_1$ to $v_2$, as shown in the first term in Fig.~\ref{fig:wavefunctionWithAnyons}. This state will be referred to as a reference state from which we build the rest of the terms in the superposition that defines $\ket{\bar{a},a;1;{\bf 1},{\bf 1}}$. We fix a gauge where the reference state has amplitude 1, ignoring overall normalization. Obviously the reference state is not an eigenstate of the Hamiltonian; demanding that we are in an eigenstate of all the $B_p$ means that the wavefunction has nonzero amplitude for configurations which represent braided fusion diagrams with net anyon charge $\bar{a}$ at $v_1$ and $a$ at $v_2$, and no other net anyon charge. The ratio of the amplitudes of any two such states is equal to the amplitude obtained when turning the first state into the second state by applying $F-$ and $R-$ moves to the braided fusion diagram. An example is given in the second term of Fig.~\ref{fig:wavefunctionWithAnyons}.

We also want to demand that $D_c = 1$ away from the endpoints of the string. We therefore must superpose over domain wall configurations, and the relative amplitude between the state with and without a given domain wall is related by nucleating and expanding the domain wall in the graphical calculus. Some examples are given in the last two terms of Fig.~\ref{fig:wavefunctionWithAnyons}. Importantly, since $\mathcal{S}^{{\bf 1},{\bf 1}}_a$ projects onto ${\bf g}_{1,2}={\bf 1}$, where ${\bf g}_i$ is the $G$ degree of freedom on cube $c_i$, $\ket{\bar{a},a;1;{\bf 1},{\bf 1}}$ only contains states where the cubes $c_1$ and $c_2$ contain identity group elements. 

\subsubsection{The surface realizes $\mathcal{C}$}

Next we want to check that the $F$ and $R$ symbols describing the algebraic properties of the deconfined surface anyons coincide with those of the UMTC $\mathcal{C}$ that was used to define the model. Here we discuss $F$ explicitly; the argument for $R$ is essentially identical.

\begin{figure*}
\centering
\subfloat[\label{fig:Fstate}]{\includegraphics[width=1.2\columnwidth]{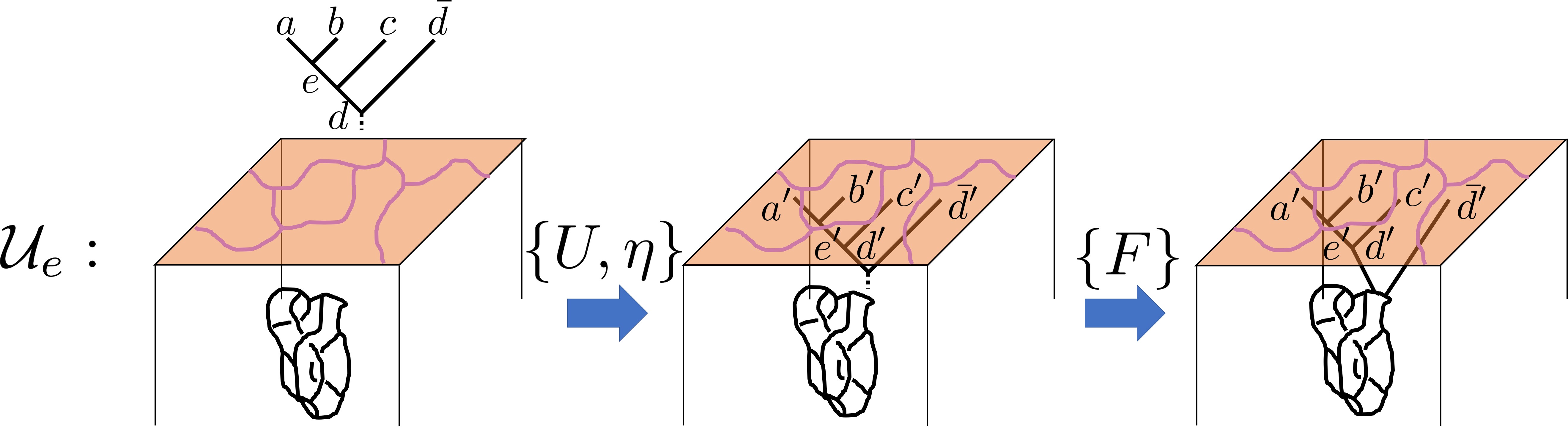}}\\
\subfloat[\label{fig:Fstate2}]{\includegraphics[width=1.2\columnwidth]{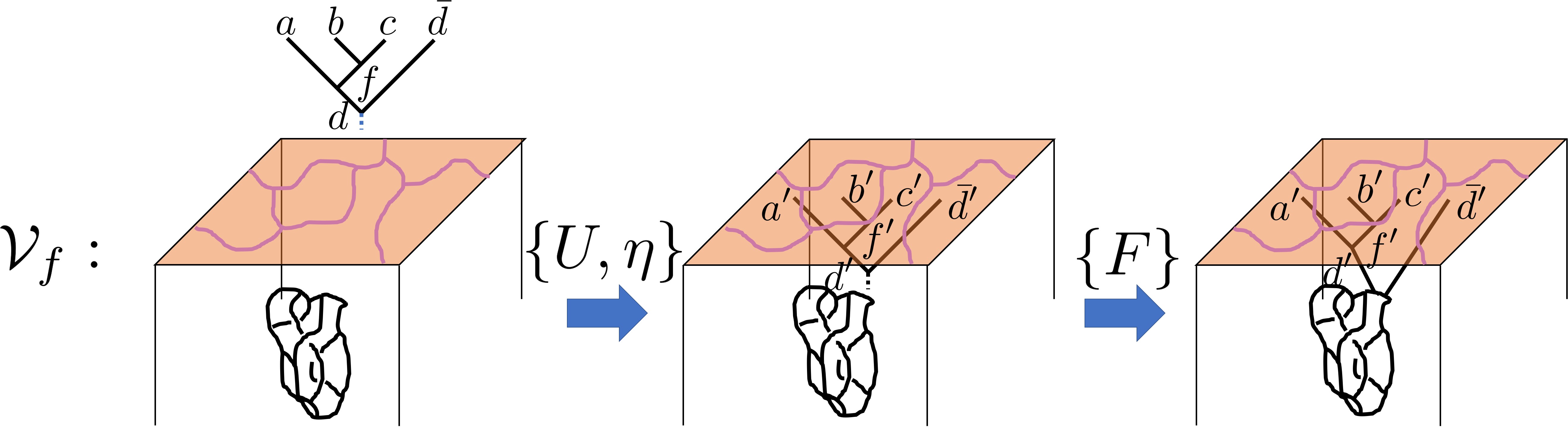}}
\caption{Graphical processes which determines the matrix elements of the operator (a) $\mathcal{U}_e$ and (b) $\mathcal{V}_f$. The appropriate fusion tree is created above the surface, then slid through the domain walls and fused into the bulk braided fusion category according to graphical calculus.  Blue are anyon lines (implicitly oriented upwards), light orange are domain walls on the surface, purple are junctions between domain walls. Primes mean the anyon type after being acted on by whatever domain walls the anyons pass through.}
\end{figure*}

We begin by defining an operator $\mathcal{U}_e$ graphically: it creates a particular braided fusion diagram far above the surface, slides it through the surface domain walls, and fuses it into the bulk braided fusion diagram. We choose the braided fusion diagram where $a,b$ fuse to $e$, $e$ and $c$ fuse to $d$, and finally $d$ fuses with $\bar{d}$ to vacuum. This process is shown in Fig.~\ref{fig:Fstate}. The matrix elements between an initial configuration $\ket{\psi}$ and $\mathcal{U}_e\ket{\psi}$ is given by performing the appropriate graphical calculus in $\mathcal{C}$ with symmetry fractionalization. We may then suggestively name a state
\begin{equation}
\ket{(((a,b;e),c; d),\bar{d};1)} = \mathcal{U}_e\ket{\Psi_0} ,
\end{equation}
where we are ignoring all of the local degrees of freedom near the endpoints of the strings involved. Similarly, we can graphically define a different operator $\mathcal{V}_f$ by fusing in a different braided fusion diagram; this diagram has $b$ and $c$ fusing to $f$, $a$ and $f$ fusing to $d$, then $d$ and $\bar{d}$ fusing to vacuum. This is shown in Fig.~\ref{fig:Fstate2}. We can then define a state
\begin{equation}
\ket{((a,(b,c;f);d),\bar{d};1)} = \mathcal{V}_f \ket{\Psi_0} .
\end{equation}

Consistency of the graphical calculus demands that
\begin{equation}
\mathcal{U}_e = \sum_f F^{abc}_{def}\mathcal{V}_f ,
\end{equation}
because we may always evaluate the fusion diagrams involved in calculating matrix elements of $\mathcal{U}_e$ by performing an $F$-move first. Therefore the states obey
\begin{equation}
\ket{(((a,b;e),c; d),\bar{d};1)} = \sum_f F^{abc}_{def}\ket{((a,(b,c;f);d),\bar{d};1)}
\end{equation}
which justifies the names we have given the states; the point-like excitations which we have called $a,b,c,\bar{d}$ indeed obey the $F$-moves given by $\mathcal{C}$. A precisely analogous argument holds for $R$-symbols; therefore the surface topological order is indeed given by $\mathcal{C}$.

We also note that projectors $\mathcal{P}_{\omega_a}$ onto fixed anyon sectors $a$ inside a loop $\omega$ can be defined graphically. In $\mathcal{C}$, such a projector is given by
\begin{equation}
\Pi_{\omega_a} = \sum_{x \in \mathcal{C}} S_{0a}S^{\ast}_{xa} W^{(\omega)}_x ,
\end{equation}
where $W^{(\omega)}_x$ is a Wilson loop operator (specifically a closed string operator) for $x$ on the loop $\omega$. We define $\mathcal{P}_{\omega_a}$ graphically by taking this same superposition of Wilson loops and fusing them onto the surface (again passing through domain walls as necessary). It is clear from consistency of the graphical calculus that this definition will also justify our labelings of states; by choosing different $\omega$, we can measure the fusion channels associated with different sets of anyons to be as we have claimed.

\subsection{Surface symmetry fractionalization}

Here we analyze the symmetry action on the surface topological order. Remarkably, we will see that the symmetry localization ansatz of Eq.~\eqref{symloc} is realized exactly, along with the symmetry fractionalization relation of Eq.~\eqref{etaDef}. Moreover, we also demonstrate how the symmetry fractionalization pattern on the surface is described by the same choice of $\{\rho, U, \eta\}$ that was used to define the model. 

\subsubsection{Local degeneracy at anyons}

First we note that each anyon at the surface carries a local degeneracy of $|G|$ states. As we will see, this degeneracy is local in the sense that local operators in the vicinity of the anyon can distinguish the different states. 

The origin of this degeneracy lies in the fact that the cube operators $D_c$ are by definition zero whenever the vertices contained in the cube have violations of the vertex operator $A_v$. The string operator $\mathcal{S}^{{\bf k},{\pmb{\ell}}}_a$ described in the previous section breaks the fusion rules at $v_{1,2}$, that is, acting on the ground state it makes a state with $A_{v_{1,2}}= 0$. If $c_i$ contains the vertex $v_i$, then we have
\begin{align}
D_{c_1} \mathcal{S}^{{\bf k},{\pmb{\ell}}}_a |\Psi_0\rangle = D_{c_2} \mathcal{S}^{{\bf k},{\pmb{\ell}}}_a |\Psi_0\rangle = 0. 
\end{align}
This is why the state $|\bar{a}, a;1; {\bf 1}, {\bf 1} \rangle$ only contains states where the group label in the cubes $c_1$ and $c_2$, at the endpoints of the string, is the identity. 

Next, let us define an operator $\tilde{D}_{c,v}^{{\bf g}}$, which is given by $D_{c}^{{\bf g}}$, but with the phase factors of $U$ involving the vertex $v$ removed. For example, if there was an anyon at the vertex where $w,w'$, and $a$ meet in Fig.~\ref{fig:plaquetteLabelings}, then $\tilde{D}_c^{{\bf g}}$ would be defined by Eq.~\eqref{eqn:cubePhase} with the factor $U_{{\bf g}}^{-1}(\act{g}{w},\acts{gg}{00'}{w'};\act{g}{a})$ removed.

It is clear by inspection that $\tilde{D}_{c,v}^{{\bf g}}$ commutes with all $A_{v'},B_p,D_{c'}$ for any vertex $v'$, cube $c' \neq c$, and plaquette $p$ which does not contain $v$. Recalling that, for $p$ containing $v$, $B_p=0$ on any state with $A_v=0$, this means that
\begin{equation}
\tilde{D}_{c,v}^{{\bf g}}H\ket{\bar{a},a,1;{\bf 1},{\bf 1}} = H\tilde{D}_{c,v}^{{\bf g}}\ket{\bar{a},a,1;{\bf 1},{\bf 1}}
\label{eqn:DtildeDegeneracy}
\end{equation}
because $A_v\ket{\bar{a},a,1;{\bf 1},{\bf 1}}=0$.

The model therefore has some degeneracy, but since $\tilde{D}_{c,v}^{{\bf g}}$ is local, the degeneracy is non-topological. Of course, $\tilde{D}_{c,v}^{{\bf g}}$ cannot change the net anyon type in cube $c$. Therefore, there are many eigenstates, related by local operators, which have the same anyon types at $v_1,v_2$. We label these states $\ket{\bar{a},a;1;{\bf k},{\pmb{\ell}}}$. The labeling scheme is as follows: $\bar{a},a$ means that the (physical) anyons $\bar{a}$ and $a$ are present on the surface, the $1$ means that $\bar{a}$ and $a$ are in the $1$ fusion channel, and the labels ${\bf k},{\pmb{\ell}}$ mean that ${\bf g}_1={\bf k}$ and ${\bf g}_2={\pmb{\ell}}$, where ${\bf g}_{1,2}$ are the group elements on $c_{1,2}$. We fix a gauge by defining
\begin{equation}
\ket{\bar{a},a;1;{\bf k},{\pmb{\ell}}}=\tilde{D}_{c_1,v_1}^{{\bf k}} \tilde{D}_{c_2,v_2}^{\pmb{\ell}}\ket{\bar{a},a;1;{\bf 1},{\bf 1}} .
\end{equation}
where $\ket{\bar{a},a;1;{\bf 1},{\bf 1}}$ was gauge-fixed earlier. The state $\ket{\bar{a},a;1;{\bf k},{\pmb{\ell}}}$ is illustrated pictorially in Fig. \ref{fig:aaWavefunction}. 

Accordingly, we can define the string operators $\mathcal{S}^{{\bf k},{\pmb{\ell}}}_a$ by 
\begin{equation}
\mathcal{S}^{{\bf k},{\pmb{\ell}}}_a = \tilde{D}_{c_1,v_1}^{{\bf k}}\tilde{D}_{c_2,v_2}^{{\pmb{\ell}}} \mathcal{S}^{{\bf 1},{\bf 1}}_a
\label{eqn:Skl}
\end{equation}
where $\mathcal{S}^{{\bf 1},{\bf 1}}_a$ is specified by Eq.~\eqref{eqn:S11}. This implies that
\begin{equation}
\mathcal{S}^{{\bf k},{\pmb{\ell}}}_a\ket{\Psi_0} = \ket{\bar{a},a;1;{\bf k},{\pmb{\ell}}}
\end{equation}

\begin{figure*}
\subfloat[\label{fig:wavefunctionWithAnyons_domains}]{\includegraphics[width=1.8\columnwidth]{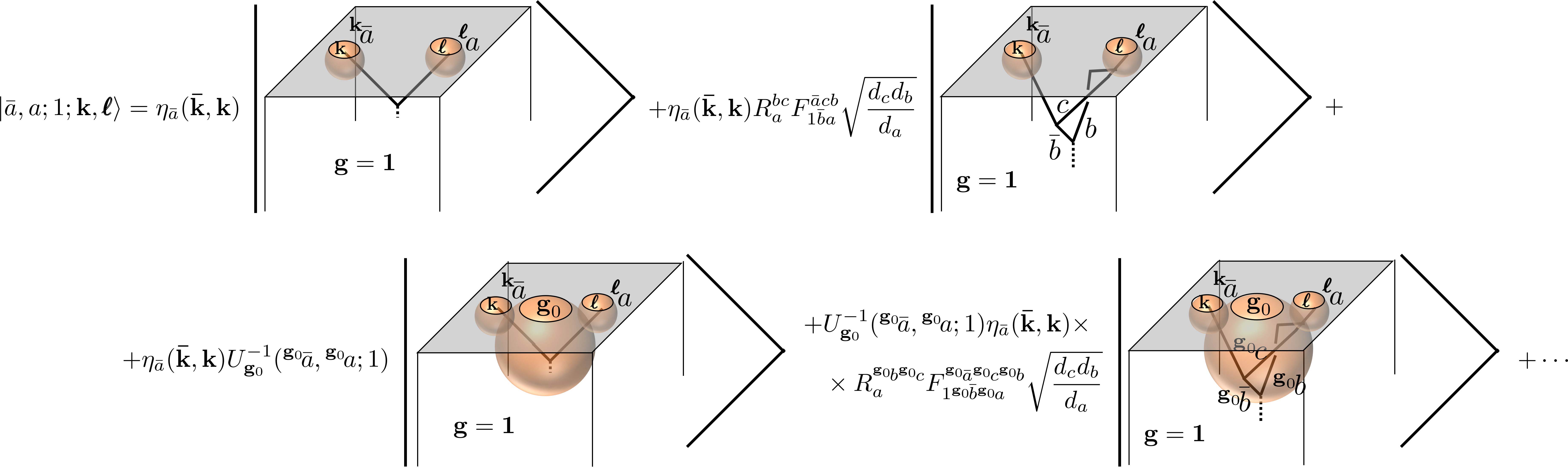}}\\
\subfloat[\label{fig:referenceConfig_phase3D}]{\includegraphics[width=\columnwidth]{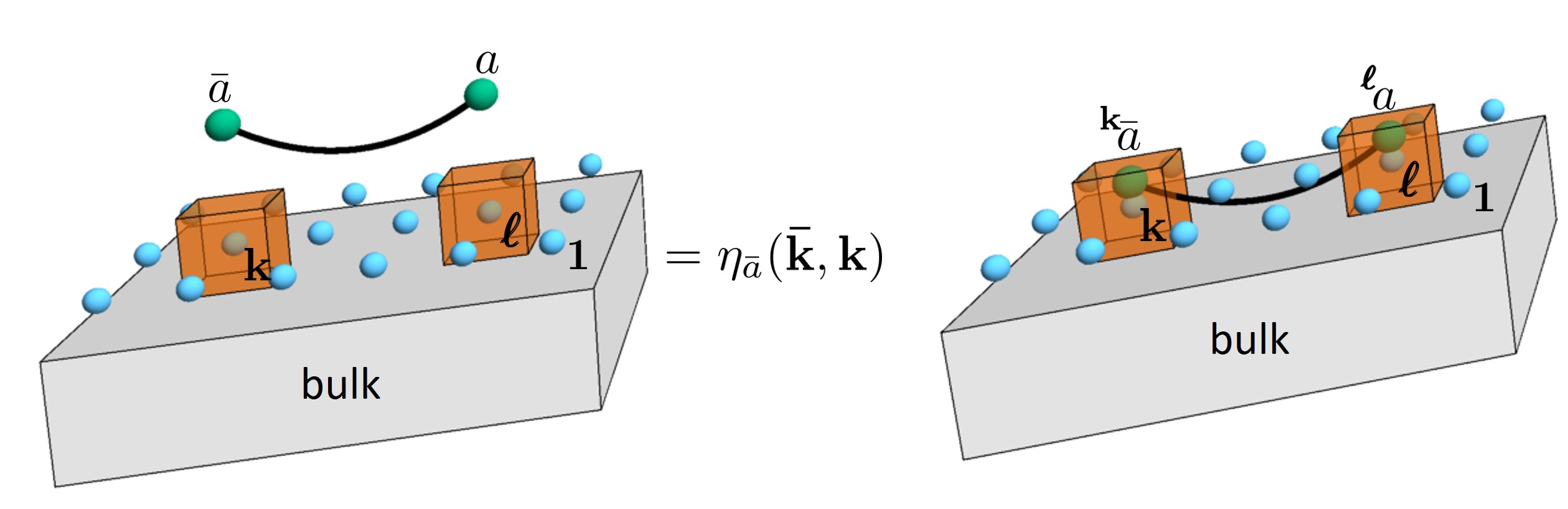}}\\
\subfloat[\label{fig:referenceConfig_phase}]{\includegraphics[width=\columnwidth]{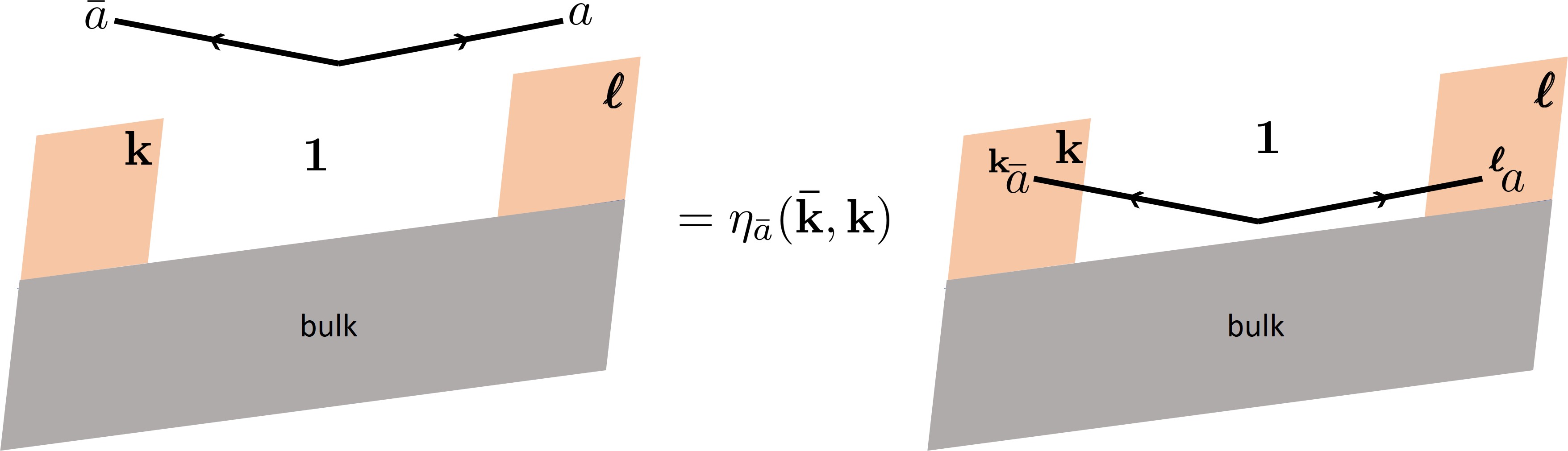}}
\caption{(a) Wavefunction of the state $\ket{\bar{a},a;1;{\bf k},\pmb{\ell}}$. Blue are anyon lines, orange are domain walls, $G$ labels label domains. The first term is the ``reference" state with an $a$-type string running from $v_1$ to $v_2$. The Hamiltonian term $B_p$ causes fluctuations from the first and third terms to the second and fourth, while $D_c$ causes fluctuations between the first two terms and the last two. The anyon labels label the string type as it intersects the surface, \textit{not} the physical anyon labels; the physical anyons are $\bar{a}$ and $a$ at $v_1$ and $v_2$, respectively. (b) Side view of the surface showing that moving a string (black) through the domain walls generates the phase $\eta_{\bar{a}}({\bf \bar{k}},{\bf k})$ appearing in the reference configuration (first term) in Fig.~\ref{fig:wavefunctionWithAnyons_domains}. There is also a factor of $\eta_a({\bf 1},{\pmb{\ell}})$ which we have gauge-fixed to 1. View is similar to Fig.~\ref{fig:surfaceTermination}. Blue spheres are group elements, all of which (in this configuration) are in the state $\ket{{\bf 1}}$ except for the labeled ones inside the domain walls (orange). Green spheres are anyons on the end of the string. (c) Same as (b), but projected into the plane of the page. Floating group elements label the domains.}
\end{figure*}

It is straightforward to check that the wave function of $\ket{\bar{a},a;1;{\bf k},{\pmb{\ell}}}$ is obtained in the same way as that of $\ket{\bar{a},a;1;{\bf 1},{\bf 1}}$ in Fig.~\ref{fig:wavefunctionWithAnyons}, but starting from a reference state where the anyons are surrounded by domain walls, as in Fig.~\ref{fig:wavefunctionWithAnyons_domains}. There is a subtlety here that the reference state should not have phase 1; instead, as shown in Fig.~\ref{fig:referenceConfig_phase3D} and more schematically in Fig.~\ref{fig:referenceConfig_phase}, moving a string through the domain walls picks up a phase of $\eta_{\bar{a}}({\bf \bar{k}},{\bf k})$. This changes the wavefunction by a global phase and thus is simply a basis choice, but it does arise from our definitions of the string operators in Eq.~\eqref{eqn:Skl}. 

Importantly, the physical anyon type $a$ cannot be changed by a local operator such as $\tilde{D}_{c,v}^{{\pmb{\ell}}}$. However, since $\tilde{D}_{c,v}^{{\pmb{\ell}}}$ still acts on the anyon labels on the links in $c$, the anyon label along the string that ends at $v_2$ changes to $\,^{\pmb{\ell}}a$. Therefore, if in the wave function a string type labeled $b$ ends at $v_2$, then the physical anyon type at $v_2$ is $a ={}^{\bar{{\bf g}}_2}b$, where ${\bf g}_2$ is the state of the $G$ degree of freedom in cube $c_2$.

Note that the local degenerate state space at each anyon excitation on the surface forms a regular representation of $G$. In general such a representation is reducible, which means that the Hamiltonian should admit $G$ symmetry-preserving perturbations that lifts this degeneracy. The understanding of the ground and excited states, and in particular the symmetry fractionalization analysis below, would then have to be modified, but the data that characterizes symmetry fractionalization should remain unchanged, up to possible gauge transformations. 

\subsubsection{Symmetry localization}

For simplicity, we consider only simple states with two anyons $a$ and $\bar{a}$ fusing to the identity; the generalization is straightforward.

We define the topological symmetry transformation
\begin{equation}
\rho_{{\bf g}}\ket{\bar{a},a;1;{\bf k},\pmb{\ell}} = U_{{\bf g}}({}^{{\bf g}}\bar{a},{}^{{\bf g}}a;1) \ket{{}^{{\bf g}}\bar{a},{}^{{\bf g}}a;1 ; {\bf k},\pmb{\ell}}
\label{eqn:HamRho}
\end{equation}
Obviously this map by definition obeys Eq.~\eqref{eqn:rhoStates} with the original $U$ that we input into the Hamiltonian. To show that we have the  correct symmetry fractionalization pattern at the surface, we first need to show that the global symmetry action $R_{{\bf g}}$ acts in a way according to Eq.~\eqref{symloc}, which we refer to as symmetry localization. Subsequently we must show that Eq.~\eqref{etaDef} is obeyed with $\rho$ defined in Eq.~\eqref{eqn:HamRho} and with $\eta$ given by the $\eta$ that we input into the Hamiltonian. 

\begin{figure*}
\subfloat[\label{fig:aaWavefunction}]{\includegraphics[width=1.7\columnwidth]{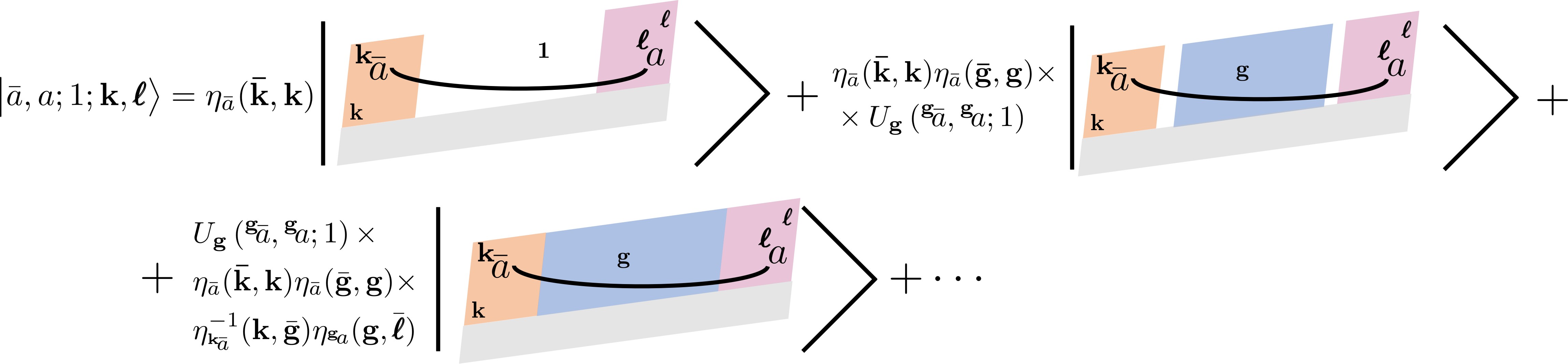}}\\
\subfloat[\label{fig:RaaWavefunction}]{\includegraphics[width=1.7\columnwidth]{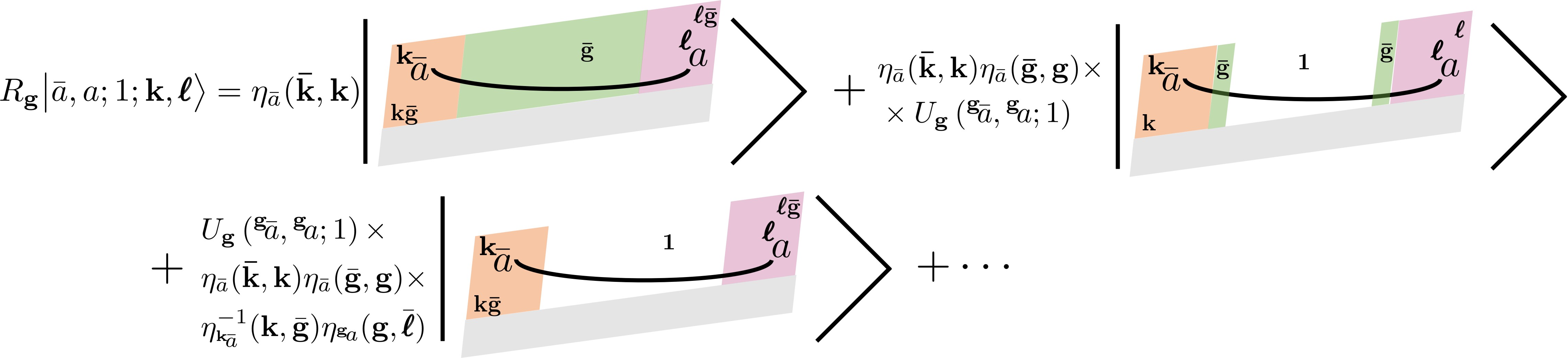}}\\
\subfloat[\label{fig:gagaWavefunction}]{\includegraphics[width=1.7\columnwidth]{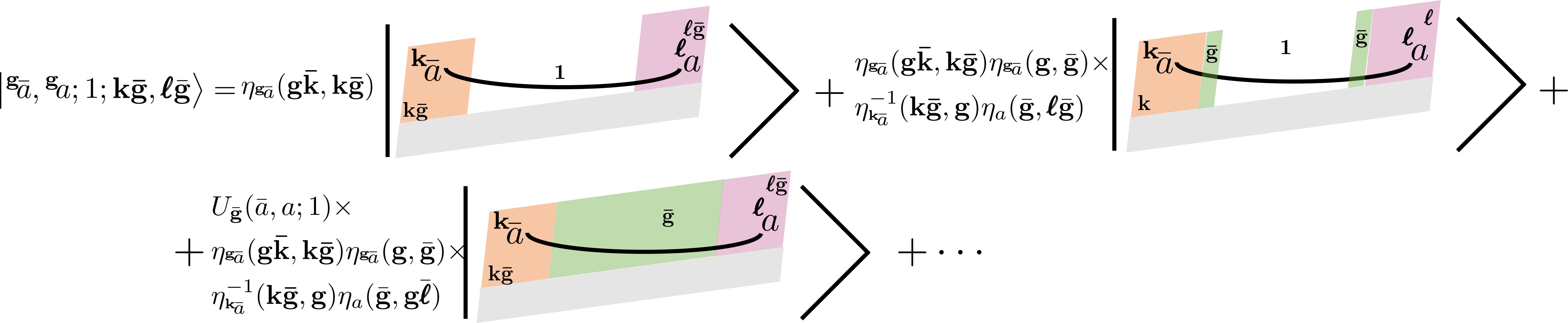}}
\caption{Some terms in the wavefunctions for symmetry-related excited states on the surface of our construction. The view is from the side in the same schematic picture as Fig.~\ref{fig:referenceConfig_phase}. Grey indicates the bulk. Colors indicate $G$ domains labeled by the ``floating" group labels; unlike in Fig.~\ref{fig:referenceConfig_phase}, we have used different colors for different domains. The anyon labels ${}^{{\bf k}}\!\bar{a}$ and ${}^{\pmb{\ell}}\!a$ label the string type which terminates at that point; the black string changes type when it passes through a domain wall.}
\label{fig:symmetryActionWavefunction}
\end{figure*}

We compute $R_{{\bf g}}\ket{\bar{a},a;1;{\bf k},\pmb{\ell}}$ by examining the wavefunction. Consider in particular the terms $\ket{\bar{a},a;1;{\bf k},\pmb{\ell}}$ shown in Fig.~\ref{fig:aaWavefunction}, where all of the group elements except for ${\bf g}_{1,2}$ are either ${\bf 1}$ or ${\bf g}$. By our normalization and gauge fixing, the first term has amplitude $\eta_{\bar{a}}({\bf \bar{k}},{\bf k})$. By nucleating a domain wall of type ${\bf g}$, as in the second term in Fig.~\ref{fig:aaWavefunction}, and expanding it, we eventually reach the third term, picking up various $U$s and $\eta$s as the domain wall expands. 

Now apply the symmetry operator $R_{{\bf g}}$ to $\ket{\bar{a},a;1;{\bf k},\pmb{\ell}}$, which right-multiplies all group elements by ${\bf {\bar g}}$ but does not change any link labels. The result is shown in Fig.~\ref{fig:RaaWavefunction}.

Separately, we may compute the wavefunction for the state $\ket{\act{g}{\bar{a}},\act{g}{a};1;{\bf k}\bar{{\bf g}},{\pmb{\ell}}\bar{{\bf g}}}$; this is shown in Fig.~\ref{fig:gagaWavefunction}. Comparing Figs.~\ref{fig:RaaWavefunction} and~\ref{fig:gagaWavefunction} term-by-term and using the consistency conditions for $U$ and $\eta$, we find that
\begin{align}
R_{{\bf g}}\ket{\bar{a},a;1;{\bf k},{\pmb{\ell}}} &= \eta_{\act{g}{\bar{a}}}({\bf g},{\bf \bar{k}})\eta_{\act{g}{a}}({\bf g},{\bf \bar{\ell}})\times \nonumber \\
&\hspace{0.5cm} \times U_{{\bf g}}(\act{g}{\bar{a}},\act{g}{a};1)\ket{\act{g}{\bar{a}},\act{g}{a};1;{\bf k}\bar{{\bf g}},\pmb{\ell}\bar{{\bf g}}} \\
&=U_{{\bf g}}^{(v_1)}U_{{\bf g}}^{(v_2)}\rho_{{\bf g}}\ket{\bar{a},a;1;{\bf k},\pmb{\ell}} \label{eqn:explicitLoc}
\end{align}
where we have defined the local operators $U^{(v_i)}_{\bf g}$ as they act on a state with anyon $a_i$ at $v_i$ by
\begin{align}
U_{{\bf g}}^{(v_i)} = \sum_{{\bf h}}\eta_{a_i}({\bf g},\bar{{\bf h}})  \ket{{\bf h}\bar{{\bf g}}}\bra{{\bf h}} \label{eqn:localU}
\end{align}
where ${\bf h}$ labels the degree of freedom on the cube $c_i$ containing the vertex $v_i$.

The topological part of the symmetry action $\rho_{{\bf g}}$ that was previously defined has indeed appeared in the actual global symmetry action of the model, and the remaining factors decompose into local factors $U_{\bf g}^{(v_i)}$, as expected.

\subsubsection{Symmetry fractionalization}

Verifying Eq.~\eqref{symloc} is now a matter of straightforward computation. Using Eq.~\eqref{eqn:etaConsistency}:
\begin{widetext}
\begin{align}
U_{\bf g}^{(v_2)}\rho_{{\bf g}} U_{{\bf h}}^{(v_2)} \rho_{{\bf g}}^{-1} \ket{\bar{a},a;1;{\bf k},\pmb{\ell}}
&= \eta_{\act{\bar{g}}{a}}({\bf h},\bar{{\pmb{\ell}}})\eta_a({\bf g},{\bf h}\bar{{\pmb{\ell}}})\ket{\bar{a},a;1;{\bf k},{\pmb{\ell}}\bar{{\bf h}}\bar{{\bf g}}}\\
&= \eta_a({\bf g},{\bf h})\eta_a({\bf gh},\bar{\pmb{\ell}})\ket{\bar{a},a;1;{\bf k},{\pmb{\ell}}\bar{{\bf h}}\bar{{\bf g}}}\\
&= \eta_a({\bf g},{\bf h})U_{{\bf gh}}^{(v_2)}\ket{\bar{a},a;1;{\bf k},{\pmb{\ell}}}
\end{align}
\end{widetext}
This demonstrates that the symmetry fractionalization at the surface is, as we have claimed, given by the $\eta$ and $U$ which were put into the theory.

As a technical aside, our definition of $\rho_{{\bf g}}$ is not invariant under independent basis rotations of the local degrees of freedom in different anyon sectors. For a simple example, we could modify the states $\ket{\bar{a},a;1,{\bf k},\pmb{\ell}}$ with a ${\bf k}$- or ${\pmb{\ell}}$-dependent phase. Such basis changes yield surface fractionalization with data that is gauge-equivalent but not identical to a given choice of $\eta$ and $U$ that defines the model.  

\subsubsection{Anti-unitary symmetries}

All of the prior analysis can be repeated for anti-unitary symmetries. We define the analog of Eq.~\eqref{eqn:HamRho}
\begin{equation}
\rho_{{\bf g}}\ket{\bar{a},a;1;{\bf k},\pmb{\ell}} = U_{{\bf g}}({}^{{\bf g}}\bar{a},{}^{{\bf g}}a;1) K^{q({\bf g})}\ket{{}^{{\bf g}}\bar{a},{}^{{\bf g}}a;1 ; {\bf k},\pmb{\ell}}.
\end{equation}
Following the same argument, the anti-unitary operator
\begin{equation}
R_{\bf g} = \prod_{c \in \text{cubes}} R_{\bf g}^{(c)}K^{q(g)}
\end{equation}
localizes according to Eq.~\eqref{eqn:explicitLoc}, with the local operators $U_{\bf g}^{(v_i)}$ still defined by Eq.~\eqref{eqn:localU}. Note that the physical surface anyon states do \textit{not} involve any domain data since the microscopic $G$ domains fluctuate. However, the Hilbert space configurations naturally carry domain information. This means that when repeating the calculation, the graphical calculus involves the data $\lbrace \tilde{F}, \tilde{R}, \tilde{U}, \tilde{\eta} \rbrace$ defined in Sec.~\ref{sec:antiunitary}, and the domain labels must be carefully tracked. The amplitudes in the anti-unitary version of Fig.~\ref{fig:symmetryActionWavefunction} are thus expressed in terms of the data with tildes, but using $G$-equivariance, they can be converted to the physical $U$ and $\eta$ data via Eqs.~\eqref{eqn:FNoTilde}-\eqref{eqn:UNoTilde}.

\section{Generalizations: spatial symmetries, continuous, infinite $G$}

So far we have defined models that possess an internal, on-site symmetry action, which may be unitary or anti-unitary. Furthermore, the state sums defined in the previous sections require finite, discrete groups. Here we briefly discuss the generalizations of the construction to spatial (crystalline) symmetries, and to continuous and/or infinite symmetry groups.

\subsection{Infinite and/or continuous $G$}

When $G$ is compact and continuous (and for simplicity of this discussion connected), there is a natural generalization of our state sum obtained by replacing the sum over group element labels by integrals, $\frac{1}{|G|^{N_0}} \sum_{ \{ {\bf g}_i \}}  \rightarrow \left( \frac{1}{\text{vol}(G)} \right)^{N_0} \prod_{i \in \mathcal{T}^0} \int d{\bf g}_i $, where $\text{vol}(G) = \int d{\bf g}_i $ is the volume of $G$ and $N_0$ is the number of 0-simplices. The integration measure is then taken to be the Haar measure on $G$. We note that such a construction necessarily still defines a $G$ SPT due to the $G$ translation invariance of the measure. Therefore our construction yields topologically invariant path integrals on non-trivial flat $G$ bundles for continuous $G$ as well. Our Hamiltonian construction furthermore also goes through straightforwardly as well. 

For infinite discrete $G$ we can define the state sum as a limit of a sequence of finite discrete groups that approach $G$ in the limit. However, if our conjecture in Sec.~\ref{sec:stateSum} is true, then the above modifications are actually all unnecessary, and our construction naturally gives a concrete way to compute the path integral on any flat $G$ bundle regardless of whether $G$ is infinite, continuous, or compact.

We note that our discussion of obtaining cohomology invariants in Sec. \ref{subsec:generateInvariants} applies only to finite groups; therefore some other method would be required to find cohomology invariants for the case of infinite or continuous $G$. 

\subsection{Spatial symmetries}

Given a symmetry group $G$, it is possible that certain elements of $G$ are spatial symmetries, meaning that they act on the quantum system by some combination of translations, rotations, and reflections in space. Here we briefly discuss some issues in incorporating such symmetries into our discussion. There are three aspects to the discussion, which we will briefly address:
\begin{enumerate}
\item Classification of spatial symmetry fractionalization and anomaly computations
\item Path integral construction
\item Hamiltonian construction: boundary and bulk symmetries
\end{enumerate}

\subsubsection{Classification of spatial symmetry fractionalization and anomaly computations}

First, let us consider classification of symmetry fractionalization and anomaly computations. It is expected that the classification of symmetry-enriched topological phases with spatial symmetries is identical to the case with internal symmetries, where spatial parity reversing symmetries are treated as anti-unitary symmetries. This expectation has already received strong direct supporting evidence from studies of SPT states\cite{song2017,huang2017,thorngren2018} and from the understanding of the relation between Lieb-Schulz-Mattis theorems and mixed anomalies between translation and on-site symmetries.\cite{cheng2016lsm} Therefore, the same data that characterizes symmetry fractionalization and the same anomaly calculations are expected to hold even in the case of spatial symmetries.

The general reason for this expectation is as follows. The low energy universal properties of these SET states is described by a TQFT enriched with a $G$ symmetry. The TQFT always has a diffeomorphism symmetry, which in particular includes the isometries of space. Any spatial symmetry of the microscopic Hamiltonian can then be taken to correspond to a combination of an internal symmetry in the field theory together with an isometry of the space. Stated differently, one can consider a combination of the spatial symmetry of the microscopic Hamiltonian, followed by an isometry of space in the field theory, which then leads to an internal symmetry acting in the field theory. Examples where the above logic is borne out are prevalent throughout the literature regarding effective quantum field theories describing quantum many-body systems. Thus classifying different ways of incorporating internal symmetry in a TQFT should also give the classification of spatial symmetry enriched topological phases of matter. 

\subsubsection{Path integral construction}

The state sum for the path integral proceeds by choosing a triangulation and a branching structure. Therefore we can consider triangulations and branching structures that are invariant under a space-time symmetry $G_s$ of interest. For example, one can consider a space-time manifold $M^4 = W^3 \times S^1$, and one can consider a triangulation and branching structure that reduces on $W^3$ to a particular (finite) space group symmetry of interest. In this case, the amplitudes that are summed over in the path integral possess a $G \times G_s$ symmetry. If we pick $G$ such that $G_s \subseteq G$, then we can consider the diagonal group elements of the form $({\bf g}, {\bf g}) \in G \times G_s$, where ${\bf g} \in G_s \subseteq G$. Then we can consider any type of space group symmetry fractionalization, and the path integral will possess the space group symmetry. 

Alternatively, one can consider a 4-manifold with boundary such that $W^3 = \partial M^4$ has a triangulation and branching structure associated with the space group symmetry of interest. We can take the triangulation of $M^4$ to correspond to just adding a single ``bulk'' 0-simplex to which all 0-simplices in the triangulation of $W^3$ are connected. The resulting wave function will be symmetric under $G \times G_s$, and again as in the previous paragraph we can consider the diagonal action $({\bf g}, {\bf g}) \in G \times G_s$ for ${\bf g} \in G_s \subseteq G$.

Therefore, the path integral construction can be easily extended to space group symmetries for which there is a triangulation and branching structure that preserves the space group symmetry. It is an interesting question whether a symmetric triangulation and branching structure exists for any given space group symmetry. 

\subsubsection{Hamiltonian construction}

Our Hamiltonian construction in general picks a preferred direction to correspond to an ``arrow of time'' in evaluating the braided fusion diagrams. As such, our Hamiltonian construction, while it preserves lattice translational symmetries, in general breaks point group symmetries. It is a non-trivial question whether our general construction can be modified to allow space group symmetries in general. 

It may be possible to consider certain classes of UMTCs that are invariant under changing the ``time'' direction. This seems to be a generalization of the ``tetrahedral'' symmetry of fusion categories to a higher dimensional version, which we may term ``hypertetrahedral'' symmetry. UMTCs that possess such a symmetry and which have self-dual anyons will then give Hamiltonians that are independent of the chosen ``time'' direction and any orientations on the edges. While the trivalently resolved cubic lattice is still incompatible with point group symmetries, we expect it is possible to consider Hamiltonians on other trivalent lattices, or to use certain gadgets that allow for sites with more than three edges, that would be compatible with any given space group symmetry.

\subsubsection{Spatial reflection symmetries mapping to anti-unitary internal symmetries}

Here we consider spatial reflection symmetries in the path integral construction and demonstrate how they can be captured using the anti-unitary on-site action. We leave for future work to study spatial reflection symmetry in the Hamiltonian formalism, and in particular to understand spatial reflection symmetry fractionalization in the (2+1)D surface SET in the Hamiltonian construction. 

Accordingly, suppose that $G$ contains symmetries that reverse spatial parity. For simplicity, we choose to examine a pure reflection ${\bf r}$, but the following discussion holds for any unitary parity-reversing element of $G$. Since
${\bf r}$ acts unitarily on the wavefunction but changes the locations of vertices, the desired $G$-action on the wavefunction is a modification of Eq.~\eqref{eqn:unitaryGAction}:
\begin{align}
Z[\partial M^4; &\lbrace {\bf g}_{{\bf r}(i)}{\bf r} \rbrace, \lbrace a_{{\bf r}(ijk)} \rbrace , \lbrace b_{{\bf r}(ijkl)} \rbrace] \nonumber \\
&= Z[\partial M^4; \lbrace {\bf g}_i \rbrace, \lbrace a_{ijk} \rbrace ,\lbrace b_{ijkl}\rbrace]
\label{eqn:reflectingGAction}
\end{align}
where, as a reminder, the $ \lbrace {\bf g}_i \rbrace, \lbrace a_{ijk} \rbrace ,\lbrace b_{ijkl}\rbrace$ are the degrees of freedom on $\partial M^4$, and $i,j,k,l$ label boundary vertices. The notation ${\bf r}(i)$ means the image under spatial reflection of the vertex $i$.

This is not enough to determine the consistency equations for the objects that appear in the state-sum. In order to do so, we need to understand the way that group multiplication by ${\bf r}$ acts \textit{onsite}. We claim that
\begin{align}
Z[\partial M^4; &\lbrace {\bf g}_{{\bf r}(i)}\rbrace, \lbrace a_{{\bf r}(ijk)} \rbrace , \lbrace b_{{\bf r}(ijkl)} \rbrace]^{\ast} \nonumber \\
&= Z[\partial M^4; \lbrace {\bf g}_i \rbrace, \lbrace a_{ijk} \rbrace ,\lbrace b_{ijkl}\rbrace]
\label{eqn:spatialIndexReflection}
\end{align}
where there is no group multiplication. To see this, note that under the action of reflection, every 4-simplex $\Delta^4$ changes its orientation. If the data associated to the 4-simplex remains unchanged, this orientation reversal simply complex conjugates the corresponding amplitude $Z^{\pm}(\Delta^4)$. Eq.~\eqref{eqn:spatialIndexReflection} therefore follows. Combining Eq.~\eqref{eqn:spatialIndexReflection} and \eqref{eqn:reflectingGAction}, we find
\begin{align}
Z[\partial M^4; &\lbrace {\bf g}_{i}{\bf r}\rbrace, \lbrace a_{ijk} \rbrace , \lbrace b_{ijkl} \rbrace]^{\ast} \nonumber \\
&= Z[\partial M^4; \lbrace {\bf g}_i \rbrace, \lbrace a_{ijk} \rbrace ,\lbrace b_{ijkl}\rbrace].
\end{align}
This is exactly the same as Eq.~\eqref{eqn:TReversingAction}; that is, the onsite action of a parity-reversing symmetry should be \textit{anti-unitary}, despite the fact that the global action is unitary. The rest of the derivation following Eq.~\eqref{eqn:TReversingAction} therefore holds for parity-reversing symmetries as well, provided the definition Eq.~\eqref{eqn:sigmaDef} of $\sigma({\bf g})$ is reinterpreted to refer to the (anti-)unitarity of the \textit{onsite} action of ${\bf g}$. In particular, we should be choosing anti-unitary $\alpha_{{\bf h}}$ despite the unitarity of the overall symmetry action.

Repeating the above argument for ${\bf g}\in G$ which is both anti-unitary and spatial parity-reversing, we find that ${\bf g}$ should have a \textit{unitary} onsite action as in Eq.~\eqref{eqn:unitaryGAction}.

\section{Discussion}

We have shown how the symmetry fractionalization data determines a (3+1)D SPT, and thus determines an element $[\coho{S}] \in \mathcal{H}^4(G, U(1))$. As we have discussed, it is also known that there is an obstruction $[\coho{O}] \in \mathcal{H}^4(G, U(1))$ in defining a $G$-crossed braided tensor category from a given symmetry fractionalization class. In mathematical terms, there is an obstruction $[\coho{O}] \in \mathcal{H}^4(G, U(1))$ in lifting a categorical $G$ action to a $G$-crossed braided tensor category. It is clear on physical grounds that $[\coho{S}] = [\coho{O}]$, since we physically understand the presence of a non-trivial bulk (3+1)D SPT as an obstruction to defining a consistent theory of symmetry defects purely at the (2+1)D surface. On mathematical grounds, we expect that the categorical $G$ action defines a single element of $\mathcal{H}^4(G, U(1))$, not two distinct elements.  Nevertheless, to have a complete mathematically rigorous theory, we must prove that $[\coho{S}] = [\coho{O}]$, which we leave for future work. 

While we have given a concrete general method to compute the anomaly, there are still many interesting and important directions left for further study. For example, it would be interesting to understand ``anomaly in-flow'' in these models. That is, to understand precisely how processes on the surface give rise to inconsistencies that are canceled by the bulk SPT. Along these same lines, it would be interesting to understand how to extract a 4-cocyle through consideration of certain processes in the (2+1)D surface theory in a way which would not require a complete computation of the path integral on closed 4-manifolds with non-trivial $G$-bundles.

A related issue is to make contact between the results here and the relative anomaly formula derived in Ref.~\onlinecite{BarkeshliRelativeAnomaly}. It was conjectured in Ref.~\onlinecite{BarkeshliRelativeAnomaly} that the relative anomaly obtained there could be canceled by a bulk (3+1)D SPT. The bulk construction provided here should in principle provide a way of demonstrating this explicitly, however we have not pursued this direction here. 

As noted in the main text, while our state sum requires a sum over group elements, we conjecture that every term in the sum over group elements is identical, and therefore that the sum over group elements is actually not required. The $G$ action would therefore enter only through the presence of non-trivial $G$ holonomies along non-contractible cycles of the space-time manifold. It would be useful for practical computations to prove this rigorously. 
 
While we have presented our construction in terms of an explicit state sum associated with a triangulation of the space-time manifold, we expect that there should be an alternative formulation where the path integral on arbitrary space-time manifolds and $G$ bundles can be obtained by gluing together path integrals on simpler manifolds via a handle decomposition. Such a formulation was developed for the Crane-Yetter state sum (i.e. the case where $G$ is trivial) in Ref. \onlinecite{walker2006}. The extension to non-trivial $G$ would require, for example, defining a vector space for 3-manifolds equipped with $G$ bundles via skein modules on $G$ bundles. A simple version of such a formulation for the case of $\mathbb{Z}_2$ space-time reflection symmetry was used in Ref.~\onlinecite{BarkeshliReflection} for computing path integrals on non-orientable manifolds. Such constructions could lead to more efficient ways of computing anomaly indicators, and we leave them for future work. 

It is clear that similar ideas should hold in all dimensions. Symmetry fractionalization in general $d$ dimensional TQFTs with $G$ symmetry should define a $d$-category where the objects (0-morphisms) correspond to group elements of $G$, and there is a $G$ action on the $d$-category. This $d$-category should then define a $(d+1)$-dimensional state sum for a $G$-SPT, which then can be used to compute the anomaly. We note that initial steps for computing anomalies in (3+1)D by defining a (4+1)D path integral have already been taken in some simple cases.\cite{kobayashi2019}

Finally, we note that our Hamiltonian constructions explicitly break most spatial symmetries. It would be interesting to develop constructions that preserve any given spatial symmetry. However we note that even the simpler Levin-Wen models for (2+1)D topological orders (and Walker-Wang models for (3+1)D topological orders) have not been generalized to take as input arbitrary unitary (braided) fusion categories in a way that is compatible with point group symmetries on the lattice. 

\begin{acknowledgments}

MB thanks Parsa Bonderson, Meng Cheng, Chao-Ming Jian, Kevin Walker, and Zhenghan Wang for previous collaborations and ongoing discussions. MB also thanks Shawn Cui for helpful conversations. This work is supported by NSF CAREER (DMR- 1753240), Alfred P. Sloan Research Fellowship, and JQI- PFC-UMD. 
  
\end{acknowledgments}

\addtocontents{toc}{\protect\setcounter{tocdepth}{1}}
\appendix

\section{Invariance properties of state sum}
\label{app:invariances}

\subsection{Invariance under Pachner moves}
\label{app:Pachner}

In this appendix, we demonstrate that our path integral is invariant under the 3-3 Pachner move. Other Pachner moves are treated similarly, but the diagrammatics are considerably more involved. Our treatment is essentially identical to that of Ref.~\onlinecite{CuiTQFT}; we reproduce the calculation here to (a) convert the calculation to our conventions and (b) make our paper self-contained.

\begin{figure*}
\subfloat[\label{fig:mergingLemma1}]{\includegraphics[width=1.8\columnwidth]{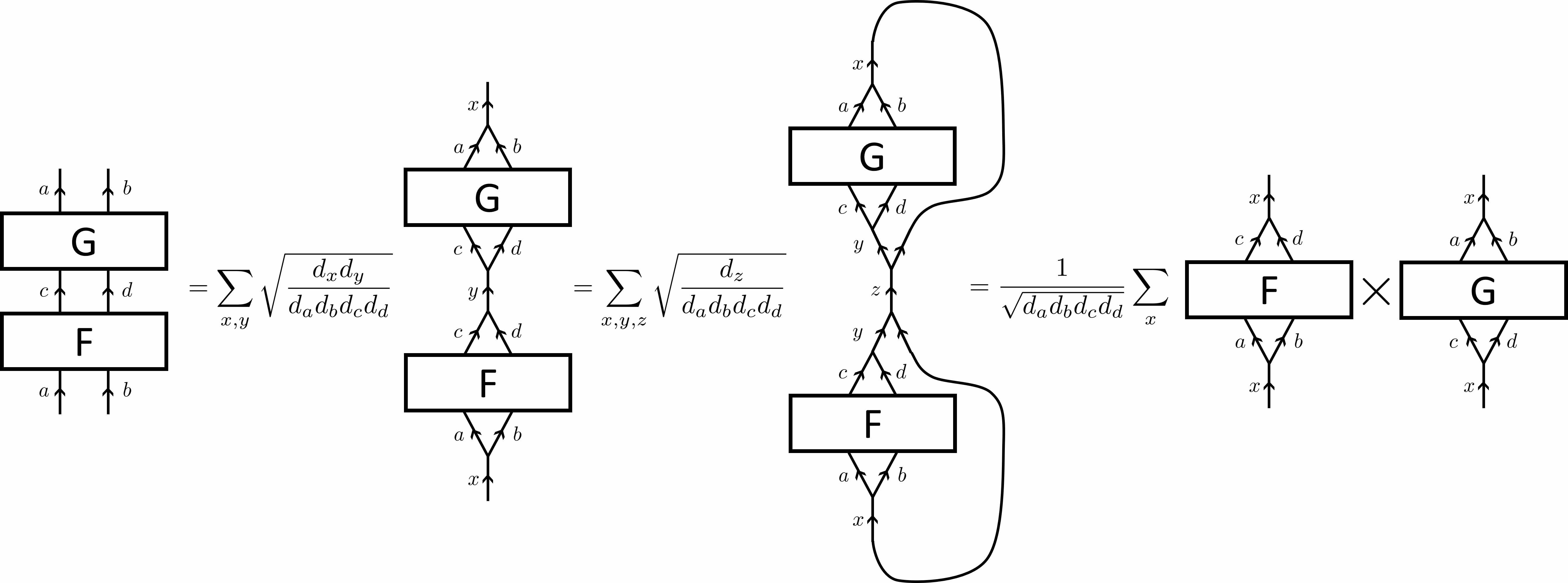}}\\
\subfloat[\label{fig:mergingLemma2}]{\includegraphics[width=1.8\columnwidth]{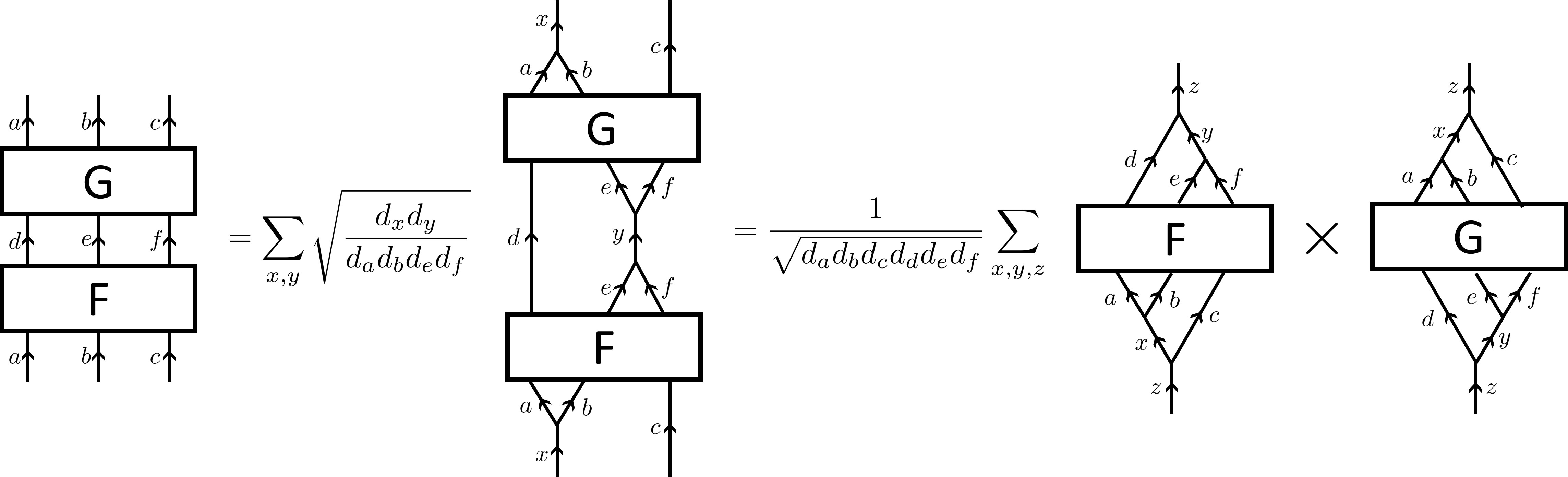}}
\caption{``Merging lemmas", with proof. Each lemma is the equality of the far left and far right-hand-sides of the graphical equation, and $F$ and $G$ are arbitrary diagrams. Dangling anyon lines at the bottom are implied to be connected to their corresponding line at the top. The last equality in (b) follows from applying the result in (a).}
\label{fig:mergingLemmas}
\end{figure*}

Before doing this, we prove two ``merging lemmas" stated in Ref.~\onlinecite{CuiTQFT}. The first is the equality of the far left- and far right-hand sides of Fig.~\ref{fig:mergingLemma1}, where the intermediate equalities are the proof of the merging lemma. The first two equalities follow from the usual UMTC diagrammatic equation
\begin{equation}
\raisebox{-0.5\height}{
\begin{pspicture}[shift=-0.65](-0.1,-0.2)(1.0,1.2)
  \small
  \psset{linewidth=0.9pt,linecolor=black,arrowscale=1.5,arrowinset=0.15}
  \psline{->}(0.25,0)(0.25,0.6)
  \psline(0.25,0)(0.25,1.0)
  \psline{->}(0.7,0)(0.7,0.6)
  \psline(0.7,0)(0.7,1.0)
  \rput[br]{0}(0.15,0.5){$a$}
  \rput[bl]{0}(0.8,0.5){$b$}
 \end{pspicture}}
=\sum_{c} \sqrt{\frac{d_c}{d_a d_b}}
\raisebox{-0.5\height}{
\begin{pspicture}[shift=-0.65](-0.4,-0.2)(1.5,1.3)
  \small
 \psset{linewidth=0.9pt,linecolor=black,arrowscale=1.5,arrowinset=0.15}
  \psline{->}(0.7,0.25)(0.7,0.7)
  \psline(0.7,0.25)(0.7,0.8)
  \psline(0.7,0.8) (0.25,1.25)
  \psline{->}(0.7,0.8)(0.3,1.2)
  \psline(0.7,0.8) (1.15,1.25)	
  \psline{->}(0.7,0.8)(1.1,1.2)
  \psline{->}(0.25,-0.3)(0.6,0.15)
  \psline(0.25,-0.3)(0.7,0.25)
  \psline{->}(1.15,-0.3)(0.8,0.15)
  \psline(1.15,-0.3)(0.7,0.25)
  \rput[bl]{0}(0.4,0.5){$c$}
  \rput[br]{0}(1.4,1.05){$b$}
  \rput[bl]{0}(0,1.05){$a$}
  \rput[bl]{0}(0,-0.2){$a$}
  \rput[br]{0}(1.4,-0.2){$b$}
  \end{pspicture}},
\end{equation}
where we have been implicitly assuming $N_{ab}^c \leq 1$ for all $a,b,c$.
Conservation of total anyon charge means that in the third diagram, the only nonzero term occurs when $z$ is the identity, producing the last equality.

We then use this first merging lemma to prove a second, shown in Fig.~\ref{fig:mergingLemma2}. 

With these lemmas in hand, we can prove invariance under the 3-3 Pachner move. Given five vertices $0,1,2,3,4,5$, a 3-3 Pachner move takes the 4-simplices $01234,01245,02345$, all positively oriented, to the 4-simplices $01235,01345,12345$, also all positively oriented. In doing so, the 2-simplex $024$ is replaced by the 2-simplex $135$ and the 3-simplices $0124,0234,0245$ are replaced by the 3-simplices $0135,1235,1345$. Our aim is to show that
\begin{widetext}
\begin{align}
\sum_{024,0124,0234,0245}\frac{d_{024}}{d_{0124}d_{0234}d_{0245}}&Z^+(01234)Z^+(01245)Z^+(02345) \nonumber \\
&= \sum_{135,0135,1235,1345}\frac{d_{135}}{d_{0135}d_{1235}d_{1345}} Z^+(01235)Z^+(01345)Z^+(12345) \label{eqn:Pachner33}
\end{align}
\end{widetext}
with fixed labels on all other simplices involved.

We first contract all of the domain walls in the diagrams and examine the symmetry fractionalization factors that arise. These are given by the factors of $U$ and $\eta$ in Eq.~\eqref{eqn:Zplusdef}. On the left-hand-side (LHS) of Eq.~\eqref{eqn:Pachner33}, we obtain
\begin{widetext}
\begin{align}
\text{LHS} \propto \frac{U_{34}(013,123,0123)}{U_{34}(023,{}^{32}012,0123)\eta_{012}(23,34)} \times \frac{U_{45}(014,124,0124)}{U_{45}(024,{}^{42}012,0124)\eta_{012}(24,45)} \times \frac{U_{45}(024,234,0234)}{U_{45}(034,{}^{43}023,0234)\eta_{023}(34,45)} 
\label{eqn:PachnerLHS}
\end{align}
\end{widetext}

\begin{figure*}
\includegraphics[width=0.95\columnwidth]{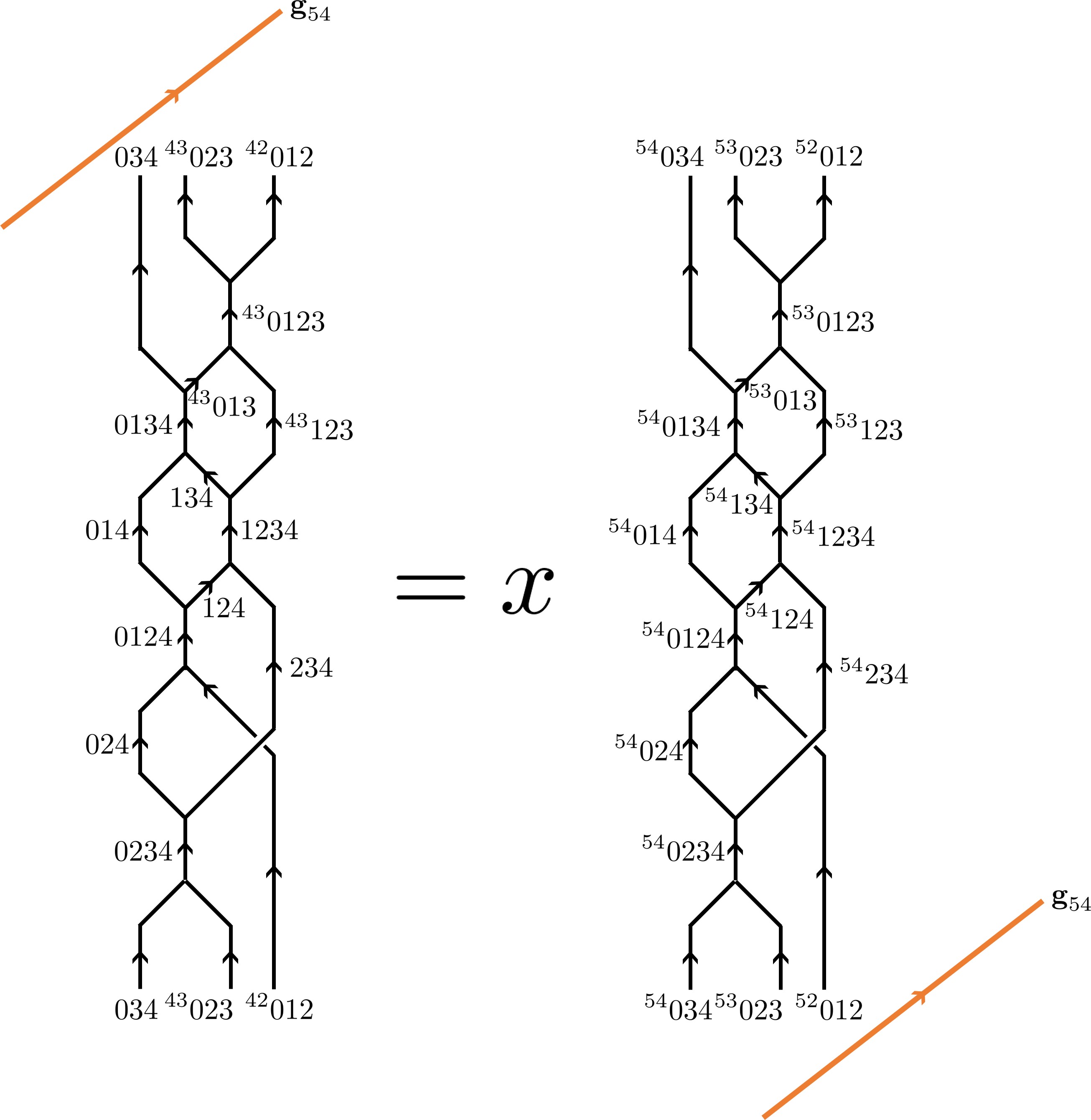}
\caption{Sweeping a ${\bf g}_{54}$ domain wall through the diagram for the 4-simplex $01234$. This is a graphical calculus version of replacing all of the $F$- and $R$-symbols in the evaluation of the diagram in Eq.~\eqref{eqn:Zplusdef} with their ${\bf g}_{54}$-transformed versions using consistency conditions. A trace is implied, i.e. open lines at the top are implied to be connected to their corresponding lines at the bottom.}
\label{fig:Pachner_sweep}
\end{figure*}

Our aim will be to use the merging lemmas in Fig.~\ref{fig:mergingLemmas} to perform the sums on $024,0124,0234,$ and $0245$ in Eq.~\eqref{eqn:Pachner33}. In order to do so, both diagrams should involve, say, $\,^{54}0124$, but as naturally written, the diagram for $01234$ involves $0124$ and the diagram for $01245$ involves $\,^{54}0124$. Hence, we sweep a ${\bf g}_{54}$ domain wall through the diagram for $01234$ as shown in Fig.~\ref{fig:Pachner_sweep}. By the consistency of the graphical calculus, we may evaluate the diagram either before or after this sweeping process; as long as we account for all of the factors of $U_{54}$ or $U_{54}^{-1}$ which arise in the sweeping process, the value of the diagram is unchanged. At the algebraic level, this process simply transforms all of the $F$- and $R$-symbols in the explicit formula Eq.~\eqref{eqn:Zplusdef} for $Z^+(01234)$ by ${\bf g}_{54}$ using the consistency condition Eq.~\eqref{eqn:UFURConsistency}. The proportionality factor $x$ in Fig.~\ref{fig:Pachner_sweep} is given by
\begin{widetext}
\begin{align}
x= \frac{U_{54}(\act{54}{024},\act{54}{234},\act{54}{0234})}{U_{54}(\act{54}{034},\act{53}{023},\act{54}{0234})} &\times 
\frac{U_{54}(\act{54}{014},\act{54}{124},\act{54}{0124})}{U_{54}(\act{54}{024},\act{53}{012},\act{54}{0124})} \times 
\frac{U_{54}(\act{53}{023},\act{52}{012},\act{53}{0123})}{U_{54}(\act{53}{013},\act{53}{123},\act{53}{0123})} \times \nonumber \\
& \times 
\frac{U_{54}(\act{54}{034},\act{53}{013},\act{54}{0134})}{U_{54}(\act{54}{014},\act{53}{134},\act{54}{0134})} \times 
\frac{U_{54}(\act{54}{134},\act{53}{123},\act{54}{1234})}{U_{54}(\act{54}{124},\act{54}{234},\act{54}{1234})}
\end{align}
\end{widetext}
After this transformation, the symmetry factors for the LHS of Eq.~\eqref{eqn:Pachner33} are $x$ times the factors in Eq.~\eqref{eqn:PachnerLHS}.

The symmetry factors for the right-hand side (RHS) of Eq.~\eqref{eqn:Pachner33} are
\begin{widetext}
\begin{align}
\text{RHS} \propto \frac{U_{34}(013,123,0123)}{U_{34}(023,{}^{32}012,0123)\eta_{012}(23,34)} \times \frac{U_{45}(014,124,0124)}{U_{45}(024,{}^{42}012,0124)\eta_{012}(24,45)} \times \frac{U_{45}(024,234,0234)}{U_{45}(034,{}^{43}023,0234)\eta_{023}(34,45)}
\end{align}
\end{widetext}

Laborious but straightforward use of the consistency conditions Eqs.~\eqref{eqn:UEtaConsistency} and \eqref{eqn:etaConsistency}  demonstrates that the symmetry factors on the LHS and the RHS are now equal.

\begin{figure*}
\includegraphics[width=1.95\columnwidth]{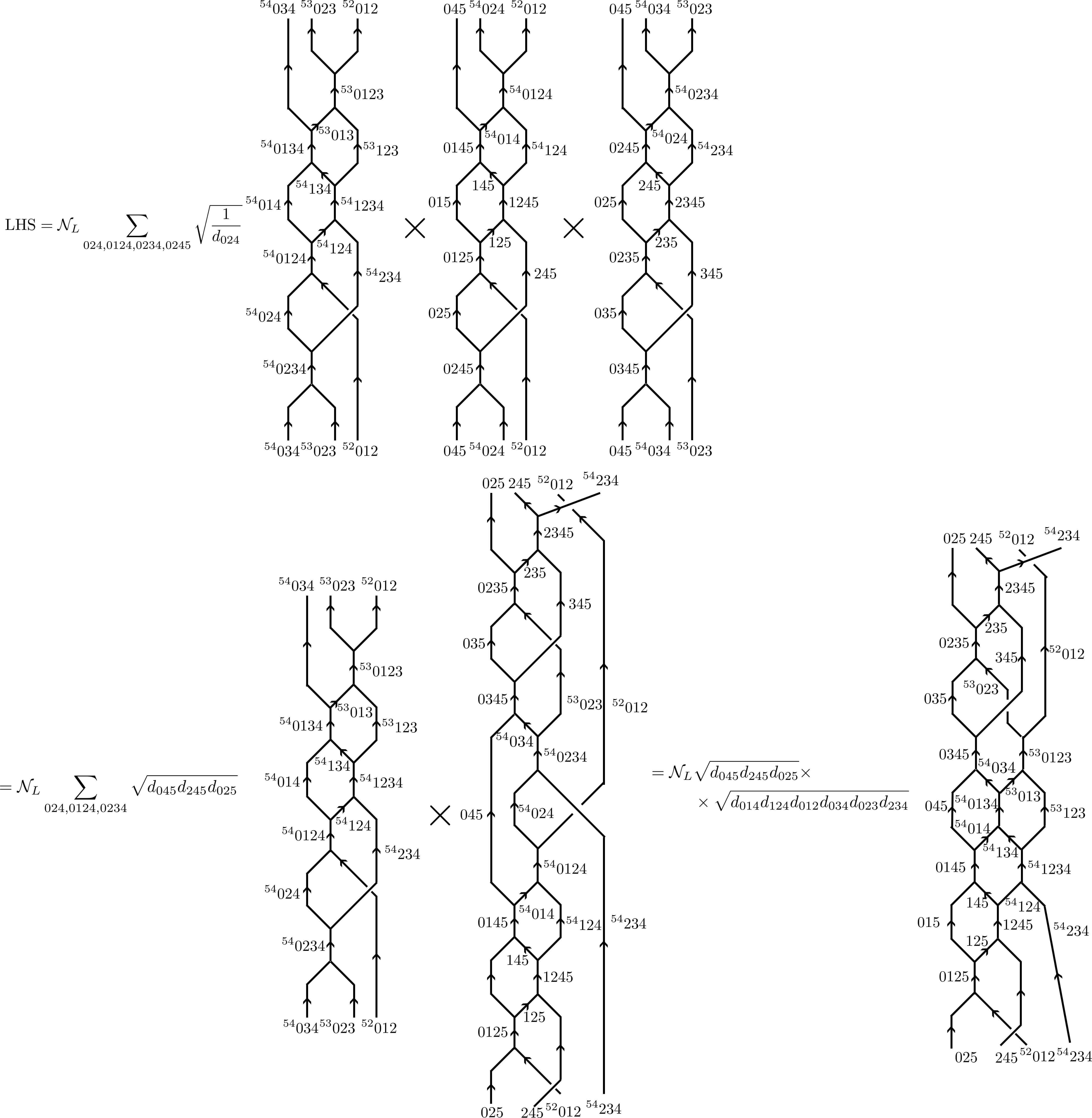}
\caption{Graphical calculus for the left-hand side of Eq.~\eqref{eqn:Pachner33} after sweeping ${\bf g}_{54}$ domain through the diagram for $01234$ (see Fig.~\ref{fig:Pachner_sweep}), excluding symmetry factors. The factor of $\mathcal{N}_L$ is defined in Eq.~\eqref{eqn:NL} and does not involve quantum dimensions of any summands. Traces are implied on all of the diagrams.}
\label{fig:Pachner_LHS}
\end{figure*}

\begin{figure*}
\includegraphics[width=1.95\columnwidth]{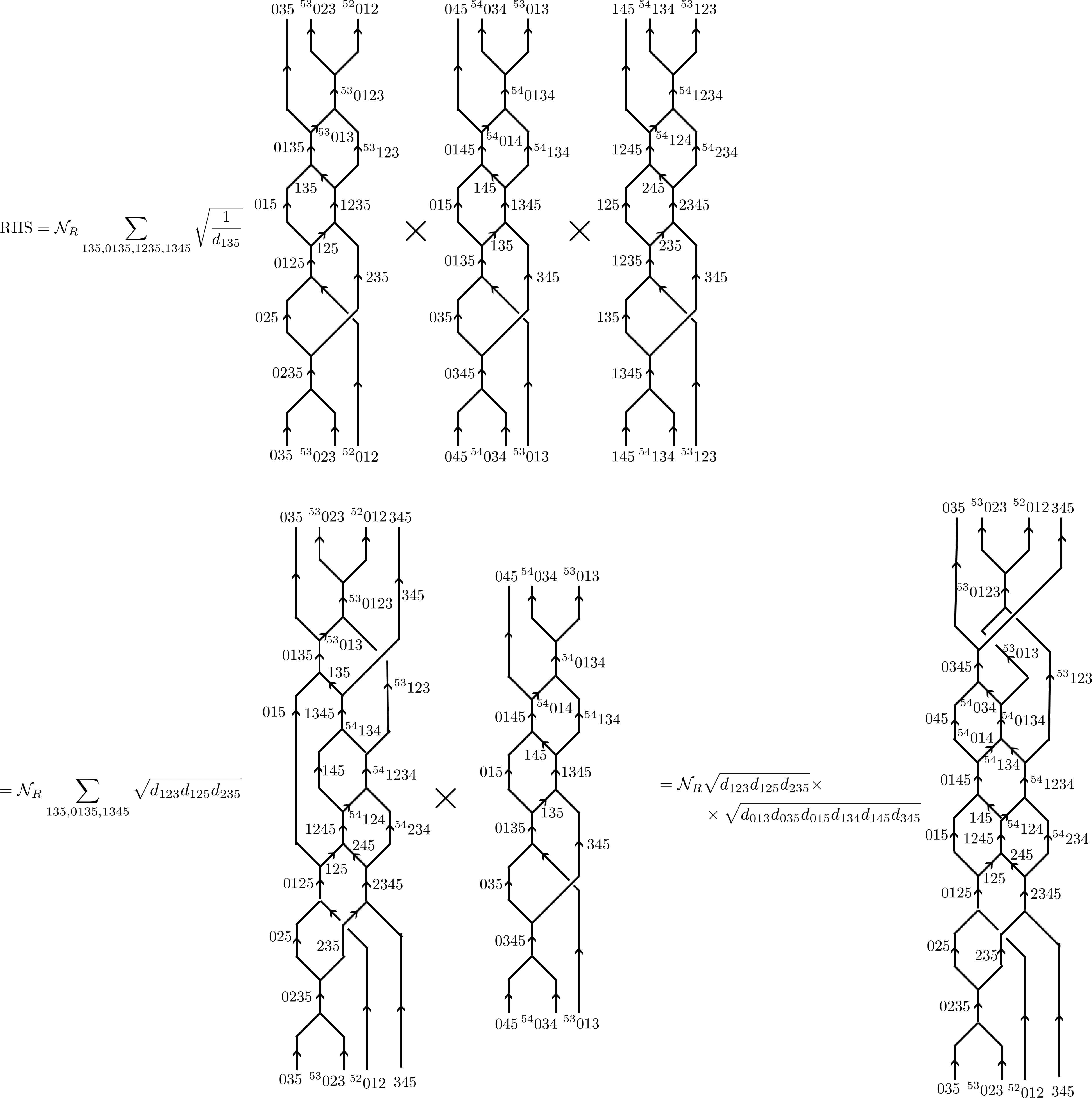}
\caption{Graphical calculus for the right-hand side of Eq.~\eqref{eqn:Pachner33}, excluding symmetry factors. The factor of $\mathcal{N}_R$ is defined in Eq.~\eqref{eqn:NR} and does not involve quantum dimensions of any summands. Traces are implied on all of the diagrams.}
\label{fig:Pachner_RHS}
\end{figure*}

Canceling these symmetry factors reduces each side of Eq.~\eqref{eqn:Pachner33} to of a product of three diagrams with no domain walls or symmetry factors. We define
\begin{align}
\frac{d_{024}}{d_{0124}d_{0234}d_{0245}}\mathcal{N}_{01234}\mathcal{N}_{01245}\mathcal{N}_{02345} &= \mathcal{N}_L\frac{1}{\sqrt{d_{024}}} \label{eqn:NL}\\
\frac{d_{135}}{d_{0135}d_{1235}d_{1345}}\mathcal{N}_{01235}\mathcal{N}_{01345}\mathcal{N}_{12345} &= \mathcal{N}_R\frac{1}{\sqrt{d_{135}}} \label{eqn:NR}
\end{align}
where $\mathcal{N}_{L,R}$ do not involve quantum dimensions of any of the summands in Eq.~\eqref{eqn:Pachner33}. We can now do graphical calculus to demonstrate the equality of those products of diagrams, as shown in Figs.~\ref{fig:Pachner_LHS} and \ref{fig:Pachner_RHS}. Careful examination of the final diagrams shows that both the diagrams and normalization factors are indeed equal, proving Eq.~\eqref{eqn:Pachner33}.

\subsection{Independence of fusion channel on 3-simplices}
In defining our path integral, we made an arbitrary choice of fusion channel for the data on the 3-simplices. Suppose we change the fusion channel; we wish to show that the path integral is unchanged. On a closed manifold, each 3-simplex appears in exactly two 4-simplices with opposite induced orientation. For concreteness we focus on the 3-simplex $0123$ which, again for concreteness, we suppose is in the 4-simplices $01234$ (+ orientation) and $01235$ (- orientation). We define the unitary basis transformation
\begin{equation}
\raisebox{-0.5\height}{
\begin{pspicture}[shift=-0.65](-0.4,-0.2)(1.5,1.3)
  \small
 \psset{linewidth=0.9pt,linecolor=black,arrowscale=1.5,arrowinset=0.15}
  \psline{->}(0.25,-0.3)(0.25,0.55)
  \psline{->}(0.25,0.65)(0.25,1.15)
  \psline(0.25,-0.3)(0.25,1.25)
  \psline{->}(1,-0.3)(1,0.1)
  \psline{->}(1,0.2)(1,1.15)
  \psline(1,-0.3)(1,1.25)
  \psline{->}(1,0.2)(0.4,0.54) 
  \psline(1,0.2)(0.25,0.65)
  \rput[bl]{0}(0.62,0.55){$e$}
  \rput[br]{0}(1.25,1.05){$b$}
  \rput[bl]{0}(0,1.05){$a$}
  \rput[bl]{0}(0,-0.2){$c$}
  \rput[br]{0}(1.25,-0.2){$d$}
  \end{pspicture}}
= \sum_f \left(F^{ab}_{cd}\right)_{ef} \raisebox{-0.5\height}{
\begin{pspicture}[shift=-0.65](-0.4,-0.2)(1.5,1.3)
  \small
 \psset{linewidth=0.9pt,linecolor=black,arrowscale=1.5,arrowinset=0.15}
  \psline{->}(0.7,0.25)(0.7,0.7)
  \psline(0.7,0.25)(0.7,0.8)
  \psline(0.7,0.8) (0.25,1.25)
  \psline{->}(0.7,0.8)(0.3,1.2)
  \psline(0.7,0.8) (1.15,1.25)	
  \psline{->}(0.7,0.8)(1.1,1.2)
  \psline{->}(0.25,-0.3)(0.6,0.15)
  \psline(0.25,-0.3)(0.7,0.25)
  \psline{->}(1.15,-0.3)(0.8,0.15)
  \psline(1.15,-0.3)(0.7,0.25)
  \rput[bl]{0}(0.35,0.43){$f$}
  \rput[br]{0}(1.4,1.05){$b$}
  \rput[bl]{0}(0,1.05){$a$}
  \rput[bl]{0}(0,-0.2){$c$}
  \rput[br]{0}(1.4,-0.2){$d$}
  \end{pspicture}}
\end{equation}
One can check that
\begin{equation}
\left(F^{ab}_{cd}\right)_{ef} = \sqrt{\frac{d_e d_f}{d_a d_d}}\left(F^{ceb}_f\right)^{\ast}_{ad}
\end{equation}
where $F^{ceb}_f$ is the usual $F$-symbol, although this form does not make the unitarity of $F^{ab}_{cd}$ manifest. It can also be checked that
\begin{equation}
\raisebox{-0.5\height}{
\begin{pspicture}[shift=-0.65](-0.4,-0.2)(1.5,1.3)
  \small
 \psset{linewidth=0.9pt,linecolor=black,arrowscale=1.5,arrowinset=0.15}
  \psline{->}(1,-0.3)(1,0.55)
  \psline{->}(1,0.65)(1,1.15)
  \psline(1,-0.3)(1,1.25)
  \psline{->}(0.25,-0.3)(0.25,0.1)
  \psline{->}(0.25,0.2)(0.25,1.15)
  \psline(0.25,-0.3)(0.25,1.25)
  \psline{->}(0.25,0.2)(0.85,0.54) 
  \psline(0.25,0.2)(1,0.65)
  \rput[bl]{0}(0.62,0.55){$e$}
  \rput[br]{0}(1.25,1.05){$d$}
  \rput[bl]{0}(0,1.05){$c$}
  \rput[bl]{0}(0,-0.2){$a$}
  \rput[br]{0}(1.25,-0.2){$b$}
  \end{pspicture}}
= \sum_f \left(F^{ab}_{cd}\right)^{\ast}_{ef} \raisebox{-0.5\height}{
\begin{pspicture}[shift=-0.65](-0.4,-0.2)(1.5,1.3)
  \small
 \psset{linewidth=0.9pt,linecolor=black,arrowscale=1.5,arrowinset=0.15}
  \psline{->}(0.7,0.25)(0.7,0.7)
  \psline(0.7,0.25)(0.7,0.8)
  \psline(0.7,0.8) (0.25,1.25)
  \psline{->}(0.7,0.8)(0.3,1.2)
  \psline(0.7,0.8) (1.15,1.25)	
  \psline{->}(0.7,0.8)(1.1,1.2)
  \psline{->}(0.25,-0.3)(0.6,0.15)
  \psline(0.25,-0.3)(0.7,0.25)
  \psline{->}(1.15,-0.3)(0.8,0.15)
  \psline(1.15,-0.3)(0.7,0.25)
  \rput[bl]{0}(0.35,0.43){$f$}
  \rput[br]{0}(1.4,1.05){$d$}
  \rput[bl]{0}(0,1.05){$c$}
  \rput[bl]{0}(0,-0.2){$a$}
  \rput[br]{0}(1.4,-0.2){$b$}
  \end{pspicture}}
\end{equation}

\begin{figure*}
\includegraphics[width=1.8\columnwidth]{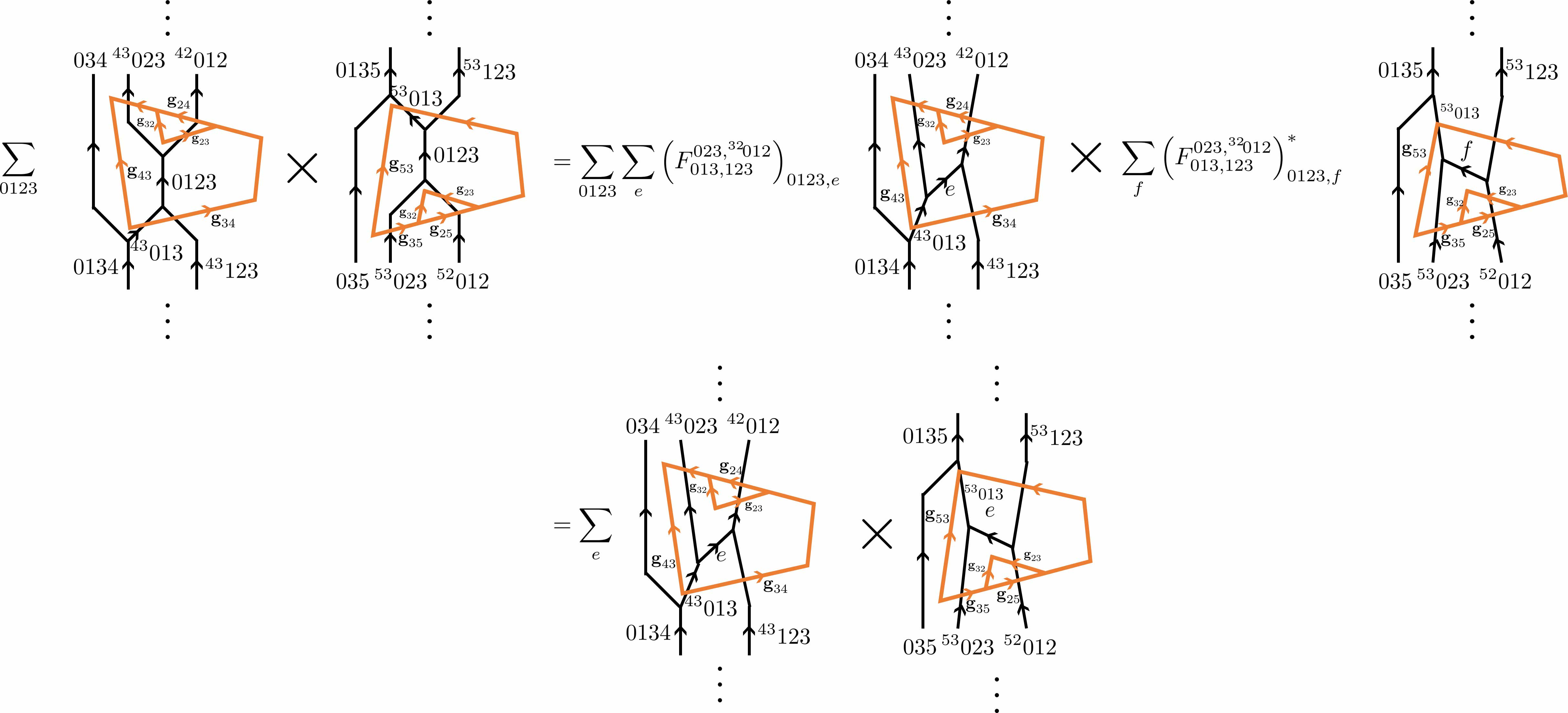}
\caption{Graphical form of Eqs.~\eqref{eqn:3simplexBasis1}-\eqref{eqn:3simplexBasis3} for the product of amplitudes for the 4-simplices $01234$ and $01235$, focusing on the portions of the diagrams involving the 3-simplex $0123$. We have assumed for simplicity that $01234$ and $01235$ have positive and negative orientations, respectively. A pair of $F$-moves changes the basis for the fusion data associated to $0123$, but the $F$-symbols cancel out.}
\label{fig:3simplexBasis}
\end{figure*}

We can relate amplitudes $Z$ in the original fusion channel to amplitudes $Z'$ in the new fusion channel by an $F$-move:
\begin{widetext}
\begin{align}
\sum_{b_{0123}} Z^+(01234;b_{0123})Z^-(01235;b_{0123})&=\sum_{b_{0123}}\sum_e  \left(F^{023,{}^{32}012}_{013,123}\right)_{0123,e}Z^{'+}(01234;e) \times \nonumber \\
&\hspace{1cm}\times\sum_f \left(F^{023,{}^{32}012}_{013,123}\right)^{\ast}_{0123,f}Z^{'-}(01235;f) \label{eqn:3simplexBasis1}\\
&= \sum_{e,f}\delta_{e,f}Z^{'+}(01234;e)Z^{'-}(01235;f) \label{eqn:3simplexBasis2}\\
&= \sum_{b'_{0123}}Z^{'+}(01234;b'_{0123})Z^{'-}(01235;b'_{0123}) \label{eqn:3simplexBasis3}
\end{align}
\end{widetext}
where the second line follows from the unitarity of the $F$-symbols and in the third line we relabeled $e \rightarrow b'_{0123}$. We have explicitly labeled the relevant 3-cell as it appears in each amplitude $Z^{\pm}$. This equation is shown graphically in Fig.~\ref{fig:3simplexBasis}.

The right-hand side is precisely the amplitude which would appear if we had chosen to construct the path integral using the new fusion basis for the 3-simplex $0123$. Therefore the choice of fusion channel does not affect the partition function for a closed manifold. The same is true for an open manifold if the 3-simplex is not on the surface. If the 3-simplex is on the surface, then there is no cancellation; instead the amplitude in the wavefunction of the initial configuration is related to the amplitude of the final configuration by an $F$-symbol, which is as expected.

\subsection{Independence of deformation of 3-simplex data}

\begin{figure*}
\includegraphics[width=1.8\columnwidth]{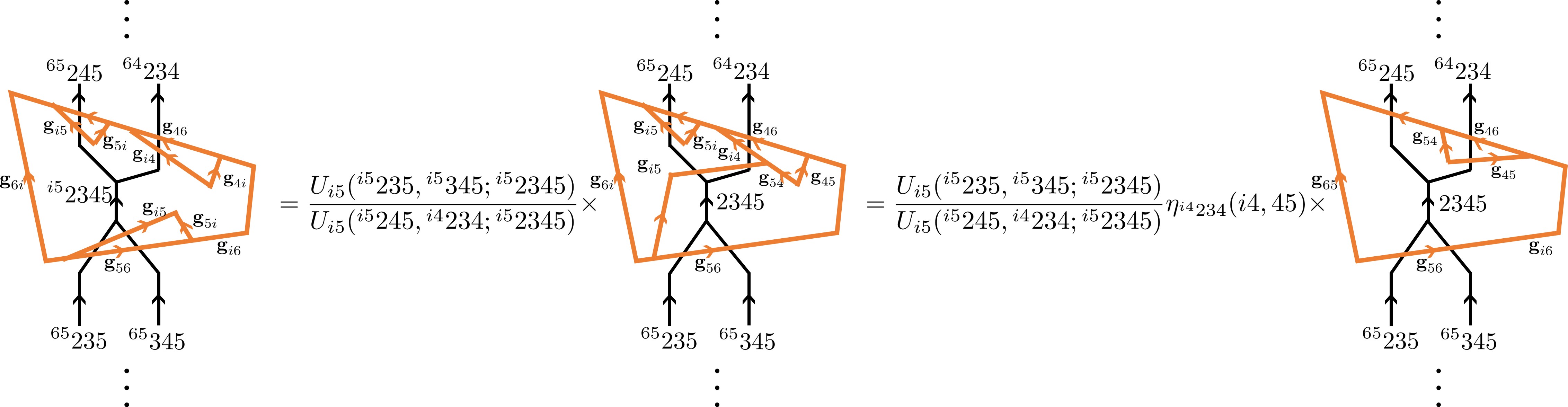}
\caption{Graphical calculus illustrating Eq.~\eqref{eqn:3simplexDeformation}, that is, relating the amplitude $Z^{'+}(23456;{}^{i5}\!b_{2345})$ (left) with the 3-simplex $2345$ deformed towards vertex $i=2,3,4$ to the amplitude $Z^+(23456;b_{2345})$ (right) with $2345$ deformed towards vertex $5$. We only show the portion of the diagram that involves the 3-simplex $2345$. The consistency condition for $\eta$ has been used in the last equality.}
\label{fig:3simplexDeformation}
\end{figure*}

We also chose a particular deformation of the data on the 3-simplices towards 0-simplices, as in Fig.~\ref{fig:tetrahedronDual}(c). Suppose that instead of deforming a $3$-simplex, call it $2345$, towards ${\bf g}_5$, we deformed it towards ${\bf g}_{i}$ instead, where $i=2,3,$ or $4$. Our claim is that the term in the state sum with data $b_{2345}$ on $2345$ in the original deformation is equal to the term in the state sum with ${}^{i5}\!b_{2345}$ on $2345$ in the new deformation. The reason is as follows. For each 4-simplex $\Delta^4$ containing $2345$ the amplitude $Z'(\Delta^4;{}^{i5}\!b_{2345})$ corresponding to $\Delta^4$ in the new deformation can be related to the amplitude $Z(\Delta^4;b_{2345})$ in the original deformation by sliding anyon lines through domain walls as shown in Fig.~\ref{fig:3simplexDeformation}. The labeling of $\Delta^4$ could be $m2345$ for $m<2$, or $2m345$ for $2<m<3$, etc. Doing the graphical calculus, shown in Fig.~\ref{fig:3simplexDeformation} for the case $\sigma=23456$ with positive orientation and checked similarly for the other cases, leads to (for all possible $\Delta^4$)
\begin{widetext}
\begin{equation}
\frac{Z'(\Delta^4;{}^{i5}\!b_{2345})}{Z(\Delta^4;b_{2345})} = \left(\frac{U_{i5}({}^{i5}\!235,{}^{i5}\!345;{}^{i5}\!2345)}{U_{i5}({}^{i5}\!245,{}^{i5}\!234;{}^{i5}\!2345)}\times \eta_{{}^{i4}\!234}(i4,45)\right)^{\epsilon(2345;\Delta^4)}
\label{eqn:3simplexDeformation}
\end{equation}
\end{widetext}

Here $\epsilon(2345;\Delta^4) = \pm$ is the relative orientation of $2345$ to the 4-simplex $\Delta^4$. If the 3-simplex is in the bulk of the spacetime manifold, then $2345$ appears in exactly two 4-simplices $\Delta^4_1$ and $\Delta^4_2$, one with each relative orientation. Hence, for every term in the state sum, the above factors coming from $\Delta^4_1$ and $\Delta^4_2$ cancel out, and the state sum is unchanged. If $2345$ is on the surface, then the aforementioned phase factor is the relative amplitude in the wavefunction between two states where the braided fusion diagrams labeling the state differ by sliding the appropriate domain walls over anyon lines, which is as we expect.

\subsection{Independence of deformation of 2-simplex data}

Another choice was a deformation of the 2-simplex data towards a 0-simplex. Consider a $2$-simplex $234$. For simplicity of presentation we assume there are no vertices $m$ such that the branching structure orders $2<m<3$ or $3<m<4$; this assumption may be relaxed at the cost of checking a few extra cases in what follows. Suppose that we deformed $234$ towards ${\bf g}_j$ for $j=2$ or $3$ instead of towards ${\bf g}_4$. We relate the term in the newly deformed state sum with ${}^{i4}\!a_{234}$ on the simplex in question to the term in the old deformation with $a_{234}$.

\begin{figure}
\includegraphics[width=0.9\columnwidth]{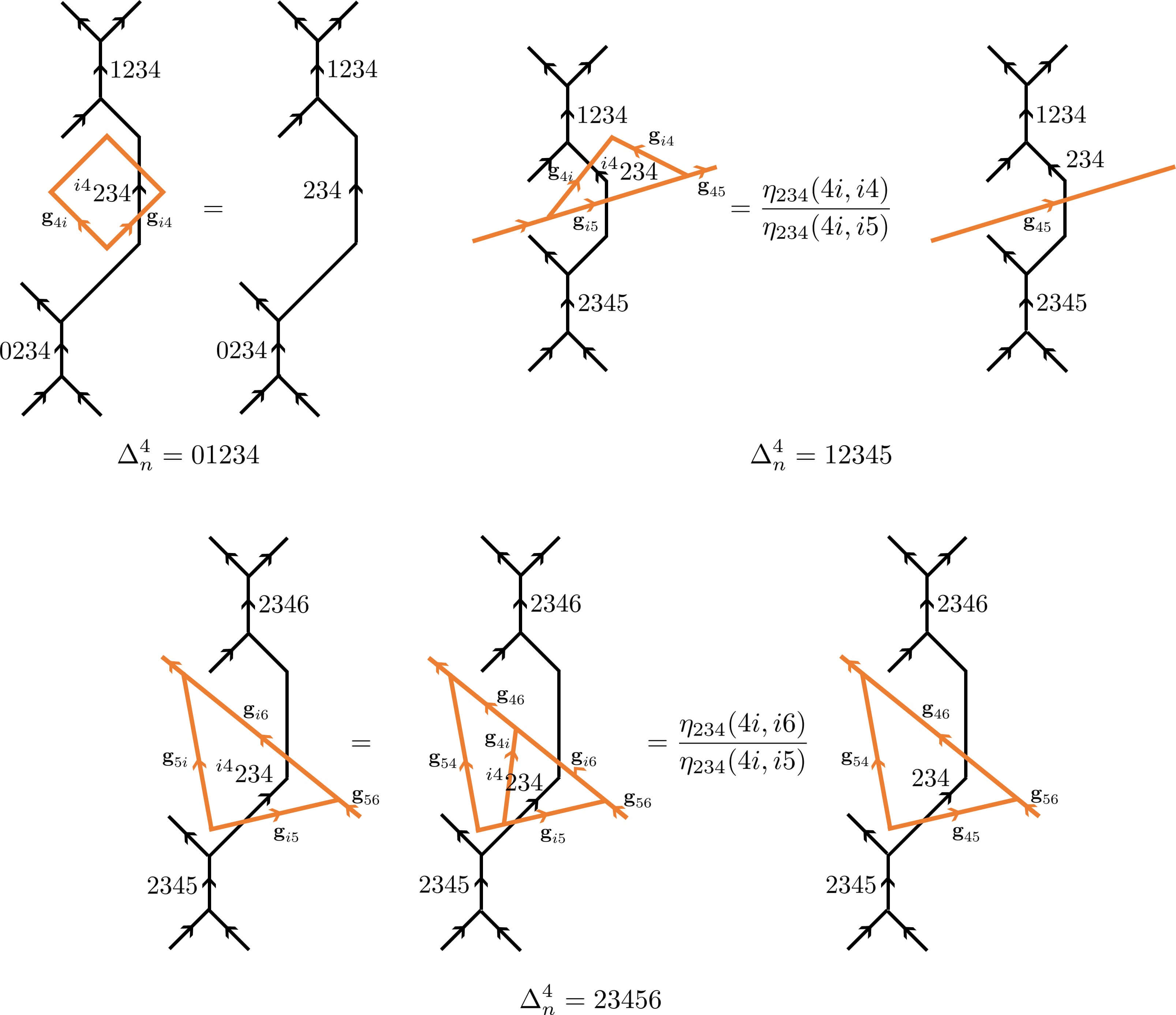}
\caption{Graphical calculus illustrating Eq.~\eqref{eqn:Zratio2Simplex}, that is, comparing the amplitude $Z'(\Delta^4_n;{}^{i4}234)$ associated to the 4-simplex $\Delta^4_n$ with the 2-simplex 234 deformed towards vertex $i=2$ or $3$ (left-hand diagram for each case) to the amplitude $Z(\Delta^4_n;234)$ with 234 deformed towards vertex 4 (right-hand diagram for each case). We only show the portion of the diagram involving 234.}
\label{fig:2simplexDeformation}
\end{figure}

Let us examine each 4-simplex $\Delta^4_n$ which contains $234$, where $n$ enumerates all such 4-simplices. There are three possibilities for $\Delta^4_n$; without loss of generality, $\Delta^4_n$ can be labeled $01234,12345,$ or $23456$, where all that matters is whether the additional vertices are smaller than 2 or larger than 4. For each of these three cases one can determine the ratio of the amplitudes $Z(\Delta^4_n;a_{234})$ and $Z'(\Delta^4_n;{}^{i4}\!a_{234})$ with the old and new deformations, respectively. The portions of the diagrams which differ for each of these cases are shown in Fig.~\ref{fig:2simplexDeformation}. Graphical calculus shows
\begin{widetext}
\begin{align}
\frac{Z'(\Delta^4_n;{}^{i2}\!a_{012})}{Z(\Delta^4_n;a_{012})} =  \begin{cases}
1 & \Delta^4_n=01234\\
\eta^{-\epsilon(\Delta^4_n)}_{234}(4i,i5)\eta^{\epsilon(\Delta^4_n)}_{234}(4i,i4) & \Delta^4_n=12345\\
\eta^{-\epsilon(\Delta^4_n)}_{234}(4i,i5)\eta^{\epsilon(\Delta^4_n)}_{234}(4i,i6) & \Delta^4_n = 23456
\end{cases}
\label{eqn:Zratio2Simplex}
\end{align}
\end{widetext}

Of course, the 4-simplex $\Delta^4_n$ always contains exactly two 3-simplices $\Delta^3_{n;m}$ which involve $234$, where $m$ enumerates the two 3-simplices. Each 3-simplex is labeled either $\Delta^3_{n;m}=m234$ or $234m$ for some $m$. Define
\begin{equation}
\zeta(\Delta^3_{n;m}) =\begin{cases}
\eta_{234}(4i,i4) & m<2\\
\eta^{-1}_{234}(4i,im) & m>4
\end{cases}
\end{equation}
Then we see that Eq.~\eqref{eqn:Zratio2Simplex} can be rewritten
\begin{equation}
\frac{Z'(\Delta^4_n;{}^{i4}\!a_{234})}{Z(\Delta^4_n;a_{234})} = \prod_{\Delta^3_{n;m} \in \Delta^4_n} \zeta(\Delta^3_{n;m})^{\epsilon(\Delta^3_{n;m};\Delta^4_n)}
\end{equation}
where the product is over all 3-simplices in $\Delta^4_n$ containing $234$ and $\epsilon(\Delta^3_{n;m};\Delta^4_n)$ is the orientation of $\Delta^3_{n;m}$ induced by $\Delta^4_n$. But every bulk 3-simplex appears in exactly two 4-simplices with opposite induced orientations. Since $\zeta$ depends only on the 3-simplex labels, it is now clear that the bulk factors of $\zeta$ all cancel pairwise, that is, for a closed manifold,
\begin{equation}
\prod_n \frac{Z'(\Delta^4_n;{}^{i4}\!a_{234})}{Z(\Delta^4_n;a_{234})} = 1
\end{equation}
Therefore, the two terms in question in the state sum are equal, as desired.

\subsection{Independence of branching structure}

In Ref.~\onlinecite{CuiTQFT}, Cui proves the invariance of a closely related TQFT under changes of branching structure. The proof constructs, for each configuration in the state-sum with the original branching structure, a corresponding configuration in the state-sum with the new branching structure, then shows that the amplitude associated to that configuration is the same with both branching structures. The construction and proof are rather involved, but apply \textit{mutatis mutandis} to our state-sum. We therefore do not reproduce the proof here.

\subsection{Gauge invariance under vertex basis gauge transformations}

Consider a basis transformation on the fusion or splitting spaces given by a unitary map  $\Gamma^{ab}_c : V^{ab}_c \rightarrow V^{ab}_c$. Our claim is that on a closed manifold $M$, the path integral is invariant under this basis transformation. On a closed manifold, each 3-simplex appears twice, once with positive orientation and once with negative orientation. Hence, for a fixed term in the state sum, each fusion vertex (say, the fusion vertex associated to a positively oriented 3-simplex) will appear exactly once and its dual splitting vertex (say, the splitting vertex associated to the same 3-simplex with opposite orientation) will also appear exactly once. Let $Z(M;\lbrace a,b, {\bf g}\rbrace, \mu,\nu)$ be the path integral amplitude with fixed anyon and group data, but we place the splitting vertex in the state $\mu$ and the fusion vertex in the state $\nu$. The gauge-transformed path integral involves sums over the terms with $\mu = \nu$, so we compute that sum:
\begin{align}
\sum_{\mu} &Z'(M;\lbrace a,b, {\bf g}\rbrace, \mu,\mu) = \nonumber \\
&=\sum_{\mu,\nu, \nu'} \left(\Gamma^{ab}_c\right)_{\mu \nu}Z(M;\lbrace a, b, {\bf g} \rbrace, \nu,\nu')\left(\Gamma^{ab}_c\right)^{\dagger}_{\nu',\mu}\\
&=\sum_{\nu, \nu'}\delta_{\nu,\nu'}Z(M;\lbrace a, b, {\bf g} \rbrace, \nu,\nu')\\
&=\sum_{\nu}Z(M;\lbrace a, b, {\bf g} \rbrace, \nu,\nu)
\end{align}
where the prime indicates the gauge-transformed amplitude. The final expression is exactly the same amplitude in the original gauge, which demonstrates the desired equality.

On an open manifold, these gauge transformations on bulk 3-simplices cancel in the same way. The gauge transformations do not cancel on surface 3-simplices, but instead transform the wavefunction in the expected way.

In the presence of anti-unitary symmetries, the gauge transformation rule needs to account for the group elements; the gauge transformation takes the form
\begin{equation}
\widehat{\ket{a,b;c;{\bf g}}} = \tilde{\Gamma}^{ab}_c({\bf g})\ket{a,b;c;{\bf g}}
\end{equation}
where the hat refers to the gauge-transformed state and the $abc$ fusion vertex is in domain ${\bf g}$. Demanding $G$-equivariance of the gauge transformations, we find
\begin{equation}
\tilde{\Gamma}^{ab}_c({\bf g}) = \left(\tilde{\Gamma}^{ab}_c\right)^{\sigma({\bf g})}({\bf 1}) \equiv \left(\Gamma^{ab}_c\right)^{\sigma({\bf g})}
\end{equation} 
where the last equality defines the shorthand $\Gamma^{ab}_c$. Using this object, the above argument for gauge invariance of the path integral goes through exactly on an orientable manifold.

\begin{figure}
\includegraphics[width=0.6\columnwidth]{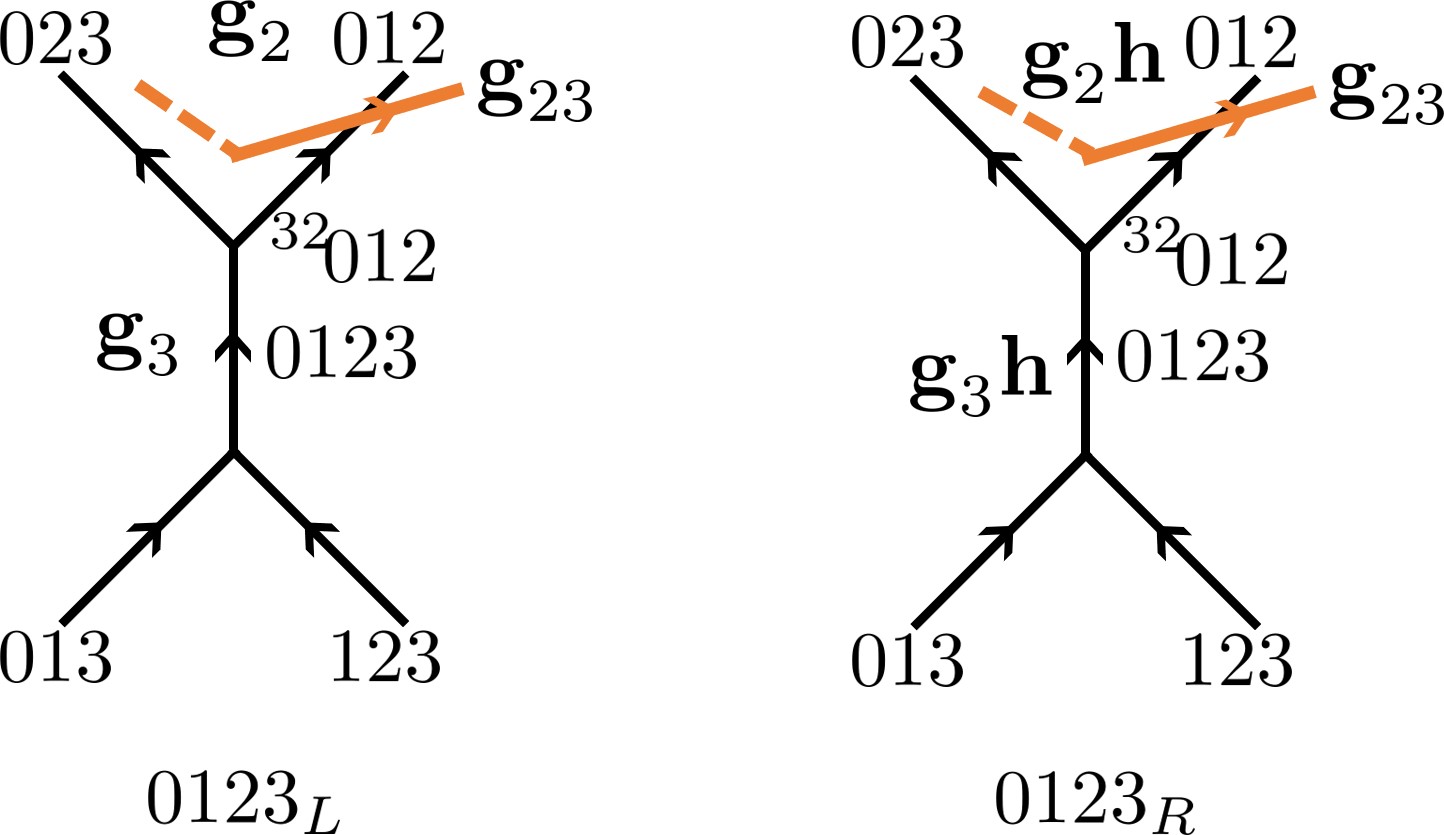}
\caption{Data associated to a 3-simplex $0123$ as it appears on the left and right sides of a cut-open, unorientable manifold. Both appearances of $0123$ have the same relative orientation, but the group elements ${\bf g_i}$ associated to it are twisted by an anti-unitary group element ${\bf h}$. Solid orange lines are domain walls and dashed orange lines indicate some way of connecting the domain wall to others that depends on the details of the 4-simplex in which the 3-simplex appears.}
\label{fig:cut3simplicesUnorientable}
\end{figure}

On an unorientable manifold, in order to define the path integral, as described in Sec.~\ref{sec:antiunitary}, we should cut open the manifold to obtain an orientable open manifold. For any 3-simplex which does not lie entirely on the cut, the argument goes through as above. However, on the cut, we identify the data on two 3-simplices with the \textit{same} relative orientation. The key difference is that the group elements on the vertices of those two 3-simplices are twisted by an anti-unitary group element ${\bf h}$. That is, if $0123$ is a 3-simplex lying on the cut and has positive induced orientation, then $0123$ appears in two diagrams in the path integral shown in Fig.~\ref{fig:cut3simplicesUnorientable}. These two diagrams pick up the gauge transformation
\begin{equation}
\tilde{\Gamma}^{ab}_c({\bf g}_3) \tilde{\Gamma}^{ab}_c({\bf g}_3{\bf h}) = \tilde{\Gamma}^{ab}_c({\bf g}_3) \left(\tilde{\Gamma}^{ab}_c({\bf g}_3)\right)^{\ast} = 1
\end{equation}
where we have used the fact that $\sigma({\bf h}) = \ast$. Hence the path integral remains gauge-invariant. The fact that the boundary conditions for the group elements are twisted by a group element ${\bf h}$ with an anti-unitary action is what ensures that the path integral is gauge-invariant on an unorientable manifold.

\subsection{Gauge invariance under symmetry action gauge transformations}

\begin{figure}
\subfloat[\label{fig:gaugeTransformCrossings}]{\includegraphics[width=0.6\columnwidth]{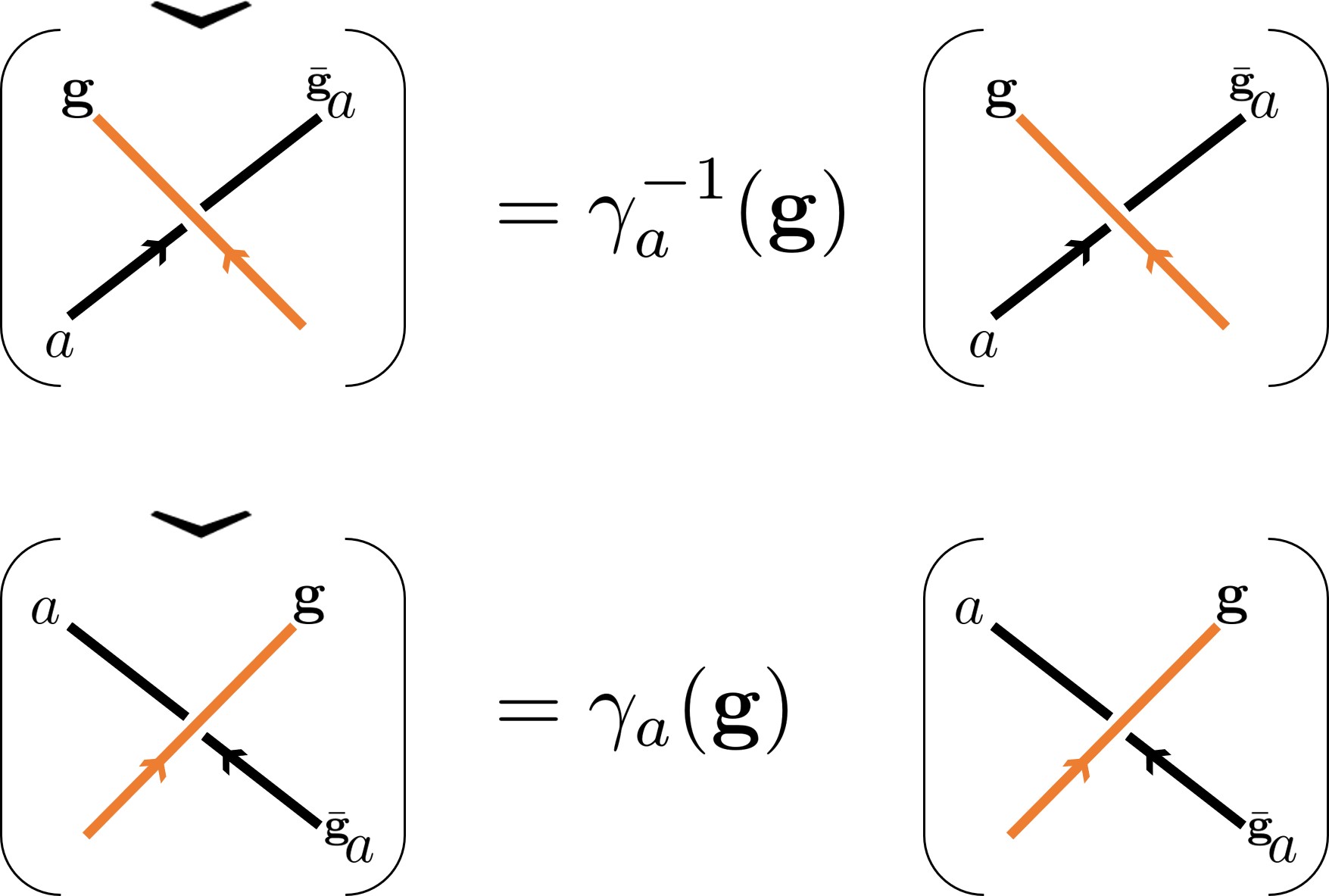}}\\
\subfloat[\label{fig:gaugeTransformCrossingsAnti}]{\includegraphics[width=0.6\columnwidth]{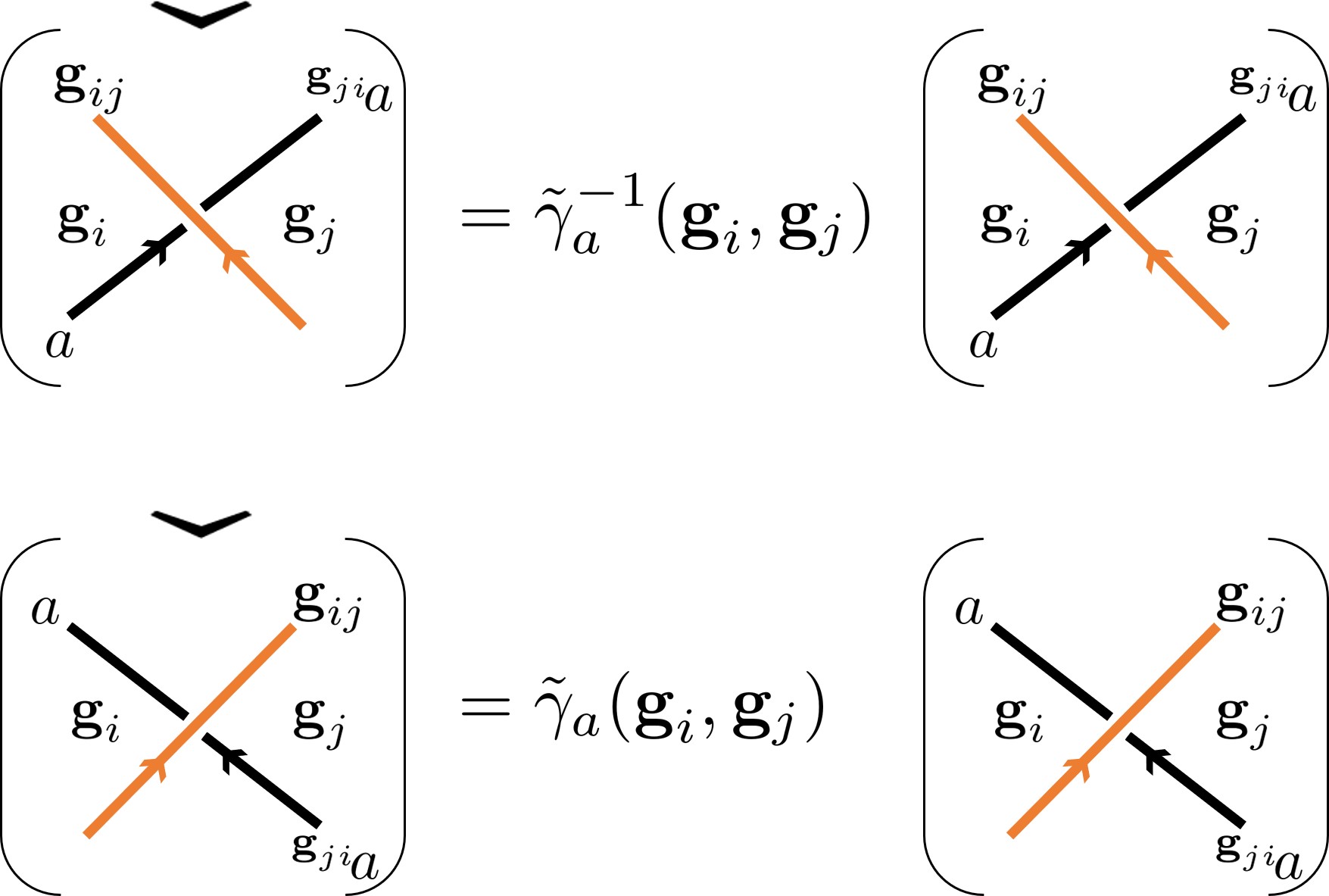}}
\caption{Gauge transformation rules for changing the factorization of the symmetry action into its topological and local parts. Anyon line is black, domain wall is orange. Checks indicate gauge-transformed diagrams. (a) Unitary case. (b) Anti-unitary case, tracking the group elements on the 0-simplices explicitly.}
\end{figure}

Gauge transforming the symmetry action changes the way that the global symmetry action factorizes into its topological and local parts according to Eq.~\eqref{symloc}. Given an anyon diagram with domain walls, the gauge transformation changes the diagram by a phase $\gamma_a({\bf g})$ for every crossing of a domain wall of type ${\bf g}$ with an anyon $a$ as shown in Fig.~\ref{fig:gaugeTransformCrossings}, where checks over diagrams represent the gauge-transformed diagram. We presently check that our state-sum is invariant under this transformation.

It is straightforward to check by inspecting Fig.~\ref{fig:diagramGluing} that the amplitudes $Z^{\pm}$ for a 4-simplex, each crossing of a domain wall over an anyon line in the full diagram can be associated to a 3-simplex, i.e. the set of crossings on the left-hand-side of Fig.~\ref{fig:diagramGluing} is in correspondence with the crossings on the right-hand-side. Using the diagrammatic rules in Fig.~\ref{fig:gaugeTransformCrossings}, we see that the crossing in a 3-simplex $ijk\ell$ (here $i,j,k,\ell$ label vertices) appearing in a 4-simplex $\Delta^4$ picks up a factor of $\gamma^{\epsilon(ijk\ell;\Delta^4)}_{ijk}({\bf g}_{k\ell})$ under gauge transformations, where $\epsilon(ijk\ell;\Delta^4)$ is the orientation of $ijk\ell$ induced by the 4-simplex $\Delta^4$. Since, on a closed manifold $M$, each 3-simplex appears in two 4-simplices, each leading to opposite induced orientation, the factors of $\gamma$ all cancel out in computing a single term of the path integral on $M$. Hence the path integral on a closed manifold is gauge invariant.

On a manifold $M$ with boundary, the gauge transformations of crossings associated to bulk 3-simplices still cancel out. However, gauge transformations of crossings involved in surface 3-simplices do not cancel. Instead, they modify the wave function by changing the value of the wave function to match the gauge-transformed value of the closed anyon diagram on the boundary of $M$, as expected.

In the presence of anti-unitary symmetries, we should modify the gauge transformation rules Fig.~\ref{fig:gaugeTransformCrossings} to those in Fig.~\ref{fig:gaugeTransformCrossingsAnti}, which allows us to define the shorthand objects $\gamma_a({\bf g}_{ij})$ by the equation
\begin{align}
\tilde{\gamma}_a({\bf g}_i,{\bf g}_j) &= \left(\tilde{\gamma}_a({\bf 1}, {\bf g}_{ji})\right)^{\sigma({\bf g})} \nonumber \\
&\equiv \left(\gamma_a({\bf g}_{ij})\right)^{\sigma({\bf g})}
\end{align}
With this definition, the only difference is that the crossings in a 3-simplex pick up the gauge transformation factor $\gamma^{(\epsilon(ijk\ell;\Delta^4)\times \sigma(c)}_{ijk}(k\ell)$. On an orientable manifold, the above argument carries through straightforwardly. 

On an unorientable manifold, we consider the same setup discussed in the previous section on vertex gauge transformations. For 3-simplices which do not lie entirely on the cut, the argument for the unitary case goes through, \textit{mutatis mutandis}. For a 3-simplex $0123$ on the cut, its two appearances are, as before, shown in the diagrams in Fig.~\ref{fig:cut3simplicesUnorientable}. These two diagrams pick up the gauge transformation factors $\tilde{\gamma}_{012}({\bf g}_i, {\bf g}_j)$ and $\tilde{\gamma}_{012}({\bf g}_i{\bf h}, {\bf g}_j{\bf h})=\tilde{\gamma}_{012}({\bf g}_i, {\bf g}_j)$ respectively, where ${\bf h}$ is some anti-unitary group element. But $\sigma(h)=\ast$, hence
\begin{equation}
\tilde{\gamma}_{012}({\bf g}_i, {\bf g}_j) \tilde{\gamma}_{012}({\bf g}_i{\bf h}, {\bf g}_j{\bf h})=\tilde{\gamma}_{012}({\bf g}_i, {\bf g}_j)\tilde{\gamma}_{012}({\bf g}_i, {\bf g}_j)^{\ast}=1
\end{equation}
Hence the path integral is gauge-invariant, even on an unorientable manifold. As for the vertex transformations, this gauge invariance fundamentally requires that the boundary conditions for the group elements are twisted by a group element ${\bf h}$ with an \textit{anti-unitary} action.

\subsection{Mirroring 3-simplices about the vertical axis}
\label{vertMirror}

Our path integral is built by assigning a fusion diagram to the 3-simplex $0123$, corresponding to the process $013 \otimes 123 \rightarrow 023 \otimes \act{32}{012}$. Mathematically this corresponds to an element of $\bigoplus_b V_{013,123}^b \otimes V_b^{023,\act{32}{012}}$ (or, in the category theory terminology, this corresponds to an element of the Hom space Hom$(013 \otimes 123, 023 \otimes \act{32}{012})$). Given an orientation of the 3-simplex, the 2-simplices $013$ and $123$ have the same induced orientation, which is why $013$ and $123$ both appear in the domain (or codomain, for the opposite orientation of the 3-simplex) of this process. However, it is not a priori clear why we should not mirror the entire diagram about the vertical axis when constructing the amplitudes associated to 4-simplices; that is, use Hom$(123 \otimes 013,\act{32}{012}\otimes 023)$ instead.

\begin{figure*}
\includegraphics[width=1.8\columnwidth]{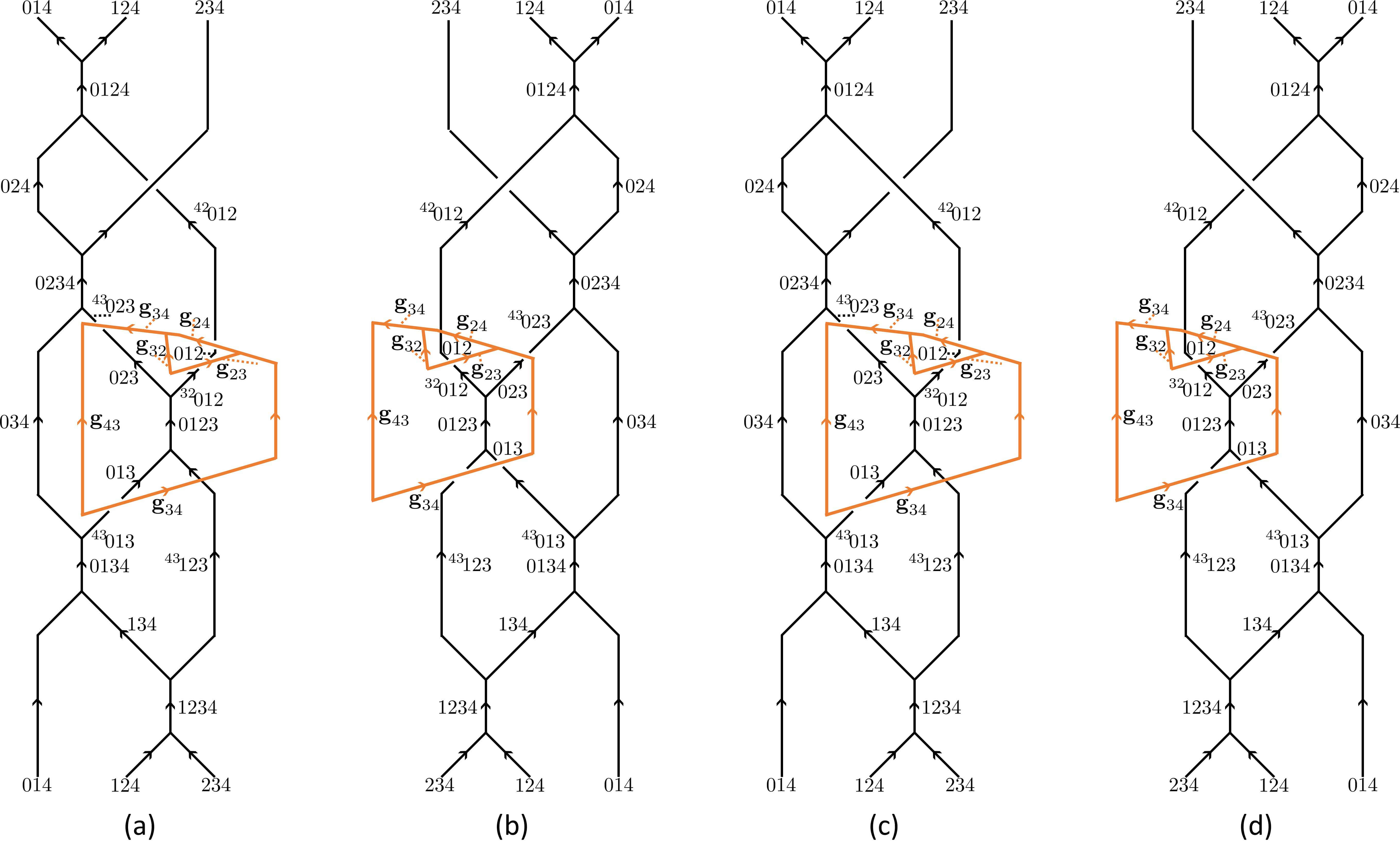}
\caption{Different possible diagrams used to define $Z^+(01234)$. On a closed manifold, (a) and (b) produce the same path integral, as do (c) and (d). (a) and (c) are related by changing the over-crossing to an under-crossing, which we conjecture complex conjugates the path integral on closed manifolds.  (a) Diagram used in the main text. (b) The 3-simplex data in (a) has been mirrored about the vertical axis and the overcrossing in (a) is switched to an undercrossing. (c) The overcrossing in (a) is switched to an undercrossing without changing the 3-simplex data. (d) The 3-simplex data in (a) has been mirrored about the vertical axis without changing the crossing.}
\label{fig:mirroredAboutVertical}
\end{figure*}

We claim that on a closed manifold, the path integral is invariant under this mirroring, provided that the over/under-crossing in the diagram is also changed. That is, constructing $Z^+$ using the diagram in Fig.~\ref{fig:mirroredAboutVertical}(a) yields the same path integral as in Fig.~\ref{fig:mirroredAboutVertical}(b), and likewise Figs.~\ref{fig:mirroredAboutVertical}(c) and (d) would yield the same path integral on a closed manifold.

Note that mirroring the diagram about the vertical axis and changing the over- to under- crossing is equivalent to rotating the entire diagram about the vertical axis by $\pi$. Thus another way of stating the result is that the path integral is invariant under rotating the diagrams associated with all 4-simplices by $\pi$ around the vertical axis. 

The idea of the proof is the following. Start with Fig.~\ref{fig:mirroredAboutVertical}(a) and shrink the domain walls, evaluating the symmetry factors as appropriate. Next, slide all of the 3-simplex lines into their positions in Fig.~\ref{fig:mirroredAboutVertical}(b). This introduces many twists into the diagram. Then, remove those twists with $R$-moves. Finally, reintroduce the domain walls as appropriate; this produces, after some isotopy, Fig.~\ref{fig:mirroredAboutVertical}(b). The whole process introduces a factor of $R^{ab}_c$ for each fusion space and a factor of $\left(R^{ab}_c\right)^{-1}$ for each splitting space in the diagram (use of the consistency of $U$ and $R$ moves is required for the $0123$ vertex inside the domain walls). Since each 3-simplex appears once with each orientation in a closed manifold, each vertex appears once as a fusion vertex and once as a splitting vertex. Hence the $R$-moves cancel.

On a non-orientable manifold, we consider, as usual, cutting the manifold to produce an orientable one. For all 3-simplices away from the cut, the above argument works identically. The only difference is for 3-simplices on the cut, which appear once on each side of the cut with the same induced orientation. However, there is a twist of the group elements on both sides of the cut; the 3-simplex lives in domain ${\bf g}$ on one side and ${\bf g\bar{h}}$ on the other side, where $\sigma({\bf h})=\ast$. Hence we pick up $\tilde{R}^{ab}_c({\bf g})\tilde{R}^{ab}_c({\bf g\bar{h}})$, which still cancels out by the $G$-equivariance of the $R$-symbols.

\subsection{Changing over-crossing to under-crossing}
\label{crossChange}

In this section, we discuss changing the crossing in the diagrams defining our state-sum (Fig.~\ref{fig:diagramGluing}) from an over-crossing to an under-crossing.

This crossing is the source of the factor of $R$ in the explicit formulas Eqs.~\eqref{eqn:Zplusdef} and \eqref{eqn:Zminusdef}, and it is easy to check that changing the crossing replaces $R^{ab}_c$ by $\left[R^{ba}_c\right]^{-1}$. By inspection of the consistency conditions for BTCs and symmetry fractionalization, if the data $\left\lbrace F, R^{ab}_c, U, \eta \right\rbrace$ defines a consistent BTC $\mathcal{C}_1$ with symmetry fractionalization, then the data $\left\lbrace F, \left[R^{ba}_c\right]^{-1}, U, \eta \right\rbrace$ also defines a consistent BTC $\mathcal{C}_2$ with symmetry fractionalization. Therefore, changing the crossing interchanges the path integrals for these two theories.

The modular $S$ and $T$ matrices, where $T$ is a diagonal matrix with
\begin{equation}
T_{aa} = \theta_a,
\end{equation} 
clearly obey $T_2 = T_1^{\ast}$, $S_2 = S_1^{\ast}$ where $S_i,T_i$ belong to the BTC $\mathcal{C}_i$.

It immediately follows that, on a closed manifold, the Crane-Yetter invariant for $\mathcal{C}_1$ and $\mathcal{C}_2$ are complex conjugates. This is because for a
UMTC $\mathcal{C}$, 
\begin{equation}
Z_{CY}(M^4) = e^{2\pi i c \sigma(M^4)/8} .
\end{equation}
for the Crane-Yetter invariant. Here $\sigma(M^4)$ is the signature of the closed 4-manifold $M^4$ and $c$ is the chiral central charge given by
\begin{equation}
 e^{2\pi i c/8} = \frac{1}{\mathcal{D}}\sum_a d_a^2 \theta_a 
\end{equation}

For our state sum with the $G$ symmetry action, it is less clear to see directly from the path integral what happens when we replace $R^{ab}_c$ with $[R^{ba}_c]^{-1}$ and thus effectively interchange $\mathcal{C}_1$ with $\mathcal{C}_2$. We will prove (subject to a mild additional assumption) that when the input to our state sum is a UMTC, this interchange complex conjugates our path integral. The technique is indirect, using the relative anomaly formalism of Ref.~\onlinecite{BarkeshliRelativeAnomaly}, but the result is borne out in the examples that we study in Sec. \ref{exampleSec} and also the example of $G = \mathbb{Z}_3\times \mathbb{Z}_3$ with the unique rank 3 Abelian UMTC with $\mathbb{Z}_3$ fusion rules.

The method of proof is to show that, given two fractionalization classes given by data $\left\lbrace F, R^{ab}_c, U, \eta^{(1)} \right\rbrace$ and $\left\lbrace F, R^{ab}_c, U, \eta^{(2)} \right\rbrace$, the path integral on any closed $4$-manifold $M$ (with a general background $G$ gauge field) obeys
\begin{equation}
\frac{Z[M;F,R^{ab}_c,U,\eta^{(2)}]}{Z[M;F,R^{ab}_c,U,\eta^{(1)}]} = \frac{Z[M;F,\left[R^{ba}_c\right]^{-1},U,\eta^{(1)}]}{Z[M;F,\left[R^{ba}_c\right]^{-1},U,\eta^{(2)}]}
\end{equation}
Rephrased in terms of anomalies, we wish to show that the anomaly of $\left\lbrace F, R^{ab}_c, U, \eta^{(2)} \right\rbrace$ relative to $\left\lbrace F, R^{ab}_c, U, \eta^{(1)} \right\rbrace$ is the inverse of the anomaly of $\left\lbrace F, \left[R^{ba}_c\right]^{-1}, U, \eta^{(2)} \right\rbrace$ relative to $\left\lbrace F, \left[R^{ba}_c\right]^{-1}, U, \eta^{(1)} \right\rbrace$. Under the additional assumption that some non-anomalous class exists for any $U$, setting $\eta^{(1)}$ to the non-anomalous class proves the desired equality, since on any manifold a non-anomalous SPT has $Z=1$.

In general, different fractionalization classes form an $\mathcal{H}^2(G,\mathcal{A})$ torsor\cite{barkeshli2019}, where $\mathcal{A}$ is the set of Abelian anyons in the theory. That is, given any two fractionalization classes $\left\lbrace F, R^{ab}_c, U, \eta^{(i)} \right\rbrace$ for $i=1,2$, there exists $\coho{t}({\bf g},{\bf h}) \in \mathcal{H}^2(G,\mathcal{A})$ such that
\begin{equation}
\eta_x^{(2)}({\bf g,h}) = M_{x,\tgh{g}{h}}\eta_x^{(1)}({\bf g,h}),
\label{eqn:relEta}
\end{equation}
where $M_{ab}$ is the double braid defined in Eq.~\eqref{doubleBraid}. The anomaly $[\coho{O}] \in \mathcal{H}^4(G,U(1))$ of $\left\lbrace F, R^{ab}_c, U, \eta^{(2)} \right\rbrace$ relative to $\left\lbrace F, R^{ab}_c, U, \eta^{(1)} \right\rbrace$ is\cite{BarkeshliRelativeAnomaly}
\begin{widetext}
\begin{align}
\coho{O}({\bf g},{\bf h},{\bf k},{\bf l}) = 
R^{\act{gh}{\tgh{k}{l}},\tgh{g}{h}}&
\eta^{(1)}_{\act{gh}{\tgh{k}{l}}}({\bf g},{\bf h})
\left[U_{\bf g}(\act{g}{\coho{t}({\bf hk},{\bf l})},\act{g}{\coho{t}({\bf h},{\bf k})})\right]^{\ast} U_{\bf g}(\act{g}{\coho{t}({\bf h},{\bf kl})},\act{gh}{\coho{t}({\bf k},{\bf l})}) \times \nonumber \\
&\times F^{\tgh{ghk}{l},\tgh{gh}{k},\tgh{g}{h}}F^{\tgh{g}{hkl},\act{g}{\tgh{hk}{l}},\act{g}{\tgh{h}{k}}}F^{\tgh{gh}{kl},\tgh{g}{h},\act{gh}{\tgh{k}{l}}} \times \nonumber \\
&\times \left[F^{\tgh{ghk}{l},\tgh{g}{hk},\act{g}{\tgh{h}{k}}}F^{\tgh{g}{hkl},\act{g}{\tgh{h}{kl}},\act{gh}{\tgh{k}{l}}}F^{\tgh{gh}{kl},\act{gh}{\tgh{k}{l}},\tgh{g}{h}}\right]^{\ast}
\label{eqn:relativeAnomaly}
\end{align}
\end{widetext}
In what follows, the factors of $U$ and $F$ will be unchanged throughout, so we suppress all of their arguments.

We now reinterpret $\coho{O}$ as a relative anomaly between theories with the $R$-symbol $\left[R^{ba}_c\right]^{-1}$. Since $\coho{t}({\bf g},{\bf h})$ is Abelian, we have
\begin{equation}
M_{x,\tgh{g}{h}} = R^{x,\tgh{g}{h}}R^{\tgh{g}{h},x}.
\end{equation}
Therefore, according to Eq.~\eqref{eqn:relEta},
\begin{equation}
R^{x,\tgh{g}{h}}\eta^{(1)}_x({\bf g},{\bf h})=\left[R^{\tgh{g}{h},x}\right]^{-1}\eta_x^{(2)}({\bf g},{\bf h})
\label{eqn:RetaReta}
\end{equation}
Setting $x=\act{gh}{\coho{t}({\bf g},{\bf h})}$ and substituting into Eq.~\eqref{eqn:relativeAnomaly}, we obtain
\begin{equation}
\coho{O} = \left[R^{\tgh{g}{h},\act{gh}{\tgh{k}{l}}}\right]^{-1}\eta_{\act{gh}{\tgh{k}{l}}}^{(2)}({\bf g},{\bf h}) U^{\ast}UFFF(FFF)^{\ast}.
\end{equation}
Comparing this expression to Eq.~\eqref{eqn:relativeAnomaly}, we directly see that we can interpret $\coho{O}$ as the anomaly of $\left\lbrace F, \left[R^{ba}_c\right]^{-1}, U, \eta' \right\rbrace$ relative to $\left\lbrace F, \left[R^{ba}_c\right]^{-1}, U, \eta^{(2)} \right\rbrace$, where $\eta'$ is obtained from $\eta^{(2)}$ using $\coho{t}$ in Eq.~\eqref{eqn:relEta} with the double braid computed with the $R$-symbol $\left[R^{ba}_c\right]^{-1}$. Explicitly,
\begin{align}
\eta'_x({\bf g,h}) &= \eta^{(2)}_x({\bf g},{\bf h})\left[R^{\tgh{g}{h},x}\right]^{-1}\left[R^{x,\tgh{g}{h}}\right]^{-1} \\
&=\eta^{(1)}_x({\bf g},{\bf h}),
\end{align}
where the last equality follows from Eq.~\eqref{eqn:RetaReta}. Hence $\eta'=\eta^{(1)}$. Furthermore, since relative anomalies form an Abelian group, the anomaly of $\left\lbrace F, \left[R^{ba}_c\right]^{-1}, U, \eta^{(1)} \right\rbrace$ relative to $\left\lbrace F, \left[R^{ba}_c\right]^{-1}, U, \eta^{(2)} \right\rbrace$ is the inverse of the anomaly of $\left\lbrace F, \left[R^{ba}_c\right]^{-1}, U, \eta^{(2)} \right\rbrace$ relative to $\left\lbrace F, \left[R^{ba}_c\right]^{-1}, U, \eta^{(1)} \right\rbrace$. Thus we have proven that $\coho{O}$, defined as the anomaly of $\left\lbrace F, R^{ab}_c, U, \eta^{(2)} \right\rbrace$ relative to $\left\lbrace F, R^{ab}_c, U, \eta^{(1)} \right\rbrace$, is also the inverse of the anomaly of $\left\lbrace F, \left[R^{ba}_c\right]^{-1}, U, \eta^{(2)} \right\rbrace$ relative to $\left\lbrace F, \left[R^{ba}_c\right]^{-1}, U, \eta^{(1)} \right\rbrace$, as claimed.

\section{Exhaustiveness of cohomology invariants}
\label{app:exhaustiveInvariants}

In Sec.~\ref{sec:anomalyIndicators}, we constructed sets of cohomology invariants for $\mathcal{H}^n(G,U(1))$ using the Smith normal form of a matrix version of the coboundary operator. In this appendix, we demonstrate that those invariants fully characterize the cohomology group. That is, following the notation in Sec.~\ref{sec:anomalyIndicators}
\begin{equation}
\mathcal{H}^n(G,U(1)) = \bigoplus_{i | D_{ii}>1} \mathbb{Z}_{D_{ii}}
\label{eqn:Hd}
\end{equation}

In the following, when no ambiguity results we will abuse terminology slightly by using ``cochain" (and ``cocycle", etc.) to interchangably mean either a function $\omega: G^n \rightarrow U(1)$ or its ``vectorized" form $v$, where $v$ is a real-valued vector of length $|G|^n$ defined modulo integers.

Recall first that $n$-cocycles are vectors which satisfy Eq.~\eqref{eqn:cocycleM}. Treating $M_n$ as a map between vector spaces over $\mathbb{R}$, it follows that all elements of $\ker M_n$ are cocycles, but not all cocycles are in $\ker M_n$. However, all $n$-coboundaries are in $\im \, M_{n-1}$ (modulo the integer ambiguity).

Next we claim that
\begin{equation}
\ker M_n = \im \, M_{n-1} .
\label{eqn:kerImM}
\end{equation}
To see this, note that by construction, $M_n$ is the matrix of the coboundary map for $\mathcal{H}^n(G,\mathbb{R})$. (Note that $M_n$ is an unambiguously defined integer matrix, while vectors which represent cocycles over $U(1)$ are only defined modulo integers.) Hence, generators of $\ker M_n/\im \, M_{n-1}$ correspond to generators of $\mathcal{H}^n(G,\mathbb{R})$. The latter, however, is known to be trivial for any finite group $G$ for $n>0$; that is, $\ker M_n/\im \, M_{n-1}$ is trivial, which proves Eq.~\eqref{eqn:kerImM}. One way to prove this latter fact is to use the universal coefficient theorem, which, for present purposes\footnote{The universal coefficient theorem is usually stated as the existence of a split exact sequence, but the fact that all the groups involved are known to be Abelian means we can rewrite it in the form Eq.~\eqref{eqn:UCT}.}, states that\cite{HatcherAT}
\begin{equation}
\mathcal{H}^n(G,\mathbb{R}) = \mbox{Hom}(\mathcal{H}_n(G,\mathbb{Z}),\mathbb{R}) \oplus \mbox{Ext}(\mathcal{H}_{n-1}(G,\mathbb{Z}),\mathbb{R})
\label{eqn:UCT}
\end{equation}
where $\mathcal{H}_k(G,\mathbb{Z})$ are the homology groups of $G$, $\mbox{Hom}(H_1,H_2)$ is the group of homomorphisms from $H_1$ to $H_2$, and $\mbox{Ext}$ is the Ext functor, whose detailed definition we do not need here.
The important properties of $\mbox{Ext}$ are that, for any finitely generated groups $G_1,G_2$, and Abelian group $H$,\cite{HatcherAT}
\begin{align}
\mbox{Ext}(G_1 \oplus G_2, H) &=\mbox{Ext}(G_1, H) \oplus \mbox{Ext}(G_2,H)\\
\mbox{Ext}(\mathbb{R},H) &= 0\\
\mbox{Ext}(\mathbb{Z}_k,H) &= H/kH
\end{align}
Using the fact that $\mathbb{R}/k\mathbb{R} = 0$, it follows that $\mbox{Ext}(\mathcal{H}_{n-1}(G,\mathbb{Z}),\mathbb{R})=0$. Further, homology groups for $n>0$ over $\mathbb{Z}$ are finite\cite{BrownCohomology}, so by inspection $\mbox{Hom}(\mathcal{H}_n(G,\mathbb{Z}),\mathbb{R})$ is trivial. Thus $\mathcal{H}^n(G,\mathbb{R})=0$, as claimed.

We now claim that if two cocycles have the same invariants $\mathcal{I}_i$ as defined in Eq. \ref{eqn:cohomologyInvariants}, then they differ by an element of $\im \, d_{n-1}$; that is, they are in the same cohomology class. This is enough to prove equality in Eq.~\eqref{eqn:Hd}.

Let $v$ and $w$ be two cocycles (over $\U$) with identical $\mathcal{I}_i$ for all $i$. Then for each $i$ with $D_{ii}>0$, 
\begin{equation}
n_i \equiv \sum_{j} B_{ij}\left(v_j -w_j\right) \in \mathbb{Z}
\end{equation}
where this equation defines $n_i$. Since $B$ has an integer inverse,
\begin{equation}
\Delta v_i \equiv \sum_{j; D_{jj}>0} (B^{-1})_{ij} n_j \in \mathbb{Z}
\end{equation}
Recalling that $v$ is ambiguous up to an integer vector, we can freely replace $v \rightarrow v - \Delta v$ without changing the cocycle. We now claim that, as a strict equality (over the reals),
\begin{equation}
M_n\left(v - \Delta v - w\right) = 0.
\end{equation}
This is readily verified by computation of
\begin{align}
\left(M_n \Delta v\right)_i &= \sum_{j,k,\ell | D_{\ell \ell} >0}A_{ij} D_{jj} B_{jk} B^{-1}_{k\ell} n_{\ell}\\
&= \sum_{\ell | D_{\ell \ell}>0}A_{i\ell}D_{\ell \ell}n_{\ell}\\
&= \sum_{\ell,m}A_{i\ell}D_{\ell \ell}B_{\ell m}\left(v_m - w_m\right)\\
&= \left(M_n \left(v - w\right)\right)_i
\end{align}
In the third line, we were free to extend the sum over $\ell$ because the summand is zero when $D_{\ell \ell}= 0$ anyway, and we substituted the definition of $n_i$.

Hence $(v - \Delta v) - w \in \ker M_n$. But we showed earlier that $\ker M_n = \im \, M_{n-1}$ Hence the cocycles represented by $v$ and $w$ are in the same cohomology class. This means that our invariants $\mathcal{I}_i$ distinguish all cohomology classes; that is,
\begin{equation}
\mathcal{H}^d(G,U(1)) \subseteq \bigoplus_{i | D_{ii}>1} \mathbb{Z}_{D_{ii}}
\label{eqn:HSubsetZ}
\end{equation}
Furthermore, as we demonstrated in Sec.~\ref{sec:anomalyIndicators}, for each possible set of values of the invariants $\mathcal{I}_i$ in Eq.~\eqref{eqn:cohomologyInvariants} (each of which is a $(D_{ii})$-th root of unity), there exists a cocycle defined in Eq.~\eqref{eqn:explicitCocycle} which realizes that set of values. Hence 
\begin{equation}
\bigoplus_{i | D_{ii}>1} \mathbb{Z}_{D_{ii}} \subseteq \mathcal{H}^d(G,U(1))
\end{equation}
which, with Eq.~\eqref{eqn:HSubsetZ}, proves Eq.~\eqref{eqn:Hd}. 

\section{Cellulations of manifolds $M \times S^1$}
\label{app:cellulations}

In this appendix, we give a general procedure for cellulating $M \times S^1$, where $M$ is a manifold of arbitrary dimension with a known cellulation. We then apply this procedure to some particular cases of $M$ to obtain cellulations used elsewhere in the paper.

\begin{figure}
\centering
\subfloat[\label{fig:D0S1cellulation}]{\includegraphics[width=0.6\columnwidth]{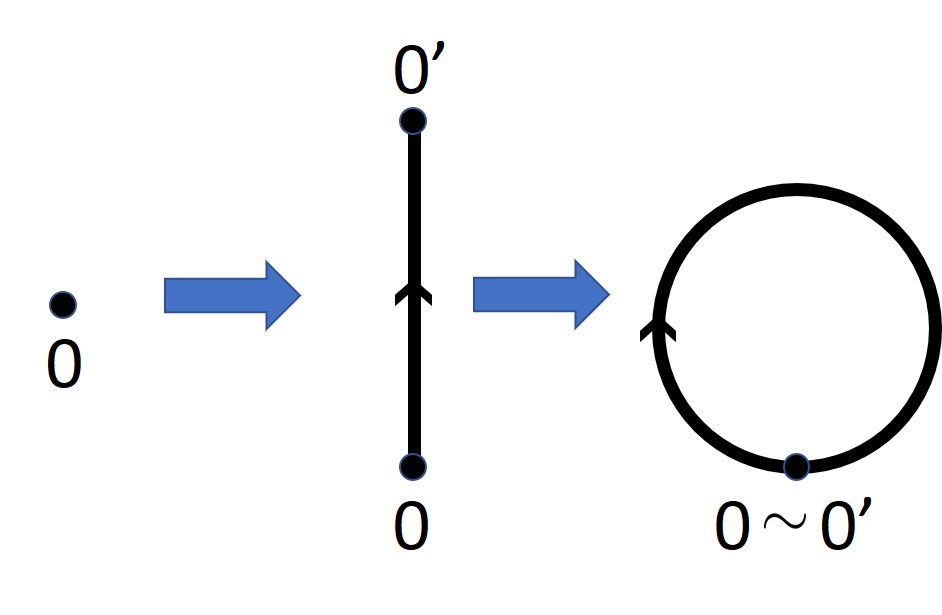}}\\
\subfloat[\label{fig:D1S1cellulation}]{\includegraphics[width=0.8\columnwidth]{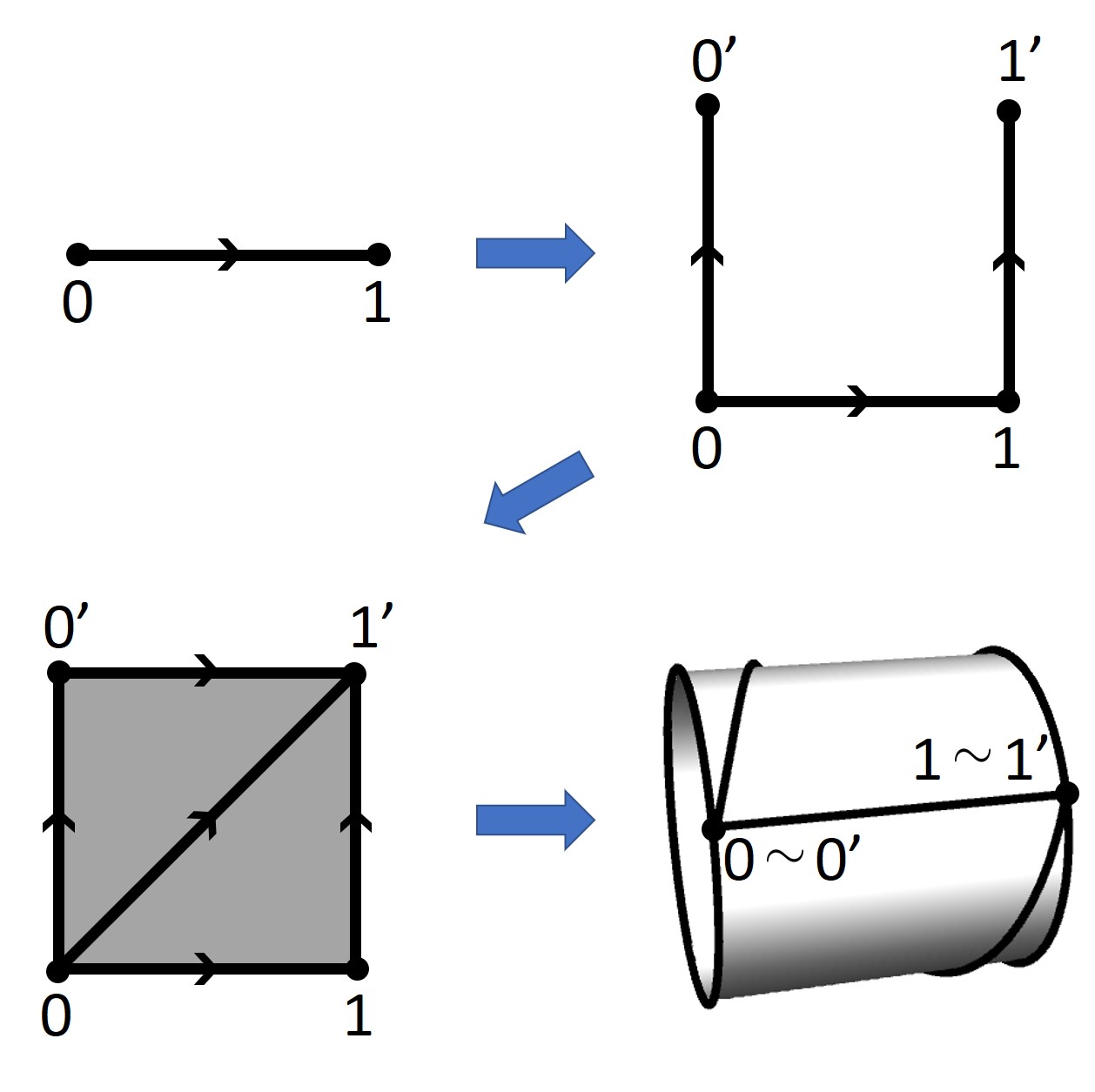}}
\caption{(a) Generating a cellulation of $D^0 \times S^1$ from a cellulation of $D^0$. (b) Generating a cellulation of $D^1 \times S^1$ from a cellulation of $D^1$.}
\end{figure}

We start with the case where $M$ consists of disjoint $d$-simplices. Start with a 0-simplex, i.e. a single vertex $0$. Obviously a point times $S^1$ is just $S^1$, which can be cellulated with a single 0-simplex $0$ and a single 1-simplex. For later purposes, imagine first adding a $0$-simplex $0'$ and a 1-simplex $00'$, then identifying $0 \sim 0'$, as shown in Fig.~\ref{fig:D0S1cellulation}.

Now cross a 1-simplex, labeled $01$, with $S^1$. As shown in Fig.~\ref{fig:D1S1cellulation}, for each of the original 0-simplices 0 and 1, we repeat the procedure for crossing a 0-simplex with $S^1$; add ``primed" vertices $0'$ and $1'$ and 1-simplices $00'$ and $11'$. Next, for the 1-simplex $01$, add a new 1-simplex $0'1'$ and two new 2-simplices $011'$,$00'1'$. Finally, identify the ``top" and ``bottom" to make the $S^1$, that is, identify $0 \sim 0$, $1 \sim 1'$, and $01 \sim 0'1'$. Even after identification, it is convenient to keep tracking $0$ and $0'$ separately (and $1$ and $1'$, etc.) in order to define the 2-simplices.

By looking at the 2-simplex $012$ crossed with $S^1$, we see how this process continues. Start by repeating the 0-simplex process for each 0-simplex in the original 2-simplex. This adds the primed vertices $0'$, $ 1'$, and $2'$, along with the 1-simplices $00'$, $11'$, $22'$. Next repeat the 1-simplex process for each 1-simplex in the original 2-simplex. This adds the 1-simplices $01'$, $12'$, and $02'$. Then, for the 2-simplex, add two new 2-simplices $012'$ and $01'2'$, and three 3-simplices $0122'$, $011'2'$, and $00'1'2'$. Finally, identify everything that consists entirely of primes with its unprimed version,e.g. $0'1'2'\sim 012$.

In general, then, given an original $k$-simplex $012\ldots n$, crossing it with $S^1$ should be done as follows. First, for each $k$-simplex in the original object, we add $k+1$ new $k$-simplices and $k+1$ new $(k+1)$-simplices to the cellulation. As above, the $k$-simplices are constructed by ``priming" some vertices starting from the highest-numbered ones, and the $(k+1)$-simplices are constructed in the same way but including both the unprimed and primed vertices for the lowest-numbered primed vertex. This crosses the original simplex with the interval. Then compactify the interval into $S^1$ by identifying simplices with all primed vertices with their unprimed counterparts; this means that in total, for each $k$-simplex in the original object, only $k$ new $k$-simplices appear after compactification.

It is now straightforward to see that given a cellulation of a general $n$-dimensional manifold $M$, one obtains a cellulation of $M \times S^1$ by performing the above process on every individual simplex in the cellulation. That is, for each $k=0,1,\ldots,n$, we find all of the $k$-simplices in the original cellulation of $M$ and add $k$ new $k$-simplices and $(k+1)$ new $(k+1)$-simplices in the manner described above.

\subsection{Dealing with identifications}
We start with a very simple example of the above procedure: cellulating $D^2 \times S^1$ given a cellulation of $S^1$. From this, we will see how to cellulate $T^3 = T^2 \times S^1$ and in the process understand how to pass from a cellulation of a manifold $M \times S^1$ to one of $M' \times S^1$, where simplices of $M$ are identified with each other to produce $M'$. This will be particularly helpful for generating a cellulation of $L(2,1)\times S^1$.

\begin{figure}
\includegraphics[width=0.4\columnwidth]{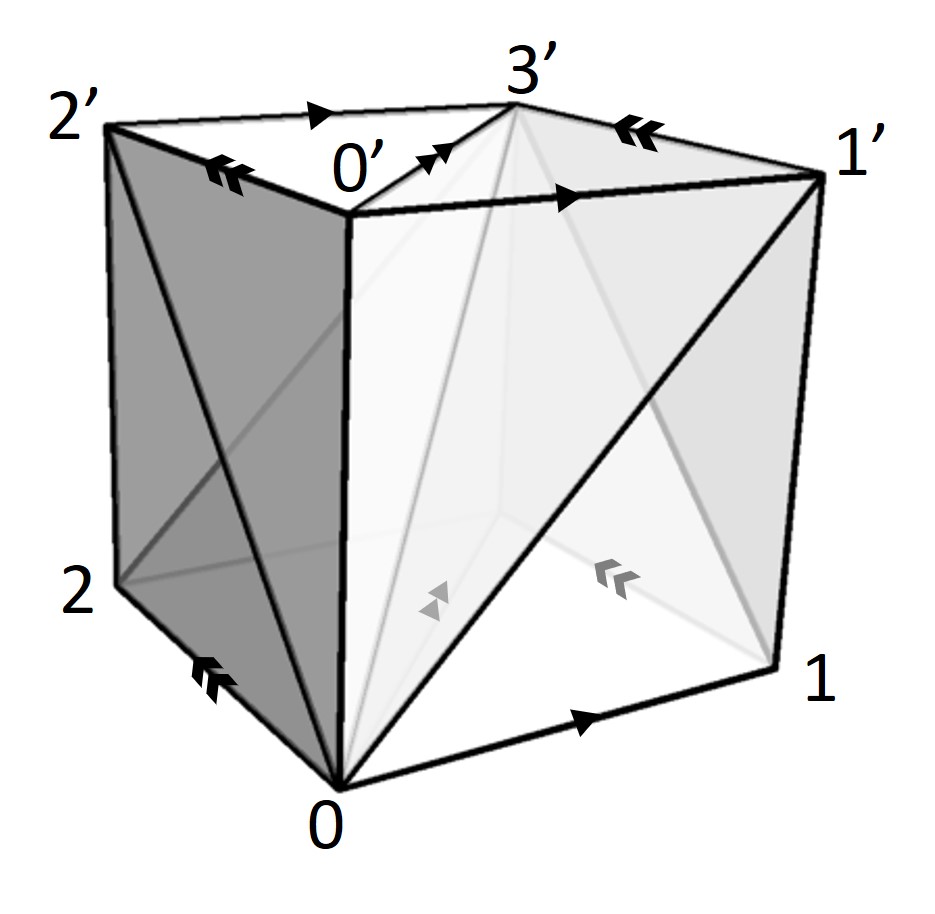}
\caption{Cellulation of $D^2 \times S^1$ produced by our procedure. Marked simplices are identified with correspondingly marked simplices, and unmarked edges are oriented from unprimed vertices to primed vertices.}
\label{fig:D2S1cellulation}
\end{figure}

\begin{figure}
\centering
\subfloat[\label{fig:D2cellulation}]{\includegraphics[width=0.4\columnwidth]{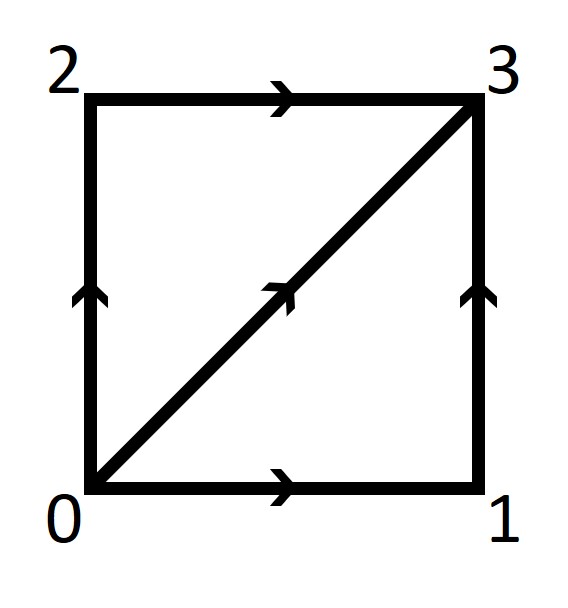}}
\subfloat[\label{fig:L21cellulation}]{\includegraphics[width=0.4\columnwidth]{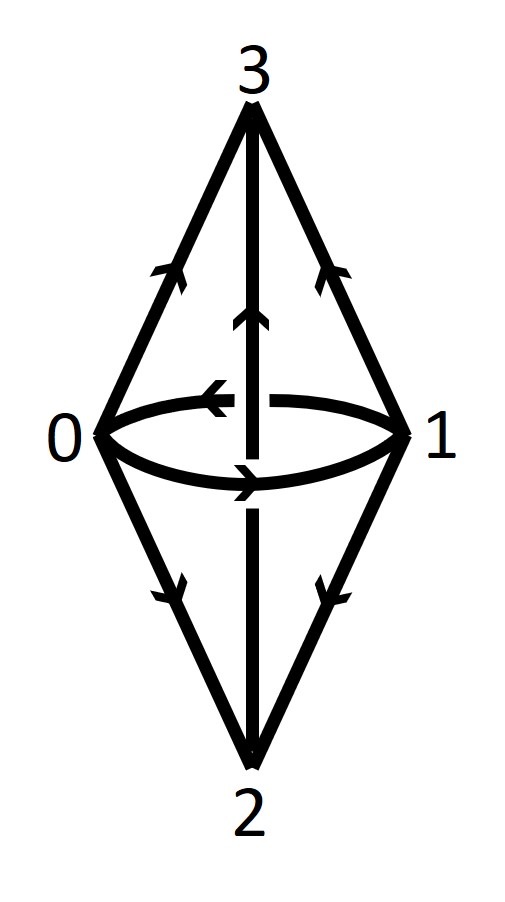}}
\caption{Cellulations of manifolds $M$ used to generate cellulations of $M \times S^1$. (a) Cellulation of $D^2$. (b) Cellulation of 3-dimensional Lens space $L(2,1)$. The front top face $013$ is identified with the back bottom face $102$ and the front bottom face $012$ is identified with the back top face $103$. The simplices for this cellulation are listed explicitly in Table~\ref{tab:L21Cellulation}.}
\end{figure}

We start with the cellulation of $D^2$ shown in Fig.~\ref{fig:D2cellulation}, which consists of four $0$-simplices $0,1,2,3$, five $1$-simplices $01,02,03,13,23$, and two $2$-simplices $013$ and $023$. According to our procedure we start with $k=0$ and add one new 1-simplex for each original 0-simplex. These are the 1-simplices $00'$, $11'$, $22'$, $33'$. For $k=1$, we add a 1-simplex for each of the original 1-simplices, i.e. $01',02',03',13',23'$. We then add 2-simplices $011',00'1',022',00'2',\ldots$. Finally, for $k=2$ we add the 2-simplices $013',01'3',023',02'3'$ and the 3-simplices $0133',011'3',00'1'3',0233',022'3',00'2'3'$. Identifying primed and unprimed vertices as usual yields a cellulation of $D^2 \times S^1$ shown in Fig.~\ref{fig:D2S1cellulation}. 

One can, of course, obtain a cellulation of $T^3$ by identifying various objects on the boundary of the cellulation of $D^2 \times S^1$.  To take a more systematic approach we start by compacting $D^2$ into $T^2$, that is, identifying $0 \sim 1 \sim 2 \sim 3$, $01 \sim 23$, and $02 \sim 13$ and ask how this affects the prior procedure.

The key point is that when two simplices in $D^2$ are identified, we should also identify their corresponding ``child" simplices. For example, $01$ generated one new $1$-simplex $01'$ and two new 2-simplices $011'$ and $00'1'$, while $23$ generated the 1-simplex $23'$ and the 2-simplices $233'$ and $22'3'$. Thus, identifying $01 \sim 23$ in $D^2$ means that we should identify $01' \sim 23'$, $011' \sim 233'$, and $00'1' \sim 22'3'$ in $D^2 \times S^1$. It is easy to check that, for each identification of the ``parent" simplices in $D^2$ to produce $T^2$, identifying ``child" simplices in $D^2 \times S^1$ in this manner is exactly how one produces the natural cellulation of $T^3$ from the given cellulation of $D^2 \times S^1$. 

To summarize, suppose we start with a manifold $M$ such that a list of simplices of $M$ can be pairwise identified to produce $M'$. To obtain a cellulation of $M' \times S^1$, first obtain a cellulation of $M \times S^1$ in the normal way. In this process, a ``parent" $k$-simplex of $M$ produces ``child" $k-$ and $(k+1)-$simplices in the cellulation of $M \times S^1$. Given two parent $k$-simplices which are identified in passing from $M$ to $M'$, then identify their corresponding child simplices as well. After this identification process, we have a cellulation of $M' \times S^1$.

\subsection{$T^4$}
We start from the cellulation of $T^3$ given in Fig.~\ref{fig:T3cellulation}. A list of all the simplices contained in it, with identifications, is given in Table~\ref{tab:T3Cellulation}.

\begin{table*}
\centering
\begin{tabular}{@{}lllllll}
\toprule[2pt]
$0$-\bf{simplices} &\phantom{a}& $1$-\bf{simplices} &\phantom{a}&$2$-\bf{simplices}&\phantom{a}&$3$-\bf{simplices}\\  \hline
$0 \sim 1 \sim 2 \sim \cdots \sim 7$ && $01 \sim 23 \sim 45 \sim 67$ && $013 \sim 457$ && $0137$ \\
 && $02 \sim 13 \sim 46 \sim 57$ && $023 \sim 467$ && $0157$ \\
 && $04 \sim 15 \sim 26 \sim 37$ && $015 \sim 237$ && $0457$\\
 && $03 \sim 47$ && $045 \sim 267$ && $0237$\\
 && $05 \sim 27$ &&$026 \sim 137$ && $0267$\\
 && $06 \sim 17$ && $046 \sim 157$ && $0467$\\
 && $07$ && $037$ && \\
 && && $047$ &&
\\	\bottomrule[2pt]	
\end{tabular}
\caption{List of simplices in a cellulation of $T^3$, with identifications.}
\label{tab:T3Cellulation}
\end{table*}

Let $\mathcal{T}_n$ be the $n$-simplices in the above cellulation of $T^3$. Then by following the above procedure we obtain a cellulation of $T^4$ consisting of the data given in Table~\ref{tab:T4Cellulation}.

\begin{table*}
\centering
\begin{tabular}{@{}lllllllll}
\toprule[2pt]
$0$-\bf{simplices} &\phantom{a}& $1$-\bf{simplices} &\phantom{a}&$2$-\bf{simplices}&\phantom{a}&$3$-\bf{simplices}&\phantom{a}&$4$-\bf{simplices}\\  \hline
$0 \sim 0' \sim 1 \sim \cdots \sim 7'$ && $00'\sim 11'\sim \cdots \sim 77'$ && $pqq'$ for $pq \in \mathcal{T}_1$ (7) && $pqrr'$ for $pqr\in \mathcal{T}_2$ (8)  && $pqrss'$ for $pqrs\in \mathcal{T}_3$ (6)\\
 && $pq$ for $pq \in \mathcal{T}_1$ (7)&& $pp'q'$ for $pq \in \mathcal{T}_1$ (7) && $pqq'r'$ for $pqr\in \mathcal{T}_2$ (8) && $pqrr's'$ for $pqrs\in \mathcal{T}_3$ (6) \\
 && $pq'$ for $pq \in \mathcal{T}_1$ (7) && $pqr$ for $pqr\in \mathcal{T}_2$ (8) && $pp'q'r'$ for $pqr\in \mathcal{T}_2$ (8) && $pqq'r's'$ for $pqrs\in \mathcal{T}_3$ (6)\\
 &&  && $pqr'$ for $pqr\in \mathcal{T}_2$ (8) && $pqrs$ for $pqrs\in \mathcal{T}_3$ (6) && $pp'q'r's'$ for $pqrs\in \mathcal{T}_3$ (6)\\
 &&  && $pq'r'$ for $pqr\in \mathcal{T}_2$ (8) && $pqrs'$ for $pqrs\in \mathcal{T}_3$ (6) &&\\
 &&  &&  && $pqr's'$ for $pqrs\in \mathcal{T}_3$ (6) &&\\
 &&  &&  && $pq'r's'$ for $pqrs\in \mathcal{T}_3$ (6) &&
\\	\bottomrule[2pt]	
\end{tabular}
\caption{List of simplices in a cellulation of $T^4$. All simplices with all vertices primed are identified with their unprimed counterparts, i.e. $p'q' \sim pq$. Numbers in parentheses indicate how many simplices are in the relevant set $\mathcal{T}_k$, which are the set of $k$-simplices in the cellulation of $T^3$ in Table ~\ref{tab:T3Cellulation}.}
\label{tab:T4Cellulation}
\end{table*}

\subsection{$L(2,1)\times S^1$}

We start with the cellulation of $L(2,1)$ shown in Fig.~\ref{fig:L21cellulation}. It consists of the simplices given in Table~\ref{tab:L21Cellulation}.

\begin{table*}
\centering
\begin{tabular}{@{}lllllll}
\toprule[2pt]
$0$-\bf{simplices} &\phantom{a}& $1$-\bf{simplices} &\phantom{a}&$2$-\bf{simplices}&\phantom{a}&$3$-\bf{simplices}\\  \hline
$0 \sim 1$ && $01 \sim 10$ && $012 \sim 103$ && $0123$ \\
$2 \sim 3$ && $02 \sim 13$ && $013 \sim 102$ && $1023$ \\
 && $03 \sim 12$ && $023$ &&  \\
 && $23$ && $123$ &&  
\\	\bottomrule[2pt]	
\end{tabular}
\caption{List of simplices in a cellulation of Lens space $L(2,1)$, with identifications.}
\label{tab:L21Cellulation}
\end{table*}

Note that we distinguish the ``front" and ``back" 2- and 3-simplices by the order in which the vertices appear, i.e. $0123$ and $1023$ contain the same vertices but are distinct 3-simplices. Letting $\mathcal{L}_k$ be the set of $k$-simplices in the above cellulation and applying the above procedure, we obtain a cellulation of $L(2,1) \times S^1$ with the data given in Table~\ref{tab:L21S1Cellulation}.

\begin{table*}
\centering
\begin{tabular}{@{}lllllllll}
\toprule[2pt]
$0$-\bf{simplices} &\phantom{a}& $1$-\bf{simplices} &\phantom{a}&$2$-\bf{simplices}&\phantom{a}&$3$-\bf{simplices}&\phantom{a}&$4$-\bf{simplices}\\  \hline
$0 \sim 0' \sim 1 \sim 1'$ && $00' \sim 11'$ && $pqq'$ for $pq \in \mathcal{L}_1$ (4) && $pqrr'$ for $pqr\in \mathcal{L}_2$ (4)  && $pqrss'$ for $pqrs\in \mathcal{L}_3$ (2)\\
$2 \sim 2' \sim 3 \sim 3'$ && $22' \sim 33'$ && $pp'q'$ for $pq \in \mathcal{L}_1$ (4) && $pqq'r'$ for $pqr\in \mathcal{L}_2$ (4) && $pqrr's'$ for $pqrs\in \mathcal{L}_3$ (2) \\
 && $pq$ for $pq \in \mathcal{L}_1$ (4) && $pqr$ for $pqr\in \mathcal{L}_2$ (4) && $pp'q'r'$ for $pqr\in \mathcal{L}_2$ (4) && $pqq'r's'$ for $pqrs\in \mathcal{L}_3$ (2)\\
 && $pq'$ for $pq \in \mathcal{L}_1$ (4) && $pqr'$ for $pqr\in \mathcal{L}_2$ (4) && $pqrs$ for $pqrs\in \mathcal{L}_3$ (2) && $pp'q'r's'$ for $pqrs\in \mathcal{L}_3$ (2)\\
 &&  && $pq'r'$ for $pqr\in \mathcal{L}_2$ (4) && $pqrs'$ for $pqrs\in \mathcal{L}_3$ (2) &&\\
 &&  &&  && $pqr's'$ for $pqrs\in \mathcal{L}_3$ (2) &&\\
 &&  &&  && $pq'r's'$ for $pqrs\in \mathcal{L}_3$ (2) &&
\\	\bottomrule[2pt]	
\end{tabular}
\caption{List of simplices in a cellulation of $L(2,1)\times S^1$. All simplices with all vertices primed are identified with their unprimed counterparts, i.e. $p'q' \sim pq$. Numbers in parentheses indicate how many simplices are in the relevant set $\mathcal{L}_k$, which are the set of $k$-simplices in the cellulation of $L(2,1)$ in Table ~\ref{tab:L21Cellulation}.}
\label{tab:L21S1Cellulation}
\end{table*}

\subsection{$S^3 \times S^1$}
A simple cellulation of $S^3$ consists of two $3$-simplices of opposite orientation glued together on their boundaries. It contains:
\begin{table*}
\centering
\begin{tabular}{@{}lllllll}
\toprule[2pt]
$0$-\bf{simplices} &\phantom{a}& $1$-\bf{simplices} &\phantom{a}&$2$-\bf{simplices}&\phantom{a}&$3$-\bf{simplices}\\  \hline
$0$ && $01$ && $012$ && $0123^{(+)}$ \\
$1$ && $02$ && $013$ && $0123^{(-)}$ \\
$2$ && $03$ && $023$ &&  \\
$3$ && $12$ && $123$ && \\
    && $13$ && &&\\
    && $23$ && &&\\ 
\\	\bottomrule[2pt]	
\end{tabular}
\caption{List of simplices in a cellulation of $S^3$. The $\pm$ superscripts refer to the orientation of the two 3-simplices with identified boundaries; this notation is needed since these two 3-simplices share the same vertices.}
\label{tab:S3Cellulation}
\end{table*}

Following the above procedure, we obtain a cellulation of $S^3 \times S^1$ with the simplices given in Table~\ref{tab:S3S1Cellulation}.

\begin{table*}
\centering
\begin{tabular}{@{}lllllllll}
\toprule[2pt]
$0$-\bf{simplices} &\phantom{a}& $1$-\bf{simplices} &\phantom{a}&$2$-\bf{simplices}&\phantom{a}&$3$-\bf{simplices}&\phantom{a}&$4$-\bf{simplices}\\  \hline
$0 \sim 0'$ && $pp'$ (4) && $pqq'$ for $p<q$ (6) && $pqrr'$ for $p<q<r$ (4)  && $01233^{'(\pm)}$ (2)\\
$1 \sim 1'$ && $pq$ for $p<q$ (6) && $pp'q'$ for $p<q$ (6) && $pqq'r'$ for $p<q<r$ (4) && $0122'3^{'(\pm)}$ (2) \\
$2 \sim 2'$ && $pq'$ for $p<q$ (6) && $pqr$ for $p<q<r$ (4) && $pp'q'r'$ for $p<q<r$ (4) && $011'2'3^{'(\pm)}$ (2)\\
$1 \sim 1'$ &&  && $pqr'$ for $p<q<r$ (4) && $0123^{(\pm)}$ (2) && $00'1'2'3^{'(\pm)}$ (2)\\
 &&  && $pq'r'$ for $p<q<r$ (4) && $0123^{'(\pm)}$ (2) &&\\
 &&  &&  && $012'3^{'(\pm)}$ (2) &&\\
 &&  &&  && $01'2'3^{'(\pm)}$ (2) &&
\\	\bottomrule[2pt]	
\end{tabular}
\caption{List of simplices in a cellulation of $S^3\times S^1$. All simplices with all vertices primed are identified with their unprimed counterparts, i.e. $p'q' \sim pq$. Numbers in parentheses indicate how many simplices the number of simplices which take that form. The labels $p,q,r$ run over $0,1,2,3$.}
\label{tab:S3S1Cellulation}
\end{table*}

\section{Partition function on $S^3 \times S^1$}
\label{app:ZS3xS1}
In this section, we explicitly compute our partition function on $S^3 \times S^1$ and show that it is always 1 in the absence of a background flux using the cellulation of $S^3 \times S^1$ containing eight 4-simplices that was constructed in the previous Appendix.

We first decompose the state-sum in the following way:
\begin{widetext}
\begin{align}
Z(S^3 \times &S^1) = \sum_g \frac{1}{|G|^4\mathcal{D}^{24}}\sum_{a,b}{}' \sum_{\substack{033',133',233',\\ 0133',1233',0233'}}Z^+(01233^{'(+)})Z^-(01233^{'(-)}) \sum_{\substack{022',122',22'3',\\ 0122',022'3',122'3'}}Z^+(0122'3^{'(+)})Z^-(0122'3^{'(-)}) \times \nonumber\\
&\times \sum_{\substack{011',11'2',11'3',\\ 011'2',011'3',11'2'3'}}Z^+(011'2'3^{'(+)})Z^-(011'2'3^{'(-)}) \sum_{\substack{00'1',00'2',00'3',\\ 00'1'2',00'1'3',00'2'3'}}Z^+(00'1'2'3^{'(+)})Z^-(00'1'2'3^{'(-)}) \nonumber \\
&\times \frac{\prod_{\alpha \in \mathcal{T}^2}d_{a_{\alpha}}}{\prod_{\tau \in \mathcal{T}^3}d_{b_{\tau}}}
\label{eqn:ZS3S1Decomp}
\end{align}
\end{widetext}
Here $\sum_{a,b}'$ means the sum on all 2-simplices and 3-simplices not appearing explicitly in Eq.~\eqref{eqn:ZS3S1Decomp}. We have chosen to group the diagrams so that the sums over certain $2$- and $3$-simplices can be separated; i.e. the 2-simplex $133'$ appears only in the 4-simplices $01233^{'(\pm)}$, so the other diagrams factor out of that sum.

Consider first the sum involving $0122'3'^{(\pm)}$. The symmetry factors (factors of $U$ and $\eta$) in each diagram are
\begin{widetext}
\begin{align}
Z^+(0122'3'^{(+)})&\propto \eta^{-1}_{012}(g_{22'},g_{2'3'})U_{2'3'}^{-1}(022',{}^{2'2}\!012;0122')U_{2'3'}(012',122';0122')\\
Z^-(0122'3'^{(-)})&\propto \eta_{012}(g_{22'},g_{2'3'})U_{2'3'}(022',{}^{2'2}\!012;0122')U^{-1}_{2'3'}(012',122';0122')
\end{align}
\end{widetext}
Clearly these factors cancel. In fact, since for any $\pm$ pair, the two diagrams involve exactly the same 2-simplices, the only way that the symmetry factors could fail to cancel is if $U$ involves different 3-simplices. In the 4-simplices $0122'3'^{(\pm)}, 011'2'3'^{(\pm)},$ and $00'1'2'3'^{(\pm)}$, the factor of $U$ acts on $0122',011'2',$ and $00'1'2'$, respectively in both the $+$ and the $-$ diagram. Therefore all of these symmetry factors cancel out. The only pair that we need to check is the $01233'^{(\pm)}$ pair, which has 
\begin{widetext}
\begin{align}
Z^+(01233'^{(+)})&\propto \eta^{-1}_{012}(g_{23},g_{33'})U_{33'}^{-1}(023,{}^{32}\!012;0123^{(+)})U_{33'}(013,123;0123^{(+)})\\
Z^-(01233'^{(-)})&\propto \eta_{012}(g_{23},g_{33'})U_{33'}(023,{}^{32}\!012;0123^{(-)})U^{-1}_{33'}(013,123;0123^{(-)})
\end{align}
\end{widetext}
Because these two 4-simplices involve $b_{0123^{(\pm)}}$, which are not a priori equal, there is not obviously a cancellation. However, because of the absence of background flux, ${\bf g}_{33'}=1$. Therefore, all of these symmetry factors are 1.

What we have shown is that, for any given term in the state-sum, the factors of $U$ and $\eta$ either cancel out or are equal to 1. At this stage, one can now check straightforwardly that each anyon always appears with the same symmetry action on it, that is, for example, $012$ always appears as ${}^{42}\!012$. Reindexing the sum (e.g. ${}^{42}\!012 \rightarrow 012$) and performing the (now-trivial) sum on ${\bf g}_i$, we obtain precisely the Crane-Yetter state-sum for $\mathcal{C}$ on $S^3 \times S^1$, which is known to be 1, that is,
\begin{equation}
Z(S^3 \times S^1) = Z_{CY}(S^3 \times S^1) = 1
\end{equation}
as desired. One can explicitly compute the Crane-Yetter state-sum on $S^3 \times S^1$; this is most conveniently done using the decomposition in Eq.~\eqref{eqn:ZS3S1Decomp} and the merging lemma in Fig.~\ref{fig:mergingLemma2}, but since the result is known, we do not reproduce the calculation here.

\bibstyle{apsrev4-1} \bibliography{references,TI}

\begin{thebibliography}{63}%
\makeatletter
\providecommand \@ifxundefined [1]{%
 \@ifx{#1\undefined}
}%
\providecommand \@ifnum [1]{%
 \ifnum #1\expandafter \@firstoftwo
 \else \expandafter \@secondoftwo
 \fi
}%
\providecommand \@ifx [1]{%
 \ifx #1\expandafter \@firstoftwo
 \else \expandafter \@secondoftwo
 \fi
}%
\providecommand \natexlab [1]{#1}%
\providecommand \enquote  [1]{``#1''}%
\providecommand \bibnamefont  [1]{#1}%
\providecommand \bibfnamefont [1]{#1}%
\providecommand \citenamefont [1]{#1}%
\providecommand \href@noop [0]{\@secondoftwo}%
\providecommand \href [0]{\begingroup \@sanitize@url \@href}%
\providecommand \@href[1]{\@@startlink{#1}\@@href}%
\providecommand \@@href[1]{\endgroup#1\@@endlink}%
\providecommand \@sanitize@url [0]{\catcode `\\12\catcode `\$12\catcode
  `\&12\catcode `\#12\catcode `\^12\catcode `\_12\catcode `\%12\relax}%
\providecommand \@@startlink[1]{}%
\providecommand \@@endlink[0]{}%
\providecommand \url  [0]{\begingroup\@sanitize@url \@url }%
\providecommand \@url [1]{\endgroup\@href {#1}{\urlprefix }}%
\providecommand \urlprefix  [0]{URL }%
\providecommand \Eprint [0]{\href }%
\providecommand \doibase [0]{http://dx.doi.org/}%
\providecommand \selectlanguage [0]{\@gobble}%
\providecommand \bibinfo  [0]{\@secondoftwo}%
\providecommand \bibfield  [0]{\@secondoftwo}%
\providecommand \translation [1]{[#1]}%
\providecommand \BibitemOpen [0]{}%
\providecommand \bibitemStop [0]{}%
\providecommand \bibitemNoStop [0]{.\EOS\space}%
\providecommand \EOS [0]{\spacefactor3000\relax}%
\providecommand \BibitemShut  [1]{\csname bibitem#1\endcsname}%
\let\auto@bib@innerbib\@empty
\bibitem [{\citenamefont {Wen}(2004)}]{wen04}%
  \BibitemOpen
  \bibfield  {author} {\bibinfo {author} {\bibfnamefont {Xiao-Gang}\
  \bibnamefont {Wen}},\ }\href@noop {} {\emph {\bibinfo {title} {Quantum Field
  Theory of Many-Body Systems}}}\ (\bibinfo  {publisher} {Oxford Univ. Press},\
  \bibinfo {address} {Oxford},\ \bibinfo {year} {2004})\BibitemShut {NoStop}%
\bibitem [{\citenamefont {Wang}(2008)}]{wang2008}%
  \BibitemOpen
  \bibfield  {author} {\bibinfo {author} {\bibfnamefont {Zhenghan}\
  \bibnamefont {Wang}},\ }\href@noop {} {\emph {\bibinfo {title} {Topological
  Quantum Computation}}}\ (\bibinfo  {publisher} {American Mathematical
  Society},\ \bibinfo {year} {2008})\BibitemShut {NoStop}%
\bibitem [{\citenamefont {Nayak}\ \emph {et~al.}(2008)\citenamefont {Nayak},
  \citenamefont {Simon}, \citenamefont {Stern}, \citenamefont {Freedman},\ and\
  \citenamefont {Sarma}}]{nayak2008}%
  \BibitemOpen
  \bibfield  {author} {\bibinfo {author} {\bibfnamefont {Chetan}\ \bibnamefont
  {Nayak}}, \bibinfo {author} {\bibfnamefont {Steven~H.}\ \bibnamefont
  {Simon}}, \bibinfo {author} {\bibfnamefont {Ady}\ \bibnamefont {Stern}},
  \bibinfo {author} {\bibfnamefont {Michael}\ \bibnamefont {Freedman}}, \ and\
  \bibinfo {author} {\bibfnamefont {Sankar~Das}\ \bibnamefont {Sarma}},\
  }\bibfield  {title} {\enquote {\bibinfo {title} {Non-abelian anyons and
  topological quantum computation},}\ }\href@noop {} {\bibfield  {journal}
  {\bibinfo  {journal} {Rev. Mod. Phys.}\ }\textbf {\bibinfo {volume} {80}},\
  \bibinfo {pages} {1083} (\bibinfo {year} {2008})}\BibitemShut {NoStop}%
\bibitem [{\citenamefont {Moore}\ and\ \citenamefont
  {Seiberg}(1989)}]{moore1989b}%
  \BibitemOpen
  \bibfield  {author} {\bibinfo {author} {\bibfnamefont {Gregory}\ \bibnamefont
  {Moore}}\ and\ \bibinfo {author} {\bibfnamefont {Nathan}\ \bibnamefont
  {Seiberg}},\ }\bibfield  {title} {\enquote {\bibinfo {title} {Classical and
  quantum conformal field theory},}\ }\href {\doibase 10.1007/BF01238857}
  {\bibfield  {journal} {\bibinfo  {journal} {Comm. Math. Phys.}\ }\textbf
  {\bibinfo {volume} {123}},\ \bibinfo {pages} {177--254} (\bibinfo {year}
  {1989})}\BibitemShut {NoStop}%
\bibitem [{\citenamefont {Witten}(1989)}]{witten1989}%
  \BibitemOpen
  \bibfield  {author} {\bibinfo {author} {\bibfnamefont {Edward}\ \bibnamefont
  {Witten}},\ }\bibfield  {title} {\enquote {\bibinfo {title} {{Quantum field
  theory and the Jones polynomial}},}\ }\href@noop {} {\bibfield  {journal}
  {\bibinfo  {journal} {Comm. Math. Phys.}\ }\textbf {\bibinfo {volume}
  {121}},\ \bibinfo {pages} {351--399} (\bibinfo {year} {1989})}\BibitemShut
  {NoStop}%
\bibitem [{\citenamefont {Barkeshli}\ \emph
  {et~al.}(2019{\natexlab{a}})\citenamefont {Barkeshli}, \citenamefont
  {Bonderson}, \citenamefont {Cheng},\ and\ \citenamefont
  {Wang}}]{barkeshli2019}%
  \BibitemOpen
  \bibfield  {author} {\bibinfo {author} {\bibfnamefont {Maissam}\ \bibnamefont
  {Barkeshli}}, \bibinfo {author} {\bibfnamefont {Parsa}\ \bibnamefont
  {Bonderson}}, \bibinfo {author} {\bibfnamefont {Meng}\ \bibnamefont {Cheng}},
  \ and\ \bibinfo {author} {\bibfnamefont {Zhenghan}\ \bibnamefont {Wang}},\
  }\bibfield  {title} {\enquote {\bibinfo {title} {Symmetry fractionalization,
  defects, and gauging of topological phases},}\ }\href {\doibase
  10.1103/PhysRevB.100.115147} {\bibfield  {journal} {\bibinfo  {journal}
  {Phys. Rev. B}\ }\textbf {\bibinfo {volume} {100}},\ \bibinfo {pages}
  {115147} (\bibinfo {year} {2019}{\natexlab{a}})},\ \Eprint
  {http://arxiv.org/abs/arXiv:1410.4540} {arXiv:1410.4540} \BibitemShut
  {NoStop}%
\bibitem [{\citenamefont {Etingof}\ \emph {et~al.}(2009)\citenamefont
  {Etingof}, \citenamefont {Nikshych},\ and\ \citenamefont {Ostrik}}]{ENO2009}%
  \BibitemOpen
  \bibfield  {author} {\bibinfo {author} {\bibfnamefont {Pavel}\ \bibnamefont
  {Etingof}}, \bibinfo {author} {\bibfnamefont {Dmitri}\ \bibnamefont
  {Nikshych}}, \ and\ \bibinfo {author} {\bibfnamefont {Victor}\ \bibnamefont
  {Ostrik}},\ }\bibfield  {title} {\enquote {\bibinfo {title} {Fusion
  categories and homotopy theory},}\ }\href@noop {} {\  (\bibinfo {year}
  {2009})},\ \Eprint {http://arxiv.org/abs/arXiv:0909.3140} {arXiv:0909.3140}
  \BibitemShut {NoStop}%
\bibitem [{\citenamefont {Gaiotto}\ \emph {et~al.}(2014)\citenamefont
  {Gaiotto}, \citenamefont {Kapustin}, \citenamefont {Seiberg},\ and\
  \citenamefont {Willett}}]{gaiotto2014}%
  \BibitemOpen
  \bibfield  {author} {\bibinfo {author} {\bibfnamefont {Davide}\ \bibnamefont
  {Gaiotto}}, \bibinfo {author} {\bibfnamefont {Anton}\ \bibnamefont
  {Kapustin}}, \bibinfo {author} {\bibfnamefont {Nathan}\ \bibnamefont
  {Seiberg}}, \ and\ \bibinfo {author} {\bibfnamefont {Brian}\ \bibnamefont
  {Willett}},\ }\bibfield  {title} {\enquote {\bibinfo {title} {Generalized
  global symmetries},}\ }\href@noop {} {\  (\bibinfo {year} {2014})},\ \Eprint
  {http://arxiv.org/abs/arXiv:1412.5148} {arXiv:1412.5148} \BibitemShut
  {NoStop}%
\bibitem [{\citenamefont {Benini}\ \emph {et~al.}(2019)\citenamefont {Benini},
  \citenamefont {C{\'o}rdova},\ and\ \citenamefont {Hsin}}]{benini2019}%
  \BibitemOpen
  \bibfield  {author} {\bibinfo {author} {\bibfnamefont {Francesco}\
  \bibnamefont {Benini}}, \bibinfo {author} {\bibfnamefont {Clay}\ \bibnamefont
  {C{\'o}rdova}}, \ and\ \bibinfo {author} {\bibfnamefont {Po-Shen}\
  \bibnamefont {Hsin}},\ }\bibfield  {title} {\enquote {\bibinfo {title} {On
  2-group global symmetries and their anomalies},}\ }\href@noop {} {\bibfield
  {journal} {\bibinfo  {journal} {Journal of High Energy Physics}\ }\textbf
  {\bibinfo {volume} {2019}},\ \bibinfo {pages} {118} (\bibinfo {year}
  {2019})}\BibitemShut {NoStop}%
\bibitem [{\citenamefont {Kapustin}\ and\ \citenamefont
  {Thorngren}(2014)}]{kapustin2014b}%
  \BibitemOpen
  \bibfield  {author} {\bibinfo {author} {\bibfnamefont {Anton}\ \bibnamefont
  {Kapustin}}\ and\ \bibinfo {author} {\bibfnamefont {Ryan}\ \bibnamefont
  {Thorngren}},\ }\bibfield  {title} {\enquote {\bibinfo {title} {Anomalies of
  discrete symmetries in various dimensions and group cohomology},}\
  }\href@noop {} {\  (\bibinfo {year} {2014})},\ \Eprint
  {http://arxiv.org/abs/arXiv:1404.3230} {arXiv:1404.3230} \BibitemShut
  {NoStop}%
\bibitem [{Note1()}]{Note1}%
  \BibitemOpen
  \bibinfo {note} {Implicit in this statement is that the symmetry acts in an
  on-site manner; that is, that the symmetry action in a lattice model
  decomposes as a tensor product of unitaries acting on each lattice site
  independently. It is possible to realize anomalous boundary theories in the
  dimension of the boundary by considering a non-on-site action of the
  symmetry.\cite {chen2013} The precise extension of the ``on-site''
  requirement for space-time symmetries has not yet been formulated, but
  presumably the requirement is that the symmetry must correspond to an on-site
  transformation combined with a classical permutation of coordinates and
  possibly complex conjugation.}\BibitemShut {Stop}%
\bibitem [{\citenamefont {Vishwanath}\ and\ \citenamefont
  {Senthil}(2013)}]{vishwanath2013}%
  \BibitemOpen
  \bibfield  {author} {\bibinfo {author} {\bibfnamefont {Ashvin}\ \bibnamefont
  {Vishwanath}}\ and\ \bibinfo {author} {\bibfnamefont {T.}~\bibnamefont
  {Senthil}},\ }\bibfield  {title} {\enquote {\bibinfo {title} {Physics of
  three-dimensional bosonic topological insulators: Surface-deconfined
  criticality and quantized magnetoelectric effect},}\ }\href {\doibase
  10.1103/PhysRevX.3.011016} {\bibfield  {journal} {\bibinfo  {journal} {Phys.
  Rev. X}\ }\textbf {\bibinfo {volume} {3}},\ \bibinfo {pages} {011016}
  (\bibinfo {year} {2013})}\BibitemShut {NoStop}%
\bibitem [{\citenamefont {Wang}\ and\ \citenamefont
  {Senthil}(2013)}]{wong2013}%
  \BibitemOpen
  \bibfield  {author} {\bibinfo {author} {\bibfnamefont {Chong}\ \bibnamefont
  {Wang}}\ and\ \bibinfo {author} {\bibfnamefont {T.}~\bibnamefont {Senthil}},\
  }\bibfield  {title} {\enquote {\bibinfo {title} {Boson topological
  insulators: A window into highly entangled quantum phases},}\ }\href
  {\doibase 10.1103/PhysRevB.87.235122} {\bibfield  {journal} {\bibinfo
  {journal} {Phys. Rev. B}\ }\textbf {\bibinfo {volume} {87}},\ \bibinfo
  {pages} {235122} (\bibinfo {year} {2013})}\BibitemShut {NoStop}%
\bibitem [{\citenamefont {Chen}\ \emph {et~al.}(2015)\citenamefont {Chen},
  \citenamefont {Burnell}, \citenamefont {Vishwanath},\ and\ \citenamefont
  {Fidkowski}}]{Chen2014}%
  \BibitemOpen
  \bibfield  {author} {\bibinfo {author} {\bibfnamefont {Xie}\ \bibnamefont
  {Chen}}, \bibinfo {author} {\bibfnamefont {F.~J.}\ \bibnamefont {Burnell}},
  \bibinfo {author} {\bibfnamefont {Ashvin}\ \bibnamefont {Vishwanath}}, \ and\
  \bibinfo {author} {\bibfnamefont {Lukasz}\ \bibnamefont {Fidkowski}},\
  }\bibfield  {title} {\enquote {\bibinfo {title} {Anomalous symmetry
  fractionalization and surface topological order},}\ }\href {\doibase
  10.1103/PhysRevX.5.041013} {\bibfield  {journal} {\bibinfo  {journal} {Phys.
  Rev. X}\ }\textbf {\bibinfo {volume} {5}},\ \bibinfo {pages} {041013}
  (\bibinfo {year} {2015})}\BibitemShut {NoStop}%
\bibitem [{\citenamefont {Senthil}(2015)}]{senthil2015}%
  \BibitemOpen
  \bibfield  {author} {\bibinfo {author} {\bibfnamefont {T.}~\bibnamefont
  {Senthil}},\ }\bibfield  {title} {\enquote {\bibinfo {title}
  {Symmetry-protected topological phases of quantum matter},}\ }\href@noop {}
  {\bibfield  {journal} {\bibinfo  {journal} {Annual Review of Condensed Matter
  Physics}\ }\textbf {\bibinfo {volume} {6}},\ \bibinfo {pages} {299--324}
  (\bibinfo {year} {2015})}\BibitemShut {NoStop}%
\bibitem [{\citenamefont {Kapustin}(2014{\natexlab{a}})}]{kapustin2014}%
  \BibitemOpen
  \bibfield  {author} {\bibinfo {author} {\bibfnamefont {Anton}\ \bibnamefont
  {Kapustin}},\ }\bibfield  {title} {\enquote {\bibinfo {title} {Symmetry
  protected topological phases, anomalies, and cobordisms: Beyond group
  cohomology},}\ }\href@noop {} {\  (\bibinfo {year} {2014}{\natexlab{a}})},\
  \Eprint {http://arxiv.org/abs/arXiv:1403.1467} {arXiv:1403.1467} \BibitemShut
  {NoStop}%
\bibitem [{\citenamefont {Kapustin}(2014{\natexlab{b}})}]{kapustin2014c}%
  \BibitemOpen
  \bibfield  {author} {\bibinfo {author} {\bibfnamefont {Anton}\ \bibnamefont
  {Kapustin}},\ }\bibfield  {title} {\enquote {\bibinfo {title} {Bosonic
  topological insulators and paramagnets: a view from cobordisms},}\
  }\href@noop {} {\  (\bibinfo {year} {2014}{\natexlab{b}})},\ \Eprint
  {http://arxiv.org/abs/arXiv:1404.6659} {arXiv:1404.6659} \BibitemShut
  {NoStop}%
\bibitem [{\citenamefont {Freed}\ and\ \citenamefont
  {Hopkins}(2016)}]{freed2016}%
  \BibitemOpen
  \bibfield  {author} {\bibinfo {author} {\bibfnamefont {Daniel~S.}\
  \bibnamefont {Freed}}\ and\ \bibinfo {author} {\bibfnamefont {Michael~J.}\
  \bibnamefont {Hopkins}},\ }\href@noop {} {} (\bibinfo {year} {2016}),\
  \Eprint {http://arxiv.org/abs/arXiv:1604.06527} {arXiv:1604.06527}
  \BibitemShut {NoStop}%
\bibitem [{\citenamefont {Chen}\ \emph {et~al.}(2013)\citenamefont {Chen},
  \citenamefont {Gu}, \citenamefont {Liu},\ and\ \citenamefont
  {Wen}}]{chen2013}%
  \BibitemOpen
  \bibfield  {author} {\bibinfo {author} {\bibfnamefont {Xie}\ \bibnamefont
  {Chen}}, \bibinfo {author} {\bibfnamefont {Zheng-Cheng}\ \bibnamefont {Gu}},
  \bibinfo {author} {\bibfnamefont {Zheng-Xin}\ \bibnamefont {Liu}}, \ and\
  \bibinfo {author} {\bibfnamefont {Xiao-Gang}\ \bibnamefont {Wen}},\
  }\bibfield  {title} {\enquote {\bibinfo {title} {Symmetry protected
  topological orders and the group cohomology of their symmetry group},}\
  }\href {\doibase 10.1103/PhysRevB.87.155114} {\bibfield  {journal} {\bibinfo
  {journal} {Phys. Rev. B}\ }\textbf {\bibinfo {volume} {87}},\ \bibinfo
  {pages} {155114} (\bibinfo {year} {2013})}\BibitemShut {NoStop}%
\bibitem [{\citenamefont {Dijkgraaf}\ and\ \citenamefont
  {Witten}(1990)}]{dijkgraaf1990}%
  \BibitemOpen
  \bibfield  {author} {\bibinfo {author} {\bibfnamefont {R.}~\bibnamefont
  {Dijkgraaf}}\ and\ \bibinfo {author} {\bibfnamefont {E.}~\bibnamefont
  {Witten}},\ }\bibfield  {title} {\enquote {\bibinfo {title} {Topological
  gauge theories and group cohomology},}\ }\href@noop {} {\bibfield  {journal}
  {\bibinfo  {journal} {Comm. Math. Phys.}\ }\textbf {\bibinfo {volume}
  {129}},\ \bibinfo {pages} {393--429} (\bibinfo {year} {1990})}\BibitemShut
  {NoStop}%
\bibitem [{\citenamefont {Song}\ \emph {et~al.}(2017)\citenamefont {Song},
  \citenamefont {Huang}, \citenamefont {Fu},\ and\ \citenamefont
  {Hermele}}]{song2017}%
  \BibitemOpen
  \bibfield  {author} {\bibinfo {author} {\bibfnamefont {Hao}\ \bibnamefont
  {Song}}, \bibinfo {author} {\bibfnamefont {Sheng-Jie}\ \bibnamefont {Huang}},
  \bibinfo {author} {\bibfnamefont {Liang}\ \bibnamefont {Fu}}, \ and\ \bibinfo
  {author} {\bibfnamefont {Michael}\ \bibnamefont {Hermele}},\ }\bibfield
  {title} {\enquote {\bibinfo {title} {Topological phases protected by point
  group symmetry},}\ }\href {\doibase 10.1103/PhysRevX.7.011020} {\bibfield
  {journal} {\bibinfo  {journal} {Phys. Rev. X}\ }\textbf {\bibinfo {volume}
  {7}},\ \bibinfo {pages} {011020} (\bibinfo {year} {2017})}\BibitemShut
  {NoStop}%
\bibitem [{\citenamefont {Huang}\ \emph {et~al.}(2017)\citenamefont {Huang},
  \citenamefont {Song}, \citenamefont {Huang},\ and\ \citenamefont
  {Hermele}}]{huang2017}%
  \BibitemOpen
  \bibfield  {author} {\bibinfo {author} {\bibfnamefont {Sheng-Jie}\
  \bibnamefont {Huang}}, \bibinfo {author} {\bibfnamefont {Hao}\ \bibnamefont
  {Song}}, \bibinfo {author} {\bibfnamefont {Yi-Ping}\ \bibnamefont {Huang}}, \
  and\ \bibinfo {author} {\bibfnamefont {Michael}\ \bibnamefont {Hermele}},\
  }\bibfield  {title} {\enquote {\bibinfo {title} {Building crystalline
  topological phases from lower-dimensional states},}\ }\href {\doibase
  10.1103/PhysRevB.96.205106} {\bibfield  {journal} {\bibinfo  {journal} {Phys.
  Rev. B}\ }\textbf {\bibinfo {volume} {96}},\ \bibinfo {pages} {205106}
  (\bibinfo {year} {2017})}\BibitemShut {NoStop}%
\bibitem [{\citenamefont {Thorngren}\ and\ \citenamefont
  {Else}(2018)}]{thorngren2018}%
  \BibitemOpen
  \bibfield  {author} {\bibinfo {author} {\bibfnamefont {Ryan}\ \bibnamefont
  {Thorngren}}\ and\ \bibinfo {author} {\bibfnamefont {Dominic~V.}\
  \bibnamefont {Else}},\ }\bibfield  {title} {\enquote {\bibinfo {title}
  {Gauging spatial symmetries and the classification of topological crystalline
  phases},}\ }\href {\doibase 10.1103/PhysRevX.8.011040} {\bibfield  {journal}
  {\bibinfo  {journal} {Phys. Rev. X}\ }\textbf {\bibinfo {volume} {8}},\
  \bibinfo {pages} {011040} (\bibinfo {year} {2018})}\BibitemShut {NoStop}%
\bibitem [{\citenamefont {Cheng}\ \emph {et~al.}(2016)\citenamefont {Cheng},
  \citenamefont {Zaletel}, \citenamefont {Barkeshli}, \citenamefont
  {Vishwanath},\ and\ \citenamefont {Bonderson}}]{cheng2016lsm}%
  \BibitemOpen
  \bibfield  {author} {\bibinfo {author} {\bibfnamefont {Meng}\ \bibnamefont
  {Cheng}}, \bibinfo {author} {\bibfnamefont {Michael}\ \bibnamefont
  {Zaletel}}, \bibinfo {author} {\bibfnamefont {Maissam}\ \bibnamefont
  {Barkeshli}}, \bibinfo {author} {\bibfnamefont {Ashvin}\ \bibnamefont
  {Vishwanath}}, \ and\ \bibinfo {author} {\bibfnamefont {Parsa}\ \bibnamefont
  {Bonderson}},\ }\bibfield  {title} {\enquote {\bibinfo {title} {Translational
  symmetry and microscopic constraints on symmetry-enriched topological phases:
  A view from the surface},}\ }\href {\doibase 10.1103/PhysRevX.6.041068}
  {\bibfield  {journal} {\bibinfo  {journal} {Phys. Rev. X}\ }\textbf {\bibinfo
  {volume} {6}},\ \bibinfo {pages} {041068} (\bibinfo {year}
  {2016})}\BibitemShut {NoStop}%
\bibitem [{\citenamefont {Crane}\ and\ \citenamefont
  {Yetter}(1993)}]{crane1993}%
  \BibitemOpen
  \bibfield  {author} {\bibinfo {author} {\bibfnamefont {Louis}\ \bibnamefont
  {Crane}}\ and\ \bibinfo {author} {\bibfnamefont {David}\ \bibnamefont
  {Yetter}},\ }\bibfield  {title} {\enquote {\bibinfo {title} {{A categorical
  construction of 4D TQFTs}},}\ }in\ \href@noop {} {\emph {\bibinfo {booktitle}
  {Quantum Topology}}},\ \bibinfo {editor} {edited by\ \bibinfo {editor}
  {\bibfnamefont {Louis}\ \bibnamefont {Kauffman}}\ and\ \bibinfo {editor}
  {\bibfnamefont {Randy}\ \bibnamefont {Baadhio}}}\ (\bibinfo  {publisher}
  {World Scientific},\ \bibinfo {address} {Singapore},\ \bibinfo {year}
  {1993})\ \Eprint {http://arxiv.org/abs/arXiv:hep-th/9301062}
  {arXiv:hep-th/9301062} \BibitemShut {NoStop}%
\bibitem [{\citenamefont {Walker}(2006)}]{walker2006}%
  \BibitemOpen
  \bibfield  {author} {\bibinfo {author} {\bibfnamefont {Kevin}\ \bibnamefont
  {Walker}},\ }\href {http://canyon23.net/math/tc.pdf} {\enquote {\bibinfo
  {title} {{TQFTs}},}\ } (\bibinfo {year} {2006})\BibitemShut {NoStop}%
\bibitem [{\citenamefont {Walker}\ and\ \citenamefont
  {Wang}(2011)}]{WalkerWang}%
  \BibitemOpen
  \bibfield  {author} {\bibinfo {author} {\bibfnamefont {Kevin}\ \bibnamefont
  {Walker}}\ and\ \bibinfo {author} {\bibfnamefont {Zhenghan}\ \bibnamefont
  {Wang}},\ }\bibfield  {title} {\enquote {\bibinfo {title} {(3$+$1)-{TQFTs}
  and topological insulators},}\ }\href {\doibase 10.1007/s11467-011-0194-z}
  {\bibfield  {journal} {\bibinfo  {journal} {Frontiers of Physics}\ }\textbf
  {\bibinfo {volume} {7}},\ \bibinfo {pages} {150--159} (\bibinfo {year}
  {2011})}\BibitemShut {NoStop}%
\bibitem [{\citenamefont {Turaev}\ and\ \citenamefont
  {Viro}(1992)}]{turaev1992}%
  \BibitemOpen
  \bibfield  {author} {\bibinfo {author} {\bibfnamefont {V.G.}\ \bibnamefont
  {Turaev}}\ and\ \bibinfo {author} {\bibfnamefont {O.Y.}\ \bibnamefont
  {Viro}},\ }\bibfield  {title} {\enquote {\bibinfo {title} {State sum
  invariants of 3-manifolds and quantum 6j-symbols},}\ }\href@noop {}
  {\bibfield  {journal} {\bibinfo  {journal} {Topology}\ }\textbf {\bibinfo
  {volume} {31}},\ \bibinfo {pages} {865--902} (\bibinfo {year}
  {1992})}\BibitemShut {NoStop}%
\bibitem [{\citenamefont {Barrett}\ and\ \citenamefont
  {Westbury}(1996)}]{barrett1996}%
  \BibitemOpen
  \bibfield  {author} {\bibinfo {author} {\bibfnamefont {John~W.}\ \bibnamefont
  {Barrett}}\ and\ \bibinfo {author} {\bibfnamefont {Bruce~W.}\ \bibnamefont
  {Westbury}},\ }\bibfield  {title} {\enquote {\bibinfo {title} {Invariants of
  piecewise-linear 3-manifolds},}\ }\href@noop {} {\bibfield  {journal}
  {\bibinfo  {journal} {Trans. Amer. Math. Soc.}\ }\textbf {\bibinfo {volume}
  {348}},\ \bibinfo {pages} {3997--4022} (\bibinfo {year} {1996})}\BibitemShut
  {NoStop}%
\bibitem [{\citenamefont {Levin}\ and\ \citenamefont {Wen}(2005)}]{levin2005}%
  \BibitemOpen
  \bibfield  {author} {\bibinfo {author} {\bibfnamefont {Michael~A.}\
  \bibnamefont {Levin}}\ and\ \bibinfo {author} {\bibfnamefont {Xiao-Gang}\
  \bibnamefont {Wen}},\ }\bibfield  {title} {\enquote {\bibinfo {title}
  {String-net condensation:\quad{}a physical mechanism for topological
  phases},}\ }\href {\doibase 10.1103/PhysRevB.71.045110} {\bibfield  {journal}
  {\bibinfo  {journal} {Phys. Rev. B}\ }\textbf {\bibinfo {volume} {71}},\
  \bibinfo {pages} {045110} (\bibinfo {year} {2005})}\BibitemShut {NoStop}%
\bibitem [{\citenamefont {Barkeshli}\ \emph
  {et~al.}(2019{\natexlab{b}})\citenamefont {Barkeshli}, \citenamefont
  {Bonderson}, \citenamefont {Cheng}, \citenamefont {Jian},\ and\ \citenamefont
  {Walker}}]{BarkeshliReflection}%
  \BibitemOpen
  \bibfield  {author} {\bibinfo {author} {\bibfnamefont {Maissam}\ \bibnamefont
  {Barkeshli}}, \bibinfo {author} {\bibfnamefont {Parsa}\ \bibnamefont
  {Bonderson}}, \bibinfo {author} {\bibfnamefont {Meng}\ \bibnamefont {Cheng}},
  \bibinfo {author} {\bibfnamefont {Chao-Ming}\ \bibnamefont {Jian}}, \ and\
  \bibinfo {author} {\bibfnamefont {Kevin}\ \bibnamefont {Walker}},\ }\bibfield
   {title} {\enquote {\bibinfo {title} {{Reflection and Time Reversal Symmetry
  Enriched Topological Phases of Matter: Path Integrals, Non-orientable
  Manifolds, and Anomalies}},}\ }\href {\doibase 10.1007/s00220-019-03475-8}
  {\bibfield  {journal} {\bibinfo  {journal} {Commun Math Phys}\ } (\bibinfo
  {year} {2019}{\natexlab{b}}),\ 10.1007/s00220-019-03475-8}\BibitemShut
  {NoStop}%
\bibitem [{\citenamefont {Heinrich}\ \emph {et~al.}(2016)\citenamefont
  {Heinrich}, \citenamefont {Burnell}, \citenamefont {Fidkowski},\ and\
  \citenamefont {Levin}}]{heinrich2016}%
  \BibitemOpen
  \bibfield  {author} {\bibinfo {author} {\bibfnamefont {Chris}\ \bibnamefont
  {Heinrich}}, \bibinfo {author} {\bibfnamefont {Fiona}\ \bibnamefont
  {Burnell}}, \bibinfo {author} {\bibfnamefont {Lukasz}\ \bibnamefont
  {Fidkowski}}, \ and\ \bibinfo {author} {\bibfnamefont {Michael}\ \bibnamefont
  {Levin}},\ }\bibfield  {title} {\enquote {\bibinfo {title} {Symmetry-enriched
  string nets: Exactly solvable models for set phases},}\ }\href {\doibase
  10.1103/PhysRevB.94.235136} {\bibfield  {journal} {\bibinfo  {journal} {Phys.
  Rev. B}\ }\textbf {\bibinfo {volume} {94}},\ \bibinfo {pages} {235136}
  (\bibinfo {year} {2016})}\BibitemShut {NoStop}%
\bibitem [{\citenamefont {Cheng}\ \emph {et~al.}(2017)\citenamefont {Cheng},
  \citenamefont {Gu}, \citenamefont {Jiang},\ and\ \citenamefont
  {Qi}}]{cheng2016}%
  \BibitemOpen
  \bibfield  {author} {\bibinfo {author} {\bibfnamefont {Meng}\ \bibnamefont
  {Cheng}}, \bibinfo {author} {\bibfnamefont {Zheng-Cheng}\ \bibnamefont {Gu}},
  \bibinfo {author} {\bibfnamefont {Shenghan}\ \bibnamefont {Jiang}}, \ and\
  \bibinfo {author} {\bibfnamefont {Yang}\ \bibnamefont {Qi}},\ }\bibfield
  {title} {\enquote {\bibinfo {title} {Exactly solvable models for
  symmetry-enriched topological phases},}\ }\href {\doibase
  10.1103/PhysRevB.96.115107} {\bibfield  {journal} {\bibinfo  {journal} {Phys.
  Rev. B}\ }\textbf {\bibinfo {volume} {96}},\ \bibinfo {pages} {115107}
  (\bibinfo {year} {2017})}\BibitemShut {NoStop}%
\bibitem [{\citenamefont {Cui}(2019)}]{CuiTQFT}%
  \BibitemOpen
  \bibfield  {author} {\bibinfo {author} {\bibfnamefont {Shawn}\ \bibnamefont
  {Cui}},\ }\bibfield  {title} {\enquote {\bibinfo {title} {{Four dimensional
  topological quantum field theories from $G$-crossed braided categories}},}\
  }\href {\doibase 10.4171/qt/128} {\bibfield  {journal} {\bibinfo  {journal}
  {Quantum Topology}\ }\textbf {\bibinfo {volume} {10}},\ \bibinfo {pages}
  {593--676} (\bibinfo {year} {2019})}\BibitemShut {NoStop}%
\bibitem [{\citenamefont {Williamson}\ and\ \citenamefont
  {Wang}(2017)}]{WilliamsonHamiltonian}%
  \BibitemOpen
  \bibfield  {author} {\bibinfo {author} {\bibfnamefont {Dominic~J.}\
  \bibnamefont {Williamson}}\ and\ \bibinfo {author} {\bibfnamefont {Zhenghan}\
  \bibnamefont {Wang}},\ }\bibfield  {title} {\enquote {\bibinfo {title}
  {Hamiltonian models for topological phases of matter in three spatial
  dimensions},}\ }\href {\doibase 10.1016/j.aop.2016.12.018} {\bibfield
  {journal} {\bibinfo  {journal} {Annals of Physics}\ }\textbf {\bibinfo
  {volume} {377}},\ \bibinfo {pages} {311--344} (\bibinfo {year}
  {2017})}\BibitemShut {NoStop}%
\bibitem [{\citenamefont {Douglas}\ and\ \citenamefont
  {Reutter}(2018)}]{douglas2018}%
  \BibitemOpen
  \bibfield  {author} {\bibinfo {author} {\bibfnamefont {Christopher~L.}\
  \bibnamefont {Douglas}}\ and\ \bibinfo {author} {\bibfnamefont {David~J.}\
  \bibnamefont {Reutter}},\ }\bibfield  {title} {\enquote {\bibinfo {title}
  {Fusion 2-categories and a state-sum invariant for 4-manifolds},}\
  }\href@noop {} {\  (\bibinfo {year} {2018})},\ \Eprint
  {http://arxiv.org/abs/1812.11933} {1812.11933} \BibitemShut {NoStop}%
\bibitem [{\citenamefont {Wang}\ \emph {et~al.}(2016)\citenamefont {Wang},
  \citenamefont {Lin},\ and\ \citenamefont {Levin}}]{wang2016b}%
  \BibitemOpen
  \bibfield  {author} {\bibinfo {author} {\bibfnamefont {Chenjie}\ \bibnamefont
  {Wang}}, \bibinfo {author} {\bibfnamefont {Chien-Hung}\ \bibnamefont {Lin}},
  \ and\ \bibinfo {author} {\bibfnamefont {Michael}\ \bibnamefont {Levin}},\
  }\bibfield  {title} {\enquote {\bibinfo {title} {Bulk-boundary correspondence
  for three-dimensional symmetry-protected topological phases},}\ }\href
  {\doibase 10.1103/PhysRevX.6.021015} {\bibfield  {journal} {\bibinfo
  {journal} {Phys. Rev. X}\ }\textbf {\bibinfo {volume} {6}},\ \bibinfo {pages}
  {021015} (\bibinfo {year} {2016})}\BibitemShut {NoStop}%
\bibitem [{\citenamefont {Barkeshli}\ and\ \citenamefont
  {Cheng}(2020)}]{BarkeshliRelativeAnomaly}%
  \BibitemOpen
  \bibfield  {author} {\bibinfo {author} {\bibfnamefont {Maissam}\ \bibnamefont
  {Barkeshli}}\ and\ \bibinfo {author} {\bibfnamefont {Meng}\ \bibnamefont
  {Cheng}},\ }\bibfield  {title} {\enquote {\bibinfo {title} {{Relative
  Anomalies in (2+1)D Symmetry Enriched Topological States}},}\ }\href
  {\doibase 10.21468/SciPostPhys.8.2.028} {\bibfield  {journal} {\bibinfo
  {journal} {SciPost Phys.}\ }\textbf {\bibinfo {volume} {8}},\ \bibinfo
  {pages} {28} (\bibinfo {year} {2020})},\ \Eprint
  {http://arxiv.org/abs/1906.10691} {1906.10691} \BibitemShut {NoStop}%
\bibitem [{\citenamefont {Wang}\ and\ \citenamefont {Levin}(2017)}]{wang2017}%
  \BibitemOpen
  \bibfield  {author} {\bibinfo {author} {\bibfnamefont {Chenjie}\ \bibnamefont
  {Wang}}\ and\ \bibinfo {author} {\bibfnamefont {Michael}\ \bibnamefont
  {Levin}},\ }\bibfield  {title} {\enquote {\bibinfo {title} {Anomaly
  indicators for time-reversal symmetric topological orders},}\ }\href
  {\doibase 10.1103/PhysRevLett.119.136801} {\bibfield  {journal} {\bibinfo
  {journal} {Phys. Rev. Lett.}\ }\textbf {\bibinfo {volume} {119}},\ \bibinfo
  {pages} {136801} (\bibinfo {year} {2017})},\ \Eprint
  {http://arxiv.org/abs/arXiv:1610.04624} {arXiv:1610.04624} \BibitemShut
  {NoStop}%
\bibitem [{\citenamefont {Tachikawa}\ and\ \citenamefont
  {Yonekura}(2016)}]{tachikawa2016b}%
  \BibitemOpen
  \bibfield  {author} {\bibinfo {author} {\bibfnamefont {Yuji}\ \bibnamefont
  {Tachikawa}}\ and\ \bibinfo {author} {\bibfnamefont {Kazuya}\ \bibnamefont
  {Yonekura}},\ }\bibfield  {title} {\enquote {\bibinfo {title} {More on
  time-reversal anomaly of 2+1d topological phases},}\ }\href@noop {} {\
  (\bibinfo {year} {2016})},\ \Eprint {http://arxiv.org/abs/arXiv:1611.01601}
  {arXiv:1611.01601} \BibitemShut {NoStop}%
\bibitem [{\citenamefont {Lee}\ and\ \citenamefont
  {Tachikawa}(2018)}]{lee2018}%
  \BibitemOpen
  \bibfield  {author} {\bibinfo {author} {\bibfnamefont {Yasunori}\
  \bibnamefont {Lee}}\ and\ \bibinfo {author} {\bibfnamefont {Yuji}\
  \bibnamefont {Tachikawa}},\ }\bibfield  {title} {\enquote {\bibinfo {title}
  {A study of time reversal symmetry of abelian anyons},}\ }\href@noop {}
  {\bibfield  {journal} {\bibinfo  {journal} {Journal of High Energy Physics}\
  }\textbf {\bibinfo {volume} {2018}},\ \bibinfo {pages} {90} (\bibinfo {year}
  {2018})}\BibitemShut {NoStop}%
\bibitem [{\citenamefont {Qi}\ \emph {et~al.}(2019)\citenamefont {Qi},
  \citenamefont {Jian},\ and\ \citenamefont {Wang}}]{qi2019}%
  \BibitemOpen
  \bibfield  {author} {\bibinfo {author} {\bibfnamefont {Yang}\ \bibnamefont
  {Qi}}, \bibinfo {author} {\bibfnamefont {Chao-Ming}\ \bibnamefont {Jian}}, \
  and\ \bibinfo {author} {\bibfnamefont {Chenjie}\ \bibnamefont {Wang}},\
  }\bibfield  {title} {\enquote {\bibinfo {title} {Folding approach to
  topological order enriched by mirror symmetry},}\ }\href {\doibase
  10.1103/PhysRevB.99.085128} {\bibfield  {journal} {\bibinfo  {journal} {Phys.
  Rev. B}\ }\textbf {\bibinfo {volume} {99}},\ \bibinfo {pages} {085128}
  (\bibinfo {year} {2019})}\BibitemShut {NoStop}%
\bibitem [{\citenamefont {Kobayashi}\ and\ \citenamefont
  {Shiozaki}(2019)}]{kobayashi2019}%
  \BibitemOpen
  \bibfield  {author} {\bibinfo {author} {\bibfnamefont {Ryohei}\ \bibnamefont
  {Kobayashi}}\ and\ \bibinfo {author} {\bibfnamefont {Ken}\ \bibnamefont
  {Shiozaki}},\ }\bibfield  {title} {\enquote {\bibinfo {title} {Anomaly
  indicator of rotation symmetry in (3+1)d topological order},}\ }\href@noop {}
  {\  (\bibinfo {year} {2019})},\ \Eprint
  {http://arxiv.org/abs/arXiv:1901.06195} {arXiv:1901.06195} \BibitemShut
  {NoStop}%
\bibitem [{\citenamefont {Mao}\ and\ \citenamefont {Wang}(2020)}]{mao2020}%
  \BibitemOpen
  \bibfield  {author} {\bibinfo {author} {\bibfnamefont {Bin-Bin}\ \bibnamefont
  {Mao}}\ and\ \bibinfo {author} {\bibfnamefont {Chenjie}\ \bibnamefont
  {Wang}},\ }\bibfield  {title} {\enquote {\bibinfo {title} {Mirror anomaly in
  fermionic topological orders},}\ }\href@noop {} {\  (\bibinfo {year}
  {2020})},\ \Eprint {http://arxiv.org/abs/arXiv:2002.07714} {arXiv:2002.07714}
  \BibitemShut {NoStop}%
\bibitem [{\citenamefont {Hermele}\ and\ \citenamefont
  {Chen}(2016)}]{hermele2016}%
  \BibitemOpen
  \bibfield  {author} {\bibinfo {author} {\bibfnamefont {Michael}\ \bibnamefont
  {Hermele}}\ and\ \bibinfo {author} {\bibfnamefont {Xie}\ \bibnamefont
  {Chen}},\ }\bibfield  {title} {\enquote {\bibinfo {title} {Flux-fusion
  anomaly test and bosonic topological crystalline insulators},}\ }\href
  {\doibase 10.1103/PhysRevX.6.041006} {\bibfield  {journal} {\bibinfo
  {journal} {Phys. Rev. X}\ }\textbf {\bibinfo {volume} {6}},\ \bibinfo {pages}
  {041006} (\bibinfo {year} {2016})}\BibitemShut {NoStop}%
\bibitem [{\citenamefont {Lapa}\ and\ \citenamefont {Levin}(2019)}]{lapa2019}%
  \BibitemOpen
  \bibfield  {author} {\bibinfo {author} {\bibfnamefont {Matthew}\ \bibnamefont
  {Lapa}}\ and\ \bibinfo {author} {\bibfnamefont {Michael}\ \bibnamefont
  {Levin}},\ }\bibfield  {title} {\enquote {\bibinfo {title} {Anomaly
  indicators for topological orders with u(1) and time-reversal symmetry},}\
  }\href@noop {} {\  (\bibinfo {year} {2019})},\ \Eprint
  {http://arxiv.org/abs/arXiv:1905.00435} {arXiv:1905.00435} \BibitemShut
  {NoStop}%
\bibitem [{\citenamefont {Cui}\ \emph {et~al.}(2016)\citenamefont {Cui},
  \citenamefont {Galindo}, \citenamefont {Plavnik},\ and\ \citenamefont
  {Wang}}]{cui2016}%
  \BibitemOpen
  \bibfield  {author} {\bibinfo {author} {\bibfnamefont {Shawn~X.}\
  \bibnamefont {Cui}}, \bibinfo {author} {\bibfnamefont {C{\'e}sar}\
  \bibnamefont {Galindo}}, \bibinfo {author} {\bibfnamefont {Julia~Yael}\
  \bibnamefont {Plavnik}}, \ and\ \bibinfo {author} {\bibfnamefont {Zhenghan}\
  \bibnamefont {Wang}},\ }\bibfield  {title} {\enquote {\bibinfo {title} {On
  gauging symmetry of modular categories},}\ }\href {\doibase
  10.1007/s00220-016-2633-8} {\bibfield  {journal} {\bibinfo  {journal} {Comm.
  Math. Phys.}\ }\textbf {\bibinfo {volume} {348}},\ \bibinfo {pages}
  {1043--1064} (\bibinfo {year} {2016})}\BibitemShut {NoStop}%
\bibitem [{\citenamefont {Fidkowski}\ and\ \citenamefont
  {Vishwanath}(2015)}]{fidkowski2015}%
  \BibitemOpen
  \bibfield  {author} {\bibinfo {author} {\bibfnamefont {Lukasz}\ \bibnamefont
  {Fidkowski}}\ and\ \bibinfo {author} {\bibfnamefont {Ashvin}\ \bibnamefont
  {Vishwanath}},\ }\bibfield  {title} {\enquote {\bibinfo {title} {Realizing
  anomalous anyonic symmetries at the surfaces of 3d gauge theories},}\
  }\href@noop {} {\  (\bibinfo {year} {2015})},\ \Eprint
  {http://arxiv.org/abs/arXiv:1511.01502} {arXiv:1511.01502} \BibitemShut
  {NoStop}%
\bibitem [{\citenamefont {Barkeshli}\ and\ \citenamefont
  {Cheng}(2018)}]{barkeshli2018}%
  \BibitemOpen
  \bibfield  {author} {\bibinfo {author} {\bibfnamefont {Maissam}\ \bibnamefont
  {Barkeshli}}\ and\ \bibinfo {author} {\bibfnamefont {Meng}\ \bibnamefont
  {Cheng}},\ }\bibfield  {title} {\enquote {\bibinfo {title} {Time-reversal and
  spatial-reflection symmetry localization anomalies in (2+1)-dimensional
  topological phases of matter},}\ }\href {\doibase 10.1103/PhysRevB.98.115129}
  {\bibfield  {journal} {\bibinfo  {journal} {Phys. Rev. B}\ }\textbf {\bibinfo
  {volume} {98}},\ \bibinfo {pages} {115129} (\bibinfo {year}
  {2018})}\BibitemShut {NoStop}%
\bibitem [{Note2()}]{Note2}%
  \BibitemOpen
  \bibinfo {note} {A non-trivial obstruction $[\protect \textswab {O}] \in
  \protect \mathcal {H}^3_{[\rho ]}(G, \protect \mathcal {A})$ can be
  alternatively interpreted as the associated TQFT possessing a non-trivial
  $2$-group symmetry, consisting of the $0$-form symmetry group $G$ and the
  $1$-form symmetry group $\protect \mathcal {A}$, with $[\protect \textswab
  {O}]$ characterizing the $2$-group \cite
  {ENO2009,barkeshli2019,benini2019}.}\BibitemShut {Stop}%
\bibitem [{\citenamefont {Essin}\ and\ \citenamefont
  {Hermele}(2013)}]{essin2013}%
  \BibitemOpen
  \bibfield  {author} {\bibinfo {author} {\bibfnamefont {Andrew~M.}\
  \bibnamefont {Essin}}\ and\ \bibinfo {author} {\bibfnamefont {Michael}\
  \bibnamefont {Hermele}},\ }\bibfield  {title} {\enquote {\bibinfo {title}
  {Classifying fractionalization: Symmetry classification of gapped
  ${\mathbb{z}}_{2}$ spin liquids in two dimensions},}\ }\href {\doibase
  10.1103/PhysRevB.87.104406} {\bibfield  {journal} {\bibinfo  {journal} {Phys.
  Rev. B}\ }\textbf {\bibinfo {volume} {87}},\ \bibinfo {pages} {104406}
  (\bibinfo {year} {2013})}\BibitemShut {NoStop}%
\bibitem [{Note3()}]{Note3}%
  \BibitemOpen
  \bibinfo {note} {We thank Shawn Cui for helpful discussions regarding this
  point.}\BibitemShut {Stop}%
\bibitem [{\citenamefont {Schommer-Pries}(2018)}]{ToriInvertibleTQFTs}%
  \BibitemOpen
  \bibfield  {author} {\bibinfo {author} {\bibfnamefont {Christopher}\
  \bibnamefont {Schommer-Pries}},\ }\bibfield  {title} {\enquote {\bibinfo
  {title} {Tori detect invertibility of topological field theories},}\ }\href
  {\doibase 10.2140/gt.2018.22.2713} {\bibfield  {journal} {\bibinfo  {journal}
  {Geometry {\&} Topology}\ }\textbf {\bibinfo {volume} {22}},\ \bibinfo
  {pages} {2713--2756} (\bibinfo {year} {2018})}\BibitemShut {NoStop}%
\bibitem [{Note4()}]{Note4}%
  \BibitemOpen
  \bibinfo {note} {Note that a $(d+1)$-dimensional TQFT is once-extended if it
  assigns data to every closed $(d+1)$-, $d$- and $(d-1)$-manifold. Our model
  assigns a complex number to each closed $4$-manifold and a quantum state to
  the boundary of every $4$-manifold with boundary. Since every 3-manifold can
  exist at the boundary of a 4-manifold, it follows that our construction
  assigns a vector space to every closed $3$-manifold. The Hamiltonian
  formulation we study later shows how our formalism also assigns states to the
  boundaries of 3-manifolds as well, which determines a UMTC with $G$ symmetry
  fractionalization data.}\BibitemShut {Stop}%
\bibitem [{Note5()}]{Note5}%
  \BibitemOpen
  \bibinfo {note} {A $d$-dimensional triangulation strictly speaking requires
  two $d$-simplices to share only one $d-1$ simplex. A cellulation does not
  require this stringent condition and therefore allows more efficient ways to
  construct manifolds by gluing together simplices. Our state sum is still
  well-defined when we relax this condition.}\BibitemShut {Stop}%
\bibitem [{Note6()}]{Note6}%
  \BibitemOpen
  \bibinfo {note} {It is interesting to consider a theory where $\protect
  \mathaccentV {tilde}07E{\rho }$ is taken to be anti-unitary, which can
  presumably lead to a consistent graphical calculus and may potentially be of
  mathematical interest. However, the natural generalization of our Hamiltonian
  construction in Sec. \ref {HamiltonianSec}-\ref {sec:surfaceTO} to this case
  is non-local and thus we do not consider this possibility here.}\BibitemShut
  {Stop}%
\bibitem [{\citenamefont {Tantivasadakarn}(2017)}]{TantivasadakarnSPT}%
  \BibitemOpen
  \bibfield  {author} {\bibinfo {author} {\bibfnamefont {Nathanan}\
  \bibnamefont {Tantivasadakarn}},\ }\bibfield  {title} {\enquote {\bibinfo
  {title} {Dimensional reduction and topological invariants of
  symmetry-protected topological phases},}\ }\href {\doibase
  10.1103/physrevb.96.195101} {\bibfield  {journal} {\bibinfo  {journal} {Phys.
  Rev. B}\ }\textbf {\bibinfo {volume} {96}},\ \bibinfo {pages} {195101}
  (\bibinfo {year} {2017})}\BibitemShut {NoStop}%
\bibitem [{\citenamefont {Tiwari}\ \emph {et~al.}(2018)\citenamefont {Tiwari},
  \citenamefont {Chen}, \citenamefont {Shiozaki},\ and\ \citenamefont
  {Ryu}}]{TiwariSPT}%
  \BibitemOpen
  \bibfield  {author} {\bibinfo {author} {\bibfnamefont {Apoorv}\ \bibnamefont
  {Tiwari}}, \bibinfo {author} {\bibfnamefont {Xiao}\ \bibnamefont {Chen}},
  \bibinfo {author} {\bibfnamefont {Ken}\ \bibnamefont {Shiozaki}}, \ and\
  \bibinfo {author} {\bibfnamefont {Shinsei}\ \bibnamefont {Ryu}},\ }\bibfield
  {title} {\enquote {\bibinfo {title} {{Bosonic topological phases of matter:
  Bulk-boundary correspondence, symmetry protected topological invariants, and
  gauging}},}\ }\href {\doibase 10.1103/physrevb.97.245133} {\bibfield
  {journal} {\bibinfo  {journal} {Phys. Rev. B}\ }\textbf {\bibinfo {volume}
  {97}},\ \bibinfo {pages} {245133} (\bibinfo {year} {2018})}\BibitemShut
  {NoStop}%
\bibitem [{\citenamefont {Wang}\ and\ \citenamefont
  {Levin}(2015)}]{WangLevinPRB}%
  \BibitemOpen
  \bibfield  {author} {\bibinfo {author} {\bibfnamefont {Chenjie}\ \bibnamefont
  {Wang}}\ and\ \bibinfo {author} {\bibfnamefont {Michael}\ \bibnamefont
  {Levin}},\ }\bibfield  {title} {\enquote {\bibinfo {title} {Topological
  invariants for gauge theories and symmetry-protected topological phases},}\
  }\href@noop {} {\bibfield  {journal} {\bibinfo  {journal} {Phys. Rev. B}\
  }\textbf {\bibinfo {volume} {91}},\ \bibinfo {pages} {165119} (\bibinfo
  {year} {2015})}\BibitemShut {NoStop}%
\bibitem [{\citenamefont {von Keyserlingk}\ \emph {et~al.}(2013)\citenamefont
  {von Keyserlingk}, \citenamefont {Burnell},\ and\ \citenamefont
  {Simon}}]{vonKeyserlingkSurfaceAnyons}%
  \BibitemOpen
  \bibfield  {author} {\bibinfo {author} {\bibfnamefont {C.~W.}\ \bibnamefont
  {von Keyserlingk}}, \bibinfo {author} {\bibfnamefont {F.~J.}\ \bibnamefont
  {Burnell}}, \ and\ \bibinfo {author} {\bibfnamefont {S.~H.}\ \bibnamefont
  {Simon}},\ }\bibfield  {title} {\enquote {\bibinfo {title} {Three-dimensional
  topological lattice models with surface anyons},}\ }\href {\doibase
  10.1103/PhysRevB.87.045107} {\bibfield  {journal} {\bibinfo  {journal} {Phys.
  Rev. B}\ }\textbf {\bibinfo {volume} {87}},\ \bibinfo {pages} {045107}
  (\bibinfo {year} {2013})}\BibitemShut {NoStop}%
\bibitem [{Note7()}]{Note7}%
  \BibitemOpen
  \bibinfo {note} {The universal coefficient theorem is usually stated as the
  existence of a split exact sequence, but the fact that all the groups
  involved are known to be Abelian means we can rewrite it in the form
  Eq.~\protect \textup {\hbox {\mathsurround \z@ \protect \normalfont
  (\ignorespaces \ref {eqn:UCT}\unskip \@@italiccorr )}}.}\BibitemShut {Stop}%
\bibitem [{\citenamefont {Hatcher}(2002)}]{HatcherAT}%
  \BibitemOpen
  \bibfield  {author} {\bibinfo {author} {\bibfnamefont {Allen}\ \bibnamefont
  {Hatcher}},\ }\href@noop {} {\emph {\bibinfo {title} {Algebraic Topology}}}\
  (\bibinfo  {publisher} {Cambridge University Press},\ \bibinfo {year}
  {2002})\BibitemShut {NoStop}%
\bibitem [{\citenamefont {Brown}(1982)}]{BrownCohomology}%
  \BibitemOpen
  \bibfield  {author} {\bibinfo {author} {\bibfnamefont {Kenneth~S.}\
  \bibnamefont {Brown}},\ }\href@noop {} {\emph {\bibinfo {title} {Cohomology
  of Groups}}}\ (\bibinfo  {publisher} {Springer-Verlag Berlin},\ \bibinfo
  {year} {1982})\BibitemShut {NoStop}%
\end{thebibliography}%

\end{document}